\documentclass[12pt]{article}%
\pdfoutput=1

\usepackage{rotating}
\usepackage{cite}
\usepackage[all]{xy}
\usepackage{amsmath,amssymb,amsthm,amsfonts,mathrsfs}
\theoremstyle{definition}

\usepackage{mathtools}
\usepackage{amsfonts}
\usepackage{amssymb}
\usepackage{authblk}
\usepackage{bbm}
\usepackage{bm}
\usepackage{caption}
\usepackage{epsfig}
\usepackage{epstopdf}
\usepackage{heck}
\usepackage[margin=1in]{geometry}
\usepackage{tabularx}
\usepackage[usenames,dvipsnames]{xcolor}
\usepackage{graphicx}
\numberwithin{equation}{section}
\usepackage{tikz}
\usetikzlibrary{decorations.pathreplacing}
\usetikzlibrary{calc}

\usepackage{stackengine}
\usepackage{verbatim}

\usepackage{hyperref}
\usepackage[nameinlink,capitalize]{cleveref}

\makeatletter
\newcommand*{\diff}{\@ifnextchar^{\DIFF}{\DIFF^{}}}
\def\DIFF^#1{\mathop{\mathrm{\mathstrut d}}\nolimits^{#1}\gobblespace}
\newcommand*{\bigdiff}{\@ifnextchar^{\BIGDIFF}{\BIGDIFF^{}}}
\def\BIGDIFF^#1{\mathop{\mathrm{\mathstrut \mathcal{D}}}\nolimits^{#1}\gobblespace}
\def\gobblespace{\futurelet\diffarg\opspace}
\def\opspace{%
	\let\DiffSpace\!%
	\ifx\diffarg(%
		\let\DiffSpace\relax
	\else
		\ifx\diffarg[%
			\let\DiffSpace\relax
		\else
			\ifx\diffarg\{%
				\let\DiffSpace\relax
			\fi
		\fi
	\fi
	\DiffSpace
}
\makeatother

\newcommand*{\T}{\mathfrak{T}}

\newcommand*{\Tp}{\mathfrak{T}{\mkern 2mu}' }

\newcommand*{\ADE}{{\textnormal{ADE}}}

\newcommand*{\Tor}{{\textnormal{Tor}}}

\usepackage{tikz}
\usetikzlibrary{arrows}
\usetikzlibrary{arrows.meta}
\usetikzlibrary{shapes.geometric,calc,arrows, positioning,shapes.misc,decorations.markings}
\tikzset{
  big arrow/.style={
    decoration={markings,mark=at position 1 with {\arrow[scale=2,#1]{>}}},
    postaction={decorate},
    shorten >=0.4pt},
  big arrow/.default=black}

\pgfdeclarelayer{edgelayer}
\pgfdeclarelayer{nodelayer}
\pgfsetlayers{edgelayer,nodelayer,main}
\tikzstyle{none}=[inner sep=0pt]

\tikzstyle{NodeCross}=[draw, shape=circle, cross out, inner sep=0pt, minimum size=6pt,line width=0.25mm]
\tikzstyle{Circle}=[draw, shape=circle, black, fill=black, inner sep=0pt, minimum size=6pt]
\tikzstyle{Star}=[draw, shape=star, fill=black, star points=8, inner sep=0pt, minimum size=8pt]
\tikzstyle{EmptyCircle}=[draw, shape=circle, inner sep=0pt,  fill=white, minimum size=7pt, line width=0.4mm]
\tikzstyle{CircleRed}=[draw=black, shape=circle, fill=black, inner sep=0pt, minimum size=12pt]
\tikzstyle{CircleRed2}=[draw=black, shape=circle, fill=black, inner sep=0pt, minimum size=8pt]

\tikzstyle{DashedLine}=[-, densely dashed, line width=0.25mm]
\tikzstyle{DottedLine}=[-, dotted, line width=0.25mm]
\tikzstyle{ThickLine}=[-, line width=0.25mm]
\tikzstyle{ArrowLineRight}=[-, -{Stealth[scale=1.75]}, line width=0.15mm, scale=5]
\tikzstyle{RedLine}=[-, draw={rgb,255: red,191; green,0; blue,0}, fill=none, line width=0.25mm]
\tikzstyle{DashedLineThin}=[-, densely dashed, line width=0.125mm, fill=none, draw=black]
\tikzstyle{ArrowLineRed}=[-, draw={rgb,255: red,191; green,0; blue,0}, -{Stealth[scale=1.75]}, line width=0.15mm, scale=5]
\tikzstyle{BlueLine}=[-, draw={rgb,255: red,0; green,0; blue,191}, fill=none, line width=0.25mm]
\tikzstyle{BlueDottedLight}=[-, dotted, draw={rgb,255: red,0; green,0; blue,191}, fill=none, line width=0.25mm]
\tikzstyle{DottedRed}=[-, dotted, draw={rgb,255: red,191; green,0; blue,0}, fill=none, dotted, line width=0.25mm]

\newcommand{\bea}{\begin{eqnarray}}
\newcommand{\eea}{\end{eqnarray}}
\newcommand{\be}{\begin{equation}}
\newcommand{\ee}{\end{equation}}
\newcommand{\ba}{\begin{aligned}}
\newcommand{\ea}{\end{aligned}}
\newcommand{\bit}{\begin{itemize}}
\newcommand{\eit}{\end{itemize}}
\newcommand{\ben}{\begin{enumerate}}
\newcommand{\een}{\end{enumerate}}

\newcommand{\lb}{\left(}
\newcommand{\rb}{\right)}
\newcommand{\lbb}{\left[}
\newcommand{\rbb}{\right]}
\newcommand{\lbbb}{\left\{}
\newcommand{\rbbb}{\right\}}

\newcommand{\Z}{{\mathbb Z}}
\newcommand{\R}{{\mathbb R}}
\newcommand{\C}{{\mathbb C}}
\newcommand{\Q}{{\mathbb Q}}
\renewcommand{\P}{{\mathbb P}}

\newcommand{\Xl}{X^{\textnormal{loc}}}

\newcommand{\XADE}{\mathscr{X}}

\begin{document}

\date{July 2023}

\title{Generalized Symmetries, \\[4mm] Gravity, and the Swampland}

\institution{PENN}{\centerline{$^{1}$Department of Physics and Astronomy, University of Pennsylvania, Philadelphia, PA 19104, USA}}
\institution{PENNmath}{\centerline{$^{2}$Department of Mathematics, University of Pennsylvania, Philadelphia, PA 19104, USA}}
\institution{MARIBOR}{\centerline{${}^{3}$Center for Applied Mathematics and Theoretical Physics, University of Maribor, Maribor, Slovenia}}

\authors{
Mirjam Cveti\v{c}\worksat{\PENN,\PENNmath,\MARIBOR}\footnote{e-mail: \texttt{cvetic@physics.upenn.edu}},
Jonathan J. Heckman\worksat{\PENN,\PENNmath}\footnote{e-mail: \texttt{jheckman@sas.upenn.edu}},\\[4mm]
Max H\"ubner\worksat{\PENN}\footnote{e-mail: \texttt{hmax@sas.upenn.edu}}, and
Ethan Torres\worksat{\PENN}\footnote{e-mail: \texttt{emtorres@sas.upenn.edu}}
}

\abstract{Generalized global symmetries are a common feature of many quantum field theories decoupled from gravity.
By contrast, in quantum gravity / the Swampland program, it is widely expected that all global symmetries are either gauged or broken,
and this breaking is in turn related to the expected completeness of the spectrum of charged states in quantum gravity.
We investigate the fate of such symmetries in the context of 7D and 5D vacua realized by compact
Calabi-Yau spaces with localized singularities in M-theory. We explicitly show how gravitational backgrounds support additional dynamical degrees of freedom which trivialize (i.e., ``break'') the higher symmetries of the local geometric models. Local compatability conditions across these different sectors lead to gluing conditions for gauging higher-form and (in the 5D case) higher-group symmetries. This also leads to a preferred global structure of the gauge group and higher-form gauge symmetries. In cases based on a genus-one fibered Calabi-Yau space, we also get an F-theory model in one higher dimension with corresponding constraints on the global form of the gauge group.}

{\small \texttt{\hfill UPR-1325-T}}

\maketitle

\setcounter{tocdepth}{2}

\tableofcontents


\section{Introduction}

Global symmetries play an important role in the study of quantum field theory (QFT).
They are helpful because they constrain correlation functions, and (via their anomalies) provide insight into
various non-perturbative phenomena (see e.g., \cite{tHooft:1979rat}). On the other hand, there is a general expectation that in quantum gravity, all such global symmetries are either gauged or explicitly broken (see e.g., \cite{Banks:1988yz, Banks:2010zn, Harlow:2018tng, McNamara:2019rup}). In this regard, the best one can hope for is that violations of a given global symmetry are exponentially suppressed.

In the context of the Swampland program \cite{Vafa:2005ui, Ooguri:2006in} (see e.g., \cite{vanBeest:2021lhn, Agmon:2022thq} for recent reviews),
this amounts to starting with the symmetry structures of a given low energy effective field theory and tracking their fate once the effects of gravity are included. The condition that there are no global symmetries is closely tied to the expectation that the spectrum of a theory is complete, i.e., there is a state of each possible representation (see e.g., \cite{Polchinski:1998rr, Banks:2010zn, Heidenreich:2021xpr, McNamara:2021cuo}). This has also been formalized as the statement that all backgrounds of quantum gravity are in fact cobordant \cite{McNamara:2019rup}.\footnote{For recent developments, see e.g., \cite{Montero:2020icj, Dierigl:2020lai, Debray:2021vob, Buratti:2021fiv, Blumenhagen:2021nmi,
Angius:2022aeq, Blumenhagen:2022mqw, Angius:2022mgh, Blumenhagen:2022bvh, Dierigl:2022reg, Debray:2023yrs, Dierigl:2023jdp}.}

Of course, instead of proceeding from the bottom up, one can also consider starting with a consistent low energy effective field theory
such as that provided by a string compactification. With an explicit UV completion in hand, one can then track the effects of decoupling gravity, and the eventual emergence of various global symmetries in the deep infrared of a given QFT sector. This is especially natural to study in the context of QFTs with generalized symmetries \cite{Gaiotto:2014kfa}.

In this regard, an interesting question is the global structure of various gauge and global symmetries which can arise in such decoupling limits.\footnote{For example, in 4D Yang-Mills theory with gauge symmetry given by the Lie algebra $\mathfrak{su}(N)$, a priori there are different choices for the global form of the gauge group, e.g., ``the electric choice'' $SU(N)$ versus ``the magnetic choice'' $SU(N) / \mathbb{Z}_N$. Gauging the electric 1-form symmetry of $SU(N)$ gauge theory takes us to the $SU(N) / \mathbb{Z}_N$ gauge theory, and gauging the magnetic 1-form symmetry of $SU(N) / \mathbb{Z}_N$ gauge theory takes us back to the original electric theory \cite{Kapustin:2014gua, Gaiotto:2014kfa}.} In each case, we get a spectrum of local and extended operators charged under various $p$-form symmetries. Gauging such symmetries (when non-anomalous) amounts to changing from one global form of the theory to another \cite{Gaiotto:2014kfa}. For a $D$-dimensional QFT, one way to understand these choices is in terms of a $(D+1)$-dimensional symmetry topological field theory (TFT) with states (i.e., boundary conditions) specifying a given global form of the $D$-dimensional QFT \cite{Freed:2012bs, Kaidi:2022cpf, Freed:2022qnc}. This can also be geometrized, with the extra dimension viewed as the ``radial direction'' of a local string compactification \cite{DelZotto:2015isa, GarciaEtxebarria:2019caf, Albertini:2020mdx, Morrison:2020ool, Apruzzi:2021nmk} (see also \cite{Aharony:1998qu, Heckman:2017uxe}). Once gravity is switched on, however, no such boundary conditions ``at infinity'' are available. Somehow, the gravitational theory ``chooses'' a global form for the theory. In practical terms, this is specified by the lightest degrees of freedom in the theory, which by convention specify an electric global form for the theory. Heavier, dual objects then specify the magnetic basis of states. The condition that the gauge group acts faithfully on the spectrum of states then fixes the global form of these gauge symmetries (see e.g., \cite{Aspinwall:1998xj, Morrison:2021wuv}). One comment here is that in practice, it is customary to restrict attention to only the massless degrees of freedom present in the low energy spectrum. To extract the full global form of the gauge group, as well as the structure of dynamical extended objects, we will find it necessary to also consider massive and finite tension objects.

While our discussion has thus far focused on the specific case of higher-form symmetries, it has by now been established that there are many generalizations of these sorts of considerations, including higher-group structures, as well as non-invertible symmetries.\footnote{The literature has in recent times grown considerably. For a partial list of recent work, see e.g.,
\cite{Gaiotto:2014kfa, Gaiotto:2010be,Kapustin:2013qsa,Kapustin:2013uxa,Aharony:2013hda,
DelZotto:2015isa,Sharpe:2015mja, Heckman:2017uxe, Tachikawa:2017gyf,
Cordova:2018cvg,Benini:2018reh,Hsin:2018vcg,Wan:2018bns,
Thorngren:2019iar,GarciaEtxebarria:2019caf,Eckhard:2019jgg,Wan:2019soo,Bergman:2020ifi,Morrison:2020ool,
Albertini:2020mdx,Hsin:2020nts,Bah:2020uev,DelZotto:2020esg,Hason:2020yqf,Bhardwaj:2020phs,
Apruzzi:2020zot,Cordova:2020tij,Thorngren:2020aph,DelZotto:2020sop,BenettiGenolini:2020doj,
Yu:2020twi,Bhardwaj:2020ymp,DeWolfe:2020uzb,Gukov:2020btk,Iqbal:2020lrt,Hidaka:2020izy,
Brennan:2020ehu,Komargodski:2020mxz,Closset:2020afy,Thorngren:2020yht,Closset:2020scj,
Bhardwaj:2021pfz,Nguyen:2021naa,Heidenreich:2021xpr,Apruzzi:2021phx,Apruzzi:2021vcu,
Hosseini:2021ged,Cvetic:2021sxm,Buican:2021xhs,Bhardwaj:2021zrt,Iqbal:2021rkn,Braun:2021sex,
Cvetic:2021maf,Closset:2021lhd,Thorngren:2021yso,Sharpe:2021srf,Bhardwaj:2021wif,Hidaka:2021mml,
Lee:2021obi,Lee:2021crt,Hidaka:2021kkf,Koide:2021zxj,Apruzzi:2021mlh,Kaidi:2021xfk,Choi:2021kmx,
Bah:2021brs,Gukov:2021swm,Closset:2021lwy,Yu:2021zmu,Apruzzi:2021nmk, Cvetic:2021sjm, Beratto:2021xmn,Bhardwaj:2021mzl,
Debray:2021vob, Wang:2021vki,
Cvetic:2022uuu,DelZotto:2022fnw,Cvetic:2022imb,DelZotto:2022joo,DelZotto:2022ras,Bhardwaj:2022yxj,Hayashi:2022fkw,
Kaidi:2022uux,Roumpedakis:2022aik,Choi:2022jqy,
Choi:2022zal,Arias-Tamargo:2022nlf,Cordova:2022ieu, Bhardwaj:2022dyt,
Benedetti:2022zbb, DelZotto:2022ohj, Bhardwaj:2022scy,Antinucci:2022eat,Carta:2022spy,
Apruzzi:2022dlm, Heckman:2022suy, Baume:2022cot, Choi:2022rfe,
Bhardwaj:2022lsg, Lin:2022xod, Bartsch:2022mpm, Apruzzi:2022rei,
GarciaEtxebarria:2022vzq, Cherman:2022eml, Heckman:2022muc, Lu:2022ver, Niro:2022ctq, Kaidi:2022cpf,
Mekareeya:2022spm, vanBeest:2022fss, Antinucci:2022vyk, Giaccari:2022xgs, Bashmakov:2022uek,Cordova:2022fhg,
GarciaEtxebarria:2022jky, Choi:2022fgx, Robbins:2022wlr, Bhardwaj:2022kot, Bhardwaj:2022maz, Bartsch:2022ytj, Gaiotto:2020iye,Agrawal:2015dbf, Robbins:2021ibx, Robbins:2021xce,Huang:2021zvu,
Inamura:2021szw, Cherman:2021nox,Sharpe:2022ene,Bashmakov:2022jtl, Lee:2022swr, Inamura:2022lun, Damia:2022bcd, Lin:2022dhv,Burbano:2021loy, Damia:2022rxw, Apte:2022xtu, Nawata:2023rdx, Bhardwaj:2023zix, Kaidi:2023maf, Etheredge:2023ler, Lin:2023uvm, Amariti:2023hev, Bhardwaj:2023wzd, Bartsch:2023pzl, Carta:2023bqn, Zhang:2023wlu, Cao:2023doz, Putrov:2023jqi, Debray:2023yrs, Davighi:2023luh, Acharya:2023bth, Debray:2023rlx, Debray:2023tdd, Dierigl:2023jdp, Cvetic:2023plv, Bashmakov:2023kwo, Chen:2023qnv, Copetti:2023mcq, Bartsch:2023wvv, Bhardwaj:2023ayw, Pasquarella:2023deo, Lawrie:2023tdz, Bah:2023ymy, Apruzzi:2023uma, vanBeest:2023dbu}, and \cite{Cordova:2022ruw, Brennan:2023mmt, Schafer-Nameki:2023jdn, Bhardwaj:2023kri, Luo:2023ive} for reviews.} In each of these cases, it is natural to see how they are explicitly gauged or broken upon coupling to gravity, and conversely, how such structures emerge when gravity is switched off.

Our aim in this paper will be to address these issues by explicitly tracking local singularities in M-theory compactified on a Calabi-Yau space. More precisely, we consider M-theory on compact K3 surfaces, and compact Calabi-Yau threefolds, resulting in 7D and 5D vacua, respectively. We shall be interested in compact models which locally support quotient singularities of the form $\mathbb{C}^{m} / \Gamma_{SU(m)}$, where $\Gamma_{SU(m)}$ is a finite subgroup of $SU(m)$. In the case of 7D vacua, these singularities result in 7D supersymmetric Yang-Mills (SYM) sectors, while in the case of 5D vacua, we generically obtain 5D superconformal field theories (SCFTs). For some discussion of coupling stringy SCFTs to gravity, see e.g., \cite{DelZotto:2014fia, Anderson:2018heq, Hayashi:2019fsa, Heckman:2019bzm}.

In the case of 7D vacua, the local QFT sectors can a priori support both line operators from M2-branes wrapped on non-compact relative 2-cycles of $\mathbb{C}^{2} / \Gamma_{SU(2)}$, as well as codimension 3 't Hooft operators from M5-branes wrapped on the same non-compact relative 2-cycles. Choosing a polarization, i.e., a set of maximally commuting fluxes, yields a global form for the associated gauge group, and in the limit where gravity is decoupled, each sector can support its own global form. Once gravity is included, however, these different choices are inevitably correlated, as dictated by a Mayer-Vietoris long exact sequence. We show that this correlation in fact persists in the deep IR (after integrating out the massive electric and magnetic states) as it specifies the data of the set of genuine Wilson and 't Hooft operators and emergent higher-form symmetries. This is accomplished by working with a compact K3 surface with local orbifold singularities, and we consider in detail different quotients of $T^4$, as well as some elliptically fibered K3 surfaces. An important feature of this analysis is that this procedure automatically selects a global form for the gauge group.

In the case of 5D vacua, local orbifolds $\mathbb{C}^3 / \Gamma_{SU(3)}$ with a collapsing divisor generate 5D SCFTs. Such systems can support higher-form symmetries, as well as higher-group symmetries  and there is by now an algorithmic procedure for reading off these properties directly from geometry (see e.g., \cite{Albertini:2020mdx, Morrison:2020ool, Apruzzi:2021vcu, Tian:2021cif, DelZotto:2022fnw, Cvetic:2022imb, DelZotto:2022joo, Acharya:2023bth}). These generalized symmetries can be extracted from the topology of the boundary space $S^5 / \Gamma_{SU(3)}$. A generic feature of many such systems is the presence of non-trivial anomalies, which at least superficially obstruct gauging of the associated discrete higher-symmetries. However, taking multiple copies of the same theory, specific subgroups are anomaly free and as such can be gauged.\footnote{At a basic level, this is guaranteed because the anomalies in question are discrete, so taking an appropriate number of copies renders any putative anomaly for the diagonal symmetry trivial.} This is precisely what happens in compact models coupled to gravity, but it also occurs in limits where gravity is decoupled but multiple 5D SCFT sectors have been glued together. This includes the gauging of 1-form symmetries, and the trivialization of 2-group structures once multiple 5D SCFTs are coupled together by gauging a common 0-form flavor symmetry. We demonstrate these general features by analyzing a few examples of quotients of $T^6$, as well as elliptically fibered Calabi-Yau threefolds.

For vacua obtained from elliptically fibered Calabi-Yau compactifications of M-theory, there is a dual description in terms of F-theory on the same background, resulting in a vacuum in one higher dimension. Now, in the case of F-theory models, there is a well-known procedure for reading off the gauge group directly from data associated with the genus-one fibration (see e.g., \cite{Grimm:2010ez, Grimm2011, Morrison:2012ei, Cvetic:2013nia, Borchmann:2013jwa, Morrison:2014era, Braun:2014oya, Mayrhofer2014, Braun:2014nva, Mayrhofer:2014laa, Cvetic:2017epq, Cvetic:2021sxm} as well as \cite{Cvetic:2018bni} and references therein). Our considerations are necessarily more topological / coarse, but are consistent with these considerations. Our analysis can be interpreted as constraining the global form of the gauge group, which in F-theory terms involves an analysis of multi-sections and additional sections. A general comment here is that in $(D+1)$-dimensional F-theory vacua, additional tuning is required in the accompanying Weierstrass model which is not required in $D$-dimensional M-theory vacua, since in $(D+1)$-dimensional terms, one can now allow field vevs which only retain Lorentz invariance in $D$ dimensions.

The rest of this paper is organized as follows. In section \ref{sec:LOCAL} we briefly summarize some of the local building blocks which comprise our QFT sectors in 7D and 5D vacua. In section \ref{sec:GLOBAL} we present a brief overview of the general procedure we will follow, namely we consider supergravity models realized by M-theory on a compact background with localized singularities. In a decoupling limit, each singularity supports a QFT sector, and the defining data of defects and symmetry operators in each local patch must consistently glue together with the other patches. After this, in section \ref{sec:7d} we turn to general features of M-theory on a K3 surface, and in section \ref{sec:7dExamp} we present some illustrative examples based on quotients of $T^4$ and elliptically fibered K3 surfaces. In section \ref{sec:5d} we discuss 5D examples as realized by M-theory on Calabi-Yau quotients of a $T^6$ and elliptic models. When available, we also comment on the dual F-theory interpretation. We present our conclusions and future directions in section \ref{sec:CONC}. Some additional details on quotients of $T^4$ and $T^6$ are collected in the Appendices.

\section{Local Ingredients} \label{sec:LOCAL}

Before proceeding to explicit compact models, let us briefly review some of the salient features of local models obtained via geometric engineering. The general setup we consider here is M-theory on a local Calabi-Yau singularity $\mathscr{X}$. This engineers a relative
QFT in
\be
D = 11 - \textnormal{dim}\, \mathscr{X}
\ee
dimensions denoted $\mathcal{T}_{\mathscr{X}}$. We assume that $\mathscr{X}$ takes the form of a conical singularity, whereby in particular $\mathscr{X}$ is topologically a cone over its boundary geometry $\partial \mathscr{X}$:
\begin{equation}
\mathscr{X} = \mathrm{Cone}(\partial \mathscr{X})\,.
\end{equation}
Further, a singularity of codimension $\textnormal{dim}\, \mathscr{X}$ sits at the tip of the cone and we clearly have a preferred ``radial direction'' with respect to which the boundary $\partial \mathscr{X}$ sits ``at infinity". Bulk modes such as those associated to a lower codimension singularity will generically fill this radial direction as well as some of the directions of $\partial \mathscr{X}$.

We now discuss heavy defects, topological symmetry operators, and their geometric realization. We then turn to the specific singularity structures we will repeatedly encounter.

\subsection{Heavy Defects and Symmetry Operators}

We begin by discussing heavy defects of the QFT $\mathcal{T}_{\mathscr{X}}$, engineered by the local geometry $\mathscr{X}$, which are in turn constructed by wrapping branes on non-compact cycles of $\mathscr{X}$. The resulting objects are formally of infinite mass\,/\,tension and realize defects in the QFT.

Given a space $\mathscr{X}$ with a singularity at the tip of the cone, we can construct defects by wrapping branes on relative cycles which stretch from the tip of the cone out to the boundary $\partial \mathscr{X}$. It can happen that the singularity is isolated, but it may also support additional singularities which stretch from the tip of the cone out to ``infinity'', and these additional singularities are physically interpreted as flavor loci.

Throughout this paper, we will be in the situation where there exists a crepant resolution $\pi: \mathscr{X}^{\prime} \rightarrow \mathscr{X}$. As such, we can explicitly enumerate the defects obtained from wrapped branes, as well as dynamical states obtained from branes wrapping collapsed cycles. We can wrap a brane on a relative cycle $H_{k+1}(\mathscr{X}^{\prime}, \partial \mathscr{X}^{\prime})$ and these defects are partially screened by branes wrapping cycles in $H^{\mathrm{cpct}}_{k+1}(\mathscr{X})$. The resulting quotient $H_{k+1}(\mathscr{X}^{\prime}, \partial \mathscr{X}^{\prime}) / H^{\mathrm{cpct}}_{k+1}(\mathscr{X}^{\prime})$ specifies a collection of unscreened charges. This is the essential physics behind the ``defect group'' \cite{DelZotto:2015isa, GarciaEtxebarria:2019caf, Albertini:2020mdx, Morrison:2020ool}.

Geometrically, the defect group capturing the spectrum of all stable\footnote{In principle, they could be stable but not BPS.} branes wrapped on non-compact relative cycles is given by
\begin{equation}
\mathbb{D} = \underset{m}{\bigoplus}\,{\mathbb{D}}^{(m)},
\end{equation}
where $m$ denotes the number of the brane worldvolume dimensions extending in spacetime, which is the dimensionality of the resulting defect, and
\begin{equation}\label{eq:DefectGp}
\mathbb{D}^{(m)} = \underset{p\mathrm{\textnormal{\;\!-}branes}}{\bigoplus}~ \underset{p - k = m}{\bigoplus} \frac{H_{k+1}(\mathscr{X}' , \partial \mathscr{X}')}{H_{k+1}^{\mathrm{cpct}}(\mathscr{X}')}\,.
\end{equation}
Picking a polarization, i.e., a maximally mutually local set of extended operators then provides us with an absolute, rather than relative theory (in the sense of \cite{Freed:2012bs}).

Symmetry operators are obtained by wrapping branes ``at infinity'', i.e., such branes are supported exclusively at the boundary $\partial \mathscr{X}$ of the internal directions. In the case of finite order discrete symmetries which act on a $p$-brane wrapped on a torsional class of \eqref{eq:DefectGp}, these symmetry operators are obtained from magnetic dual $q$-branes wrapped on torsional cycles of $H_{l}(\partial \mathscr{X})$ which link with the heavy defect in both the spacetime and $\partial \mathscr{X}$ (see \cite{Apruzzi:2022rei, GarciaEtxebarria:2022vzq, Heckman:2022muc} as well as \cite{Heckman:2022xgu}), while in the case of continuous symmetries, these symmetry operators are obtained from magnetic dual flux $(q+1)$-branes (see e.g., \cite{Costa:2001ifa, Gutperle:2001mb, Emparan:2001gm}) wrapped on free cycles of $H_{l+1}(\partial \mathscr{X})$ which link in the spacetime, but intersect in $\partial \mathscr{X}$ with the heavy defect \cite{Cvetic:2023plv} (see also \cite{Garcia-Valdecasas:2023mis}).

An important feature of this perspective is that the topological sector of these branes automatically descends (upon reduction on the internal cycles) to a topological field theory attached to the symmetry operator. This gives a general algorithm for producing non-trivial symmetry operators and extracting the corresponding fusion rules directly from the brane construction.

\subsection{Local Models}

We now discuss in more detail some properties of the local singular geometries $\mathscr{X}$ we shall encounter. We shall mainly focus on the special case of orbifold singularities
\be
\mathscr{X}=\mathbb{C}^{m} / \Gamma_{SU(m)}\,, \qquad \partial \mathscr{X}=S^{2m-1}/\Gamma_{SU(m)}\,,
\ee
with $\Gamma_{SU(m)}$ a finite subgroup of $SU(m)$. Many of the global orbifolds we consider will be modelled patchwise on such finite quotients. In the case of elliptically fibered Calabi-Yau spaces, similar considerations will apply, up to the small subtlety of including the data of the elliptic fiber.

In the case $\mathscr{X}=\mathbb{C}^{2} / \Gamma_{SU(2)}$, there is an ADE classification of possible singularities. It is well-known that M-theory on such spaces gives rise to a 7D supersymmetric Yang-Mills theory with Lie algebra $\mathfrak{g}_{{\ADE}}$ of ADE type.\footnote{We do not consider frozen singularities of M-theory \cite{Witten:1997bs, Tachikawa:2015wka}.} The global form of the 7D gauge group depends on the boundary conditions for torsional fluxes ``at infinity'' along the asymptotic boundary $\partial\mathscr{X}$, and this in turn is implicitly dictated by the spectrum of extended operators one allows in the theory.

The associated defect group $\mathbb{D}$ \cite{DelZotto:2015isa, GarciaEtxebarria:2019caf, Albertini:2020mdx, Morrison:2020ool, Apruzzi:2021nmk} is obtained from M2- and M5-branes wrapped on relative 2-cycles which extend along the radial direction of $\mathbb{C}^{2} / \Gamma_{SU(2)}$ terminating on torsional 1-cycles of the boundary geometry $S^{3} / \Gamma_{SU(2)}$. They are catalogued by elements of
\be
H_{2}(\mathbb{C}^2/\Gamma_{SU(2)},S^3 / \Gamma_{SU(2)}) \cong H_{1}(S^3 / \Gamma_{SU(2)}) \cong \mathrm{Ab}[\Gamma_{SU(2)}] \equiv \mathcal{Z}\,,
\ee
which is the Abelianization of $\Gamma_{SU(2)}$, or equivalently, the center of the simply connected Lie group with Lie algebra $\mathfrak{g}_{\ADE}$. The defect group is then
\begin{equation}
\mathbb{D} = \mathcal{Z}^{(1)}_{M2} \oplus \mathcal{Z}^{(4)}_{M5},
\end{equation}
where the superscript gives the defect dimension and therefore the degree of the corresponding $p$-form symmetry. The choice of boundary conditions for torsional fluxes then dictates the global form of the gauge group.

For example, taking $\Gamma_{SU(2)} = \mathbb{Z}_N$ we get $\mathfrak{g}_{\ADE} = \mathfrak{su}(N)$ and the global form of the group may take the form $SU(N) / \mathbb{Z}_K$ for some $K$ dividing $N$. Topological symmetry operators acting on $\mathcal{Z}^{(1)}_{M2}$ and $\mathcal{Z}^{(4)}_{M5}$ are constructed from M5-branes and M2-branes wrapped on torsional 1-cycles of the lens space boundary $S^{3} / \mathbb{Z}_N$ respectively \cite{Heckman:2022muc}.

The other building blocks we shall frequently encounter are orbifold singularities of the form $\mathscr{X}=\mathbb{C}^{3} / \Gamma_{SU(3)}$. In this case, we get a 5D superconformal field theory (SCFT) provided the resolution of the singularity includes a collapsing 4-cycle \cite{Douglas:1996xp, Intriligator:1997pq} and, depending on the choice of $\Gamma_{SU(3)}$ as well as group action, we can get many different choices for the singularity. In particular, it can happen that in addition to a codimension 6 singularity (at the tip of a cone), there can also be codimension 4 singularities which extend out to the boundary $\partial \mathscr{X}=S^{5} / \Gamma_{SU(3)}$, associated with 7D SYM sectors.

In the 5D SCFT, such sectors specify non-Abelian flavor symmetry factors, namely 0-form symmetries. Neglecting such factors, the contributions to the defect group are obtained from wrapping M2-branes on relative cycles
\be
H_{2}(\mathbb{C}^{3} / \Gamma_{SU(3)} , S^{5} / \Gamma_{SU(3)})\cong H_{1}( S^{5} / \Gamma_{SU(3)})\cong  \textnormal{Ab}(\Gamma/H)\equiv \mathcal{Z}
\ee
where $H$ is the normal subgroup of $\Gamma$ generated by all elements with fixed points on $S^{5}$ as determined by a theorem of Armstrong \cite{armstrong_1968} on the fundamental group of orbit spaces. Dually, M5-branes are wrapped on non-compact 4-cycles also contributing a copy of $\mathcal{Z}$ to the defect group, overall we have the unscreened line and surface defects captured by:
\begin{equation}
 \mathcal{Z}^{(1)}_{M2} \oplus \mathcal{Z}^{(2)}_{M5}\subset \mathbb{D}\,.
\end{equation}
Specific algorithms are now available for extracting these structures (see e.g., \cite{Albertini:2020mdx,Morrison:2020ool,Tian:2021cif,DelZotto:2022fnw, Cvetic:2022imb}). There are corresponding magnetic dual branes which wrap 3-cycles and 1-cycles at infinity, and these produce the corresponding topological symmetry operators associated with these higher-form symmetries \cite{Heckman:2022muc}.

In addition to these higher-form symmetries, there can also be non-trivial entwinement of these structures via higher-group structures. For 2-groups, this is an entwinement between 0-form and 1-form symmetries, and amounts to a failure of associativity in forming various junctions for the topological 0-form symmetry operators \cite{Benini:2018reh}. Such 2-group symmetries have been investigated in 5D SCFTs, for example, in \cite{Apruzzi:2021vcu, DelZotto:2022fnw, Cvetic:2022imb, DelZotto:2022joo}.

Working in an electric polarization for our 5D SCFT, the statement that we have a 2-group requires a specific four-term long exact sequence which is not split:
\begin{equation}\label{eq:2gpSequence}
0 \rightarrow \mathcal{A} \rightarrow \widetilde{\mathcal{A}} \rightarrow \widetilde{G} \rightarrow G \rightarrow 1,
\end{equation}
with $ \widetilde{\mathcal{A}}$ being the ``naive'' 1-form symmetry as obtained by excising from the boundary $S^{5} / \Gamma_{SU(3)}$ all singularities. In terms of the geometric content of the generalized lens space, we have:
\be
\mathcal{A} \cong  (\mathcal{Z}^{(1)}_{M2})^\vee \cong \mathrm{Ab}(\Gamma/H)^\vee\,, \qquad \widetilde{\mathcal{A}}\cong \mathrm{Ab}(\Gamma)^\vee\,,
\ee
where $Z^\vee$ denotes the Pontryagin dual of an Abelian group $Z$:
\begin{equation}
Z^{\vee} = \mathrm{Hom}(Z \rightarrow U(1)).
\end{equation}
Similarly, $G$ is the actual flavor symmetry group and $\widetilde{G}$ is the simply connected ``naive'' answer.
In practice, this is captured in geometry by a specific condition on non-splitness of a short exact sequence associated with the 1-form
symmetries:
\begin{equation}
 0\rightarrow \mathcal{A} \xrightarrow[]{~\iota~} \widetilde{\mathcal{A}} \rightarrow \textnormal{coker}\,\iota \subset Z(\widetilde{G}) \rightarrow 1,
\end{equation}
where $Z(\widetilde{G})$ is the center of $\widetilde{G}$.

Having reviewed some general features of how to engineer generalized symmetries in M-theory backgrounds, we now turn to their fate in models
coupled to gravity.

\section{Global Models: General Considerations} \label{sec:GLOBAL}

We now present a general overview to our procedure of coupling local models to gravity. Our aim will be to
track the fate of various generalized symmetries as a single local model is ``glued'' to other sectors of a string compactification. In subsequent sections we consider specific examples, mainly in the context of 7D and 5D M-theory backgrounds, and when applicable, their lifts to F-theory backgrounds.

In this paper we focus on QFT sectors obtained from M-theory on local geometric singularities.\footnote{In the case of spacetime filling brane probes of a singularity, tadpole cancellation of the associated flux introduces additional constraints.} We obtain light degrees of freedom from branes wrapped on collapsed cycles of the singularity. Our main interest here will be to track the impact of coupling this theory to gravity. It could happen that there is no such embedding available, and this is one of the core issues which the Swampland program aims to address. Additionally, it could also happen that there is more than one way to recouple a local model to gravity.

To bypass these issues, we shall instead start the other way around, beginning with a supergravity theory $\mathcal{S}_X$ as specified by M-theory compactified on a compact Calabi-Yau manifold $X$. This gives rise to a $D = 11 - \mathrm{dim}\, X $ low energy effective field theory on $\mathbb{R}^{D-1,1}$ with Newton's constant set by $G_{N} \sim \mathrm{Vol}(X)^{-1}$ in 11D Planck units. In the special case where $X$ is genus-one fibered, shrinking the fiber class to zero size results in a dual F-theory background on $S^1 \times X$ with the radius of the $S^1$ inversely related to (a power of) the volume of the elliptic fiber \cite{Vafa:1996xn, Morrison:1996na, Morrison:1996pp}. For $D > 4$ supersymmetric models, we generically have a moduli space of vacua so we can tune the geometric moduli by freezing the asymptotic vacuum expectation values of these moduli. In particular, we can work in a tuned limit where the Calabi-Yau $X$ supports various singularities.

For $X$ a K3 surface, the singularities we shall encounter are codimension 4. In this case, there is a clear geometric interpretation of each localized singularity $\mathscr{X}_i$ coupled together by the ``bulk'' K3 surface which is the complement of some neighborhoods containing the singularities. For each such singularity we can speak of a corresponding local QFT sector $\mathcal{T}_{i}$. In the process of gluing these singularities together, we can expect to pick up additional dynamical states as well as modes of the compact geometry $X$. For example, in compact models there can be additional $U(1)$ gauge group factors which are ``shared'' across multiple singularities. We shall see explicit examples of this. Additionally, we can expect various discrete symmetries to either be gauged or broken as we glue the local models back together.

For $X$ a higher-dimensional Calabi-Yau, more general phenomena are possible. Indeed, in this case we can expect singularities of various codimension. In particular, we shall be interested in singularities which are codimension $\mathrm{dim}\, X$, namely they are ``pointlike'' in $X$. For $X$ a K3 surface, this is essentially all that can occur, but for Calabi-Yau threefolds consistency of the background\,/\,local model will often require the participation of lower codimension subspaces.

For example, in engineering a 5D SCFT with a continuous flavor symmetry, we shall often have to contend with a codimension 6 singularity which arises as an enhancement within non-compact singular loci of codimension 4. In such situations, after gluing, the codimension 4 loci are necessarily compactified and could be ``shared'' between multiple codimension 6 singularities.

Let us then label each of the codimension $\mathrm{dim}\,X$ sectors as specifying a local Calabi-Yau geometry $\mathscr{X}_{i}$ and an associated theory $\mathcal{T}_i$. The process of taking a decoupling limit can thus take various stages: One can first decouple gravity by sending $\mathrm{Vol}(X) \rightarrow \infty$. In this limit, it could still happen that various $\mathcal{T}_i$ are still coupled together by lower codimension singularities. Decompactifying these as well, we can reach a local model in which the interacting degrees of freedom support a theory $\mathcal{T}_i$ and other singularities are either decoupled, or interpreted as non-dynamical ``bulk modes'' of a higher-dimensional theory.\footnote{For a recent discussion on such examples, see reference \cite{Acharya:2023bth}.} Let us note that these further decoupling limits are not unique; there are various ways one could consider gluing together a subset of the $\mathcal{T}_i$.

One of the general features of this approach is that starting from our supergravity theory $\mathcal{S}_X$, the resulting QFTs obtained from decoupling from gravity will often have emergent global symmetries. We can clearly track these from the structure of the boundary topology $\partial \mathscr{X}_{i}$, although in principle there could be other contributions such as isometries of the extra dimensions, or non-geometric features of the associated QFT. One of our tasks will be to identify such emergent symmetries.

The important feature of each local model $\mathcal{T}_i$ is that we can fix the global form of the theory by specifying boundary conditions on $\partial \mathscr{X}_i$. When non-anomalous, we can also gauge these symmetries and change polarization to a different global form of the theory. On the other hand, once we start gluing together these different local geometries, some of these gaugings will automatically occur since the boundary conditions on distinct local models such as $\partial \mathscr{X}_i$ and $\partial \mathscr{X}_j$ will now end up being identified. One can also see this from the perspective of the heavy defects of a local model: Previously infinite tension objects can now stretch from the singularity in $\mathscr{X}_i$ to the singularity in $\mathscr{X}_j$ along finite volume cycles. In the low energy effective field theory this amounts to introducing additional dynamical states into the theory. As for the symmetry operators, branes ``wrapped at infinity'' will now wrap finite volume subspaces, which in general will not be of minimal volume. This is consistent with the fact that we are gauging some symmetries, and breaking\,/\,screening others by adding extra states.

At this point it is natural to ask in what sense the supergravity theory $\mathcal{S}_X$ specifies a canonical choice of gauge group both for the 0-form as well as higher-form symmetries of a given model. The main point is that the lightest degrees of freedom in the theory implicitly specify the structure of the gauge group. For example, in the context of Maxwell theory one can invariantly define an ``electron'' as the lightest charged state of the gauge theory. In the context of M-theory backgrounds, the M2-branes are the electric degrees of freedom whereas the M5-branes are their dual magnetic counterparts. Provided, however, that the resulting collection of states obtained after compactification are mutually local (i.e., there is no flux non-commutativity), then both serve to determine the global form of the gauge group. As such, specifying the global form of the gauge group in a supergravity model in M-theory amounts to determining how M2- and M5-branes stretch between various local patches $\mathscr{X}_i$. This, in turn, is specified by the cutting and gluing rules associated with the Mayer-Vietoris long exact sequence.

In more detail, recall that in the Mayer-Vietoris sequence, one considers subspaces $U$ and $V$ of $X$ such that the interiors of $U$ and $V$ cover $X$. Then, we have the long exact sequence which relates the homologies of the local patches to those of the global geometry:
\begin{equation}
... \rightarrow H_{n+1}(X) \rightarrow H_{n}(U \cap V) \rightarrow H_{n}(U) \oplus H_{n}(V) \rightarrow H_{n}(X) \rightarrow ...
\end{equation}
in the obvious notation. Refining each subspace to consist of the various local models, it follows that we can read off the spectrum of charged states from wrapped branes, and thus implicitly the global form of the gauge groups in the M-theory model.

In the related context of F-theory on $X$ a genus-one fibered Calabi-Yau manifold, the global form of the
gauge group is dictated by the Mordell-Weil group of the Calabi-Yau, $\mathrm{MW}(X)$, which in turn requires determining the group law for rational sections of the elliptic fiber (see e.g., \cite{Grimm:2010ez, Grimm2011, Morrison:2012ei, Cvetic:2013nia, Borchmann:2013jwa, Morrison:2014era, Braun:2014oya, Mayrhofer2014, Braun:2014nva, Mayrhofer:2014laa, Cvetic:2017epq, Cvetic:2021sxm} as well as \cite{Cvetic:2018bni} and references therein). This is known to be a somewhat involved procedure which depends in a rather delicate way on the structure of the associated Weierstrass model for the genus-one fibration. Our method can be viewed as a complementary approach which can be extracted based on comparatively ``coarse'' data of $X$. Indeed, our method applies even when there is no globally defined genus-one fibration. That being said, it is important to note that the relation between the M-theory and F-theory backgrounds involves a further circle compactification. The generic appearance of $U(1)$ factors in the M-theory background compared with the tuning required in F-theory models arises because we can now allow possibly position dependent vevs of the 6D background which only preserve 5D Lorentz invariance.

\paragraph{Iterative Gluing vs Spectral Sequences}\mbox{}\medskip \\ In the context of geometries with singularities of different codimensions, the process of ``cutting and gluing'' includes some additional subtleties. The first challenge in analyzing higher-dimensional geometries $X$ lies in the plethora of possibilities for singular strata. We need to specify how we cut up $X$ or equivalently which localized field theory sectors we identify as physical building blocks in our construction.

One option is to consider the finest decomposition possible including tubular neighborhoods for all singular substrata individually. These can then be glued together democratically using the Mayer-Vietoris spectral sequence (see e.g., chapter VII, section 4 of \cite{Brown}).

A second option consists of iteratively gluing ``more'' singular substrata to ``less'' singular substrata and updating our notion of local model in the process. The penultimate step then consists of studying the interactions between local field theory sectors $\mathcal{T}$ associated with connected components of the singular locus within the global model.

To frame the discussion of these two approaches, let us introduce some notation. An orbifold $X$ admits a canonical filtration with strata consisting of points with isomorphic isotropy groups \cite{Caramello2019IntroductionTO, ademleidaruan2007}. The isotropy group of a point is the finite group $\Gamma_x$ folding the singularity at $x\in X$. More precisely there exists a chart, centered on $x$, mapping from the contractible patch $U_x\subset X$ and modeled on $\R^n/\Gamma_x$ where $n=\dim_{\;\!\R} X$. The strata $X_{\Gamma_I} \subset X$ are then defined as the subspace containing points with isotropy groups isomorphic to $\Gamma_I$ labelled by $I$. For global quotients $X=Y/\Gamma$ strata are labeled by subgroups $\Gamma_I\subset \Gamma$.

The strata $X_{\Gamma_I}$ are (possibly disjoint) manifolds without boundaries and connected components of well-defined dimension. We denote the union of strata of codimension $m$ or higher as $X_m$ with complement $X_m^\circ=X\setminus X_m$ and tubular neighborhood $X_m^{\textnormal{loc}}$ consisting of the union of small contractible sets $U_x$ for $x\in X_m$. The set of regular points is the open manifold $X_1^\circ$. Clearly, for $m_1\geq m_2$, we have the inclusions
\be\label{eq:nesting}
X_{m_1}\subseteq X_{m_2}\,, \qquad  X_{m_1}^{\textnormal{loc}}\subseteq X_{m_2}^{\textnormal{loc}}\,, \qquad X_{m_2}^\circ \subseteq X_{m_1}^\circ\,.
\ee
Note that $X_m^{\textnormal{loc}}$ deformation retracts to $X_m$ while the intersection $X_m^{\textnormal{loc}} \cap X_m^\circ$ deformation retracts to both $\partial X_m^{\textnormal{loc}}$ and $\partial X_m^\circ$. We now have the covering
\be \label{eq:Cover}
X=\lb X_{d}^{\textnormal{loc}} \cup X_d^\circ\rb \cup \dots \cup \lb X_{1}^{\textnormal{loc}} \cup X_{1}^\circ\rb
\ee
which is the starting point for the analysis based on the Mayer-Vietoris spectral sequence.\footnote{For example for singular K3 surfaces we have the ADE locus $X_4$ with local model $X_{4}^{\textnormal{loc}}$ consisting of a disjoint collection of balls, one centered on each point in $X_4$. The covering is then simply $X=X_4^\circ\cup \Xl_4$. In this case the two approaches presented here coincide.}

However, we will mainly take the second approach which carries more physical intuition. We begin by considering connected components $\textnormal{Sing}(X)_i$ of the singular set
\be
\textnormal{Sing}(X)=\bigcup_{\Gamma_I\neq 1} X_{\Gamma_I} = \coprod_i \textnormal{Sing}(X)_i\,.
\ee
Similarly we attach an index to the codimension $m$ substrata yielding disjoint connected components $ X_{m,i}^{\textnormal{loc}} , X_{m,i}^\circ$
\be
X_{m}^{\textnormal{loc}}=\coprod_i X_{m,i}^{\textnormal{loc}}\,,\qquad X_{m}^\circ=\coprod_i X_{m,i}^\circ
\ee
We then focus on each connected component, i.e., fixed $i$, and glue the local model for $\textnormal{Sing}(X)_i$ from the local models of its singular substrata of distinct codimension $m$.

In more detail, we iteratively glue $X_{n,i}^{\textnormal{loc}} $ to $X_{n-1,i}^{\textnormal{loc}} $ for fixed $i$ and this gives a space we denote as $X_{n,n-1,i}^{\textnormal{loc}} $ which is the local model of singularities of codimension $n$ and $n-1$. Next, this space is then glued to $X_{n-2,i}^{\textnormal{loc}} $ and so on. At the $k$-th iteration of this gluing procedure the space is denoted $X_{n, n-k, \,i}^{\textnormal{loc}} $. The largest value of $k$ for which we do not produce the total space $X$ is the final local model and contains a compact singular locus. In particular this local model has a smooth boundary.

One advantage of this second approach lies in noting at which iteration certain symmetry structures dissolve. For example, $2$-group symmetries as studied in \cite{Cvetic:2022imb, DelZotto:2022joo} can arise when local models exhibit non-compact singular loci. These are necessarily compactified in the above construction prior to the iteration where local models glue to a global model. The gauging and breaking of 2-group symmetries therefore occurs purely upon gluing local models to local models in this context.\footnote{One can also find new 2-group structures once one has coupled to gravity, we exhibit some examples of this sort in our discussion of 5D vacua.}

Having spelled out some general considerations for how to analyze the contributions from different singularities, we now turn to explicit examples. We begin with M-theory on K3 surfaces, where all the singularities are codimension 4. We then proceed to some examples of M-theory on a Calabi-Yau threefold, where we encounter examples with codimension 6 as well as codimension 4 singularities.

\section{M-theory on a K3 with Singularities} \label{sec:7d}

In this section we consider M-theory on the background  $\mathbb{R}^{6,1} \times X$, with $X$ a Calabi-Yau twofold, i.e., a K3 surface. Even though there is a single K3 surface, tuning the metric moduli will result in distinct low energy behavior. Our aim in this section will be to develop some general aspects of how 7D generalized symmetries recouple to gravity. In section \ref{sec:7dExamp} we turn to some illustrative examples.

At low energies, M-theory on a K3 surface results in a 7D theory with sixteen supercharges. Allowing degenerations in the metric of $X$, we obtain various codimension 4 singularities, and as such, 7D supersymmetric Yang-Mills theory sectors. In addition to these localized gauge theory degrees of freedom, we also have Abelian vector multiplets which are ``delocalized'' across the geometry. In what follows we focus on torsional contributions to homology which can be tracked in terms of defects / symmetry operators obtained from wrapped branes. In particular, we neglect contributions from ``accidental'' discrete isometries of the geometry. The full gauge algebra of the model thus takes the form:
\begin{equation}\label{eq:algebra}
\mathfrak{g}_{\mathrm{full}} = \underset{i}{\bigoplus} \,\mathfrak{g}_i \oplus \mathfrak{u}(1)^b,
\end{equation}
where the $\mathfrak{g}_i$ come from local singularities $\mathscr{X}_i \cong \mathbb{C}^2 / \Gamma_{i}$, and the $\mathfrak{u}(1)$ factors depend on global data of the model and $b$ is some integer. We denote tubular neighbourhoods centered on each singularity by $U_i\subset X$. The $U_i$ are disjoint and topologically modelled on balls in $\mathscr{X}_i$ containing the origin. For each such gauge theory, there is a corresponding simply connected gauge group $\widetilde{G}_{i}$ of ADE type. The global structure of the gauge group depends on the choice of boundary conditions on each $\mathscr{X}_{i}$. Once we start gluing into the rest of the compact geometry, the different choices of consistent boundary conditions will necessarily be correlated.

One of our tasks will be to track the light degrees of freedom stretched between these different singularities. Doing so, we will be able to determine the global form of the gauge group in the supergravity model. This will be accomplished in stages. First, we determine the contribution from the localized non-abelian gauge groups:
\begin{equation}
G_{\mathrm{loc}} = \widetilde{G}_{\mathrm{loc}} / \mathcal{C}_{\mathrm{loc}} \,\,\, \text{with} \,\,\, \widetilde{G}_{\mathrm{loc}} = \underset{i}{\prod} \widetilde{G}_{i}.
\end{equation}
With this in place, there can be an extra quotient which also acts on the remaining centers of $G_{\mathrm{loc}}$ and the $U(1)$ factors:
\begin{equation}\label{eq:quotient}
G_{\mathrm{full}} = \frac{G_{\mathrm{loc}} \times U(1)^{b}}{\mathcal{C}_{\mathrm{Extra}}}\,,
\end{equation}
for some integer $b$. In the context of 7D vacua obtained from M-theory on a K3 surface, there are no finite gauge groups which arise from this procedure.\footnote{A priori, however, there could be discrete isometries. We ignore these contributions in what follows.}. Another goal of our analysis will be to study how the higher-form symmetries are either gauged or broken\,/\,screened in the full compact model and how different heavy defects become dynamical.

\subsection{An Exact Sequence for K3 Surfaces}

\begin{figure}
\centering
\scalebox{0.7}{\begin{tikzpicture}
	\begin{pgfonlayer}{nodelayer}
		\node [style=none] (0) at (-4, -4) {};
		\node [style=none] (1) at (-3, -1) {};
		\node [style=none] (2) at (-1, -3) {};
		\node [style=none] (3) at (-3, 1) {};
		\node [style=none] (4) at (-1, 3) {};
		\node [style=none] (5) at (1, 3) {};
		\node [style=none] (6) at (3, 1) {};
		\node [style=none] (7) at (3, -1) {};
		\node [style=none] (8) at (1, -3) {};
		\node [style=none] (9) at (4, -4) {};
		\node [style=none] (10) at (4, 4) {};
		\node [style=none] (11) at (-4, 4) {};
		\node [style=none] (12) at (-1.75, 2.25) {};
		\node [style=none] (13) at (-2.25, 1.75) {};
		\node [style=none] (14) at (1.75, 2.25) {};
		\node [style=none] (15) at (2.25, 1.75) {};
		\node [style=none] (16) at (2.25, -1.75) {};
		\node [style=none] (17) at (1.75, -2.25) {};
		\node [style=none] (18) at (-1.75, -2.25) {};
		\node [style=none] (19) at (-2.25, -1.75) {};
		\node [style=Circle] (20) at (-4, 4) {};
		\node [style=Circle] (21) at (4, 4) {};
		\node [style=Circle] (22) at (4, -4) {};
		\node [style=Circle] (23) at (-4, -4) {};
		\node [style=none] (24) at (8, -4) {};
		\node [style=none] (25) at (9, -1) {};
		\node [style=none] (26) at (11, -3) {};
		\node [style=none] (27) at (9, 1) {};
		\node [style=none] (28) at (11, 3) {};
		\node [style=none] (29) at (13, 3) {};
		\node [style=none] (30) at (15, 1) {};
		\node [style=none] (31) at (15, -1) {};
		\node [style=none] (32) at (13, -3) {};
		\node [style=none] (33) at (16, -4) {};
		\node [style=none] (34) at (16, 4) {};
		\node [style=none] (35) at (8, 4) {};
		\node [style=none] (36) at (10.25, 2.25) {};
		\node [style=none] (37) at (9.75, 1.75) {};
		\node [style=none] (38) at (13.75, 2.25) {};
		\node [style=none] (39) at (14.25, 1.75) {};
		\node [style=none] (40) at (14.25, -1.75) {};
		\node [style=none] (41) at (13.75, -2.25) {};
		\node [style=none] (42) at (10.25, -2.25) {};
		\node [style=none] (43) at (9.75, -1.75) {};
		\node [style=Circle] (44) at (8, 4) {};
		\node [style=Circle] (45) at (16, 4) {};
		\node [style=Circle] (46) at (16, -4) {};
		\node [style=Circle] (47) at (8, -4) {};
		\node [style=none] (48) at (0, -5.5) {\Large $(1):\textnormal{Trivialization of Symmetry Operators }$};
		\node [style=none] (49) at (12, -5.5) {\Large $(2):\textnormal{Compactification of Defect Operators }$};
		\node [style=none] (50) at (-4, 4.5) {\large $x_i$};
		\node [style=none] (51) at (4, 4.5) {\large $x_j$};
		\node [style=none] (52) at (4, -4.5) {\large $x_k$};
		\node [style=none] (54) at (0, 0) {\large $\Sigma_{ijk\dots}$};
		\node [style=none] (55) at (12, 0) {\large $\Sigma_{ ijk\dots}$};
		\node [style=none] (56) at (-1.5, 2.5) {\large $\gamma_i$};
		\node [style=none] (58) at (-4, -4.5) {\large $\dots$};
		\node [style=none] (59) at (1.5, 2.5) {\large $\gamma_j$};
		\node [style=none] (60) at (1.5, -2.5) {\large $\gamma_k$};
		\node [style=none] (62) at (8, 4.5) {\large $x_i$};
		\node [style=none] (63) at (16, 4.5) {\large $x_j$};
		\node [style=none] (64) at (16, -4.5) {\large $x_k$};
		\node [style=none] (65) at (12, -6) {};
		\node [style=none] (66) at (10.5, 2.5) {\large ${\sigma}_i$};
		\node [style=none] (67) at (8, -4.5) {\large $\dots$};
		\node [style=none] (68) at (13.5, 2.5) {\large ${\sigma}_j$};
		\node [style=none] (69) at (13.5, -2.5) {\large ${\sigma}_k$};
        \node [style=none] (70) at (-3, 2.875) {\large $U_i$};
		\node [style=none] (71) at (3, 2.875) {\large $U_j$};
		\node [style=none] (72) at (3, -2.875) {\large $U_k$};
        \node [style=none] (74) at (3, -6.5) {};
	\end{pgfonlayer}
	\begin{pgfonlayer}{edgelayer}
		\draw [style=ThickLine] (11.center) to (4.center);
		\draw [style=ThickLine] (5.center) to (10.center);
		\draw [style=ThickLine] (10.center) to (6.center);
		\draw [style=ThickLine] (11.center) to (3.center);
		\draw [style=ThickLine] (1.center) to (0.center);
		\draw [style=ThickLine] (0.center) to (2.center);
		\draw [style=ThickLine] (8.center) to (9.center);
		\draw [style=ThickLine] (7.center) to (9.center);
		\draw [bend right=45] (4.center) to (3.center);
		\draw [bend left=45] (5.center) to (6.center);
		\draw [bend left=45] (7.center) to (8.center);
		\draw [bend left=45] (2.center) to (1.center);
		\draw [bend right=45] (3.center) to (4.center);
		\draw [bend right=45] (5.center) to (6.center);
		\draw [bend left=315] (7.center) to (8.center);
		\draw [bend left=45, looseness=1.25] (1.center) to (2.center);
		\draw [style=RedLine, bend left=45] (12.center) to (13.center);
		\draw [style=RedLine, bend left=45] (19.center) to (18.center);
		\draw [style=RedLine, bend left=45] (17.center) to (16.center);
		\draw [style=RedLine, bend right=45] (14.center) to (15.center);
		\draw [style=RedLine, bend left=45] (13.center) to (12.center);
		\draw [style=RedLine, bend left=45] (14.center) to (15.center);
		\draw [style=RedLine, bend left=45] (16.center) to (17.center);
		\draw [style=RedLine, bend right=45] (19.center) to (18.center);
		\draw [style=DottedLine, bend left=15] (5.center) to (4.center);
		\draw [style=DottedLine, bend right=15] (6.center) to (7.center);
		\draw [style=DottedLine, bend left=15] (2.center) to (8.center);
		\draw [style=DottedLine, bend left=15] (3.center) to (1.center);
		\draw [style=DottedRed, bend right=45, looseness=1.25] (12.center) to (14.center);
		\draw [style=DottedRed, bend right=45, looseness=1.25] (15.center) to (16.center);
		\draw [style=DottedRed, bend left=315, looseness=1.25] (17.center) to (18.center);
		\draw [style=DottedRed, bend right=45, looseness=1.25] (19.center) to (13.center);
		\draw [style=ThickLine] (35.center) to (28.center);
		\draw [style=ThickLine] (29.center) to (34.center);
		\draw [style=ThickLine] (34.center) to (30.center);
		\draw [style=ThickLine] (35.center) to (27.center);
		\draw [style=ThickLine] (25.center) to (24.center);
		\draw [style=ThickLine] (24.center) to (26.center);
		\draw [style=ThickLine] (32.center) to (33.center);
		\draw [style=ThickLine] (31.center) to (33.center);
		\draw [bend right=45] (28.center) to (27.center);
		\draw [bend left=45] (29.center) to (30.center);
		\draw [bend left=45] (31.center) to (32.center);
		\draw [bend left=45] (26.center) to (25.center);
		\draw [bend right=45] (27.center) to (28.center);
		\draw [bend right=45] (29.center) to (30.center);
		\draw [bend left=315] (31.center) to (32.center);
		\draw [bend left=45, looseness=1.25] (25.center) to (26.center);
		\draw [style=DottedLine, bend left=15] (29.center) to (28.center);
		\draw [style=DottedLine, bend right=15] (30.center) to (31.center);
		\draw [style=DottedLine, bend left=15] (26.center) to (32.center);
		\draw [style=DottedLine, bend left=15] (27.center) to (25.center);
		\draw [style=BlueLine, bend right=45] (37.center) to (36.center);
		\draw [style=BlueLine, bend right=45] (38.center) to (39.center);
		\draw [style=BlueLine, bend left=45] (40.center) to (41.center);
		\draw [style=BlueLine, bend right=45] (43.center) to (42.center);
		\draw [style=BlueLine, bend left=45] (37.center) to (36.center);
		\draw [style=BlueLine, bend left=45] (38.center) to (39.center);
		\draw [style=BlueLine, bend right=45] (40.center) to (41.center);
		\draw [style=BlueLine, bend left=45] (43.center) to (42.center);
		\draw [style=BlueLine] (43.center) to (47);
		\draw [style=BlueLine] (47) to (42.center);
		\draw [style=BlueLine] (41.center) to (46);
		\draw [style=BlueLine] (40.center) to (46);
		\draw [style=BlueLine] (38.center) to (45);
		\draw [style=BlueLine] (39.center) to (45);
		\draw [style=BlueLine] (44) to (36.center);
		\draw [style=BlueLine] (37.center) to (44);
		\draw [style=BlueDottedLight, bend right=45] (36.center) to (38.center);
		\draw [style=BlueDottedLight, bend right=45] (39.center) to (40.center);
		\draw [style=BlueDottedLight, bend left=315] (41.center) to (42.center);
		\draw [style=BlueDottedLight, bend right, looseness=1.25] (43.center) to (37.center);
	\end{pgfonlayer}
\end{tikzpicture}
}
\caption{Both subfigures (1) and (2) show a singular K3 surface $X$ and within it the tubular neighbourhoods $U_i\subset X$ as black cones. Subfigure (1) shows cycles $\gamma_i\in H_1(\partial U_i)$ supporting symmetry operators (red) which are bounded by a bulk 2-chain $\Sigma_{ijk\dots}$. Subfigure (2) shows cycles $\sigma_i\in H_2(U_i,\partial U_i)$ with boundaries compactified by the same 2-chain $\Sigma_{ijk\dots}$ to a compact curve in $X$.}
\label{fig:Comp}
\end{figure}

To address this question, note that in the compact geometry $X$ the localized 7D SYM sectors $\mathcal{T}_{\mathscr{X}_i}$
are coupled together by the embeddings $U_i\hookrightarrow X$. At the level of homology two effects occur. First of all, cycles in:
\be
\bigoplus_{i} H_2(\mathscr{X}_i,\partial \mathscr{X}_i)\cong \bigoplus_{i}  H_2(U_i,\partial U_i)\,,
\ee
which locally support defects, are compactified to curves by gluing in 2-chains of the bulk K3 surface. Second of all, cycles in
\be
\bigoplus_{i}  H_1(\partial \mathscr{X}_i)\cong \bigoplus_{i}  H_1(\partial U_i)
\ee
which locally support symmetry operators, are trivialized by 2-chains in the bulk K3 which bound the 1-cycles in the local model. For ADE singularities we have\footnote{It is important to note that here we are stating the relative homology group for the singular space $\mathscr{X}_{i}$ as opposed to its resolution $\mathscr{X}_{i}^{\prime}$. Because of this, there is no need to ``quotient by the compact cycles''.} $H_2(\mathscr{X}_i,\partial \mathscr{X}_i)\cong H_1(\partial \mathscr{X}_i)$ and the relevant 2-chains are identical for both effects, see figure \ref{fig:Comp}.

Recall that wrapping M2-/M5-branes on $H_2(\mathscr{X}_i,\partial \mathscr{X}_i)$ give heavy defects, while wrapping branes over cycles in $H_1(\partial \mathscr{X}_i)$ give topological symmetry operators (which act on the heavy defects). Completion in a global model $X$ determines which collection of defects combine to dynamical objects in $\mathcal{S}_X$ while trivialization effects determine which symmetries are gauged in $\mathcal{S}_X$. Symmetries not gauged are broken, as will follow from a canonical (linking / intersection-theoretic) pairing in geometry.

To track these two effects, we introduce the union of local patches and their complement
\be
\Xl=\cup_{i\!\;} U_i\,, \qquad X^\circ=X\setminus \Xl
\ee
where $i$ runs over all ADE singularities. Consequently the bulk geometry away from these patches is smooth and does not contain singular strata.

Coupling the $\mathcal{T}_{i}$ is now determined by how $\Xl$ glues to $X^\circ$. Therefore, we next consider the covering $X=\Xl\cup X^\circ$ for the K3 surface and its associated Mayer-Vietoris sequence. This long exact sequence contains the exact subsequence
\be \label{eq:MVS}\ba
0~&\xrightarrow[\text{}]{} ~H_2(X^\circ )  ~\xrightarrow[\text{}]{\;\jmath_2\;}  ~H_2(X) ~\xrightarrow[\text{}]{\;\partial_2\;}  ~   H_1(\partial \Xl)\cong \oplus_i \,H_1(\partial U_i) ~\xrightarrow[\text{}]{\;\imath_1\;} ~H_1(X^\circ)  ~\xrightarrow[\text{}]{}  ~0
\ea \ee
cut out by $H_2(\partial U_i)=H_1(X)=0$. Recall that in each local patch, we have an ADE singularity, a corresponding simply connected ADE Lie group $\widetilde{G}_{i}$. This serves to specify a ``naive'' local contribution to the gauge group, which we write as:
\begin{equation}
\widetilde{G}_{\mathrm{loc}} = \underset{i}{\prod} \widetilde{G}_i.
\end{equation}
Our task will be to determine the overall global gauge group, which will include $U(1)$ factors, as well as quotients by finite subgroups, as specified by the Mayer-Vietoris sequence. The interpretation of each term in this sequence is shown in figure \ref{Fig:Summary}.

\begin{figure}
\centering
\scalebox{0.8}{
\begin{tikzpicture}
	\begin{pgfonlayer}{nodelayer}
		\node [style=none] (0) at (0, 0) {\Large $0~\rightarrow~ H_2( X^\circ) ~\rightarrow~ H_2(X) ~\rightarrow~ H_1(\partial X^{\textnormal{loc}}) ~\rightarrow~ H_1( X^\circ) ~\rightarrow~0$~~~};
		\node [style=none] (1) at (-5.25, 0.5) {};
		\node [style=none] (2) at (-5.25, 1.5) {};
		\node [style=none] (3) at (-2.25, 0.5) {};
		\node [style=none] (4) at (-2.25, 1.5) {};
		\node [style=none] (5) at (1.25, 0.5) {};
		\node [style=none] (6) at (1.25, 1.5) {};
		\node [style=none] (7) at (5, 0.5) {};
		\node [style=none] (8) at (5, 1.5) {};
		\node [style=none] (9) at (-5.25, 2) {Extra $U(1)$'s};
		\node [style=none] (10) at (1, 2.5) {Local Model Defect,};
		\node [style=none] (11) at (1, 2) {Symmetry Operators};
		\node [style=none] (12) at (-2.25, 2.05) {Matter};
		\node [style=none] (13) at (5.25, 2.525) {Broken\,/\,Emergent};
		\node [style=none] (14) at (5.25, 2.025) {Global Symmetry};
		\node [style=none] (15) at (-3.625, -1.75) {};
		\node [style=none] (16) at (-3.625, -0.75) {};
		\node [style=none] (17) at (-3.675, -2.25) {$U(1)$ Charge };
		\node [style=none] (18) at (-3.675, -2.75) {Normalization};
		\node [style=none] (19) at (-0.625, -1.75) {};
		\node [style=none] (20) at (-0.625, -0.75) {};
		\node [style=none] (25) at (-0.25, -2.25) {Gauged\,/\,Trivialized};
		\node [style=none] (26) at (-0.25, -2.8) {Global Symmetry};
        \node [style=none] (26) at (-0.625, -3.25) {};
		\node [style=none] (27) at (-2.25, 2.55) {Massive};
		\node [style=none] (28) at (-1, -4.25) {};
  \node [style=none] (29) at (3.5, -2.215) {Bulk Relations};
    \node [style=none] (30) at (3.5, -2.75) {and Redundancies};
		\node [style=none] (31) at (3.3, -3.2) {};
  \node [style=none] (32) at (3.25, -1.75) {};
		\node [style=none] (33) at (3.25, -0.75) {};
	\end{pgfonlayer}
	\begin{pgfonlayer}{edgelayer}
		\draw [style=DottedLine] (2.center) to (1.center);
		\draw [style=DottedLine] (4.center) to (3.center);
		\draw [style=DottedLine] (6.center) to (5.center);
		\draw [style=DottedLine] (8.center) to (7.center);
		\draw [style=DottedLine] (16.center) to (15.center);
		\draw [style=DottedLine] (20.center) to (19.center);
        \draw [style=DottedLine] (32.center) to (33.center);
	\end{pgfonlayer}
\end{tikzpicture}
}
\caption{Simplified interpretation of the key long exact Mayer-Vietoris subsequences for a 7D M-theory vacuum from a K3 surface $X$. The distinction between broken and emergent global symmetries is a question of perspective: $H_1(X^\circ)$ characterizes symmetries broken by massive states in the UV and consequently gives the global symmetries emerging in the IR. The distinction of gauged\,/\,trivialized symmetries depends on whether an element in the image is mapped onto by the torsional\,/\,free elements respectively.}
\label{Fig:Summary}
\end{figure}

Our plan in the remainder of this subsection will be to elaborate on the interpretation of each of these contributions. In what follows, $G_{\mathrm{full}}$ refers to the full gauge group of the 7D supergravity model; we shall often separately analyze the non-abelian and abelian factors. We now proceed to analyze how the various local models glue into the bulk geometry:
\begin{itemize}

\item $H_1(\partial X^{\mathrm{loc}}) \cong \oplus_i \,H_1(\partial U_i)$: This homology group tracks the ``naive'' center of the simply connected Lie group:
    \be
    \widetilde{G}_{\mathrm{loc}} \equiv \underset{i}{\prod} \widetilde{G}_i\,.
    \ee
    This is because $\partial X^{\mathrm{loc}}$ is the disjoint union of lens spaces around all of the ADE singularities in $X$ and $H_1$ for a given lens space labelled by $i$ is equal to $Z(\widetilde{G}_i)$. As such, we have:
    \begin{equation}
    H_1(\partial X^{\mathrm{loc}}) \cong \oplus_i \,H_1(\partial U_i) \cong \oplus_i \,Z(\widetilde{G}_i)= Z(\widetilde{G}_{\mathrm{loc}})
    \end{equation}

\item $ \iota_1:H_1(\partial \Xl)\rightarrow H_1(X^\circ)$ is the lift of the embedding $\iota:\partial \Xl\hookrightarrow X^\circ$ to homology. Its kernel determines which collection of 1-cycles in the boundaries of individual local models trivialize when embedded into the bulk $X^\circ$. The associated collection of symmetry operators therefore acts trivially on the matter constructed from M2-branes wrapped on cycles in $H_2(X)$. We find
   \be\label{eq:globalform}
\mathcal{C}_{\mathrm{full}}\cong \textnormal{ker}\,\iota_1 \cong \textnormal{Im}\,\partial_2\,,
\ee
where $\mathcal{C}_{\mathrm{full}}$ refers to a center subgroup of $\widetilde{G}_{\mathrm{full}} = \widetilde{G}_{\mathrm{loc}} \times U(1)^{b}$ which we will quotient out. Combining with the map $\partial_2:H_2(X)\rightarrow H_1(\Xl)$, we later show that this quotienting center sits in the short exact sequence:
\begin{equation}
0~\rightarrow ~ \mathcal{C}_{\mathrm{loc}}  ~\rightarrow ~ \mathcal{C}_{\mathrm{full}} ~\rightarrow ~  \mathcal{C}_\textnormal{Extra} ~\rightarrow ~ 0\,.
\end{equation}

\item $H_1(X^\circ) $ is isomorphic to $\mathrm{Im}\; \iota_1$, and thus $\mathcal{C}_{\mathrm{loc}}$, because $\iota_1$ is surjective. As alluded to in the item above, $H_1(X^\circ)$ should also be equal to the center of the non-Abelian part of the gauge group $G_{\textnormal{full}}$. This follows from the duality relation:
\be\label{eq:TorRelation}
\big(\textnormal{Tor}\, H_2(X)\big)^\vee \cong \Tor\,H_1(X^\circ),
\ee
which we derive later. This implies that the matter obtained from $\textnormal{Tor}\, H_2(X)$ is acted on by the center subgroup isomorphic to $\Tor\,H_1(X^\circ)$ and therefore broken upon compactification.

\item $H_2(X^\circ )$ determines Abelian gauge symmetries. This follows because M2-brane states that wrap vanishing cycles are necessarily uncharged under the Abelian gauge symmetries appearing in \eqref{eq:algebra}. Otherwise, such $\mathfrak{u}(1)$'s would participate in gauge symmetry enhancement and not split off as a direct summand. The associated cycles therefore can be deformed away from the singular locus and are contained purely in $X^\circ$. Further, we have $\textnormal{rank}\,H_2(X^\circ )=b_2$, where $b_2\equiv b_2(X)$ is the second Betti-number of the singular K3 surface $X$. This follows from $H_1(\partial \Xl)$ being pure torsion. The free part is therefore associated with so-called ``extra'' $\mathfrak{u}(1)$ factors. In \eqref{eq:algebra} we therefore have $b=b_2$. In principle $H_2(X^\circ )$ could contain a torsion subgroup which would parameterize extra discrete Abelian gauge symmetries. However, for K3 surfaces $\textnormal{Tor}\,H_2(X^\circ)=0$.

\item $ j_2:H_2(X^\circ)\rightarrow H_2(X)$ is the lift of the embedding $j_2:X^\circ \hookrightarrow X$ to homology and its cokernel determines the normalization for Abelian gauge symmetry generators. Given $H_2(X^\circ)\cong \Z^{b_2}$ and
\be
\textnormal{coker}\,j_2 / \textnormal{Tor}\,H_2(X)\cong  \Z_{n_1}\oplus \dots \oplus \Z_{n_{b_2}}
\ee
then the correct normalization for the $k$-th $U(1)$ factor is $1/n_k$.
We argue this in section \ref{sec:Ab} from the viewpoint of the resolved geometry and give a characterization for elliptically fibered K3 surfaces (following Shioda's treatment) in section \ref{sec:EllipticK3}.

\item $H_2(X)$ determines the completion of the spectrum by positive mass\,/\,tensions states obtained from M2- and M5-brane wrappings. $H_2(X)$ receives no contributions from vanishing cycles. The homology group contains both curves in the bulk $X^\circ$ and curves stretching between multiple singularities. Restricting curves of $H_2(X)$ to $\Xl$ results in a collection of defects. See Figure \ref{fig:Comp}.

\item $ \partial_2:H_2(X)\rightarrow H_1(\Xl)$ is the boundary map and determines the charges of the electric and magnetic states, discussed in the previous bullet point, some of which are charged under $U(1)$ factors and some are not. To understand this, consider the decomposition into naive Abelian and non-Abelian parts
\be
\widetilde{G}_{\textnormal{full}}\equiv\widetilde{G}_{\mathrm{loc}} \times U(1)^{b_2}=\prod_i \widetilde{G}_i \times U(1)^{b_2}\,.
\ee
We then further split the image $\textnormal{Im}\,\partial_2$ into torsional and free parts
\be
\mathcal{C}_{\mathrm{loc}}\cong \textnormal{Im}\lb \partial_2|_{\textnormal{Tor}\,H_2(X)} \rb \cong \mathrm{Tor}\; H_2(X)\,, \qquad \mathcal{C}_{\textnormal{Extra}}\cong \textnormal{Im}\lb \partial_2|_{\textnormal{Free}\,H_2(X)} \rb\,.
\ee
Here the free part of $H_2(X)$ is defined as $\textnormal{Free}\,H_2(X)=H_2(X)/\textnormal{Tor}\,H_2(X)$.
Abelian symmetries are only associated with the free part, and the group $\textnormal{Im}\lb \partial_2|_{\textnormal{Free}\,H_2(X)} \rb$ carries the data of how $U(1)$ charges are correlated with $Z(\widetilde{G}_{\mathrm{loc}})$, or more precisely, $Z(\widetilde{G}_{\mathrm{loc}}/\mathcal{C}_{\mathrm{loc}})$ charges. We therefore have that
\be\label{eq:globalform2}
G_{\textnormal{full}}=\frac{\widetilde{G}_{{\mathrm{loc}}}/\mathcal{C}_{\mathrm{loc}} \times U(1)^{b_2}}{\mathcal{C}_{\textnormal{Extra}}}\,
\ee
which is related to the form of $G_{\textnormal{full}} = \widetilde{G}_{\mathrm{full}} / \mathcal{C}_{\mathrm{full}}$
via the short exact sequence:
\be
0~\rightarrow ~ \mathcal{C}_{\mathrm{loc}}  ~\rightarrow ~ \mathcal{C}_{\mathrm{full}} ~\rightarrow ~  \mathcal{C}_\textnormal{Extra} ~\rightarrow ~ 0\,.
\ee
We discuss this further in section \ref{sec:NonAb} and \ref{sec:Ab} from the view point of the resolved geometry. We give an equivalent, more combinatorial, point of view in Appendices \ref{app:AdditionalCyclicStuff} and \ref{app:morequotients} where we do not require resolutions. Both make completely explicit how $\textnormal{Im}\,\partial_2$ is embedded into the center of $\widetilde{G}_{\mathrm{full}}$.
\end{itemize}

We now expand on a few of the items listed above. In particular, we establish (\ref{eq:TorRelation}),
i.e., that $\big(\textnormal{Tor}\, H_2(X)\big)^\vee \cong \Tor\,H_1(X^\circ)$. This will prove crucial
in establishing that all global symmetries of the local models are either gauged or broken.

Along these lines, denote the location of the ADE singularities by $x_i \in X$, and then start with the identification
\be\label{eq:startingpoint}
H_2(X)\cong H_2(X, \lbbb x_i \rbbb)
\ee
which follows from the long exact sequence in relative homology for the pair $(X,  \lbbb x_i \rbbb)$. The collection of neighborhoods $\Xl$ deformation retract to $ \lbbb x_i \rbbb$. We therefore have
\be
H_2(X, \lbbb x_i \rbbb)\cong  H_2(X,\Xl)
\ee
via a deformation equivalence. By excision and Poincar\'e-Lefchetz duality we then have
\be
 H_2(X,\Xl)\cong H_2(X^\circ, \partial  X^\circ)\cong H^2(X^\circ)\,.
\ee
Equation \eqref{eq:TorRelation} then follows from the universal coefficient theorem. In particular there is a canonical linking pairing
\be
\textnormal{Tor}\, H_2(X) \times  \textnormal{Tor}\, H_1(X^\circ) ~\rightarrow~\mathbb{Q}/\mathbb{Z}\,.
\ee
This linking is defined much as the standard linking form on smooth spaces. Observe that one of the groups makes reference to $X^\circ$ and intersection computations are therefore always away from the singular locus.

Let us further comment that we are evaluating the sequence \eqref{eq:MVS} on a singular K3 surface $X$, and do not, in principle, require a crepant resolution for computation. Also, the sequence and the above analysis relies purely on singular homology and applies regardless of additional structure such as an elliptic fibration.

Additionally, one may wonder whether ordinary integer-valued homology is the correct generalized homology theory of conserved M-theory charges that one should consider, similar to how K-theory is a better approximation of Type IIA charges than ordinary cohomology. This subtlety can be mostly ignored for K3 compactifications due to the fact that even for singular K3's, the homology is entirely concentrated in even degrees. This implies that any Atiyah-Hirzebruch spectral sequence with respect to some generalized homology theory $\mathbb{E}_*$ collapses at the second page
\begin{equation}\label{eq:ahspecseq}
  E_{p,q}^2=H_p(X,\mathbb{E}_q(\mathrm{pt}))\implies \mathbb{E}_{p+q}(X)\,.
\end{equation}
It now follows from
\be
H_p(X,\mathbb{E}_q(\mathrm{pt}))\cong H_p(X,\mathbb{Z})\otimes_{\mathbb{Z}} \mathbb{E}_q(\mathrm{pt})
\ee
that such a putative generalized homology theory for M-theory charges on K3 is essentially captured by the group $H_p(X,\mathbb{Z})$ and we remain agnostic as to the possible constraints or additions introducing by tensoring with $\mathbb{E}_*$.\footnote{For a suggestion on what the dual $\mathbb{E}^*$ might be, see for example \cite{Sati:2021uhj,Fiorenza:2019ckz} which proposes that $\mathbb{E}^*$ is a twisted cohomotopy theory.}

Finally, recall that in a $D$-dimensional QFT, whenever a $n$-form symmetry is gauged, a dual quantum $m$-form symmetry emerges, $m=D-n-1$. For gauge theories, the gauging of a 1-form symmetry leads to a change in gauge group topology and is captured by quotients
\be \label{eq:gauginginFieldtheory}
\widetilde{G}_{\textnormal{loc}}/\mathcal{C}_\textnormal{loc}
\ee
which leads to a dual $(d-3)$-form symmetry isomorphic to $\mathcal{C}_{\mathrm{loc}}$. Further, there is an additional quotient by $\mathcal{C}_{\textnormal{Extra}}$ mixing Abelian and non-Abelian structures. In the present context where we have coupled to gravity, this quotient is not to be interpreted as a gauging but simply as an identification of redundancies. Consequently the quotient by $\mathcal{C}_{\textnormal{Extra}}$ does not give rise to an emergent dual symmetry.

At this point, let us compare our results with expectations from various quantum gravity / Swampland conjectures. As one might have anticipated, we find no global symmetries (see e.g., \cite{Banks:1988yz, Banks:2010zn, Harlow:2018tng, McNamara:2019rup}); all symmetries are either gauged or broken. From the perspective of the sequence \eqref{eq:MVS} the absence of 1-form and dual 4-form symmetries is a consequence of its exactness.

Let us demonstrate this explicitly by tracing through the field theory manipulations prescribed by the geometry. Our starting point are the local disjoint ADE geometries with their associated 7D SYM theories. We consider purely electric polarizations, then we have a 1-form symmetry group isomorphic to $\textnormal{Tor}\,H_1(\partial \Xl)$. By gluing in $X^\circ$ we add compact cycles, this introduces massive matter which breaks a 1-form symmetry subgroup isomorphic to Tor\,$H_1(X^\circ)$ explicitly. We further add additional abelian gauge symmetries which trivialize center symmetries by absorbing these as subgroups. With this the only center symmetries not accounted for are a isomorphic to Tor\,$H_2(X)$. These are gauged, giving dual magnetic 4-form symmetries, which are broken by M5-branes wrapped over classes in Tor\,$H_2(X)$.

Additionally, we observe that the completeness hypothesis (see e.g., \cite{Polchinski:1998rr, Banks:2010zn}) is satisfied, but in a rather subtle way; in passing from local to global models, we observe that a preferred polarization has been selected by the compact geometry $X$, and that there are suitable mutually local states which populate the relevant charge lattices. Finally, we observe that in these models the gauge coupling of the $U(1)$ factors is set by the volumes of curves which traverse the bulk geometry $X$. As such, decompactifying gravity also makes these $U(1)$ factors non-dynamical, in accordance with the weak gravity conjecture \cite{Arkani-Hamed:2006emk}.

\subsection{Resolutions to Smooth K3 Surfaces}
\label{sec:Resolved}

Given a singular K3 surface $X$ with ADE singularities, it is often convenient to perturb within the metric moduli space onto the smooth locus. In this case resolution divisors are small compared to all other curves of the geometry and we preserve the notion of local models and localized 7D supersymmetric Yang-Mills theory sectors, albeit on the Coulomb branch obtained from adjoint Higgsing. We denote such crepant resolutions of all ADE loci to smooth K3 surfaces by $\pi: X'\rightarrow X$. Throughout we use primes to denote data of the resolved geometry.

Many definitions carry over straightforwardly. Let us introduce the crepant resolutions of the local models $\pi_i\,: U_i' \rightarrow U_i$ where now $U_i'$ are modeled on the relevant ALE spaces \cite{Kronheimer}. Following the reasoning above we are again lead to consider the Mayer-Vietoris sequence for the smooth space $X'$ which is formulated with respect to the covering
\be
(\Xl)'=\cup_{i\!\;} U_i'\,, \qquad (X^\circ)' \equiv X'\setminus (\Xl)' \,.
\ee
Here, we remark that strictly speaking, there are no singularities in $X^\circ$ (we already excised them all), but by abuse of notation we continue to use a prime to denote the corresponding space obtained by deleting the local models (be they singular or resolved). The resolution preserves topology away from the singularities and we have
\be
H_n(\partial (\Xl)') =H_n(\partial \Xl)\,, \qquad H_n((X^\circ)')=H_n( X^\circ)\,.
\ee
With this we find the Mayer-Vietors sequence for the covering $X'=(\Xl)'\cup (X^\circ)'$ to contain the exact subsequence:\footnote{Here we use the fact that $H_2(\partial U_i;\Z)=0$ and K3 surfaces are simply connected.}
\be \label{eq:MVS2}\ba
0~&\xrightarrow[\text{}]{} ~H_2( X^\circ)\oplus L_E  ~\xrightarrow[\text{}]{\;\jmath_2'\;}  ~\Gamma_{3,19}~\xrightarrow[\text{}]{\;\partial_2'\;}  ~  H_1(\partial \Xl )= \oplus_{i\,} H_1(\partial U_i')  ~\xrightarrow[\text{}]{\;\imath_1\;} ~H_1( X^\circ)  ~\xrightarrow[\text{}]{}  ~0\,.
\ea \ee
Here $L_E$ denotes the lattice of exceptional curves contracted by the blowdown\,/\,projection $\pi: X^{\prime} \rightarrow X$ and $\Gamma_{3,19}$ is the K3 lattice of signature (3,19). As a point of notation, later on we shall need to make reference to $\overline{L}_E$, the minimal primitive sublattice of $H_2(X')\cong \Gamma_{3,19}$ containing $L_E$, that is, its saturation.

We now proceed to extract the global form of the non-Abelian gauge group by piecing together the local data of each singularity. We then show how to include contributions from ``delocalized'' $U(1)$ factors.

\subsubsection{Non-Abelian Gauge Symmetry}
\label{sec:NonAb}

\begin{figure}
\centering
\scalebox{0.85}{
\begin{tikzpicture}
	\begin{pgfonlayer}{nodelayer}
		\node [style=EmptyCircle] (0) at (-1, 0) {};
		\node [style=EmptyCircle] (1) at (0, 0) {};
		\node [style=EmptyCircle] (2) at (1, 0) {};
		\node [style=EmptyCircle] (3) at (3, 0) {};
		\node [style=EmptyCircle] (4) at (-3, 0) {};
		\node [style=EmptyCircle] (5) at (-3, -3) {};
		\node [style=EmptyCircle] (8) at (0.75, -3) {};
		\node [style=EmptyCircle] (9) at (1.75, -3) {};
		\node [style=EmptyCircle] (10) at (2.5, -2.25) {};
		\node [style=EmptyCircle] (11) at (2.5, -3.75) {};
		\node [style=EmptyCircle] (18) at (0, -8) {};
		\node [style=EmptyCircle] (19) at (0, -9) {};
		\node [style=EmptyCircle] (20) at (1, -9) {};
		\node [style=EmptyCircle] (21) at (2, -9) {};
		\node [style=EmptyCircle] (22) at (-1, -9) {};
		\node [style=EmptyCircle] (23) at (-2, -9) {};
		\node [style=EmptyCircle] (24) at (0, -11) {};
		\node [style=EmptyCircle] (25) at (0, -12) {};
		\node [style=EmptyCircle] (26) at (1, -12) {};
		\node [style=EmptyCircle] (27) at (2, -12) {};
		\node [style=EmptyCircle] (29) at (-1, -12) {};
		\node [style=EmptyCircle] (30) at (-2, -12) {};
		\node [style=EmptyCircle] (31) at (-3, -12) {};
		\node [style=none] (40) at (-5.5, 0) {$A_{n}\,:$};
		\node [style=none] (41) at (-5.5, -3) {$D_{2n}\,:$};
		\node [style=none] (42) at (-5.5, -6) {$D_{2n+1}\,:$};
		\node [style=none] (43) at (-5.5, -9) {$E_6\,:$};
		\node [style=none] (44) at (-5.5, -12) {$E_7\,:$};
		\node [style=none] (46) at (-2.25, 0) {};
		\node [style=none] (47) at (-1.75, 0) {};
		\node [style=none] (48) at (1.75, 0) {};
		\node [style=none] (49) at (2.25, 0) {};
		\node [style=none] (50) at (-2.25, -3) {};
		\node [style=none] (51) at (0, -3) {};
		\node [style=none] (54) at (-3, -0.5) {\footnotesize $1$};
		\node [style=none] (55) at (3, -0.5) {\footnotesize$n$};
		\node [style=none] (56) at (-1, -0.5) {\footnotesize$k-1$};
		\node [style=none] (57) at (0, -0.5) {\footnotesize$k$};
		\node [style=none] (58) at (1, -0.5) {\footnotesize$k+1$};
		\node [style=none] (59) at (-3, -3.5) {\footnotesize$1$};
		\node [style=none] (60) at (0, -3) {};
		\node [style=EmptyCircle] (61) at (-3, -6) {};
		\node [style=EmptyCircle] (62) at (0.75, -6) {};
		\node [style=EmptyCircle] (63) at (1.75, -6) {};
		\node [style=EmptyCircle] (64) at (2.5, -5.25) {};
		\node [style=EmptyCircle] (65) at (2.5, -6.75) {};
		\node [style=none] (66) at (-2.25, -6) {};
		\node [style=none] (67) at (0, -6) {};
		\node [style=none] (68) at (-3, -6.5) {\footnotesize$1$};
		\node [style=none] (69) at (0, -6) {};
		\node [style=none] (70) at (3.25, -2.25) {\footnotesize$\,2n-1$};
		\node [style=none] (71) at (3, -3.75) {\footnotesize$2n$};
		\node [style=none] (72) at (3, -5.25) {\footnotesize$2n$};
		\node [style=none] (73) at (3.25, -6.75) {\footnotesize$\,2n+1$};
		\node [style=none] (74) at (0.75, -3.5) {\footnotesize$2n-3$};
		\node [style=none] (75) at (0.75, -6.5) {\footnotesize$ 2n-2$};
		\node [style=none] (76) at (2.5, -6) {\footnotesize$\,2n-1$};
		\node [style=none] (77) at (2.5, -3) {\footnotesize$\,2n-2$};
		\node [style=none] (78) at (-2, -9.5) {\footnotesize$1$};
		\node [style=none] (79) at (-1, -9.5) {\footnotesize$2$};
		\node [style=none] (80) at (0, -9.5) {\footnotesize$3$};
		\node [style=none] (81) at (1, -9.5) {\footnotesize$4$};
		\node [style=none] (82) at (2, -9.5) {\footnotesize$5$};
		\node [style=none] (83) at (0.5, -8) {\footnotesize$6$};
		\node [style=none] (84) at (-3, -12.5) {\footnotesize$1$};
		\node [style=none] (85) at (-2, -12.5) {\footnotesize$2$};
		\node [style=none] (86) at (-1, -12.5) {\footnotesize$3$};
		\node [style=none] (87) at (0, -12.5) {\footnotesize$4$};
		\node [style=none] (88) at (1, -12.5) {\footnotesize$5$};
		\node [style=none] (89) at (2, -12.5) {\footnotesize$6$};
		\node [style=none] (90) at (0.5, -11) {\footnotesize$7$};
		\node [style=none] (91) at (7.48, 0) {$\T_{A_n}=\frac{1}{n}\sum_{k=1}^{n-1} ke_{k}$};
		\node [style=none] (92) at (8.175, -3.5) {$\T_{D_{2n}}^{(s)}=\frac{1}{2}e_{2n}+\frac{1}{2}\sum_{k=1}^{n-1} e_{2k-1}$};
		\node [style=none] (93) at (7.525, -2.5) {$\T_{D_{2n}}^{(c)}=\frac{1}{2}\sum_{k=1}^n e_{2k-1}$};
		\node [style=none] (94) at (8.85, -6) {$\T_{D_{2n+1}}=\frac{1}{4}e_{2n+1}+\frac{3}{4}e_{2n}+\frac{1}{2}\sum_{k=1}^{n} e_{2k-1}$};
		\node [style=none] (95) at (8.425, -9) {$\T_{E_6}=\frac{1}{3}e_{1}+\frac{2}{3}e_{2}+\frac{1}{3}e_{4}+\frac{2}{3}e_{5}$};
		\node [style=none] (96) at (7.875, -12) {$\T_{E_7}=\frac{1}{2}e_{1}+\frac{1}{2}e_{3}+\frac{1}{2}e_{7}$};
		\node [style=none] (97) at (3, -13.25) {};
	\end{pgfonlayer}
	\begin{pgfonlayer}{edgelayer}
		\draw [style=DottedLine] (46.center) to (47.center);
		\draw [style=DottedLine] (48.center) to (49.center);
		\draw [style=DottedLine] (50.center) to (51.center);
		\draw (4) to (46.center);
		\draw (47.center) to (0);
		\draw (0) to (1);
		\draw (1) to (2);
		\draw (2) to (48.center);
		\draw (49.center) to (3);
		\draw (5) to (50.center);
		\draw (8) to (9);
		\draw (9) to (10);
		\draw (9) to (11);
		\draw (18) to (19);
		\draw (19) to (22);
		\draw (22) to (23);
		\draw (19) to (20);
		\draw (20) to (21);
		\draw (24) to (25);
		\draw (25) to (26);
		\draw (26) to (27);
		\draw (29) to (25);
		\draw (30) to (29);
		\draw (31) to (30);
		\draw (60.center) to (8);
		\draw [style=DottedLine] (66.center) to (67.center);
		\draw (61) to (66.center);
		\draw (62) to (63);
		\draw (63) to (64);
		\draw (63) to (65);
		\draw (69.center) to (62);
	\end{pgfonlayer}
\end{tikzpicture}
}
\caption{We list the ADE Dynkin diagrams and their thimbles. We list compact representatives, dropping the subscript `$(c)$'. For $D_{2n}$ Dynkin diagrams, we have two thimbles corresponding to the spinor and cospinor representations. For the $E_8$ Dynkin diagram the thimbles are trivial and this case is therefore not considered. The $e_i$ are the exceptional curves introduced by the resolution of the ADE singularity to an ALE space.}
\label{fig:Thimbles}
\end{figure}
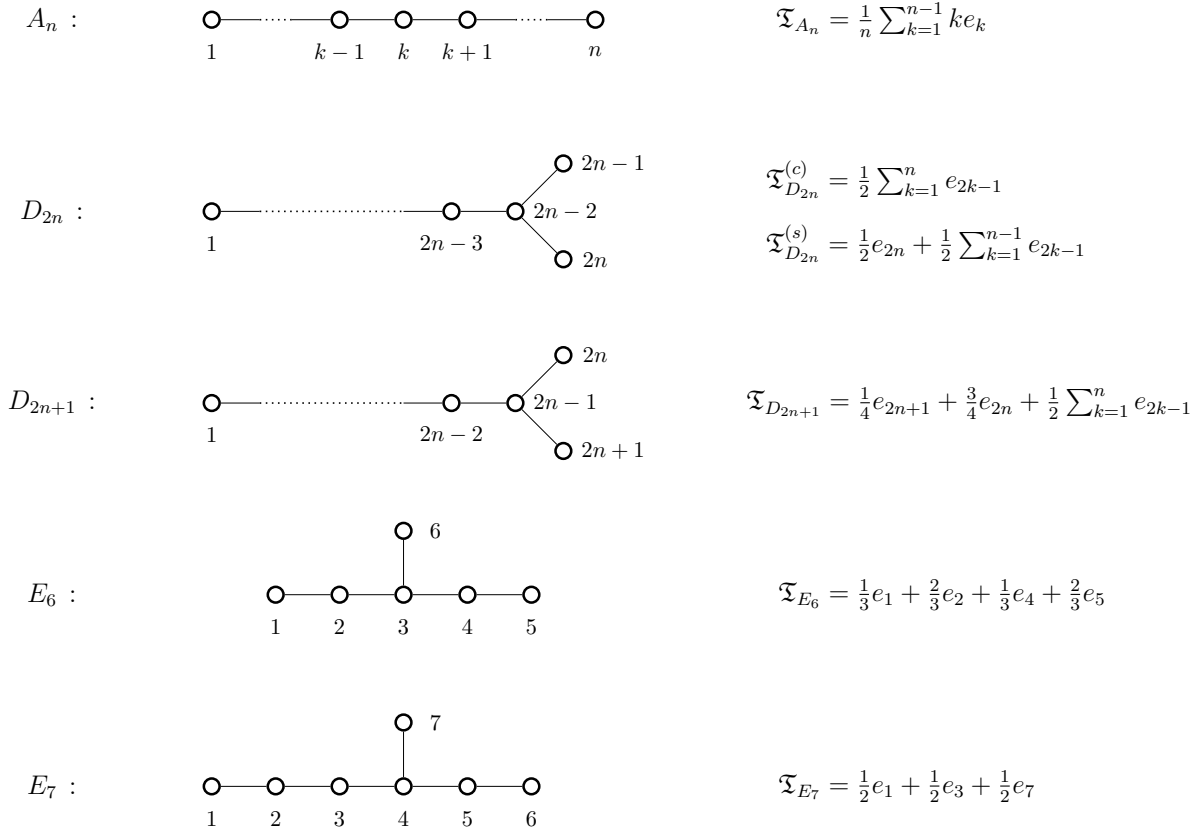

We now determine the global form of the non-Abelian gauge group:
\be\label{eq:GADE}
G_{\mathrm{loc}}= \widetilde{G}_{\mathrm{loc}} / \mathcal{C}_{\mathrm{loc}} \,\,\, \text{with \,\,\,} \widetilde{G}_{\mathrm{loc}}=\prod_i{ \widetilde{G}}_{i}\,,
\ee
where $\widetilde{G}_i$ and is the simply connected ADE group with algebra $ \mathfrak{g}_i$ and center subgroup $Z(\widetilde{G}_i)$. Here $\mathcal{C}_{\mathrm{loc}}$ is a central subgroup of $\Pi_i{\widetilde{G}}_{i}$ acting trivially on the spectrum.

In order to characterize $\mathcal{C}_{\mathrm{loc}}$ in terms of resolution data of $\pi: X' \rightarrow X$ we require the notion of a thimble and their compact representatives \cite{Hubner:2022kxr} which we briefly review now. The action of individual centers is described using the associated compact representatives and this allows us to solve for the group $\mathcal{C}_{\mathrm{loc}}$ with overall trivial action.
\medskip

\noindent \underline{\bf Thimbles}\,: The thimbles $\mathfrak{T}$ of an ADE singularity $\XADE=\C^2/\Gamma$, with associated Lie algebra $\mathfrak{g}$, are the relative 2-cycles generating $H_2(\XADE,\partial \XADE)$. Resolving the ADE singularity to an ALE space $\XADE'$, thimbles are identified with the generators of
\be \label{eq:ResolvedThimble}
\big\langle \Tp\big\rangle =H_2(\XADE',\partial \XADE')/H_2(\XADE')\,.
\ee
In the context of the smooth ALE space thimbles admit a presentation in terms of a rational linear combination of compact curves with coefficients in $\Q/\Z$, that is
\be\label{eq:ThimblesADE}
\Tp_{\!c}=\sum_{k=1}^r q^k e_{k}\,, \qquad q^k\in \Q/\Z
\ee
with exceptional curves $e_k$ associated with nodes in the Dynkin diagram of $\mathfrak{g}$ and intersection matrix given by the Cartan matrix of $\mathfrak{g}$, and $r=\textnormal{rank}\,\mathfrak{g}$. We list all compact representatives in figure \ref{fig:Thimbles}. The compact representatives \eqref{eq:ThimblesADE} have the following properties:
\begin{enumerate}
\item Intersections: Let $\Tp{}'_{\!\!c}$ denote a lift of $\Tp_{\!c}$ to some rational linear combination of exceptional curves such that $\Tp{}'_{\!\!c}=\Tp_{\!c}$ modulo 1. With respect to the ordinary intersection product we have $\Tp{}'_{\!\!c}\cdot C \in \Z$ for all curves $C\in H_2(\XADE')$. More precisely $w_k= \Tp{}'_{\!\!c}\cdot C_k$ belongs to the lattice generated by the weight vector for some specific representation. The relevant representations are the fundamental ($A$-type), spinor ($D_{\mathrm{odd}}$-type), spinor and cospinor ($D_{\mathrm{even}}$-type), ${\bf 27}$ ($E_6$-type), ${\bf 56}$ ($E_7$-type). In particular $\Tp{}'_{\!\!c}\cdot \Sigma =0$ modulo 1 for all curves of the ALE geometry $\mathscr{X}'$.

\item Linkings: Following point (1) thimbles $\Tp_{\!c}\in H_2(\XADE',\Q/\Z)$ have vanishing intersection with all compact curves modulo 1. Intersections modulo 1 between thimbles therefore give a well-defined pairing. This pairing is perfect and an extension of the linking pairing of the boundary $\partial \XADE'$ to the bulk $\XADE'$. Given two thimbles $\Tp_{\;\!\!1}, \Tp_{\;\!\!2}$ with boundaries $\gamma_1,\gamma_2$ and compact representatives $\Tp_{\!c,1},\Tp_{\!c,2}$ we have
\be
\ell(\gamma_1,\gamma_2)=\Tp_{\!c,1}\cdot\Tp_{\!c,2}
\ee
with linking form $\ell : \Tor\, H_1(\partial \XADE') \times \Tor\, H_1(\partial \XADE') \rightarrow \Q/\Z$ on $\partial \XADE'=S^3/\Gamma$.

\item Computation: Let $\Xi$ be the intersection matrix of the ALE space (i.e., minus the Cartan matrix of the Lie algebra $\mathfrak{g}$) and denote by $S,A,B$ three integral matrices such that $A \Xi B=S$ where $S$ is the Smith-normal form of $\Xi$. The matrix $S$ is diagonal with integral entries which are either 1 or the order of the simple factors of the center subgroup of the associated simply connected ADE Lie group. The columns of $A$ determine the expansion of thimbles in compact curves (see figure \ref{fig:Thimbles}).

\end{enumerate}

\noindent It is often convenient to consider local neighborhoods of ADE singularities which are not modeled on $\XADE=\C^2/\Gamma$ but rather on the elliptic fibration $\mathbb{E}\hookrightarrow \XADE_{\mathbb{E}} \rightarrow \C$ with a single Kodaira type singularity at a marked point of the base. We have an isomorphism of thimbles
\be
\Tor\, H_2(\XADE,\partial \XADE)\cong \Tor\, H_2(\XADE_{\mathbb{E}} ,\partial \XADE_{\mathbb{E}} )
\ee
whenever geometries engineer identical ADE gauge algebras. Thimbles of $\XADE_{\mathbb{E}} $ are referred to as Kodaira thimbles and constructed from vanishing cycles of the elliptic fibration modulo identifications due to monodromy (see e.g., \cite{Hubner:2022kxr, Cvetic:2022imb} for further discussion). Their compact representatives are identical to those given in figure \ref{fig:Thimbles}.\medskip

To see how the compact representatives of thimbles enter our previous discussion, we begin by making the homology groups of the singular geometry $X$ explicit using the resolved data. One immediate consequence of comparing \eqref{eq:MVS} with \eqref{eq:MVS2} is
\be\label{eq:TorsionMechanism}
\frac{H_2(X')}{L_E}\cong H_2(X)\,,
\ee
as this quotient is invariant under the contraction of the lattice $L_E$ and therefore computes the second homology groups of the singular geometry. In particular, denoting by $\overline{L}_E$ the minimal primitive sublattice of $H_2(X')\cong \Gamma_{3,19}$ containing $L_E$ (its saturation), we have:
\be\label{eq:Saturation}
\overline{L}_E/L_E = \textnormal{Tor}\,\overline{L}_E/L_E\cong \Tor\,H_2(X)\,.
\ee
Let $e_\alpha$ denote the exceptional curves of $\pi: X'\rightarrow X$ then we can give $\overline{L}_E$ as the collection of all rational linear combinations of exceptional curves contained in the integral K3 lattice
\be
\overline{L}_E=\lbbb^{\;\!} p^\alpha e_\alpha\in \Gamma_{3,19}\,|\, e_\alpha \in L_E\,, p^\alpha \in \Q \rbbb
\ee
from which we now derive the presentation
\be\label{eq:Lattice}
\Tor\,H_2(X)\cong \overline{L}_E \otimes \Q/\Z = \lbbb  q^\alpha e_\alpha \,|\, p^\alpha e_\alpha\in  \overline{L}_E\,, \,q^\alpha =\big[ p^\alpha\big] \in \Q/\Z \rbbb\,.
\ee
The non-trivial point is that \eqref{eq:Lattice} is generated by linear combinations of thimbles $\Tp_{\!c,i}$, i.e., only specific rational linear combinations (mod 1) occur in \eqref{eq:Lattice}. This is geometrically clear, namely the combinations occuring are such that the corresponding $\T_{i}$ glue as shown in the right subfigure of figure \ref{fig:Comp} to an element of $\Tor\,H_2(X)$.

Next we express via thimbles the center action of $\Pi_i \widetilde{G}_i$ whose center is the product of the centers $Z(\widetilde{G}_i)$ running over all ADE singularities. A state obtained from an M2-brane wrapped on the curve $C\in H_2(X')$ with restriction $\sigma_i=C\cap  U_i'$ is rotated by $Z(\widetilde{G}_i)$ with phase
\be\label{eq:Linking}
\exp\lbb 2 \pi i  \,\ell_{i}(\gamma_i, \partial \sigma_i ) \rbb
\ee
where $\ell_i$ is the linking pairing on the boundary $\partial U_i' = S^3/\Gamma_i$ where $\Gamma_i$ is the finite group folding the $i$-th ADE singularity. By property (2) of thimbles this phase is equal to
 \be\label{eq:Intersection}
\exp\big[ 2 \pi i\, \Tp_{\!c,i} \cdot C  \big]
 \ee
where $\Tp_{\!c,i}$ is the compact representative of the thimble $\Tp_{\;\!\!i}$ of the neighborhood $U_i$.

Compact representatives $\Tp_{\!c,i}$ have vanishing intersection with exceptional curves modulo 1 by property (1) of thimbles and therefore we see that elements of $\mathcal{C}_{\mathrm{loc}}$ are geometrically characterized as linear combinations of thimbles which have vanishing intersections with all curves in $ X'$ modulo 1.

With this characterization we now show that:
\be \label{eq:somecharacterization}
\mathcal{C}_{\mathrm{loc}}\cong \textnormal{Tor}\, H_2(X)
\ee
by showing that one side embeds into the other and vice versa. We begin with showing that $\textnormal{Tor}\,H_2(X)$ is a subgroup of $\mathcal{C}_{\mathrm{loc}}$. Consider the class $[\Sigma]\in H_2(X')/L_E\cong \textnormal{Tor}\,H_2(X)$, which is a sum of thimbles, and an arbitrary representative $\Sigma$ thereof, which is an integral element of the K3 lattice. By integrality $\Sigma \cdot C \in \Z $ for any representative and therefore $[\Sigma]\cdot C=0$ modulo 1 for any curve $C$ in the K3 lattice.

Conversely, consider a linear combination of thimbles with vanishing intersection for all curves $C$ modulo 1. Consider some lift of these thimbles from coefficients in $\mathbb{Q}/\mathbb{Z}$ to $\mathbb{Q}$, resulting in a linear combination of exceptional curves with rational coefficients. Inserting the latter into the intersection pairing we obtain a linear form on the K3 lattice mapping into $\Z$. By self-duality of the K3 lattice the inserted linear combination of exceptional curves with rational coefficients must have been an integral element of the K3 lattice. The initial collection of thimbles therefore describes an element in $\overline{L}_E/L_E\cong\textnormal{Tor}\, H_2(X)$.

Summarizing, we have now established that for the non-Abelian factors of the gauge group $G_{\mathrm{full}} = (G_{\mathrm{loc}} \times U(1)^{b_2}) / \mathcal{C}_{\mathrm{Extra}}$, the contribution from the localized ADE singularities takes the form:
\begin{equation}
G_{\mathrm{loc}} = \underset{i}{\prod} \widetilde{G}_i \,\Big/\, \mathcal{C}_{\mathrm{loc}},
\end{equation}
where $\mathcal{C}_{\mathrm{loc}}$ is given by:
\begin{equation}
\mathcal{C}_{\mathrm{loc}} = \mathrm{Tor} H_{2}(X) = \overline{L}_E / L_E.
\end{equation}

\subsubsection{Abelian Gauge Symmetries}
\label{sec:Ab}

We now determine the global form of the gauge group
\be
G_\textnormal{full}=(G_{\mathrm{loc}}\times U(1)^{b_2})/\mathcal{C}_{\textnormal{Extra}}\,,
\ee
via resolution data. Here $\mathcal{C}_{\textnormal{Extra}}$ is a central subgroup of ${G}_{\mathrm{loc}}\times U(1)^{b_2}$ acting trivially on the spectrum and we denote the Lie algebra of ${G}_{\ADE}$ by $\mathfrak{g}_{\ADE}$.

The generators of $ \mathfrak{u}(1)^{b_2}$ are associated in the singular space $X$ with equivalence classes of free 2-cycles generating the homology quotient
\be\label{eq:SingQuotient}
H_2(X)\big/\Tor\, H_2(X) \cong \Z^{b_2}\,.
\ee
Blowing up, we find by \eqref{eq:TorsionMechanism} and \eqref{eq:Saturation}, this quotient extends as
\be\label{eq:classes}
H_2(X')\big/\;\!\overline{L}_E\cong H_2(X)\big/\;\!\Tor\, H_2(X)\,.
\ee
In order to identify the generators of $ \mathfrak{u}(1)^{b_2}$ in the smooth geometry note that the contraction $X'\rightarrow X$ produces the gauge algebra $\mathfrak{g}_{\mathrm{loc}}\times \mathfrak{u}(1)^{b_2}$ only if the states enhancing gauge symmetry are uncharged under $\mathfrak{u}(1)^{b_2}$. These states result from M2-branes wrapped on the exceptional lattice $L_E$. Therefore, the supergravity 3-form expanded along an irreducible curve $\Pi$ generates an Abelian subalgebra of $\mathfrak{u}(1)^{b_2}$, denoted as $ \mathfrak{u}(1)_{\Pi}$, precisely when
\be \label{eq:u(1)}
\Pi \in L_E^\perp\, \equiv \{\ell \in H_{2}(X)\,\,\, \mathrm{such\, that}\,\,\, \ell \cdot L_{E} = 0\},
\ee
i.e., the collection of classes in $H_2(X)$ which do not intersect any of the exceptional curves obtained from blowups. So, consider two generators $\Pi, \Pi'\in  L_E^\perp$ equivalent in $H_2(X')\big/\;\!\overline{L}_E$. Then the difference $\Pi-\Pi'$ is an element of $\overline L_E \cap L_E^\perp$ and by the non-degeneracy of the pairing on $L_E$ we conclude $ \Pi=\Pi' $. In every class $[\Pi] \in H_2(X')\big/\;\!\overline{L}_E$ there thus exists a unique irreducible curve $\Pi \in [\Pi] \cap L_E^\perp$.

We can make the generators of $\mathfrak{u}(1)^{b_2}$ more explicit noting the tautology $H_2((X^\circ)')\subset L_E^\perp$. This follows simply from the fact that the exceptional curves are confined to the subset excised from $X'$ and therefore cannot intersect cycles in $(X^\circ)'$. However, generally we only have
\be
H_2((X^\circ)')\subset  L_E^\perp
\ee
and the failure of equality is measured by the cokernel of the map $\jmath_2'$ taken as a map into $H_2(X')/\overline{L}_E$, i.e., restricted such that the mapping is $\Z^{b_2}\rightarrow \Z^{b_2}$. The appropriate geometrization of the generators of $\mathfrak{u}(1)^{b_2}$ are therefore rescaled linear combinations of the generators of $H_2((X^\circ)')$. If the cokernel is isomorphic to
\be
\Z_{n_1}\oplus \dots \oplus \Z_{n_{b_2}}
\ee
then the appropriate rescaling is by $1/n_k$ for various generators, the resulting fractional cycles need not be contained in the K3 lattice. Given these rescaled generators we can now perform a further change of basis, if necessary, yielding fractional cycles which we denote by $\pi_k$ which do not intersect pairwise, only intersect the exceptional curves and generate $\mathfrak{u}(1)^{b_2}$. Here the index runs as $k=1,\dots,b_2$.

We constructed $\pi_k$ from $H_2(X^\circ)=H_2((X^\circ)')$ and can therefore consider it in both the smooth and singular geometry. First, consider $X'$ and note that there are no integral homology representatives for a $\pi_k$ in the K3 lattice, it is not a cycle in the smooth geometry, however $n_k$ copies thereof is. Second, consider $X$ and note that $\pi_k$ is a cycle, albeit with the feature that all its representatives contain orbifold points modelled on (real) surface singularities $\mathbb{C}/\Gamma_i$ with $i\in I_k$ for the index set $I_k$ running over all ADE singularities contained in $\pi_k$. These cycles are stuck at the ADE loci and can not be deformed away from these. In the singular geometry we have that restricting to local models gives a collection of thimbles
\be\label{eq:PiRestrict}
\pi_k|_{\Xl}=-\sum_{i\in I_k} n_i \mathfrak{T}_i\,.
\ee
Upon resolving $X'\rightarrow X$, it therefore follows that (with coefficients in $\mathbb{Q}/\mathbb{Z}$)
\be\label{eq:PiRestrict2}
 \bar \pi_k + \sum_{i\in I_k} n_i \Tp_{\!c,i}
 \ee
 has vanishing intersection with all curves of the K3 lattice. Here $ \bar \pi_k$ is the proper transform of $\pi_k$. This $\mathbb{Q}/\mathbb{Z}$ combination of cycles does not belong to the K3 lattice, however it nonetheless specifies a $U(1)$ rotation and an element in the center of $\Pi_i \widetilde{G}_i$ acting trivially on all states of the spectrum. This combination therefore belongs to $\mathcal{C}_{\textnormal{Extra}}$ and further we find overall
 \be\label{eq:normalizations}
 \mathcal{C}_{\textnormal{Extra}}\cong \textnormal{coker}\,\jmath_2'\,\Big/\,\overline{L}_E \cong  \textnormal{coker}\,\jmath_2\,\big/\,\textnormal{Tor}\,H_2(X)\,.
 \ee

\subsection{Decoupling Gravity and Emergent Global Symmetries}\label{ssec:Emergent}

Having embedded various QFT sectors in a model coupled to 7D supergravity,
it is instructive to consider the opposite limit where we now
decouple gravity, namely we send the 7D Newton's constant $G_{N} \sim \mathrm{Vol}(X)^{-1} \rightarrow 0$. In this limit, we observe that the various bulk $U(1)$ symmetries obtained from reduction of the 3-form potential on compact 2-cycles also become non-dynamical, and as such can formally be identified with ``global symmetries''. More precisely, these $U(1)$ factors are delocalized across all of $X$, and as such they can be viewed as free vector multiplets.

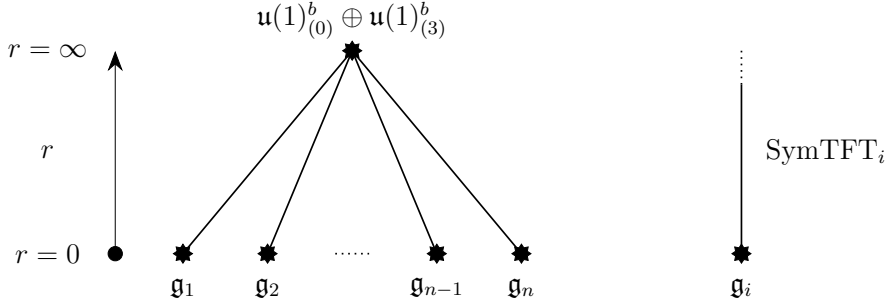
\begin{figure}
\centering
\scalebox{0.9}{
\begin{tikzpicture}
	\begin{pgfonlayer}{nodelayer}
        \node [style=none] (0) at (0.5, 2) {};
		\node [style=none] (3) at (0.5, 2.45) {$\mathfrak{u}(1)^b_{(0)}\oplus \mathfrak{u}(1)^b_{(3)}$};
		\node [style=none] (4) at (-2, -1) {};
		\node [style=none] (5) at (-0.75, -1) {};
		\node [style=none] (6) at (0.25, -1) {};
		\node [style=none] (7) at (0.75, -1) {};
		\node [style=none] (8) at (1.75, -1) {};
		\node [style=none] (9) at (3, -1) {};
		\node [style=none] (10) at (-3, -1) {};
		\node [style=none] (11) at (-3, 2) {};
		\node [style=none] (12) at (-4, 0.5) {$r$};
		\node [style=none] (13) at (-4, -1) {$r=0$};
		\node [style=none] (14) at (-4, 2) {$r=\infty$};
		\node [style=Star] (15) at (-2, -1) {};
		\node [style=Star] (16) at (-0.75, -1) {};
		\node [style=Star] (17) at (1.75, -1) {};
		\node [style=Star] (18) at (3, -1) {};
		\node [style=none] (19) at (-2, -1.5) {$\mathfrak{g}_1$};
		\node [style=none] (20) at (-0.75, -1.5) {$\mathfrak{g}_2$};
		\node [style=none] (21) at (1.75, -1.5) {$\mathfrak{g}_{n-1}$};
		\node [style=none] (22) at (3, -1.5) {$\mathfrak{g}_n$};
		\node [style=none] (23) at (6.25, -1) {};
		\node [style=none] (24) at (6.25, 1.5) {};
		\node [style=none] (25) at (6.25, 2) {};
		\node [style=Star] (26) at (6.25, -1) {};
		\node [style=none] (27) at (6.25, -1.5) {$\mathfrak{g}_i$};
		\node [style=none] (28) at (7.5, 0.5) {SymTFT$_i$};
        \node [style=none] (29) at (1, -2.25) {};
        \node [style=Circle] (30) at (-3, -1) {};
        \node [style=Star] (31) at (0.5, 2) {};
	\end{pgfonlayer}
	\begin{pgfonlayer}{edgelayer}
		\draw [style=ThickLine] (4.center) to (0.center);
		\draw [style=ThickLine] (0.center) to (5.center);
		\draw [style=ThickLine] (0.center) to (8.center);
		\draw [style=ThickLine] (0.center) to (9.center);
		\draw [style=DottedLine] (6.center) to (7.center);
		\draw [style=ArrowLineRight] (10.center) to (11.center);
		\draw [style=ThickLine] (24.center) to (23.center);
		\draw [style=DottedLine] (25.center) to (24.center);
	\end{pgfonlayer}
\end{tikzpicture}
}
\caption{The IR symmetry TFT is a junction of the symmetry TFTs of the local models with partially non-topological boundary conditions at the junction. The boundary conditions are therefore physical boundary conditions associated with the 7D SYM degrees of freedom localized at the ADE singularities and labled by Lie algebras $\mathfrak{g}_i$ and partially physical boundary conditions determined by the Abelian bulk physics of $X^\circ$. On the right we give our sketch for the symmetry TFT associated with a connected component of $\Xl$.}
\label{fig:SymFT}
\end{figure}

In addition to these delocalized contributions, we of course also observe the appearance of various local singularities, as obtained from the individual $\mathscr{X}_i$ local models. Observe that for each of these local geometries, we have a boundary $\partial \mathscr{X}_i$, and our analysis of gluing into the bulk has implicitly specified a bulk profile for the 11D fields as they reach this boundary. In particular, the specification of these boundary conditions means that the decoupling limit has already singled out a ``preferred'' polarization for the defect group, and a global form for the gauge group.

For each such local model there is a corresponding symmetry topological field theory (SymTFT) which captures the various generalized symmetries and their anomalies (see e.g., \cite{Freed:2012bs, Kaidi:2022cpf, Freed:2022qnc}). In the context of a string construction, one can start from the topological terms of a higher-dimensional theory and perform a reduction on the boundary geometry \cite{Apruzzi:2021nmk}.

In the present case where we have multiple local sectors, we can again straightforwardly determine the symmetry topological field theory (TFT) for the localized sectors from the local models $\Xl$ following the procedure laid out in \cite{Apruzzi:2021nmk}. This results in a disjoint union of symmetry TFTs ordinarily associated to 7D SYM theories. The 8D TFT action has a term of the schematic form $B_2 \wedge dC_{5}$, which controls the braiding statistics for wrapped M2- and M5-branes. There are additional cubic terms which also mix with a continuous $U(1)$ 2-form symmetry associated with the instanton density (as obtained from reduction of the 11D Chern-Simons term). We refer the interested reader to \cite{Apruzzi:2021nmk} for additional discussion.

Observer, however, that when completing $\Xl$ to the global model $X$ any obvious notion of a symmetry TFT breaks down. That being said, we can also consider an intermediate limit where we retain some of the data of $X$ as obtained from gluing all the $\mathscr{X}_i$ to a common boundary associated with $X^{\circ}$.

The first step lies in identifying the asymptotic boundary of $\Xl$. Initially these are a collection of disjoint lens spaces, however, to capture the IR emergent symmetries we should view $\Xl$ as a subset of $X$. We can enlarge $\Xl$ until it gives a partition of $X$ and now the lens spaces partially overlap and are shared between the different components of $\Xl$. The asymptotic boundary is thus more appropriately viewed to be the 3D deformation retract of $X^\circ$, or equivalently (at the level of topological structures), as simply $X^\circ$.

The bulk geometry $X^\circ$ supports the extra Abelian $U(1)$ factors and therefore constitutes a partially non-topological boundary condition for the symmetry TFT formulated with respect to $\Xl$. In particular, this means that our path integral will also include a sum over possible configurations for these $U(1)$ fields. Observe that in passing from the global to the local model, the map $H_2(X) \rightarrow H_{1}(X^{\mathrm{loc}})$ of line \eqref{eq:MVS} specifies how these $U(1)$ factors will descend to torsional elements associated with the defect group of a single local singularity.

Now, as $\Xl$ is disjoint, the 8D bulk data of the latter is simply the direct sum of TFTs for each localized sector (see figure \ref{fig:SymFT}). However, because all components of $\partial \Xl$ embed into $X^\circ$ these disjoint unions share a common 7D boundary at ``infinity'' and overall we are left with a junction of symmetry TFTs.

\section{M-theory on Specific K3 Surfaces: Illustrative Examples} \label{sec:7dExamp}

To illustrate some of the general considerations just presented, we now turn to some illustrative examples for M-theory on a K3 surface. Indeed, by tuning the hyperkahler moduli of the K3 surface, we can reach examples with various localized singularities. It is then instructive to ask how specific local models embed in global examples. We begin with a treatment of $T^{4} / \mathbb{Z}_2$, and then briefly turn to further examples of quotients of $T^4$; additional details on these examples are deferred to the Appendices. As another class of examples, we also consider elliptically fibered K3 surfaces. In these cases we can also consider F-theory on the same background, which results in an 8D vacuum.

\subsection{$X = T^4 / \mathbb{Z}_2$} \label{sec:Kummer}

To illustrate these general considerations, we now treat in detail the case $X = T^{4} / \mathbb{Z}_2$. Resolving the local $A_1$ singularities of this geometry produces the Kummer surface. The involution generating $\Z_2$ acts on each of the four circles of
\be\label{eq:MasterEquation}
T^4=S^1_1\times S^1_2 \times S^1_3 \times S^1_4
\ee
with two fixed points by reflection along an axis. In total there are therefore $2^4=16$ fixed points which in $X$ project to an $A_1^{16}$ ADE singularity.

\begin{figure}
\centering
\scalebox{0.75}{
\begin{tikzpicture}
	\begin{pgfonlayer}{nodelayer}
		\node [style=none] (0) at (0, 1.75) {};
		\node [style=none] (1) at (-3.25, 0) {};
		\node [style=none] (2) at (0, -1.75) {};
		\node [style=none] (3) at (3.25, 0) {};
		\node [style=NodeCross] (4) at (-1, 0.75) {};
		\node [style=NodeCross] (5) at (1, -0.75) {};
		\node [style=NodeCross] (6) at (2.25, 0) {};
		\node [style=NodeCross] (7) at (-2.25, 0) {};
		\node [style=none] (8) at (-1, 0.25) {$I_0^*$};
		\node [style=none] (9) at (-2.25, -0.5) {$I_0^*$};
		\node [style=none] (10) at (1, -1.25) {$I_0^*$};
		\node [style=none] (11) at (2.25, -0.5) {$I_0^*$};
		\node [style=none] (12) at (-2.25, 4.25) {};
		\node [style=none] (13) at (-2.25, 3.25) {};
		\node [style=none] (14) at (-1.75, 3.75) {};
		\node [style=none] (15) at (-2.75, 3.75) {};
		\node [style=none] (16) at (-1.5, 4) {};
		\node [style=none] (17) at (-1.5, 3.5) {};
		\node [style=none] (18) at (-1.25, 3.75) {};
		\node [style=none] (19) at (-1.75, 3.75) {};
		\node [style=none] (20) at (-2.25, 4.75) {};
		\node [style=none] (21) at (-2.25, 4.25) {};
		\node [style=none] (22) at (-2, 4.5) {};
		\node [style=none] (23) at (-2.5, 4.5) {};
		\node [style=none] (24) at (-2.25, 3.25) {};
		\node [style=none] (25) at (-2.25, 2.75) {};
		\node [style=none] (26) at (-2, 3) {};
		\node [style=none] (27) at (-2.5, 3) {};
		\node [style=none] (28) at (-3, 4) {};
		\node [style=none] (29) at (-3, 3.5) {};
		\node [style=none] (30) at (-2.75, 3.75) {};
		\node [style=none] (31) at (-3.25, 3.75) {};
		\node [style=none] (32) at (-1, 6.25) {};
		\node [style=none] (33) at (-1, 5.25) {};
		\node [style=none] (34) at (-0.5, 5.75) {};
		\node [style=none] (35) at (-1.5, 5.75) {};
		\node [style=none] (36) at (-0.25, 6) {};
		\node [style=none] (37) at (-0.25, 5.5) {};
		\node [style=none] (38) at (0, 5.75) {};
		\node [style=none] (39) at (-0.5, 5.75) {};
		\node [style=none] (40) at (-1, 6.75) {};
		\node [style=none] (41) at (-1, 6.25) {};
		\node [style=none] (42) at (-0.75, 6.5) {};
		\node [style=none] (43) at (-1.25, 6.5) {};
		\node [style=none] (44) at (-1, 5.25) {};
		\node [style=none] (45) at (-1, 4.75) {};
		\node [style=none] (46) at (-0.75, 5) {};
		\node [style=none] (47) at (-1.25, 5) {};
		\node [style=none] (48) at (-1.75, 6) {};
		\node [style=none] (49) at (-1.75, 5.5) {};
		\node [style=none] (50) at (-1.5, 5.75) {};
		\node [style=none] (51) at (-2, 5.75) {};
		\node [style=none] (52) at (1, 4.25) {};
		\node [style=none] (53) at (1, 3.25) {};
		\node [style=none] (54) at (1.5, 3.75) {};
		\node [style=none] (55) at (0.5, 3.75) {};
		\node [style=none] (56) at (1.75, 4) {};
		\node [style=none] (57) at (1.75, 3.5) {};
		\node [style=none] (58) at (2, 3.75) {};
		\node [style=none] (59) at (1.5, 3.75) {};
		\node [style=none] (60) at (1, 4.75) {};
		\node [style=none] (61) at (1, 4.25) {};
		\node [style=none] (62) at (1.25, 4.5) {};
		\node [style=none] (63) at (0.75, 4.5) {};
		\node [style=none] (64) at (1, 3.25) {};
		\node [style=none] (65) at (1, 2.75) {};
		\node [style=none] (66) at (1.25, 3) {};
		\node [style=none] (67) at (0.75, 3) {};
		\node [style=none] (68) at (0.25, 4) {};
		\node [style=none] (69) at (0.25, 3.5) {};
		\node [style=none] (70) at (0.5, 3.75) {};
		\node [style=none] (71) at (0, 3.75) {};
		\node [style=none] (72) at (2.25, 6.25) {};
		\node [style=none] (73) at (2.25, 5.25) {};
		\node [style=none] (74) at (2.75, 5.75) {};
		\node [style=none] (75) at (1.75, 5.75) {};
		\node [style=none] (76) at (3, 6) {};
		\node [style=none] (77) at (3, 5.5) {};
		\node [style=none] (78) at (3.25, 5.75) {};
		\node [style=none] (79) at (2.75, 5.75) {};
		\node [style=none] (80) at (2.25, 6.75) {};
		\node [style=none] (81) at (2.25, 6.25) {};
		\node [style=none] (82) at (2.5, 6.5) {};
		\node [style=none] (83) at (2, 6.5) {};
		\node [style=none] (84) at (2.25, 5.25) {};
		\node [style=none] (85) at (2.25, 4.75) {};
		\node [style=none] (86) at (2.5, 5) {};
		\node [style=none] (87) at (2, 5) {};
		\node [style=none] (88) at (1.5, 6) {};
		\node [style=none] (89) at (1.5, 5.5) {};
		\node [style=none] (90) at (1.75, 5.75) {};
		\node [style=none] (91) at (1.25, 5.75) {};
		\node [style=none] (92) at (-1, 4.5) {};
		\node [style=none] (93) at (2.25, 4.5) {};
		\node [style=none] (94) at (1, 2.5) {};
		\node [style=none] (95) at (-2.25, 2.5) {};
		\node [style=none] (96) at (-2.25, 0.25) {};
		\node [style=none] (97) at (-1, 1) {};
		\node [style=none] (98) at (1, -0.5) {};
		\node [style=none] (99) at (2.25, 0.25) {};
		\node [style=none] (100) at (9.25, 1.75) {};
		\node [style=none] (101) at (6, 0) {};
		\node [style=none] (102) at (9.25, -1.75) {};
		\node [style=none] (103) at (12.5, 0) {};
		\node [style=NodeCross] (104) at (8.5, 0.75) {};
		\node [style=NodeCross] (105) at (10, -0.75) {};
		\node [style=NodeCross] (106) at (11.5, 0) {};
		\node [style=NodeCross] (107) at (7, 0) {};
		\node [style=none] (112) at (7, 4) {};
		\node [style=none] (113) at (7, 3) {};
		\node [style=none] (114) at (7.5, 3.5) {};
		\node [style=none] (115) at (6.5, 3.5) {};
		\node [style=none] (119) at (7.5, 3.5) {};
		\node [style=none] (121) at (7, 4) {};
		\node [style=none] (124) at (7, 3) {};
		\node [style=none] (130) at (6.5, 3.5) {};
		\node [style=none] (132) at (8.5, 6) {};
		\node [style=none] (133) at (8.5, 5) {};
		\node [style=none] (134) at (9, 5.5) {};
		\node [style=none] (135) at (8, 5.5) {};
		\node [style=none] (139) at (9, 5.5) {};
		\node [style=none] (141) at (8.5, 6) {};
		\node [style=none] (144) at (8.5, 5) {};
		\node [style=none] (150) at (8, 5.5) {};
		\node [style=none] (152) at (10, 4) {};
		\node [style=none] (153) at (10, 3) {};
		\node [style=none] (154) at (10.5, 3.5) {};
		\node [style=none] (155) at (9.5, 3.5) {};
		\node [style=none] (159) at (10.5, 3.5) {};
		\node [style=none] (161) at (10, 4) {};
		\node [style=none] (164) at (10, 3) {};
		\node [style=none] (170) at (9.5, 3.5) {};
		\node [style=none] (172) at (11.5, 6) {};
		\node [style=none] (173) at (11.5, 5) {};
		\node [style=none] (174) at (12, 5.5) {};
		\node [style=none] (175) at (11, 5.5) {};
		\node [style=none] (179) at (12, 5.5) {};
		\node [style=none] (181) at (11.5, 6) {};
		\node [style=none] (184) at (11.5, 5) {};
		\node [style=none] (190) at (11, 5.5) {};
		\node [style=none] (192) at (8.5, 4.25) {};
		\node [style=none] (193) at (11.5, 4.25) {};
		\node [style=none] (194) at (10, 2.25) {};
		\node [style=none] (195) at (7, 2.25) {};
		\node [style=none] (196) at (7, 0.25) {};
		\node [style=none] (197) at (8.5, 1) {};
		\node [style=none] (198) at (10, -0.5) {};
		\node [style=none] (199) at (11.5, 0.25) {};
		\node [style=Circle] (200) at (8.5, 6) {};
		\node [style=Circle] (201) at (9, 5.5) {};
		\node [style=Circle] (202) at (8.5, 5) {};
		\node [style=Circle] (203) at (8, 5.5) {};
		\node [style=Circle] (204) at (7, 4) {};
		\node [style=Circle] (205) at (7.5, 3.5) {};
		\node [style=Circle] (206) at (7, 3) {};
		\node [style=Circle] (207) at (6.5, 3.5) {};
		\node [style=Circle] (208) at (10, 4) {};
		\node [style=Circle] (209) at (9.5, 3.5) {};
		\node [style=Circle] (210) at (10.5, 3.5) {};
		\node [style=Circle] (211) at (10, 3) {};
		\node [style=Circle] (212) at (11.5, 5) {};
		\node [style=Circle] (213) at (11, 5.5) {};
		\node [style=Circle] (214) at (12, 5.5) {};
		\node [style=Circle] (215) at (11.5, 6) {};
		\node [style=none] (216) at (0, -2.5) {$(T^4/\Z_2)'$};
		\node [style=none] (217) at (9.25, -2.5) {$T^4/\Z_2$};
		\node [style=none] (240) at (-4, 0) {$\P^1_\alpha\,:$};
		\node [style=none] (241) at (5.25, 0) {$\P^1_\alpha\,:$};
		\node [style=NodeCross] (242) at (0, 0) {};
		\node [style=none] (243) at (0, 0.25) {};
		\node [style=none] (244) at (0, 2.25) {};
		\node [style=none] (245) at (0, 2.75) {$T^2_\beta$};
	\end{pgfonlayer}
	\begin{pgfonlayer}{edgelayer}
		\draw [style=ThickLine, in=-180, out=90] (1.center) to (0.center);
		\draw [style=ThickLine, in=90, out=0] (0.center) to (3.center);
		\draw [style=ThickLine, in=0, out=-90] (3.center) to (2.center);
		\draw [style=ThickLine, in=-90, out=180] (2.center) to (1.center);
		\draw [style=ThickLine, in=-180, out=90] (15.center) to (12.center);
		\draw [style=ThickLine, in=90, out=0] (12.center) to (14.center);
		\draw [style=ThickLine, in=0, out=-90] (14.center) to (13.center);
		\draw [style=ThickLine, in=-90, out=-180] (13.center) to (15.center);
		\draw [style=ThickLine, in=-180, out=90] (19.center) to (16.center);
		\draw [style=ThickLine, in=90, out=0] (16.center) to (18.center);
		\draw [style=ThickLine, in=0, out=-90] (18.center) to (17.center);
		\draw [style=ThickLine, in=-90, out=-180] (17.center) to (19.center);
		\draw [style=ThickLine, in=-180, out=90] (23.center) to (20.center);
		\draw [style=ThickLine, in=90, out=0] (20.center) to (22.center);
		\draw [style=ThickLine, in=0, out=-90] (22.center) to (21.center);
		\draw [style=ThickLine, in=-90, out=-180] (21.center) to (23.center);
		\draw [style=ThickLine, in=-180, out=90] (27.center) to (24.center);
		\draw [style=ThickLine, in=90, out=0] (24.center) to (26.center);
		\draw [style=ThickLine, in=0, out=-90] (26.center) to (25.center);
		\draw [style=ThickLine, in=-90, out=-180] (25.center) to (27.center);
		\draw [style=ThickLine, in=-180, out=90] (31.center) to (28.center);
		\draw [style=ThickLine, in=90, out=0] (28.center) to (30.center);
		\draw [style=ThickLine, in=0, out=-90] (30.center) to (29.center);
		\draw [style=ThickLine, in=-90, out=-180] (29.center) to (31.center);
		\draw [style=ThickLine, in=-180, out=90] (35.center) to (32.center);
		\draw [style=ThickLine, in=90, out=0] (32.center) to (34.center);
		\draw [style=ThickLine, in=0, out=-90] (34.center) to (33.center);
		\draw [style=ThickLine, in=-90, out=-180] (33.center) to (35.center);
		\draw [style=ThickLine, in=-180, out=90] (39.center) to (36.center);
		\draw [style=ThickLine, in=90, out=0] (36.center) to (38.center);
		\draw [style=ThickLine, in=0, out=-90] (38.center) to (37.center);
		\draw [style=ThickLine, in=-90, out=-180] (37.center) to (39.center);
		\draw [style=ThickLine, in=-180, out=90] (43.center) to (40.center);
		\draw [style=ThickLine, in=90, out=0] (40.center) to (42.center);
		\draw [style=ThickLine, in=0, out=-90] (42.center) to (41.center);
		\draw [style=ThickLine, in=-90, out=-180] (41.center) to (43.center);
		\draw [style=ThickLine, in=-180, out=90] (47.center) to (44.center);
		\draw [style=ThickLine, in=90, out=0] (44.center) to (46.center);
		\draw [style=ThickLine, in=0, out=-90] (46.center) to (45.center);
		\draw [style=ThickLine, in=-90, out=-180] (45.center) to (47.center);
		\draw [style=ThickLine, in=-180, out=90] (51.center) to (48.center);
		\draw [style=ThickLine, in=90, out=0] (48.center) to (50.center);
		\draw [style=ThickLine, in=0, out=-90] (50.center) to (49.center);
		\draw [style=ThickLine, in=-90, out=-180] (49.center) to (51.center);
		\draw [style=ThickLine, in=-180, out=90] (55.center) to (52.center);
		\draw [style=ThickLine, in=90, out=0] (52.center) to (54.center);
		\draw [style=ThickLine, in=0, out=-90] (54.center) to (53.center);
		\draw [style=ThickLine, in=-90, out=-180] (53.center) to (55.center);
		\draw [style=ThickLine, in=-180, out=90] (59.center) to (56.center);
		\draw [style=ThickLine, in=90, out=0] (56.center) to (58.center);
		\draw [style=ThickLine, in=0, out=-90] (58.center) to (57.center);
		\draw [style=ThickLine, in=-90, out=-180] (57.center) to (59.center);
		\draw [style=ThickLine, in=-180, out=90] (63.center) to (60.center);
		\draw [style=ThickLine, in=90, out=0] (60.center) to (62.center);
		\draw [style=ThickLine, in=0, out=-90] (62.center) to (61.center);
		\draw [style=ThickLine, in=-90, out=-180] (61.center) to (63.center);
		\draw [style=ThickLine, in=-180, out=90] (67.center) to (64.center);
		\draw [style=ThickLine, in=90, out=0] (64.center) to (66.center);
		\draw [style=ThickLine, in=0, out=-90] (66.center) to (65.center);
		\draw [style=ThickLine, in=-90, out=-180] (65.center) to (67.center);
		\draw [style=ThickLine, in=-180, out=90] (71.center) to (68.center);
		\draw [style=ThickLine, in=90, out=0] (68.center) to (70.center);
		\draw [style=ThickLine, in=0, out=-90] (70.center) to (69.center);
		\draw [style=ThickLine, in=-90, out=-180] (69.center) to (71.center);
		\draw [style=ThickLine, in=-180, out=90] (75.center) to (72.center);
		\draw [style=ThickLine, in=90, out=0] (72.center) to (74.center);
		\draw [style=ThickLine, in=0, out=-90] (74.center) to (73.center);
		\draw [style=ThickLine, in=-90, out=-180] (73.center) to (75.center);
		\draw [style=ThickLine, in=-180, out=90] (79.center) to (76.center);
		\draw [style=ThickLine, in=90, out=0] (76.center) to (78.center);
		\draw [style=ThickLine, in=0, out=-90] (78.center) to (77.center);
		\draw [style=ThickLine, in=-90, out=-180] (77.center) to (79.center);
		\draw [style=ThickLine, in=-180, out=90] (83.center) to (80.center);
		\draw [style=ThickLine, in=90, out=0] (80.center) to (82.center);
		\draw [style=ThickLine, in=0, out=-90] (82.center) to (81.center);
		\draw [style=ThickLine, in=-90, out=-180] (81.center) to (83.center);
		\draw [style=ThickLine, in=-180, out=90] (87.center) to (84.center);
		\draw [style=ThickLine, in=90, out=0] (84.center) to (86.center);
		\draw [style=ThickLine, in=0, out=-90] (86.center) to (85.center);
		\draw [style=ThickLine, in=-90, out=-180] (85.center) to (87.center);
		\draw [style=ThickLine, in=-180, out=90] (91.center) to (88.center);
		\draw [style=ThickLine, in=90, out=0] (88.center) to (90.center);
		\draw [style=ThickLine, in=0, out=-90] (90.center) to (89.center);
		\draw [style=ThickLine, in=-90, out=-180] (89.center) to (91.center);
		\draw [style=DottedLine] (93.center) to (99.center);
		\draw [style=DottedLine] (92.center) to (97.center);
		\draw [style=DottedLine] (94.center) to (98.center);
		\draw [style=DottedLine] (95.center) to (96.center);
		\draw [style=ThickLine, in=-180, out=90] (101.center) to (100.center);
		\draw [style=ThickLine, in=90, out=0] (100.center) to (103.center);
		\draw [style=ThickLine, in=0, out=-90] (103.center) to (102.center);
		\draw [style=ThickLine, in=-90, out=180] (102.center) to (101.center);
		\draw [style=ThickLine, in=-180, out=90] (115.center) to (112.center);
		\draw [style=ThickLine, in=90, out=0] (112.center) to (114.center);
		\draw [style=ThickLine, in=0, out=-90] (114.center) to (113.center);
		\draw [style=ThickLine, in=-90, out=-180] (113.center) to (115.center);
		\draw [style=ThickLine, in=-180, out=90] (135.center) to (132.center);
		\draw [style=ThickLine, in=90, out=0] (132.center) to (134.center);
		\draw [style=ThickLine, in=0, out=-90] (134.center) to (133.center);
		\draw [style=ThickLine, in=-90, out=-180] (133.center) to (135.center);
		\draw [style=ThickLine, in=-180, out=90] (155.center) to (152.center);
		\draw [style=ThickLine, in=90, out=0] (152.center) to (154.center);
		\draw [style=ThickLine, in=0, out=-90] (154.center) to (153.center);
		\draw [style=ThickLine, in=-90, out=-180] (153.center) to (155.center);
		\draw [style=ThickLine, in=-180, out=90] (175.center) to (172.center);
		\draw [style=ThickLine, in=90, out=0] (172.center) to (174.center);
		\draw [style=ThickLine, in=0, out=-90] (174.center) to (173.center);
		\draw [style=ThickLine, in=-90, out=-180] (173.center) to (175.center);
		\draw [style=DottedLine] (193.center) to (199.center);
		\draw [style=DottedLine] (192.center) to (197.center);
		\draw [style=DottedLine] (194.center) to (198.center);
		\draw [style=DottedLine] (195.center) to (196.center);
		\draw [style=DottedLine] (244.center) to (243.center);
	\end{pgfonlayer}
\end{tikzpicture}
}
\caption{The quotient $T^4/\Z_2$ and its resolution presented as a fibration with base $T^2/\Z_2\cong \P^1_\alpha$. The exceptional fibers are four copies of the resolution of a Kodaira type $I_0^*$ fiber. The external nodes of the associated $D_4$ Dynkin diagram are exceptional curves. Blowing these down results in an $A_1^4$ singularity marked with black dots (right). }
\label{fig:T4Z2}
\end{figure}
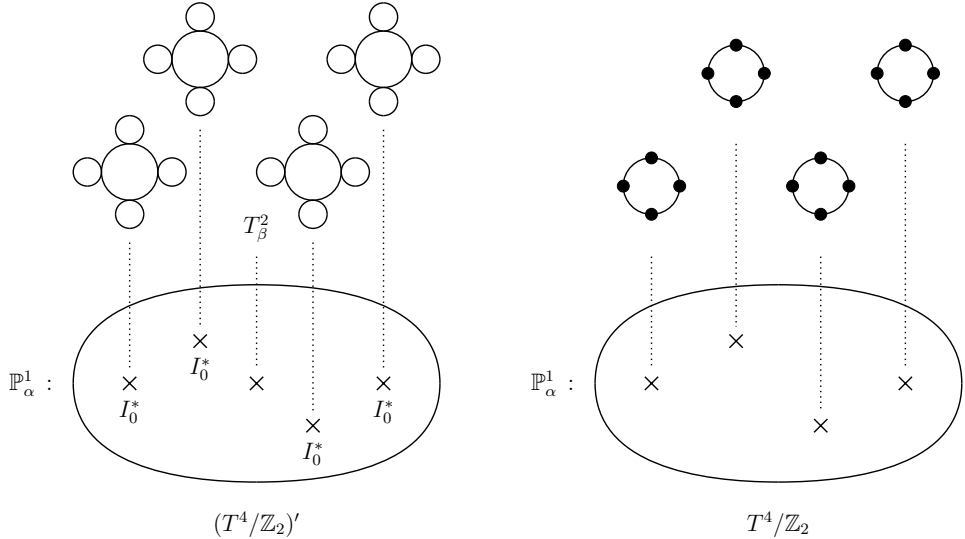

We first lay out a convenient parameterization of the problem and review the relevant topological structures. We introduce the labels
\be\ba
\alpha&=\lb \alpha_1,\alpha_2\rb\in \lbbb \lb 1,2\rb, \lb 1,3\rb,\lb 1,4\rb,\lb2,3\rb,\lb2,4\rb,\lb3,4\rb \rbbb \\
 \beta&=\lb \beta_1,\beta_2\rb\,, \quad \beta_1 < \beta_2\,, \quad       \alpha_1,\alpha_2,\beta_1,\beta_2 \; \; \textnormal{all distinct}
\ea \ee
which determine a decomposition of $T^4$ into two two-tori
\be \ba
T^2_\alpha &= S^1_{\alpha_1} \times S^1_{\alpha_2}\\
T^2_\beta &=  S^1_{\beta_1} \times S^1_{\beta_2}\,.
\ea \ee
For each choice of $\alpha$ we obtain an elliptic pencil
\be\ba
 T^4/\Z_2 ~ \rightarrow ~T^2_\alpha/\Z_2&\equiv \P^1_\alpha
\ea \ee
with generic fiber $T^2_\beta$ which at four points of $\P^1_\alpha$ degenerates to a copy of $T^2_\beta/\Z_2 \equiv \P^1_\beta$. See the right panel in figure \ref{fig:T4Z2}. We also introduce the label
\be
I\in \Z_2^4=\lbbb \lb i,j,k,l\rb \,|\, i,j,k,l \in \Z_2\,, ~i,j,k,l=0,1\rbbb
\ee
which gives a parameterization of the fixed points. Given an elliptic pencil $\pi_\alpha$ we label the fixed points by pairs $I_\alpha,I_\beta \in \Z_2^2$. Given $I_\alpha,I_\beta$ we construct the associated value for $I$ by prescribing its entries at position $\alpha_i,\beta_i$ to be the $i$-th value of $I_\alpha,I_\beta $ respectively. For example, if $\alpha=( 1,2 )$ and $I_\alpha=(0,0)$ and $I_\beta=(1,1)$ then the associated index is $I=(0,0,1,1)$. The index $I_\alpha$ runs over the four orbifold points in $\P^1_\alpha$, while $I_\beta$ runs over the fixed points projecting to the point labelled by $I_\alpha$.

\begin{figure}
\centering

\begin{tikzpicture}
	\begin{pgfonlayer}{nodelayer}
		\node [style=none] (1) at (0, 0.75) {};
		\node [style=none] (2) at (0, -0.75) {};
		\node [style=none] (3) at (0.75, 0) {};
		\node [style=none] (4) at (-0.75, 0) {};
		\node [style=none] (5) at (1.5, 0.75) {};
		\node [style=none] (6) at (1.5, -0.75) {};
		\node [style=none] (7) at (2.25, 0) {};
		\node [style=none] (8) at (0.75, 0) {};
		\node [style=none] (9) at (0, 2.25) {};
		\node [style=none] (10) at (0, 0.75) {};
		\node [style=none] (11) at (0.75, 1.5) {};
		\node [style=none] (12) at (-0.75, 1.5) {};
		\node [style=none] (13) at (0, -0.75) {};
		\node [style=none] (14) at (0, -2.25) {};
		\node [style=none] (15) at (0.75, -1.5) {};
		\node [style=none] (16) at (-0.75, -1.5) {};
		\node [style=none] (17) at (-1.5, 0.75) {};
		\node [style=none] (18) at (-1.5, -0.75) {};
		\node [style=none] (19) at (-0.75, 0) {};
		\node [style=none] (20) at (-2.25, 0) {};
		\node [style=none] (21) at (0, 0) {$F_{I_\alpha}$};
		\node [style=none] (22) at (1.5, 0) {$e_{I_\alpha,01}$};
		\node [style=none] (23) at (0, 1.5) {$e_{I_\alpha,00}$};
		\node [style=none] (24) at (-1.5, 0) {$e_{I_\alpha,10}$};
		\node [style=none] (25) at (0, -1.5) {$e_{I_\alpha,11}$};
	\end{pgfonlayer}
	\begin{pgfonlayer}{edgelayer}
		\draw [style=ThickLine, in=-180, out=90] (4.center) to (1.center);
		\draw [style=ThickLine, in=90, out=0] (1.center) to (3.center);
		\draw [style=ThickLine, in=0, out=-90] (3.center) to (2.center);
		\draw [style=ThickLine, in=-90, out=-180] (2.center) to (4.center);
		\draw [style=ThickLine, in=-180, out=90] (8.center) to (5.center);
		\draw [style=ThickLine, in=90, out=0] (5.center) to (7.center);
		\draw [style=ThickLine, in=0, out=-90] (7.center) to (6.center);
		\draw [style=ThickLine, in=-90, out=-180] (6.center) to (8.center);
		\draw [style=ThickLine, in=-180, out=90] (12.center) to (9.center);
		\draw [style=ThickLine, in=90, out=0] (9.center) to (11.center);
		\draw [style=ThickLine, in=0, out=-90] (11.center) to (10.center);
		\draw [style=ThickLine, in=-90, out=-180] (10.center) to (12.center);
		\draw [style=ThickLine, in=-180, out=90] (16.center) to (13.center);
		\draw [style=ThickLine, in=90, out=0] (13.center) to (15.center);
		\draw [style=ThickLine, in=0, out=-90] (15.center) to (14.center);
		\draw [style=ThickLine, in=-90, out=-180] (14.center) to (16.center);
		\draw [style=ThickLine, in=-180, out=90] (20.center) to (17.center);
		\draw [style=ThickLine, in=90, out=0] (17.center) to (19.center);
		\draw [style=ThickLine, in=0, out=-90] (19.center) to (18.center);
		\draw [style=ThickLine, in=-90, out=-180] (18.center) to (20.center);
	\end{pgfonlayer}
\end{tikzpicture}

\caption{Exceptional $I_0^*$ fiber labelled by $I_\alpha\in \Z_2^2$. The label $I_\beta\in \Z_2^2$ runs over all values.}
\label{fig:ExceptionalFibers}
\end{figure}
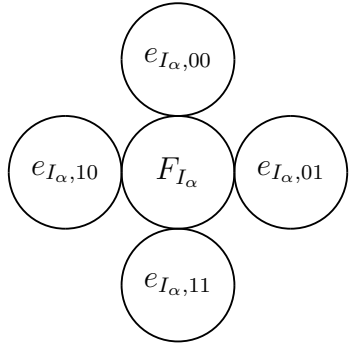

The pencils lifts to the smooth surface $X'$ and the labelling of fixed points lifts to a labelling of exceptional curves. For a given $\alpha$ we obtain the so-called double Kummer pencil (see figure \ref{fig:T4Z2}),
\be\ba
\pi_\alpha\,: ~X'~ \rightarrow ~T^2_\alpha/\Z_2&\equiv \P^1_\alpha \\
\pi_\beta \,: ~X'~ \rightarrow ~T^2_\beta/\Z_2&\equiv \P^1_\beta \,.
\ea \ee
Overall we have three double Kummer pencils collecting a total of six distinct elliptic pencils. The resolution $X'\rightarrow X$ replaces the degenerate fibers $\P^1_\beta,\P^1_\alpha$ by Kodaira type $I_0^*$ fibers, that is a collection of five rational curves with adjacency matrix given by minus the Cartan matrix of the affine $D_4$ Dynkin diagram. The external nodes are exceptional curves. With respect to the elliptic pencil $\pi_\alpha$ we label the 16 exceptional curves as
\be
e_{I_\alpha,I_\beta}\,, \qquad I_\alpha,I_\beta\in \Z_2^2\,.
\ee
The proper transform of the central node of the $I_0^*$ fiber is denoted $F_{I_\alpha}$ and projects to the point in $\P^1_\alpha$ determined by $I_\alpha$, see figure \ref{fig:ExceptionalFibers}.

By the $\Z_2$ quotient we find the generic fiber $T^2_\beta$ to be a double covering of the degenerate fibers  $\P^1_\beta$. In the resolved geometry the exceptional curves extend this relation to \cite{shioda_inose_1977}
\be\ba\label{eq:KeyRelation}
2F_{I_\alpha} + \sum_{I_\beta \;\! \in \;\! \Z_2^2} e_{I_\alpha,I_\beta} = T^2_\beta \,.
\ea\ee
There are $6\times 4=24$ such relations parameterized by $\alpha$ and $I_\alpha$. The 16 exceptional curves $e_{I_\alpha,I_\beta}$ and 24 curves $F_{I_\alpha}$ generate the K3 lattice. Here the curve $F_{I_\alpha}$ is the proper transform of the fiber projecting to the fixed point in $T^2_\alpha/\Z_2$ labelled by $I_\alpha$ (see figure \ref{fig:ExceptionalFibers}).

A crucial consequence of the relation \eqref{eq:KeyRelation} is that fractional linear combinations of exceptional curves are also integral classes of the K3 lattice. This follows by taking the difference of \eqref{eq:KeyRelation} for two different values of $I_\alpha$ for fixed $\alpha$ (and therefore fixed $\beta$). We have
\be\ba\label{eq:KeyRelation2}
F_{I_\alpha} - F_{I_\alpha'}= \frac{1}{2} \lb \sum_{I_\beta' \;\! \in \;\! \Z_2^2} e_{I_\alpha',I_\beta'} - \sum_{I_\beta \;\! \in \;\! \Z_2^2} e_{I_\alpha,I_\beta}  \rb \in \Gamma_{3,19} \,,
\ea\ee
which is argued for from a different point of view by Nikulin in \cite{Nikulin_1975}. This construction gives $6\times {4 \choose 2}=36$ fractional linear combinations of 8 exceptional curves which are actually integral classes in the K3 lattice.

\begin{figure}
\centering
\scalebox{0.8}{
\begin{tikzpicture}
	\begin{pgfonlayer}{nodelayer}
		\node [style=none] (0) at (0.75, 2.75) {};
		\node [style=none] (1) at (0.75, -3.25) {};
		\node [style=none] (2) at (2.25, 2.75) {};
		\node [style=none] (3) at (2.25, -3.25) {};
		\node [style=none] (4) at (3.75, 2.75) {};
		\node [style=none] (5) at (3.75, -3.25) {};
		\node [style=none] (6) at (5.25, 2.75) {};
		\node [style=none] (7) at (5.25, -3.25) {};
		\node [style=none] (8) at (0.25, 2.25) {};
		\node [style=none] (9) at (6.25, 2.25) {};
		\node [style=none] (10) at (0.25, 0.75) {};
		\node [style=none] (11) at (6.25, 0.75) {};
		\node [style=none] (12) at (0.25, -0.75) {};
		\node [style=none] (13) at (6.25, -0.75) {};
		\node [style=none] (14) at (0.25, -2.25) {};
		\node [style=none] (15) at (6.25, -2.25) {};
		\node [style=none] (16) at (1.5, 2.5) {};
		\node [style=none] (17) at (0.5, 1.5) {};
		\node [style=none] (18) at (0.5, 0) {};
		\node [style=none] (19) at (1.5, 1) {};
		\node [style=none] (20) at (0.5, -1.5) {};
		\node [style=none] (21) at (1.5, -0.5) {};
		\node [style=none] (22) at (2, -1.5) {};
		\node [style=none] (23) at (3, -0.5) {};
		\node [style=none] (24) at (3, 2.5) {};
		\node [style=none] (25) at (2, 1.5) {};
		\node [style=none] (26) at (4.5, 2.5) {};
		\node [style=none] (27) at (3.5, 1.5) {};
		\node [style=none] (28) at (6, 2.5) {};
		\node [style=none] (29) at (5, 1.5) {};
		\node [style=none] (30) at (3, 1) {};
		\node [style=none] (31) at (2, 0) {};
		\node [style=none] (32) at (4.5, 1) {};
		\node [style=none] (33) at (3.5, 0) {};
		\node [style=none] (34) at (6, 1) {};
		\node [style=none] (35) at (5, 0) {};
		\node [style=none] (36) at (6, -0.5) {};
		\node [style=none] (37) at (5, -1.5) {};
		\node [style=none] (38) at (4.5, -0.5) {};
		\node [style=none] (39) at (3.5, -1.5) {};
		\node [style=none] (40) at (1.5, -2) {};
		\node [style=none] (41) at (0.5, -3) {};
		\node [style=none] (42) at (3, -2) {};
		\node [style=none] (43) at (2, -3) {};
		\node [style=none] (44) at (4.5, -2) {};
		\node [style=none] (45) at (3.5, -3) {};
		\node [style=none] (46) at (6, -2) {};
		\node [style=none] (47) at (5, -3) {};
		\node [style=none] (48) at (7, 2.25) {$F_{\lbbb 00\rbbb_\alpha }$};
		\node [style=none] (49) at (7, 0.75) {$F_{\lbbb 01\rbbb_\alpha }$};
		\node [style=none] (50) at (7, -0.75) {$F_{\lbbb 10\rbbb_\alpha }$};
		\node [style=none] (51) at (7, -2.25) {$F_{\lbbb 11\rbbb_\alpha }$};
i		\node [style=none] (56) at (1.375, 1.625) {$e_{0000}$};
		\node [style=none] (57) at (2.875, 1.625) {$e_{0001}$};
		\node [style=none] (58) at (4.375, 1.625) {$e_{0010}$};
		\node [style=none] (59) at (5.875, 1.625) {$e_{0011}$};
		\node [style=none] (60) at (1.375, 0.125) {$e_{0100}$};
		\node [style=none] (61) at (2.875, 0.125) {$e_{0101}$};
		\node [style=none] (62) at (4.375, 0.125) {$e_{0110}$};
		\node [style=none] (63) at (5.875, 0.125) {$e_{0111}$};
		\node [style=none] (64) at (1.375, -1.375) {$e_{1000}$};
		\node [style=none] (65) at (2.875, -1.375) {$e_{1001}$};
		\node [style=none] (66) at (4.375, -1.375) {$e_{1010}$};
		\node [style=none] (67) at (5.875, -1.375) {$e_{1011}$};
		\node [style=none] (68) at (1.375, -2.875) {$e_{1100}$};
		\node [style=none] (69) at (2.875, -2.875) {$e_{1101}$};
		\node [style=none] (70) at (4.375, -2.875) {$e_{1110}$};
		\node [style=none] (71) at (5.875, -2.875) {$e_{1111}$};
		\node [style=none] (72) at (0.625, 2.25) {};
		\node [style=none] (73) at (0.875, 2.25) {};
		\node [style=none] (74) at (2.125, 2.25) {};
		\node [style=none] (75) at (2.375, 2.25) {};
		\node [style=none] (76) at (3.625, 2.25) {};
		\node [style=none] (77) at (3.875, 2.25) {};
		\node [style=none] (78) at (5.125, 2.25) {};
		\node [style=none] (79) at (5.375, 2.25) {};
		\node [style=none] (80) at (0.625, 0.75) {};
		\node [style=none] (81) at (0.875, 0.75) {};
		\node [style=none] (82) at (2.125, 0.75) {};
		\node [style=none] (83) at (2.375, 0.75) {};
		\node [style=none] (84) at (3.625, 0.75) {};
		\node [style=none] (85) at (3.875, 0.75) {};
		\node [style=none] (86) at (5.125, 0.75) {};
		\node [style=none] (87) at (5.375, 0.75) {};
		\node [style=none] (88) at (0.625, -0.75) {};
		\node [style=none] (89) at (0.875, -0.75) {};
		\node [style=none] (90) at (2.125, -0.75) {};
		\node [style=none] (91) at (2.375, -0.75) {};
		\node [style=none] (92) at (3.625, -0.75) {};
		\node [style=none] (93) at (3.875, -0.75) {};
		\node [style=none] (94) at (5.125, -0.75) {};
		\node [style=none] (95) at (5.375, -0.75) {};
		\node [style=none] (96) at (0.625, -2.25) {};
		\node [style=none] (97) at (0.875, -2.25) {};
		\node [style=none] (98) at (2.125, -2.25) {};
		\node [style=none] (99) at (2.375, -2.25) {};
		\node [style=none] (100) at (3.625, -2.25) {};
		\node [style=none] (101) at (3.875, -2.25) {};
		\node [style=none] (102) at (5.125, -2.25) {};
		\node [style=none] (103) at (5.375, -2.25) {};
		\node [style=none] (104) at (0.75, -3.75) {$F_{\lbbb 00\rbbb_\beta }$};
		\node [style=none] (105) at (2.25, -3.75) {$F_{\lbbb 01\rbbb_\beta }$};
		\node [style=none] (106) at (3.75, -3.75) {$F_{\lbbb 10\rbbb_\beta }$};
		\node [style=none] (107) at (5.25, -3.75) {$F_{\lbbb 11\rbbb_\beta }$};
	\end{pgfonlayer}
	\begin{pgfonlayer}{edgelayer}
		\draw [style=ThickLine] (0.center) to (1.center);
		\draw [style=ThickLine] (2.center) to (3.center);
		\draw [style=ThickLine] (4.center) to (5.center);
		\draw [style=ThickLine] (6.center) to (7.center);
		\draw (17.center) to (16.center);
		\draw (25.center) to (24.center);
		\draw (27.center) to (26.center);
		\draw (29.center) to (28.center);
		\draw (33.center) to (32.center);
		\draw (22.center) to (23.center);
		\draw (20.center) to (21.center);
		\draw (18.center) to (19.center);
		\draw (31.center) to (30.center);
		\draw (35.center) to (34.center);
		\draw (39.center) to (38.center);
		\draw (37.center) to (36.center);
		\draw (47.center) to (46.center);
		\draw (44.center) to (45.center);
		\draw (43.center) to (42.center);
		\draw (40.center) to (41.center);
		\draw [style=ThickLine] (8.center) to (72.center);
		\draw [style=ThickLine] (73.center) to (74.center);
		\draw [style=ThickLine] (75.center) to (76.center);
		\draw [style=ThickLine] (77.center) to (78.center);
		\draw [style=ThickLine] (79.center) to (9.center);
		\draw [style=ThickLine] (11.center) to (87.center);
		\draw [style=ThickLine] (86.center) to (85.center);
		\draw [style=ThickLine] (84.center) to (83.center);
		\draw [style=ThickLine] (82.center) to (81.center);
		\draw [style=ThickLine] (80.center) to (10.center);
		\draw [style=ThickLine] (12.center) to (88.center);
		\draw [style=ThickLine] (89.center) to (90.center);
		\draw [style=ThickLine] (91.center) to (92.center);
		\draw [style=ThickLine] (93.center) to (94.center);
		\draw [style=ThickLine] (95.center) to (13.center);
		\draw [style=ThickLine] (14.center) to (96.center);
		\draw [style=ThickLine] (97.center) to (98.center);
		\draw [style=ThickLine] (99.center) to (100.center);
		\draw [style=ThickLine] (101.center) to (102.center);
		\draw [style=ThickLine] (103.center) to (15.center);
	\end{pgfonlayer}
\end{tikzpicture}
}
\caption{Double Kummer pencil for the elliptic pencils $T^4/\Z_2 \rightarrow \P^1_\alpha$ and $T^4/\Z_2 \rightarrow \P^1_\beta$. The choice of $\alpha$ determines $\beta$, there are six choices for $\alpha$. We display intersections for the choice $\alpha=\lbbb 1,2\rbbb$  and $\beta=\lbbb 3,4\rbbb$.  }
\label{fig:dblKummer}
\end{figure}
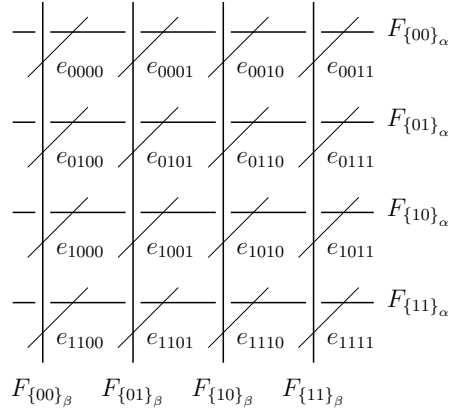

We next give the intersection matrices for various 2-cycles. The curves of the double Kummer pencil intersect as
\be \ba
F_{I_\alpha} \cdot  e_{I_\alpha',I_\beta'} =\delta_{I_\alpha, I_\alpha' }\,, \qquad  F_{I_\beta} \cdot  e_{I_\alpha',I_\beta'} =\delta_{I_\beta, I_\beta' },
\ea \ee
following Shioda and Inose \cite{shioda_inose_1977}. The exceptional curves and the curves $ F_{I_\alpha},F_{I_\beta} $ of the pencils have self-intersections
\be
e_{I_\alpha,I_\beta}\cdot e_{I_\alpha,I_\beta}=F_{I_\alpha} \cdot F_{I_\alpha}=F_{I_\beta}  \cdot F_{I_\beta} =-2
\ee
and the generic torus fibers have intersections
\be
 F_{I_\alpha} \cdot T^2_\alpha =1\,, \qquad  F_{I_\beta} \cdot T^2_\beta =1\,, \qquad T^2_\alpha \cdot T^2_\beta=2\,.
\ee
See figure \ref{fig:dblKummer} for a pictorial presentation for an example with fixed $\alpha$.

This completes the framing of the problem. We now compute the continuous non-Abelian gauge group $G_{\mathrm{loc}}$ of M-theory compactified on $T^4/\Z_2$. Geometrically, this translates into computing $\Tor\,H_2(X)$, and applying \eqref{eq:somecharacterization} and \eqref{eq:GADE} with $\widetilde{G}_\mathrm{loc} = SU(2)^{16}$.

Different aspects of this computation are discussed from the viewpoint of affine geometry (see chapter VIII of \cite{Barth}) and from a lattice perspective (see chapter 14.3 of \cite{huybrechts_2016}). In the physics literature there is the treatment of references \cite{Wendland:2000ry, Nahm:1999ps} which centers on studying orbifold conformal field theories with K3 target space.

We present a method of computation suitable for generalizations away from the Kummer case. We note that line \eqref{eq:Lattice} can simply be computed by determining the number of linearly independent fractional combinations \eqref{eq:KeyRelation2} over $\Z_2$. This rather simple perspective generalizes very concretely and straightforwardly to other examples, see Appendix \ref{app:AdditionalCyclicStuff}.

For this we need to consider the coefficient matrix of the righthand side of \eqref{eq:KeyRelation2}, whose double is
\be\label{eq:CoeffMatrix}
\scalebox{0.7}{ $\textnormal{{\Large $M_{T^4/\mathbb{Z}_2}$}}=\left(
\begin{array}{cccccccccccccccc}
 1 & 1 & 1 & 1 & -1 & -1 & -1 & -1 & 0 & 0 & 0 & 0 & 0 & 0 & 0 & 0 \\
 1 & 1 & -1 & -1 & 1 & 1 & -1 & -1 & 0 & 0 & 0 & 0 & 0 & 0 & 0 & 0 \\
 1 & -1 & 1 & -1 & 1 & -1 & 1 & -1 & 0 & 0 & 0 & 0 & 0 & 0 & 0 & 0 \\
 1 & 1 & -1 & -1 & 0 & 0 & 0 & 0 & 1 & 1 & -1 & -1 & 0 & 0 & 0 & 0 \\
 1 & -1 & 1 & -1 & 0 & 0 & 0 & 0 & 1 & -1 & 1 & -1 & 0 & 0 & 0 & 0 \\
 1 & -1 & 0 & 0 & 1 & -1 & 0 & 0 & 1 & -1 & 0 & 0 & 1 & -1 & 0 & 0 \\
 1 & 1 & 1 & 1 & 0 & 0 & 0 & 0 & -1 & -1 & -1 & -1 & 0 & 0 & 0 & 0 \\
 1 & 1 & 0 & 0 & 1 & 1 & 0 & 0 & -1 & -1 & 0 & 0 & -1 & -1 & 0 & 0 \\
 1 & 0 & 1 & 0 & 1 & 0 & 1 & 0 & -1 & 0 & -1 & 0 & -1 & 0 & -1 & 0 \\
 1 & 1 & 0 & 0 & -1 & -1 & 0 & 0 & 1 & 1 & 0 & 0 & -1 & -1 & 0 & 0 \\
 1 & 0 & 1 & 0 & -1 & 0 & -1 & 0 & 1 & 0 & 1 & 0 & -1 & 0 & -1 & 0 \\
 1 & 0 & -1 & 0 & 1 & 0 & -1 & 0 & 1 & 0 & -1 & 0 & 1 & 0 & -1 & 0 \\
 1 & 1 & 1 & 1 & 0 & 0 & 0 & 0 & 0 & 0 & 0 & 0 & -1 & -1 & -1 & -1 \\
 1 & 1 & 0 & 0 & 1 & 1 & 0 & 0 & 0 & 0 & -1 & -1 & 0 & 0 & -1 & -1 \\
 1 & 0 & 1 & 0 & 1 & 0 & 1 & 0 & 0 & -1 & 0 & -1 & 0 & -1 & 0 & -1 \\
 1 & 1 & 0 & 0 & 0 & 0 & -1 & -1 & 1 & 1 & 0 & 0 & 0 & 0 & -1 & -1 \\
 1 & 0 & 1 & 0 & 0 & -1 & 0 & -1 & 1 & 0 & 1 & 0 & 0 & -1 & 0 & -1 \\
 1 & 0 & 0 & -1 & 1 & 0 & 0 & -1 & 1 & 0 & 0 & -1 & 1 & 0 & 0 & -1 \\
 0 & 0 & 0 & 0 & 1 & 1 & 1 & 1 & -1 & -1 & -1 & -1 & 0 & 0 & 0 & 0 \\
 0 & 0 & 1 & 1 & 0 & 0 & 1 & 1 & -1 & -1 & 0 & 0 & -1 & -1 & 0 & 0 \\
 0 & 1 & 0 & 1 & 0 & 1 & 0 & 1 & -1 & 0 & -1 & 0 & -1 & 0 & -1 & 0 \\
 0 & 0 & 1 & 1 & -1 & -1 & 0 & 0 & 0 & 0 & 1 & 1 & -1 & -1 & 0 & 0 \\
 0 & 1 & 0 & 1 & -1 & 0 & -1 & 0 & 0 & 1 & 0 & 1 & -1 & 0 & -1 & 0 \\
 0 & 1 & -1 & 0 & 0 & 1 & -1 & 0 & 0 & 1 & -1 & 0 & 0 & 1 & -1 & 0 \\
 0 & 0 & 0 & 0 & 1 & 1 & 1 & 1 & 0 & 0 & 0 & 0 & -1 & -1 & -1 & -1 \\
 0 & 0 & 1 & 1 & 0 & 0 & 1 & 1 & 0 & 0 & -1 & -1 & 0 & 0 & -1 & -1 \\
 0 & 1 & 0 & 1 & 0 & 1 & 0 & 1 & 0 & -1 & 0 & -1 & 0 & -1 & 0 & -1 \\
 0 & 0 & 1 & 1 & 0 & 0 & -1 & -1 & 0 & 0 & 1 & 1 & 0 & 0 & -1 & -1 \\
 0 & 1 & 0 & 1 & 0 & -1 & 0 & -1 & 0 & 1 & 0 & 1 & 0 & -1 & 0 & -1 \\
 0 & 1 & 0 & -1 & 0 & 1 & 0 & -1 & 0 & 1 & 0 & -1 & 0 & 1 & 0 & -1 \\
 0 & 0 & 0 & 0 & 0 & 0 & 0 & 0 & 1 & 1 & 1 & 1 & -1 & -1 & -1 & -1 \\
 0 & 0 & 0 & 0 & 0 & 0 & 0 & 0 & 1 & 1 & -1 & -1 & 1 & 1 & -1 & -1 \\
 0 & 0 & 0 & 0 & 0 & 0 & 0 & 0 & 1 & -1 & 1 & -1 & 1 & -1 & 1 & -1 \\
 0 & 0 & 0 & 0 & 1 & 1 & -1 & -1 & 0 & 0 & 0 & 0 & 1 & 1 & -1 & -1 \\
 0 & 0 & 0 & 0 & 1 & -1 & 1 & -1 & 0 & 0 & 0 & 0 & 1 & -1 & 1 & -1 \\
 0 & 0 & 1 & -1 & 0 & 0 & 1 & -1 & 0 & 0 & 1 & -1 & 0 & 0 & 1 & -1 \\
\end{array}
\right) $\qquad }
\ee

\noindent with 16 columns associated to exceptional curves and 36 rows expanding the fractional combinations in \eqref{eq:KeyRelation2} with respect to these. Next, compute the Smith normal form (SNF) which is block diagonal with blocks (in the obvious notation):
\be
\textnormal{SNF}(M_{T^4/\Z_2}) = \lb M_2 ~ {\mathbf 0}_{16\times 20}  \rb^{T} \,, \qquad M_2 = \diag(\textnormal{Id}_{5\times 5}, 2\, \textnormal{Id}_{5\times 5}, \mathbf{0}_{6 \times 6}),
\ee
and recall that $M_{T^4/\mathbb{Z}_2}$ is double the coefficient matrix and that we are working with mod 1.
Therefore the object of interest is $(1/2)\textnormal{SNF}(M_{T^4/\Z_2})$ mod 1 which has $5$ non-trivial entries. We conclude that:
\be
\overline{L}_E/L_E  = \Tor\,H_2(X)\cong \Z_2^5\,.
\ee
The global form of the non-Abelian continuous gauge symmetry is therefore
\be
G_\mathrm{loc}= SU(2)^{16} / \Z_2^5 \,,
\ee
where the embedding of $\Z_2^5\hookrightarrow \Z_2^{16}$ is described (in a redundant characterization)
by the rows of the coefficient matrix $M_{T^4/\mathbb{Z}_2}$. It is interesting to note that $G_\mathrm{loc}$ appears in the dual description of Heterotic string on $T^3$ \cite{Fraiman:2021soq}.\footnote{We leave a detailed matching between our global gauge group results and the heterotic calculations for future work.}

Now we turn to compute the global form of the full continuous gauge group including Abelian factors. For this we make the generators and maps in the exact sequence \eqref{eq:MVS} explicit. We begin by characterizing the homology groups. Let us denote the five linearly independent, half integral exceptional generators of the K3 lattice descending to generators $\overline{L}_E/L_E$ as $t_k$ with $k=1,\dots, 5$. Denote by $F_\alpha$ the curves $F_{I_\alpha}$ for fixed $I_\alpha$ and varying $\alpha$. Then, as was already demonstrated in \cite{Taimanov_2018} explicitly by Taimanov
\be
\Gamma_{3,19}=\langle F_\alpha, e_I, t_k \rangle
\ee
and contracting the exceptional lattice we have
\be
H_2(X)\cong \langle F_\alpha, t_k \rangle \cong \Z^6 \oplus \Z^5_2\,.
\ee
The groups $H_2(X^\circ )$ and $H_1(X^\circ )\cong \Z_2^5$ were computed by Spanier \cite{10.2307/2033261} and in the conventions laid out here note that the former is generated by the invariant 2-cycles $T^2_\alpha$
\be
H_2(X^\circ )\cong \langle T^2_\alpha \rangle \cong \Z^6\,.
\ee
Next consider $H_1(\partial \Xl)$ whose 1-cycles $\gamma_I =\partial U_I \cap \mathfrak{T}_I$ are in correspondence with ADE thimbles whose compact representatives are $\Tp_{\!c,I}=e_{I}/2$, so by the listing in figure \ref{fig:Thimbles} we have:
\be\label{eq:ref}
 \bigoplus_{I\;\! \in \;\! \Z_2^4} \langle \gamma_I \rangle  \cong  \bigoplus_{I\;\! \in \;\! \Z_2^4} \langle  \Tp_{\!c,I} \rangle \cong \bigoplus_{I\;\! \in \;\! \Z_2^4} \Big\langle \frac{1}{2}e_I \Big\rangle_{\Q/\Z} \cong \Z_2^{16}\,.
\ee

Now we discuss the maps. The map $\jmath_2$ is simply given by decomposing generic torus fiber following \eqref{eq:KeyRelation}. Let us note here that exactness of the sequence then follows from noting that torsional generators in degree two stretch between ADE singularities as depicted in figure \ref{fig:Comp}. The map $\partial_2|_{\Tor\, H_2(X;\Z)}$ is given by \eqref{eq:KeyRelation2} and is extended to the map $\partial_2$ by generalizing to intersection with the ADE boundaries $\partial U_i$.

Let us next make the discussion in section \ref{sec:Ab} on the Abelian gauge symmetries explicit in this example. We have, in the notation introduced there,
\be
\Pi_\alpha=T^2_\alpha\,, \qquad \pi_\alpha=F_{I_\alpha}\,, \qquad  \pi_\alpha|_{\Xl}=-\frac{1}{2} \sum_{I_\beta\;\! \in \;\! \Z_2^2}e_{I_\alpha, I_\beta}
\ee
and the correctly normalized $U(1)$ generators are
\be \label{eq:U1Generator}
F_{I_\alpha}+\frac{1}{2} \sum_{I_\beta\;\! \in \;\! \Z_2^2}e_{I_\alpha, I_\beta}=\frac{1}{2}T^2_\alpha \,.
\ee
Next note that the double of \eqref{eq:U1Generator} for two different values of $I_\alpha$ for fixed $\alpha$ lie in the same homology class by \eqref{eq:KeyRelation} and consequently the number of $U(1)$ factors is $6=b_2(X)$ as counted by $\alpha$. The map $j_2:\mathbb{Z}^6\rightarrow \mathbb{Z}^6\oplus \mathbb{Z}_2^5$ does not map into $ \mathbb{Z}_2^5$ and is multiplication by 2 otherwise. Consequently $U(1)$ charges are quantized with $q=1/2$.

We summarize the above discussion. The sequence \eqref{eq:MVS} reads
\be \ba
0~&\xrightarrow[\text{}]{\;\imath_2\;} ~\big \langle T^2_\alpha \big\rangle  ~\xrightarrow[\text{}]{\;\jmath_2\;}  ~\langle F_\alpha, t_k \rangle ~\xrightarrow[\text{}]{\; \partial_2 \;}  ~   \bigoplus_{I\;\! \in \;\! \Z_2^4} \Big\langle \frac{1}{2}e_I \Big\rangle_{\Q/\Z} ~\xrightarrow[\text{}]{\;\imath_1\;} ~H_1(X^\circ)  ~\xrightarrow[\text{}]{\;\jmath_1\;}  ~0
\ea \ee
which gives the exact sequence of groups
\be \ba
0~&\xrightarrow[\text{}]{\;\imath_2\;} ~\Z^6  ~\xrightarrow[\text{}]{\;\jmath_2\;}  ~\Z^6\oplus \Z^5_2 ~\xrightarrow[\text{}]{\; \partial_2 \;}  ~  \Z_2^{16} ~\xrightarrow[\text{}]{\;\imath_1\;} ~\Z_2^5  ~\xrightarrow[\text{}]{\;\jmath_1\;}  ~0\,.
\ea \ee
The continuous gauge group is therefore:
\be
G_\textnormal{full}=\frac{(SU(2)^{16}/\Z_2^5)\times U(1)^6}{\Z_2^6}
\ee
The embedding of $\mathbb{Z}_2^6$ follows from \eqref{eq:U1Generator}. For each $U(1)$ factor (which are labelled by $\alpha$) there is a corresponding $\mathbb{Z}_2$ quotient by the diagonal center of four $SU(2)$ factors labelled by $I_\beta$ and $-1\in U(1)_\alpha$. 

\subsection{Further Orbifold Examples} \label{sec:FurtherExamples}

We now turn to some further examples of orbifolds of tori. Here, we specify some further orbifold group actions, and specify the resulting gauge group for these models. Additional details of these computations are deferred to the Appendices.

We now consider the orbifolds $X=T^4/\Z_n$ with $n=2,3,4,6$. To frame the following discussion we begin by introducing notation. We split $T^4=T^2_1\times T^2_2$, and we specify $T^2_{k}$ as a quotient $\mathbb{C} / \Lambda_{\tau_k}$, with $\Lambda_{\tau_k}$ is the lattice generated by $1$ and $\tau_k$, the complex structure modulus of the $T^2_{k}$. Denote by $z_k$ the local coordinate on $T^{2}_{k}$. Consider the $\Z_n$ orbifold action $ (z_1,z_2)\mapsto (\omega z_1, \omega^{-1} z_2)$ with $\omega$ a primitive $n$-th root of unity. This group action is well-defined precisely when $\tau_i=\tau$ for $n=3,4,6$ (for $n=2$ there are no constraints) and
\be
\tau =\left\{
  \begin{aligned}
   ~e^{2\pi i/6}& && (n=3) \\
    ~e^{2\pi i/4}&  && (n=4)\\
    ~e^{2\pi i/6}&\quad  && (n=6)
    \end{aligned}
  \right.  \smallskip
 \ee
which implies $\tau^l=\bar\tau=0$ mod $\tau, 1$ for all integers $l\in \Z$.

\begin{figure}
\centering
\scalebox{0.6}{
\begin{tikzpicture}
	\begin{pgfonlayer}{nodelayer}
		\node [style=none] (0) at (0, 0) {};
		\node [style=none] (1) at (0, 4) {};
		\node [style=none] (2) at (7, 0) {};
		\node [style=none] (3) at (2, 3) {};
		\node [style=none] (4) at (4, 0) {};
		\node [style=none] (5) at (6, 3) {};
		\node [style=CircleRed] (6) at (2, 1) {};
		\node [style=CircleRed] (7) at (4, 2) {};
		\node [style=CircleRed] (8) at (0, 0) {};
		\node [style=none] (9) at (-0.875, 3.5) {\Large Im\,$z_k$};
		\node [style=none] (10) at (6.5, -0.5) {\Large Re\,$z_k$};
		\node [style=none] (11) at (9, 0) {};
		\node [style=none] (12) at (9, 4) {};
		\node [style=none] (13) at (13, 0) {};
		\node [style=none] (14) at (9, 3) {};
		\node [style=none] (15) at (12, 0) {};
		\node [style=none] (16) at (12, 3) {};
		\node [style=CircleRed] (17) at (9, 1.5) {};
		\node [style=CircleRed] (18) at (10.5, 1.5) {};
		\node [style=CircleRed] (19) at (9, 0) {};
		\node [style=none] (20) at (8.125, 3.5) {\Large Im\,$z_k$};
		\node [style=none] (21) at (12.5, -0.5) {\Large Re\,$z_k$};
		\node [style=none] (22) at (15, 0) {};
		\node [style=none] (23) at (15, 4) {};
		\node [style=none] (24) at (22, 0) {};
		\node [style=none] (25) at (17, 3) {};
		\node [style=none] (26) at (19, 0) {};
		\node [style=none] (27) at (21, 3) {};
		\node [style=CircleRed] (28) at (17, 1) {};
		\node [style=CircleRed] (29) at (19, 2) {};
		\node [style=CircleRed] (30) at (15, 0) {};
		\node [style=none] (31) at (14.125, 3.5) {\Large Im\,$z_k$};
		\node [style=none] (32) at (21.5, -0.5) {\Large Re\,$z_k$};
		\node [style=CircleRed] (33) at (10.5, 0) {};
		\node [style=CircleRed] (34) at (16, 1.5) {};
		\node [style=CircleRed] (35) at (17, 0) {};
		\node [style=CircleRed] (36) at (18, 1.5) {};
		\node [style=none] (41) at (3.5, -1.5) {\Large $n=3$};
		\node [style=none] (42) at (11, -1.5) {\Large $n=4$};
		\node [style=none] (43) at (18.5, -1.5) {\Large $n=6$};
		\node [style=none] (44) at (0, 0) {\footnotesize \color{white} $3$};
		\node [style=none] (45) at (2, 1) {\footnotesize \color{white} $3$};
		\node [style=none] (46) at (4, 2) {\footnotesize \color{white} $3$};
		\node [style=none] (47) at (9, 0) {\footnotesize \color{white} $4$};
		\node [style=none] (48) at (10.5, 1.5) {\footnotesize \color{white} $4$};
		\node [style=none] (49) at (10.5, 0) {\footnotesize \color{white} $2$};
		\node [style=none] (50) at (9, 1.5) {\footnotesize \color{white} $2$};
		\node [style=none] (51) at (15, 0) {\footnotesize \color{white} $6$};
		\node [style=none] (52) at (17, 0) {\footnotesize \color{white} $2$};
		\node [style=none] (53) at (18, 1.5) {\footnotesize \color{white} $2$};
		\node [style=none] (54) at (16, 1.5) {\footnotesize \color{white} $2$};
		\node [style=none] (55) at (17, 1) {\footnotesize \color{white} $3$};
		\node [style=none] (56) at (19, 2) {\footnotesize \color{white} $3$};
		\node [style=none] (57) at (11, -2) {};
	\end{pgfonlayer}
	\begin{pgfonlayer}{edgelayer}
		\draw [style=ArrowLineRight] (0.center) to (1.center);
		\draw [style=ArrowLineRight] (0.center) to (2.center);
		\draw [style=ThickLine] (0.center) to (3.center);
		\draw [style=ThickLine] (4.center) to (5.center);
		\draw [style=ThickLine] (5.center) to (3.center);
		\draw [style=ArrowLineRight] (11.center) to (12.center);
		\draw [style=ArrowLineRight] (11.center) to (13.center);
		\draw [style=ThickLine] (11.center) to (14.center);
		\draw [style=ThickLine] (15.center) to (16.center);
		\draw [style=ThickLine] (16.center) to (14.center);
		\draw [style=ArrowLineRight] (22.center) to (23.center);
		\draw [style=ArrowLineRight] (22.center) to (24.center);
		\draw [style=ThickLine] (22.center) to (25.center);
		\draw [style=ThickLine] (26.center) to (27.center);
		\draw [style=ThickLine] (27.center) to (25.center);
	\end{pgfonlayer}
\end{tikzpicture}
}
\caption{Fundamental domains for $T^2_k/\Z_n$ for $n=2,3,6$. Points fixed by some subgroups $\Z_m\subset \Z_n$ with $m\,|\, n$ are marked black and labelled by $m\geq 2$. The group action $\Z_n$ groups points labelled by $m$ into orbits containing $n/m$ points.}
\label{fig:FundamentalDomains}
\end{figure}
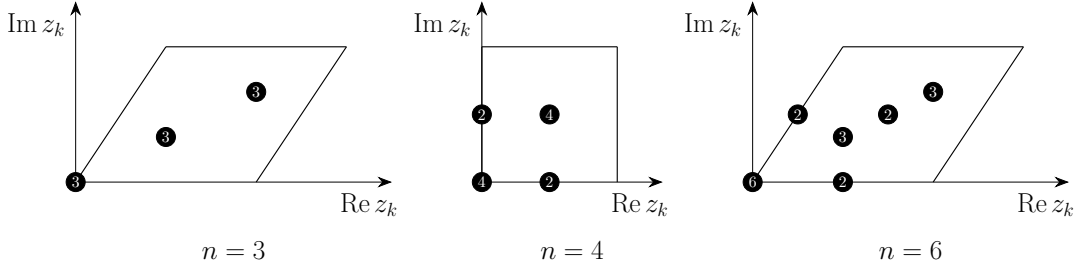

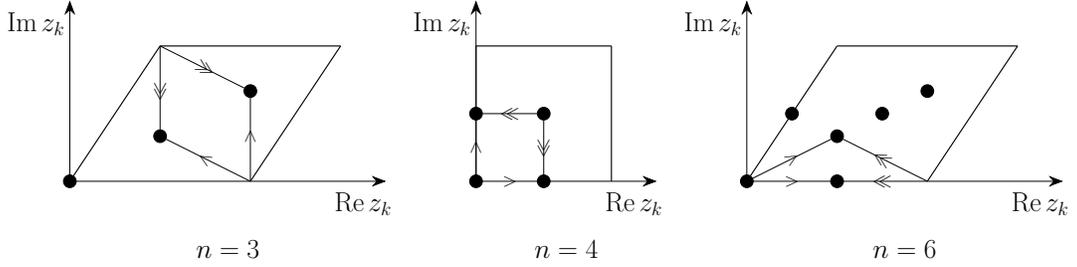
\begin{figure}
\centering
\scalebox{0.6}{
\begin{tikzpicture}
	\begin{pgfonlayer}{nodelayer}
		\node [style=none] (0) at (-8, 0) {};
		\node [style=none] (1) at (-8, 4) {};
		\node [style=none] (2) at (-1, 0) {};
		\node [style=none] (3) at (-6, 3) {};
		\node [style=none] (4) at (-4, 0) {};
		\node [style=none] (5) at (-2, 3) {};
		\node [style=CircleRed2] (6) at (-6, 1) {};
		\node [style=CircleRed2] (7) at (-4, 2) {};
		\node [style=CircleRed2] (8) at (-8, 0) {};
		\node [style=none] (9) at (-8.75, 3.5) {\Large Im\,$z_k$};
		\node [style=none] (10) at (-1.5, -0.5) {\Large Re\,$z_k$};
		\node [style=none] (11) at (1, 0) {};
		\node [style=none] (12) at (1, 4) {};
		\node [style=none] (13) at (5, 0) {};
		\node [style=none] (14) at (1, 3) {};
		\node [style=none] (15) at (4, 0) {};
		\node [style=none] (16) at (4, 3) {};
		\node [style=CircleRed2] (17) at (1, 1.5) {};
		\node [style=CircleRed2] (18) at (2.5, 1.5) {};
		\node [style=CircleRed2] (19) at (1, 0) {};
		\node [style=none] (20) at (0.25, 3.5) {\Large Im\,$z_k$};
		\node [style=none] (21) at (4.5, -0.5) {\Large Re\,$z_k$};
		\node [style=none] (22) at (7, 0) {};
		\node [style=none] (23) at (7, 4) {};
		\node [style=none] (24) at (14, 0) {};
		\node [style=none] (25) at (9, 3) {};
		\node [style=none] (26) at (11, 0) {};
		\node [style=none] (27) at (13, 3) {};
		\node [style=CircleRed2] (28) at (9, 1) {};
		\node [style=CircleRed2] (29) at (11, 2) {};
		\node [style=CircleRed2] (30) at (7, 0) {};
		\node [style=none] (31) at (6.25, 3.5) {\Large Im\,$z_k$};
		\node [style=none] (32) at (13.5, -0.5) {\Large Re\,$z_k$};
		\node [style=CircleRed2] (33) at (2.5, 0) {};
		\node [style=CircleRed2] (34) at (8, 1.5) {};
		\node [style=CircleRed2] (35) at (9, 0) {};
		\node [style=CircleRed2] (36) at (10, 1.5) {};
		\node [style=none] (37) at (-4.5, -1.5) {\Large $n=3$};
		\node [style=none] (38) at (3, -1.5) {\Large $n=4$};
		\node [style=none] (39) at (10.5, -1.5) {\Large $n=6$};
		\node [style=none] (53) at (3, -2) {};
		\node [style=none] (54) at (1.75, 0) {$>$};
		\node [style=none] (55) at (1, 0.75) {$\rotatebox{90}{$>$}$};
		\node [style=none] (56) at (2.5, 0.75) {\rotatebox{90}{$\ll$}};
		\node [style=none] (57) at (1.75, 1.5) {$\ll$};
		\node [style=none] (58) at (-5, 0.5) {\rotatebox{-30}{$<$}};
		\node [style=none] (59) at (-6, 2) {\rotatebox{90}{$\ll$}};
		\node [style=none] (62) at (8, 0.5) {\rotatebox{30}{$>$}};
		\node [style=none] (63) at (10, 0.5) {\rotatebox{-30}{$\ll$}};
		\node [style=none] (64) at (10, 0) {$\ll$};
		\node [style=none] (65) at (8, 0) {$>$};
		\node [style=none] (66) at (-4, 1) {\rotatebox{90}{$>$}};
		\node [style=none] (67) at (-5, 2.5) {\rotatebox{-30}{$\gg$}};
	\end{pgfonlayer}
	\begin{pgfonlayer}{edgelayer}
		\draw [style=ArrowLineRight] (0.center) to (1.center);
		\draw [style=ArrowLineRight] (0.center) to (2.center);
		\draw [style=ThickLine] (0.center) to (3.center);
		\draw [style=ThickLine] (4.center) to (5.center);
		\draw [style=ThickLine] (5.center) to (3.center);
		\draw [style=ArrowLineRight] (11.center) to (12.center);
		\draw [style=ArrowLineRight] (11.center) to (13.center);
		\draw [style=ThickLine] (11.center) to (14.center);
		\draw [style=ThickLine] (15.center) to (16.center);
		\draw [style=ThickLine] (16.center) to (14.center);
		\draw [style=ArrowLineRight] (22.center) to (23.center);
		\draw [style=ArrowLineRight] (22.center) to (24.center);
		\draw [style=ThickLine] (22.center) to (25.center);
		\draw [style=ThickLine] (26.center) to (27.center);
		\draw [style=ThickLine] (27.center) to (25.center);
		\draw (17) to (18);
		\draw (18) to (33);
		\draw (3.center) to (6);
		\draw (6) to (4.center);
		\draw (30) to (28);
		\draw (28) to (26.center);
		\draw (3.center) to (7);
		\draw (7) to (4.center);
	\end{pgfonlayer}
\end{tikzpicture}
}
\caption{Fundamental domains for $T^{\;\!2}_{\;\!k}/\Z_n$ for $n=3,4,6$. We give the sides to be identified by labelling these as standard with $>$ and $\gg$. Topologically $T^{\;\!2}_{\;\!k}/\Z_n$ is a sphere.}
\label{fig:T4Zk}
\end{figure}

Next, we determine the subset $F_k\subset T^2_k$ with non-trivial stabilizers. The points in $F_k$ with stabilizer subgroup $\Z_m\subset \Z_n$ where $m\,|\, n$ are grouped into orbits by the group action $\Z_n/\Z_m\cong \Z_{n/m}$. Each orbit is mapped to a single point in $ T^2_k/\Z_n$. With this the subset $F\subset T^4$ containing all points with non-trivial stabilizers takes the form:
\be
F=\lbbb (f_1,f_2) \,|\, f_i\in F_i \rbbb\subset T^4
\ee
and again points with identical stabilizer group are grouped into orbits mapped to a single orbifold point of A-type in $T^4/\Z_n$. The rank of the singularity is determined by the stabilizer group while the number of singular points is determined by the number of orbits. We have
\be
  \begin{aligned}
  ~F_k&=\lbbb 0,\frac{1}{2}, \frac{\tau_i}{2} ,\frac{1+\tau_i}{2} \rbbb  && (n=2) \\[5pt]
  ~F_k&=\lbbb 0,\frac{e^{\pi i /6}}{\sqrt{3}}, \frac{2e^{\pi i /6}}{\sqrt{3}} \rbbb  && (n=3) \\[5pt]
  ~F_k&=\lbbb 0,\frac{1}{2},\frac{i}{2},\frac{1+i}{2} \rbbb  && (n=4) \\[5pt]
  ~F_k&=\lbbb 0,\frac{e^{\pi i /6}}{\sqrt{3}}, \frac{2e^{\pi i /6}}{\sqrt{3}}, \frac{1}{2},\frac{e^{\pi i /3}}{2}, \frac{1+e^{\pi i /3}}{2}  \rbbb  \quad  && (n=6)
  \end{aligned}
\ee
which we depict for $n=3,4,6$ in figure \ref{fig:FundamentalDomains}. In fact the stabilizer subgroups of each point are determined uniquely by the integer $m$. Straightforwardly we now determine the fundamental domain of $T^2_k/\Z_n$ as a subset of the fundamental domain of $T^2_k$ (see figure \ref{fig:T4Zk}). It follows that
\be
T^2_k/\Z_n\cong \P^1_k
\ee
is topologically a sphere with $4,3,3,3$ orbifold points for $k=2,3,4,6$ respectively. Grouping elements in $F$ with non-trivial stabilizer subgroups into orbits (see Appendix \ref{app:AdditionalCyclicStuff} for details),
we next determine the A-type singularities of $T^4/\Z_n$ to be:
\be\label{eq:ADESingularities}
\oplus_{i\:\! } \mathfrak{g}_{i }=\left\{
  \begin{aligned}
   ~&A_1^{16} && (n=2) \\
    ~&A_2^9&& (n=3)\\
    ~& A_3^4\oplus A_1^6  && (n=4)    \\
    ~& A_5\oplus A_2^4\oplus A_1^5 \quad && (n=6)\,.
    \end{aligned}
  \right.
 \ee

\begin{figure}
\centering
\scalebox{0.6}{
\begin{tikzpicture}
	\begin{pgfonlayer}{nodelayer}
		\node [style=none] (0) at (0, 0) {};
		\node [style=none] (1) at (1, 1) {};
		\node [style=none] (2) at (1, -1) {};
		\node [style=none] (3) at (-1, -1) {};
		\node [style=none] (4) at (-1, 1) {};
		\node [style=none] (5) at (5, -1) {};
		\node [style=none] (6) at (4, -1) {};
		\node [style=none] (7) at (3, -1) {};
		\node [style=none] (8) at (6, -1) {};
		\node [style=none] (9) at (7, -1) {};
		\node [style=none] (10) at (5, 1) {};
		\node [style=none] (11) at (5, 0) {};
		\node [style=none] (12) at (12, -1) {};
		\node [style=none] (13) at (11, -1) {};
		\node [style=none] (14) at (10, -1) {};
		\node [style=none] (15) at (13, -1) {};
		\node [style=none] (16) at (14, -1) {};
		\node [style=none] (17) at (15, -1) {};
		\node [style=none] (18) at (12, 0) {};
		\node [style=none] (19) at (9, -1) {};
		\node [style=none] (20) at (22, -1) {};
		\node [style=none] (21) at (21, -1) {};
		\node [style=none] (22) at (20, -1) {};
		\node [style=none] (23) at (23, -1) {};
		\node [style=none] (24) at (24, -1) {};
		\node [style=none] (25) at (17, -1) {};
		\node [style=none] (26) at (22, 0) {};
		\node [style=none] (27) at (19, -1) {};
		\node [style=none] (28) at (18, -1) {};
		\node [style=CircleRed] (29) at (-1, 1) {};
		\node [style=CircleRed] (30) at (0, 0) {};
		\node [style=CircleRed] (31) at (1, 1) {};
		\node [style=CircleRed] (32) at (1, -1) {};
		\node [style=CircleRed] (33) at (-1, -1) {};
		\node [style=CircleRed] (34) at (24, -1) {};
		\node [style=CircleRed] (35) at (23, -1) {};
		\node [style=CircleRed] (36) at (22, -1) {};
		\node [style=CircleRed] (37) at (22, 0) {};
		\node [style=CircleRed] (38) at (21, -1) {};
		\node [style=CircleRed] (39) at (20, -1) {};
		\node [style=CircleRed] (40) at (19, -1) {};
		\node [style=CircleRed] (41) at (18, -1) {};
		\node [style=CircleRed] (42) at (17, -1) {};
		\node [style=CircleRed] (43) at (9, -1) {};
		\node [style=CircleRed] (44) at (10, -1) {};
		\node [style=CircleRed] (45) at (11, -1) {};
		\node [style=CircleRed] (46) at (12, -1) {};
		\node [style=CircleRed] (47) at (12, 0) {};
		\node [style=CircleRed] (48) at (13, -1) {};
		\node [style=CircleRed] (49) at (14, -1) {};
		\node [style=CircleRed] (50) at (15, -1) {};
		\node [style=CircleRed] (51) at (3, -1) {};
		\node [style=CircleRed] (52) at (4, -1) {};
		\node [style=CircleRed] (53) at (5, -1) {};
		\node [style=CircleRed] (54) at (6, -1) {};
		\node [style=CircleRed] (55) at (7, -1) {};
		\node [style=CircleRed] (56) at (5, 0) {};
		\node [style=CircleRed] (57) at (5, 1) {};
		\node [style=none] (58) at (0, 2) {\Large $D_4$};
		\node [style=none] (59) at (5, 2) {\Large $E_6$};
		\node [style=none] (60) at (12, 2) {\Large $E_7$};
		\node [style=none] (61) at (20.5, 2) {\Large $E_8$};
		\node [style=none] (62) at (-1, 1) {\footnotesize\color{white} $1$};
		\node [style=none] (63) at (1, 1) {\footnotesize\color{white} $1$};
		\node [style=none] (64) at (-1, -1) {\footnotesize\color{white} $1$};
		\node [style=none] (65) at (1, -1) {\footnotesize\color{white} $1$};
		\node [style=none] (66) at (0, 0) {\footnotesize\color{white} $2$};
		\node [style=none] (67) at (5, -1) {\footnotesize\color{white} $3$};
		\node [style=none] (68) at (4, -1) {\footnotesize\color{white} $2$};
		\node [style=none] (69) at (3, -1) {\footnotesize\color{white} $1$};
		\node [style=none] (70) at (6, -1) {\footnotesize\color{white} $2$};
		\node [style=none] (71) at (7, -1) {\footnotesize\color{white} $1$};
		\node [style=none] (72) at (5, 0) {\footnotesize\color{white} $2$};
		\node [style=none] (73) at (5, 1) {\footnotesize\color{white} $1$};
		\node [style=none] (74) at (9, -1) {\footnotesize\color{white} $1$};
		\node [style=none] (75) at (10, -1) {\footnotesize\color{white} $2$};
		\node [style=none] (76) at (11, -1) {\footnotesize\color{white} $3$};
		\node [style=none] (77) at (12, -1) {\footnotesize\color{white} $4$};
		\node [style=none] (78) at (13, -1) {\footnotesize\color{white} $3$};
		\node [style=none] (79) at (14, -1) {\footnotesize\color{white} $2$};
		\node [style=none] (80) at (15, -1) {\footnotesize\color{white} $1$};
		\node [style=none] (81) at (12, 0) {\footnotesize\color{white} $2$};
		\node [style=none] (82) at (17, -1) {\footnotesize\color{white} $1$};
		\node [style=none] (83) at (18, -1) {\footnotesize\color{white} $2$};
		\node [style=none] (84) at (19, -1) {\footnotesize\color{white} $3$};
		\node [style=none] (85) at (20, -1) {\footnotesize\color{white} $4$};
		\node [style=none] (86) at (21, -1) {\footnotesize\color{white} $5$};
		\node [style=none] (87) at (22, 0) {\footnotesize\color{white} $3$};
		\node [style=none] (88) at (22, -1) {\footnotesize\color{white} $6$};
		\node [style=none] (89) at (23, -1) {\footnotesize\color{white} $4$};
		\node [style=none] (90) at (24, -1) {\footnotesize\color{white} $2$};
		\node [style=none] (94) at (2.75, -1.5) {};
		\node [style=none] (95) at (3.5, -1.75) {};
		\node [style=none] (96) at (4.25, -1.5) {};
		\node [style=none] (97) at (5.5, 1.25) {};
		\node [style=none] (98) at (5.75, 0.5) {};
		\node [style=none] (99) at (5.5, -0.25) {};
		\node [style=none] (103) at (5.75, -1.5) {};
		\node [style=none] (104) at (6.5, -1.75) {};
		\node [style=none] (105) at (7.25, -1.5) {};
		\node [style=none] (106) at (8.75, -1.5) {};
		\node [style=none] (107) at (10, -1.75) {};
		\node [style=none] (108) at (11.25, -1.5) {};
		\node [style=none] (109) at (12.75, -1.5) {};
		\node [style=none] (110) at (14, -1.75) {};
		\node [style=none] (111) at (15.25, -1.5) {};
		\node [style=none] (112) at (16.75, -1.5) {};
		\node [style=none] (113) at (19, -1.75) {};
		\node [style=none] (114) at (21.25, -1.5) {};
		\node [style=none] (115) at (22.75, -1.5) {};
		\node [style=none] (116) at (23.5, -1.75) {};
		\node [style=none] (117) at (24.25, -1.5) {};
		\node [style=none] (118) at (0.625, 0.05) {$\P^1$};
		\node [style=none] (119) at (22, -1.625) {$\P^1$};
		\node [style=none] (120) at (12, -1.625) {$\P^1$};
		\node [style=none] (121) at (5, -1.625) {$\P^1$};
		\node [style=none] (122) at (-1.625, 1) {$A_1$};
		\node [style=none] (123) at (1.625, 1) {$A_1$};
		\node [style=none] (124) at (1.625, -1) {$A_1$};
		\node [style=none] (125) at (-1.625, -1) {$A_1$};
		\node [style=none] (126) at (6.25, 0.5) {$A_2$};
		\node [style=none] (127) at (6.5, -2.25) {$A_2$};
		\node [style=none] (128) at (3.5, -2.25) {$A_2$};
		\node [style=none] (129) at (12, 0.625) {$A_1$};
		\node [style=none] (130) at (10, -2.25) {$A_3$};
		\node [style=none] (131) at (14, -2.25) {$A_3$};
		\node [style=none] (132) at (19, -2.25) {$A_5$};
		\node [style=none] (133) at (23.5, -2.25) {$A_2$};
		\node [style=none] (134) at (22, 0.625) {$A_1$};
		\node [style=none] (135) at (6.5, -3) {};
	\end{pgfonlayer}
	\begin{pgfonlayer}{edgelayer}
		\draw [style=ThickLine] (4.center) to (0.center);
		\draw [style=ThickLine] (0.center) to (1.center);
		\draw [style=ThickLine] (0.center) to (2.center);
		\draw [style=ThickLine] (0.center) to (3.center);
		\draw [style=ThickLine] (10.center) to (5.center);
		\draw [style=ThickLine] (7.center) to (9.center);
		\draw [style=ThickLine] (19.center) to (17.center);
		\draw [style=ThickLine] (18.center) to (12.center);
		\draw [style=ThickLine] (26.center) to (20.center);
		\draw [style=ThickLine] (24.center) to (25.center);
		\draw [in=90, out=-90] (94.center) to (95.center);
		\draw [in=-90, out=90] (95.center) to (96.center);
		\draw [in=-180, out=0] (97.center) to (98.center);
		\draw [in=0, out=-180] (98.center) to (99.center);
		\draw [in=90, out=-90] (103.center) to (104.center);
		\draw [in=-90, out=90] (104.center) to (105.center);
		\draw [in=90, out=-90, looseness=0.75] (106.center) to (107.center);
		\draw [in=-90, out=90, looseness=0.75] (107.center) to (108.center);
		\draw [in=90, out=-90] (109.center) to (110.center);
		\draw [in=-90, out=90] (110.center) to (111.center);
		\draw [in=90, out=-90, looseness=0.50] (112.center) to (113.center);
		\draw [in=-90, out=90, looseness=0.50] (113.center) to (114.center);
		\draw [in=90, out=-90] (115.center) to (116.center);
		\draw [in=-90, out=90] (116.center) to (117.center);
	\end{pgfonlayer}
\end{tikzpicture}
}
\caption{Extended ADE Dynkin diagrams of type $D_4,E_6,E_7,E_8$ and their Kac labels. The node marked $\P^1$ is the proper transform of the exceptional fibers in $T^4/\Z_n \rightarrow T^2_k/\Z_n$. Contracting exceptional curves is equivalent of collapsing the $A$-type subgraphs marked in the figure which for $D,E$ type results in a sphere with $4,3,3,3$ singular points respectively. The Kac label of $\P^1$ is geometrized as $n$.}
\label{fig:ExtendedDynkin}
\end{figure}
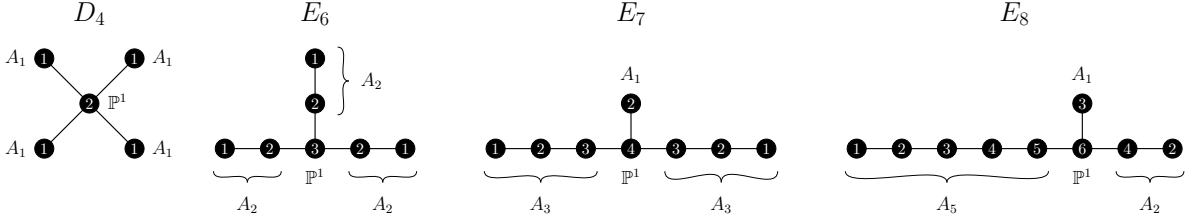

Next view $T^4/\Z_n$ as a fibration. Consider the double elliptic pencil
\be\label{eq:fib}
\pi_k\,:~ T^4/\Z_n~\rightarrow ~T^2_k/\Z_n\equiv \P^1_k
\ee
with generic two-torus fiber $\mathcal{F}_k$ which at orbifold points $z$ of the base is folded to the sphere $\mathcal{F}_{k,z}=\pi_k^{-1}(z)$. We collect such exceptional fibers into the set $S_k=\lbbb \mathcal{F}_{k,z}\rbbb$ which in the crepantly resolved geometry lift to $S_k'$ containing collections of rational curves whose adjacency matrix is captured by an extended Dynkin diagram, explicitly\footnote{For example consider the $E_8$ summand for $n=6$ which projects to $0 \in T^2_k/\Z_6$. In the singular limit it contracts to a sphere $\P^1$ with three orbifold points of type $A_5,A_2,A_1$. The minimal resolution correspondingly attaches chains of $5,2,1$ spheres to this central $\P^1$ as shown on the right of figure \ref{fig:ExtendedDynkin}. }
\be\label{eq:ExceptionalCurves}
{S}'_k\cong \left\{
  \begin{aligned}
   ~&D_4\oplus D_4 \oplus D_4 \oplus D_4 = {\rm I}_0^*\oplus {\rm I}_0^* \oplus {\rm I}_0^* \oplus {\rm I}_0^* \qquad  && (n=2) \\
    ~&E_6\oplus E_6 \oplus E_6 ={\rm IV}^*\oplus {\rm IV}^* \oplus {\rm IV}^*&& (n=3)\\
    ~ &E_7\oplus E_7\oplus D_4 = \rm{III}^*\oplus\rm{III}^*\oplus {\rm I}_0^*\qquad && (n=4) \\
     ~&E_8\oplus E_6 \oplus D_4=  {\rm II}^* \oplus {\rm IV}^* \oplus {\rm I}_0^* \qquad && (n=6)\,.
    \end{aligned}
  \right.
 \ee
In figure \ref{fig:ExtendedDynkin} we depict the relevant extended Dynkin diagrams and mark the collection of nodes resulting from resolutions of the A-type singularities listed in \eqref{eq:ADESingularities}.

Let us now turn to the explicit form of the gauge groups for different choices of orbifolds. We collect much of the discussion in the appendices, including combinatorial perspectives based on affine structures, Appendix \ref{app:AdditionalCyclicStuff} and \ref{app:morequotients}, and equivariant cohomology computations, Appendix \ref{app:nonAbelian}. The equivariant cohomology computations are readily applicable to non-Abelian group actions which we analyze in detail in Appendix \ref{app:nonAbelian}. We note here, that the analysis based on equivariant cohomology carries less physical intuition and obscures data contained in the maps of the key exact sequence \eqref{eq:MVS}. However, often this approach is computationally more convenient.

\medskip

\noindent {\bf Case $X=T^4/\Z_2$\,:} We find the exact sequence
\be \ba
0~&\xrightarrow[\text{}]{\;\imath_2\;} ~\Z^6  ~\xrightarrow[\text{}]{\;\jmath_2\;}  ~\Z^6\oplus \Z^5_2 ~\xrightarrow[\text{}]{\; \partial_2 \;}  ~  \Z_2^{16} ~\xrightarrow[\text{}]{\;\imath_1\;} ~\Z_2^5  ~\xrightarrow[\text{}]{\;\jmath_1\;}  ~0\,.
\ea \ee
The continuous gauge group is
\be \label{eq:SugraGroup}
G_{\textnormal{full}}=\frac{\big(SU(2)^{16}/\Z_{2}^5\:\! \big)\times U(1)^6}{{\Z_2^6}}
\ee
The quotient by $\mathbb{Z}_2^5$ on the non-Abelian group $\widetilde{G}_{\mathrm{loc}}=SU(2)^{16}$ is determined by the rows of the matrix $M_{T^4/\Z_2}$ given in \eqref{eq:CoeffMatrix}. The quotient by $\mathbb{Z}_2^6$ including the Abelian factors follows from \eqref{eq:KeyRelation}.

\medskip

\noindent {\bf Case $X=T^4/\Z_3$\,:} We find the exact sequence
\be \ba
0~&\xrightarrow[\text{}]{\;\imath_2\;} ~\Z^4  ~\xrightarrow[\text{}]{\;\jmath_2\;}  ~\Z^4\oplus \Z^3_3 ~\xrightarrow[\text{}]{\; \partial_2 \;}  ~  \Z_3^{9} ~\xrightarrow[\text{}]{\;\imath_1\;} ~\Z_3^3  ~\xrightarrow[\text{}]{\;\jmath_1\;}  ~0\,.
\ea \ee
The continuous gauge group is
\be \label{eq:SugraGroup2}
G_{\textnormal{full}}=\frac{\big(SU(3)^{9}/\Z_3^3\big)\times U(1)^4}{\Z_3^3}\,,
\ee
The quotient by $\Z_3^3$ on the non-Abelian group $\widetilde{G}_\mathrm{loc}=SU(3)^9$ is determined by the rows of the matrix $M_{T^4/\Z_3}$ given in \eqref{eq:T4Z3Matrix}. The quotient by $\mathbb{Z}_3^3$ including the Abelian factors follows from a relation analogous to \eqref{eq:KeyRelation}.

\medskip

\noindent {\bf Case $X=T^4/\Z_4$\,:} We find the exact sequence
\be \ba
0~&\xrightarrow[\text{}]{\;\imath_2\;} ~\Z^4  ~\xrightarrow[\text{}]{\;\jmath_2\;}  ~\Z^4\oplus \Z_4\oplus \Z_2^2 ~\xrightarrow[\text{}]{\; \partial_2 \;}  ~   \Z^4_4\oplus \Z^6_2~\xrightarrow[\text{}]{\;\imath_1\;} ~ \Z_4\oplus \Z_2^2  ~\xrightarrow[\text{}]{\;\jmath_1\;}  ~0\,.
\ea \ee
The continuous gauge group is
\be
G_{\textnormal{full}}=\frac{([SU(4)^4\times SU(2)^{6}]/\Z_4\times \Z_2^2)\times U(1)^4}{ \Z_4^2\times \Z_2^2}.
\ee
The quotient by $\Z_4 \times \mathbb{Z}_{2}^{2}$ on the non-Abelian group $\widetilde{G}_\mathrm{loc}=SU(4)^4\times SU(2)^{6}$ is determined by the rows of the matrix $M_{T^4/\Z_4}$ given in \eqref{eq:T4Z4Matrix}. Explicit computations show that this quotient actually only acts on the $SU(4)^4$ factor (see Appendix \ref{app:nonAbelian}). The quotient by $\Z_4^2\times \Z_2^2$ including the Abelian factors follows from a relation analogous to \eqref{eq:KeyRelation}.

\medskip

\noindent {\bf Case $X=T^4/\Z_6$\,:} We find the exact sequence
\be \ba
0~&\xrightarrow[\text{}]{\;\imath_2\;} ~\Z^4  ~\xrightarrow[\text{}]{\;\jmath_2\;}  ~\Z^4\oplus \Z_3\oplus \Z_2 ~\xrightarrow[\text{}]{\; \partial_2 \;}  ~ \Z_6\oplus \Z_3^4 \oplus \Z_2^5 ~\xrightarrow[\text{}]{\;\imath_1\;} ~ \Z_6  ~\xrightarrow[\text{}]{\;\jmath_1\;}  ~0\,.
\ea \ee
The continuous gauge group is
\be
G_{\textnormal{full}}=\frac{([SU(6) \times SU(3)^{4} \times SU(2)^{5}]/\Z_3\times \Z_2)\times U(1)^4}{\Z_6^3\times \Z_2}.
\ee
The quotient by $\Z_3 \times \Z_2$ on the non-Abelian group $\widetilde{G}_\mathrm{loc} =SU(6) \times SU(3)^{4} \times SU(2)^{5}$ follows from the rows of the matrix $M_{T^4/\Z_6}$ given in \eqref{eq:T4Z6Matrix}. The quotient by $\Z_6^3\times \Z_2$ including the Abelian factors follows from a relation analogous to \eqref{eq:KeyRelation}.

\subsection{Mordell-Weil Group and the Shioda Map}
\label{sec:EllipticK3}

The next class of examples we consider are elliptically fibered K3 surfaces $X\rightarrow \P^1$ with section $\sigma$ and generic fiber $\mathbb{E}$. The crepant resolution $X' \rightarrow X$ gives the fibration $\pi: X' \rightarrow \P^1$.

In the limit where the volume of the elliptic fiber degenerates to zero size, we obtain a dual description in terms of F-theory on $S^1 \times X$, with the $S^1$ decompactifying, namely an 8D background. Our considerations are topological, and thus somewhat more coarse than the delicate factorization conditions which need to be imposed in an F-theory model to ensure the existence of various $U(1)$ factors (additional sections of the Mordell-Weil group) and finite quotients (multi-sections of the genus-one model). We can view our M-theory analysis as providing a set of candidate enhancement points which might be reached by a further tuning of the associated Weierstrass model. Indeed, from the perspective of the 8D model compactified on a circle, we are allowing position dependent scalar vevs which only retain 7D Lorentz invariance. As such, less tuning is required.

It is well-known in the F-theory literature that the global form of the gauge group is determined by the Mordell-Weil group ${\rm MW}(\pi)$ \cite{Grimm:2010ez, Grimm2011, Morrison:2012ei, Cvetic:2013nia, Borchmann:2013jwa, Morrison:2014era, Braun:2014oya, Mayrhofer2014, Braun:2014nva, Mayrhofer:2014laa, Cvetic:2017epq, Cvetic:2021sxm} (for reviews see \cite{Cvetic:2018bni, Weigand:2018rez}). Instead of discussing explicit elliptic examples we show how the formalism laid out there arises as a specialization of section \ref{sec:NonAb}. In particular, we explain how this data is captured by the exact sequence \eqref{eq:MVS}.

The argument proceeds via the group homomorphism $\Phi$ mapping from the Mordell-Weil group into the N\'eron-Severi group of the fibration
\be
\Phi\,:\quad \textnormal{MW}(\pi) ~ \rightarrow~  \textnormal{NS}(X')_{\mathbb{Q}}\,.
\ee
known as the Shioda map \cite{10.3792/pjaa.65.268}. The group $ \textnormal{NS}(X')$  is the group of divisors modulo algebraic equivalence and intersections between divisors gives $  \textnormal{NS}(X') $ the structure of an integral lattice, $  \textnormal{NS}(X')_\mathbb{Q}\equiv   \textnormal{NS}(X')\otimes \mathbb{Q}$ denotes the extension of this lattice to the rational numbers.

The fibration and resolution determine the so-called trivial sublattice $T$ of $ \textnormal{NS}(X')$ which is generated by the zero section and all irreducible components of the elliptic fibers, that is the exceptional curves, and the fiber itself
\be
T=\langle \sigma , \mathbb{E} \rangle \oplus L_E\subset  \textnormal{NS}(X') \,.
\ee
The Shioda map satisfies $\textnormal{Im}\,\Phi \perp T$ \cite{10.3792/pjaa.65.268} and Shioda further proved the splitting $ \textnormal{NS}(X')_{\mathbb{Q}} \cong T \oplus_\perp \textnormal{Im}\,\Phi$ resulting in the identification
\be
\textnormal{Free\,MW}(\pi) \cong  \textnormal{Im}\,\Phi\cong   \textnormal{NS}(X')_\mathbb{Q} /T_\mathbb{Q} \,.
\ee
Here $\textnormal{Free}\,\textnormal{MW}(\pi)=\textnormal{MW}(\pi)/  \textnormal{Tor}\,\textnormal{MW}(\pi) $. Let us therefore consider $\textnormal{Im}\,\Phi $ mod $T$ more closely. Shioda explicitly defined \smallskip
\be
\Phi(s)=s-\sigma-\lb s\cdot \sigma -\sigma\cdot \sigma\rb \mathbb{E}-\sum_i \lb \mathbf{e}_i^t\rb A_i^{-1} \lb s\cdot \mathbf{e}_i\rb
\ee
where $i$ runs over all exceptional reducible fibers, each with $n_i+1$ components, which are labelled by the indices $\alpha,\beta=0,\dots,n_i$ and where the $0$-th label identifies the affine node of the Dynkin diagram associated with the $i$-th exceptional fiber, further
\be\ba
\mathbf{e}_i&=(e_{i1},\dots, e_{in_i})\\
s\cdot \mathbf{e}_i&=(s\cdot e_{i1},\dots, s\cdot e_{in_i})\\
(A_i)_{\alpha\beta}&=e_{i\alpha}\cdot e_{i\beta}\,, \qquad \alpha,\beta\neq 0\,,
\ea \ee
where $e_{i\alpha}$ is the $\alpha$-th exceptional curve of the $i$-th exceptional fiber and $A_i$ is minus the Cartan matrix associated with the $i$-th exceptional fiber, i.e., the intersection matrix of $e_{i\alpha}$ for fixed index $i$. The image of the Shioda map modulo the trivial lattice is therefore\smallskip
\be
\Phi(s)=s-\sum_i \lb \mathbf{e}_i^t\rb A_i^{-1} \lb s\cdot \mathbf{e}_i\rb=s-\sum_i \sum_{\alpha\beta}  e_{i\beta}  \lb A_i^{-1} \rb_{\alpha\beta}  s\cdot e_{i\alpha}\quad \textnormal{mod }T
\ee
where mod $T$ determines the over all coefficients multiplying the exceptional curves $e_{i\beta}$ to be mod 1. The torsion subgroup of the Mordell-Weil group is in the kernel of the Shioda map, as it is a group homomorphism over $\mathbb{Q}$, and therefore we have for $t\in \textnormal{Tor\,MW}(\pi)$
\be\label{eq:ShiodaTorsion}
t=\sum_i \lb \mathbf{e}_i^t\rb A_i^{-1} \lb s\cdot  \mathbf{e}_i\rb=\sum_i \sum_{\alpha\beta}  e_{i\beta}  \lb A_i^{-1} \rb_{\alpha\beta}  s\cdot e_{i\alpha}\quad \textnormal{mod }1\,.
\ee
Next, we compute the inverse matrices $A_i^{-1}$ mod 1. For example, we have for Kodaira type IV$^*$ fibers with Lie algebra $E_6$ the inverse
\be
A_{E_6}^{-1}=\left(
\begin{array}{cccccc}
 \frac{2}{3} & \frac{1}{3} & 0 & \frac{2}{3} & \frac{1}{3} & 0 \\
 \frac{1}{3} & \frac{2}{3} & 0 & \frac{1}{3} & \frac{2}{3} & 0 \\
 0 & 0 & 0 & 0 & 0 & 0 \\
 \frac{2}{3} & \frac{1}{3} & 0 & \frac{2}{3} & \frac{1}{3} & 0 \\
 \frac{1}{3} & \frac{2}{3} & 0 & \frac{1}{3} & \frac{2}{3} & 0 \\
 0 & 0 & 0 & 0 & 0 & 0 \\
\end{array}
\right)\qquad \textnormal{mod}\,1\,.
\ee
By inspection we clearly see that rows and columns are multiples of the thimble
\be
\T_{E_6}=\frac{1}{3}e_{1}+\frac{2}{3}e_{2}+\frac{1}{3}e_{4}+\frac{2}{3}e_{5}
\ee
with coefficients mod 1. For example, the first row and column are  $2\mathfrak{T}_{E_6}$ mod 1, and the second row and column are $\mathfrak{T}_{E_6}$.
Here we follow the numbering conventions laid out in figure \ref{fig:Thimbles}. This holds true for all other ADE-types across all ranks and we derive that $t\in \textnormal{Tor}\,\textnormal{MW}(\pi)$ is related by the Shioda map with an integral linear combinations of thimbles
\be\label{eq:sumofthimbles}
t=\sum_i m_{(t),i}^{} \Tp_{\!c,i} \qquad \textnormal{mod}\,1\,, ~ m_{(t),i}^{} \in \mathbb{Z}_{N_i}\,.
\ee
With this we find compact representatives of ADE thimbles to determine, in each fiber, the possible fractional linear combination of exceptional curves the Shioda map can map onto. Clearing denominators we find that the collection of integers $m_{(t),i}$ are valued in $\mathbb{Z}_{N_i}$ where $N_i$ is the smallest positive integer such that $N_{i} \Tp_{\!c,i} =0$ mod 1.

With this we further conclude $ \textnormal{Tor}\,\textnormal{MW}(\pi)=\overline L_E/L_E$, replicating \eqref{eq:Saturation} and making contact with our previous formalism via \eqref{eq:Lattice}. We leave a purely F-theoretic analysis of the full Mordell-Weil group, including the free part, as an exercise to the reader.

Let us note how the two sequences
\be \label{eq:TwoSequences}\ba
0~&\xrightarrow[\text{}]{} ~H_2(X^\circ )  ~\xrightarrow[\text{}]{\;\jmath_2\;}   ~H_2(X) ~\xrightarrow[\text{}]{}  ~  \textnormal{coker}\,\jmath_2  ~\xrightarrow[\text{}]{}  ~0\\
0~&\xrightarrow[\text{}]{} ~  \textnormal{Free}\,\textnormal{MW}(\pi)  ~\xrightarrow[\text{}]{}  ~ \textnormal{MW}(\pi) ~\xrightarrow[\text{}]{\;|_{\textnormal{Tor}} \;}  ~  \textnormal{Tor}\,\textnormal{MW}(\pi)  ~\xrightarrow[\text{}]{}  ~0
\ea \ee
differ. First, the rank of $H_2(X^\circ)$ sets an upper bound for the number of Abelian symmetries that can be tuned in the F-theory model. Second, let us consider the mapping of individual generators when a tuning has occurred. In this case we have a class in $H_2(X^\circ)$ which contains a representative that is horizontal with respect to the fibration $\pi$. Both groups clearly count ``extra" $U(1)$'s. However, $\jmath_2$ maps into $H_2(X)$ with a cokernel generically larger than $ \textnormal{Tor}\,\textnormal{MW}(\pi)$ and we recall that $\textnormal{coker}\,\jmath_2$ contains information on how $U(1)$ charges are to be normalized, see section \ref{sec:NonAb}.

To explain this distinction from the F-theory perspective, consider a rational section $s$ generating a free class of the Mordell-Weil group and an exceptional fiber $F_i$ of $X'$. The section $s$ can then intersect $F_i$ in an exceptional curve which is not the affine node. An integer multiple $ns$ however can intersect a different exceptional curve and the smallest positive integer $N_i$ such that $N_is$ intersects the affine node then results in a cycle which, when the exceptional curves are contracted, does not pass through the $i$-th singularity. Taking the lowest common multiple $N=\textnormal{lcm}_{i}(N_i)$ we find that $Ns$ passes through no ADE singularity giving an element of $H_2(\partial X^\circ)$. The cokernel of $\jmath_2$ modulo $\textnormal{Tor}\,H_2(X)$ then determines the list of all such integers $N$ via the order of its subgroups.

Shioda, realizing this structure, introduced in  \cite{10.3792/pjaa.65.268}  the group $\textnormal{MW}(\pi)^0$, the subgroup of the Mordell-Weil group of finite index consisting of those sections which pass through the same irreducible component in each fiber as the zero section $\sigma$. He further introduced the essential sublattice of NS$(X')$ as the orthogonal complement of the trivial lattice  $T$
\be
\textnormal{Ess}(\pi)=T^\perp\,.
\ee
With this lattice we have $\textnormal{MW}(\pi)^0\cong \textnormal{Ess}(\pi)$ and we have that the normalization data \eqref{eq:normalizations}, restricted to the tuned Abelian gauge symmetries, takes the form
in our notation
\be
\textnormal{Ess}(\pi)^*\big/\,\textnormal{Ess}(\pi)
\ee
where $\textnormal{Ess}(\pi)^*$ is the lattice dual to $\textnormal{Ess}(\pi)$.

\subsection{Example: Sen Limits for $T^4/\Z_2$}

The metric moduli space of K3 surfaces is connected and we can therefore pass from the Kummer surface $T^4/\Z_2$ to its resolution and then to an elliptic degeneration limit described by Sen \cite{Sen:1996vd, Sen:1997gv}. We track these steps in topology and demonstrate how our formalism is smoothly deformed onto elliptic data in this example. We use the notation of section \ref{sec:Kummer}.

Consider the fibration $\pi_\alpha :X\rightarrow \P^1_\alpha$ which has 4 sections $F_{I_\beta}$ satisfying $\pi_\alpha(F_{I_\beta})=\P^1_\alpha$ for all $I_\beta \in \Z_2^2$. Consider the resolution $X'\rightarrow \P^1_\alpha$ then the proper transforms of $ F_{I_\beta}$ intersects 4 exceptional curves $e_{I_\alpha,I_\beta}$ which project to 4 distinct points in $\P^1_\alpha$ labelled by $I_\alpha \in \Z_2^2$.

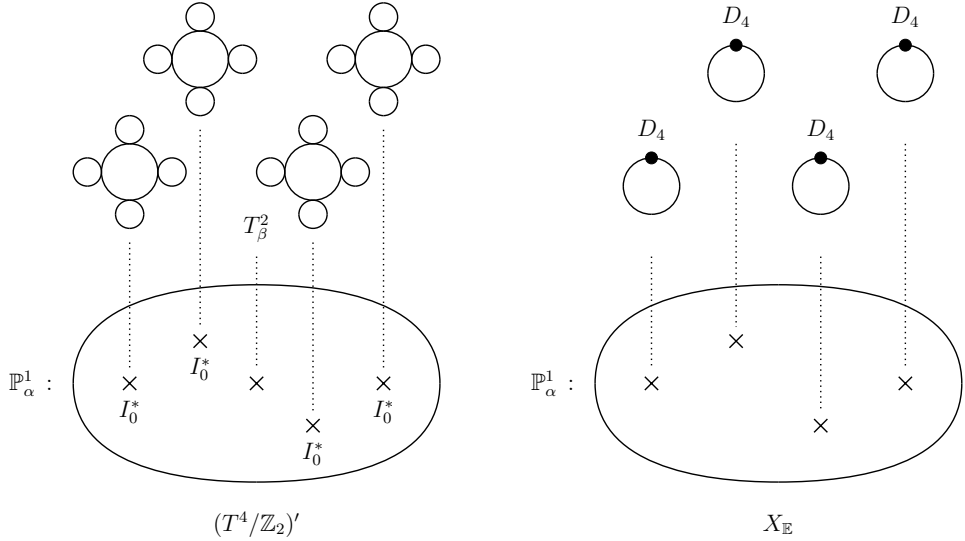
\begin{figure}
\centering
\scalebox{0.75}{
\begin{tikzpicture}
	\begin{pgfonlayer}{nodelayer}
		\node [style=none] (0) at (0, 1.75) {};
		\node [style=none] (1) at (-3.25, 0) {};
		\node [style=none] (2) at (0, -1.75) {};
		\node [style=none] (3) at (3.25, 0) {};
		\node [style=NodeCross] (4) at (-1, 0.75) {};
		\node [style=NodeCross] (5) at (1, -0.75) {};
		\node [style=NodeCross] (6) at (2.25, 0) {};
		\node [style=NodeCross] (7) at (-2.25, 0) {};
		\node [style=none] (8) at (-1, 0.25) {$I_0^*$};
		\node [style=none] (9) at (-2.25, -0.5) {$I_0^*$};
		\node [style=none] (10) at (1, -1.25) {$I_0^*$};
		\node [style=none] (11) at (2.25, -0.5) {$I_0^*$};
		\node [style=none] (12) at (-2.25, 4.25) {};
		\node [style=none] (13) at (-2.25, 3.25) {};
		\node [style=none] (14) at (-1.75, 3.75) {};
		\node [style=none] (15) at (-2.75, 3.75) {};
		\node [style=none] (16) at (-1.5, 4) {};
		\node [style=none] (17) at (-1.5, 3.5) {};
		\node [style=none] (18) at (-1.25, 3.75) {};
		\node [style=none] (19) at (-1.75, 3.75) {};
		\node [style=none] (20) at (-2.25, 4.75) {};
		\node [style=none] (21) at (-2.25, 4.25) {};
		\node [style=none] (22) at (-2, 4.5) {};
		\node [style=none] (23) at (-2.5, 4.5) {};
		\node [style=none] (24) at (-2.25, 3.25) {};
		\node [style=none] (25) at (-2.25, 2.75) {};
		\node [style=none] (26) at (-2, 3) {};
		\node [style=none] (27) at (-2.5, 3) {};
		\node [style=none] (28) at (-3, 4) {};
		\node [style=none] (29) at (-3, 3.5) {};
		\node [style=none] (30) at (-2.75, 3.75) {};
		\node [style=none] (31) at (-3.25, 3.75) {};
		\node [style=none] (32) at (-1, 6.25) {};
		\node [style=none] (33) at (-1, 5.25) {};
		\node [style=none] (34) at (-0.5, 5.75) {};
		\node [style=none] (35) at (-1.5, 5.75) {};
		\node [style=none] (36) at (-0.25, 6) {};
		\node [style=none] (37) at (-0.25, 5.5) {};
		\node [style=none] (38) at (0, 5.75) {};
		\node [style=none] (39) at (-0.5, 5.75) {};
		\node [style=none] (40) at (-1, 6.75) {};
		\node [style=none] (41) at (-1, 6.25) {};
		\node [style=none] (42) at (-0.75, 6.5) {};
		\node [style=none] (43) at (-1.25, 6.5) {};
		\node [style=none] (44) at (-1, 5.25) {};
		\node [style=none] (45) at (-1, 4.75) {};
		\node [style=none] (46) at (-0.75, 5) {};
		\node [style=none] (47) at (-1.25, 5) {};
		\node [style=none] (48) at (-1.75, 6) {};
		\node [style=none] (49) at (-1.75, 5.5) {};
		\node [style=none] (50) at (-1.5, 5.75) {};
		\node [style=none] (51) at (-2, 5.75) {};
		\node [style=none] (52) at (1, 4.25) {};
		\node [style=none] (53) at (1, 3.25) {};
		\node [style=none] (54) at (1.5, 3.75) {};
		\node [style=none] (55) at (0.5, 3.75) {};
		\node [style=none] (56) at (1.75, 4) {};
		\node [style=none] (57) at (1.75, 3.5) {};
		\node [style=none] (58) at (2, 3.75) {};
		\node [style=none] (59) at (1.5, 3.75) {};
		\node [style=none] (60) at (1, 4.75) {};
		\node [style=none] (61) at (1, 4.25) {};
		\node [style=none] (62) at (1.25, 4.5) {};
		\node [style=none] (63) at (0.75, 4.5) {};
		\node [style=none] (64) at (1, 3.25) {};
		\node [style=none] (65) at (1, 2.75) {};
		\node [style=none] (66) at (1.25, 3) {};
		\node [style=none] (67) at (0.75, 3) {};
		\node [style=none] (68) at (0.25, 4) {};
		\node [style=none] (69) at (0.25, 3.5) {};
		\node [style=none] (70) at (0.5, 3.75) {};
		\node [style=none] (71) at (0, 3.75) {};
		\node [style=none] (72) at (2.25, 6.25) {};
		\node [style=none] (73) at (2.25, 5.25) {};
		\node [style=none] (74) at (2.75, 5.75) {};
		\node [style=none] (75) at (1.75, 5.75) {};
		\node [style=none] (76) at (3, 6) {};
		\node [style=none] (77) at (3, 5.5) {};
		\node [style=none] (78) at (3.25, 5.75) {};
		\node [style=none] (79) at (2.75, 5.75) {};
		\node [style=none] (80) at (2.25, 6.75) {};
		\node [style=none] (81) at (2.25, 6.25) {};
		\node [style=none] (82) at (2.5, 6.5) {};
		\node [style=none] (83) at (2, 6.5) {};
		\node [style=none] (84) at (2.25, 5.25) {};
		\node [style=none] (85) at (2.25, 4.75) {};
		\node [style=none] (86) at (2.5, 5) {};
		\node [style=none] (87) at (2, 5) {};
		\node [style=none] (88) at (1.5, 6) {};
		\node [style=none] (89) at (1.5, 5.5) {};
		\node [style=none] (90) at (1.75, 5.75) {};
		\node [style=none] (91) at (1.25, 5.75) {};
		\node [style=none] (92) at (-1, 4.5) {};
		\node [style=none] (93) at (2.25, 4.5) {};
		\node [style=none] (94) at (1, 2.5) {};
		\node [style=none] (95) at (-2.25, 2.5) {};
		\node [style=none] (96) at (-2.25, 0.25) {};
		\node [style=none] (97) at (-1, 1) {};
		\node [style=none] (98) at (1, -0.5) {};
		\node [style=none] (99) at (2.25, 0.25) {};
		\node [style=none] (100) at (9.25, 1.75) {};
		\node [style=none] (101) at (6, 0) {};
		\node [style=none] (102) at (9.25, -1.75) {};
		\node [style=none] (103) at (12.5, 0) {};
		\node [style=NodeCross] (104) at (8.5, 0.75) {};
		\node [style=NodeCross] (105) at (10, -0.75) {};
		\node [style=NodeCross] (106) at (11.5, 0) {};
		\node [style=NodeCross] (107) at (7, 0) {};
		\node [style=none] (112) at (7, 4) {};
		\node [style=none] (113) at (7, 3) {};
		\node [style=none] (114) at (7.5, 3.5) {};
		\node [style=none] (115) at (6.5, 3.5) {};
		\node [style=none] (119) at (7.5, 3.5) {};
		\node [style=none] (121) at (7, 4) {};
		\node [style=none] (124) at (7, 3) {};
		\node [style=none] (130) at (6.5, 3.5) {};
		\node [style=none] (132) at (8.5, 6) {};
		\node [style=none] (133) at (8.5, 5) {};
		\node [style=none] (134) at (9, 5.5) {};
		\node [style=none] (135) at (8, 5.5) {};
		\node [style=none] (139) at (9, 5.5) {};
		\node [style=none] (141) at (8.5, 6) {};
		\node [style=none] (144) at (8.5, 5) {};
		\node [style=none] (150) at (8, 5.5) {};
		\node [style=none] (152) at (10, 4) {};
		\node [style=none] (153) at (10, 3) {};
		\node [style=none] (154) at (10.5, 3.5) {};
		\node [style=none] (155) at (9.5, 3.5) {};
		\node [style=none] (159) at (10.5, 3.5) {};
		\node [style=none] (161) at (10, 4) {};
		\node [style=none] (164) at (10, 3) {};
		\node [style=none] (170) at (9.5, 3.5) {};
		\node [style=none] (172) at (11.5, 6) {};
		\node [style=none] (173) at (11.5, 5) {};
		\node [style=none] (174) at (12, 5.5) {};
		\node [style=none] (175) at (11, 5.5) {};
		\node [style=none] (179) at (12, 5.5) {};
		\node [style=none] (181) at (11.5, 6) {};
		\node [style=none] (184) at (11.5, 5) {};
		\node [style=none] (190) at (11, 5.5) {};
		\node [style=none] (192) at (8.5, 4.25) {};
		\node [style=none] (193) at (11.5, 4.25) {};
		\node [style=none] (194) at (10, 2.25) {};
		\node [style=none] (195) at (7, 2.25) {};
		\node [style=none] (196) at (7, 0.25) {};
		\node [style=none] (197) at (8.5, 1) {};
		\node [style=none] (198) at (10, -0.5) {};
		\node [style=none] (199) at (11.5, 0.25) {};
		\node [style=Circle] (200) at (8.5, 6) {};
		\node [style=none] (201) at (9, 5.5) {};
		\node [style=none] (202) at (8.5, 5) {};
		\node [style=none] (203) at (8, 5.5) {};
		\node [style=Circle] (204) at (7, 4) {};
		\node [style=none] (205) at (7.5, 3.5) {};
		\node [style=none] (206) at (7, 3) {};
		\node [style=none] (207) at (6.5, 3.5) {};
		\node [style=Circle] (208) at (10, 4) {};
		\node [style=none] (209) at (9.5, 3.5) {};
		\node [style=none] (210) at (10.5, 3.5) {};
		\node [style=none] (211) at (10, 3) {};
		\node [style=none] (212) at (11.5, 5) {};
		\node [style=none] (213) at (11, 5.5) {};
		\node [style=none] (214) at (12, 5.5) {};
		\node [style=Circle] (215) at (11.5, 6) {};
		\node [style=none] (216) at (0, -2.5) {$(T^4/\Z_2)'$};
		\node [style=none] (217) at (9.25, -2.5) {$X_\mathbb{E}$};
		\node [style=none] (240) at (-4, 0) {$\P^1_\alpha\,:$};
		\node [style=none] (241) at (5.25, 0) {$\P^1_\alpha\,:$};
		\node [style=NodeCross] (242) at (0, 0) {};
		\node [style=none] (243) at (0, 0.25) {};
		\node [style=none] (244) at (0, 2.25) {};
		\node [style=none] (245) at (0, 2.75) {$T^2_\beta$};
		\node [style=none] (246) at (11.5, 6.5) {$D_4$};
		\node [style=none] (208) at (10, 4.5) {$D_4$};
		\node [style=none] (208) at (7, 4.5) {$D_4$};
		\node [style=none] (200) at (8.5, 6.5) {$D_4$};
	\end{pgfonlayer}
	\begin{pgfonlayer}{edgelayer}
		\draw [style=ThickLine, in=-180, out=90] (1.center) to (0.center);
		\draw [style=ThickLine, in=90, out=0] (0.center) to (3.center);
		\draw [style=ThickLine, in=0, out=-90] (3.center) to (2.center);
		\draw [style=ThickLine, in=-90, out=180] (2.center) to (1.center);
		\draw [style=ThickLine, in=-180, out=90] (15.center) to (12.center);
		\draw [style=ThickLine, in=90, out=0] (12.center) to (14.center);
		\draw [style=ThickLine, in=0, out=-90] (14.center) to (13.center);
		\draw [style=ThickLine, in=-90, out=-180] (13.center) to (15.center);
		\draw [style=ThickLine, in=-180, out=90] (19.center) to (16.center);
		\draw [style=ThickLine, in=90, out=0] (16.center) to (18.center);
		\draw [style=ThickLine, in=0, out=-90] (18.center) to (17.center);
		\draw [style=ThickLine, in=-90, out=-180] (17.center) to (19.center);
		\draw [style=ThickLine, in=-180, out=90] (23.center) to (20.center);
		\draw [style=ThickLine, in=90, out=0] (20.center) to (22.center);
		\draw [style=ThickLine, in=0, out=-90] (22.center) to (21.center);
		\draw [style=ThickLine, in=-90, out=-180] (21.center) to (23.center);
		\draw [style=ThickLine, in=-180, out=90] (27.center) to (24.center);
		\draw [style=ThickLine, in=90, out=0] (24.center) to (26.center);
		\draw [style=ThickLine, in=0, out=-90] (26.center) to (25.center);
		\draw [style=ThickLine, in=-90, out=-180] (25.center) to (27.center);
		\draw [style=ThickLine, in=-180, out=90] (31.center) to (28.center);
		\draw [style=ThickLine, in=90, out=0] (28.center) to (30.center);
		\draw [style=ThickLine, in=0, out=-90] (30.center) to (29.center);
		\draw [style=ThickLine, in=-90, out=-180] (29.center) to (31.center);
		\draw [style=ThickLine, in=-180, out=90] (35.center) to (32.center);
		\draw [style=ThickLine, in=90, out=0] (32.center) to (34.center);
		\draw [style=ThickLine, in=0, out=-90] (34.center) to (33.center);
		\draw [style=ThickLine, in=-90, out=-180] (33.center) to (35.center);
		\draw [style=ThickLine, in=-180, out=90] (39.center) to (36.center);
		\draw [style=ThickLine, in=90, out=0] (36.center) to (38.center);
		\draw [style=ThickLine, in=0, out=-90] (38.center) to (37.center);
		\draw [style=ThickLine, in=-90, out=-180] (37.center) to (39.center);
		\draw [style=ThickLine, in=-180, out=90] (43.center) to (40.center);
		\draw [style=ThickLine, in=90, out=0] (40.center) to (42.center);
		\draw [style=ThickLine, in=0, out=-90] (42.center) to (41.center);
		\draw [style=ThickLine, in=-90, out=-180] (41.center) to (43.center);
		\draw [style=ThickLine, in=-180, out=90] (47.center) to (44.center);
		\draw [style=ThickLine, in=90, out=0] (44.center) to (46.center);
		\draw [style=ThickLine, in=0, out=-90] (46.center) to (45.center);
		\draw [style=ThickLine, in=-90, out=-180] (45.center) to (47.center);
		\draw [style=ThickLine, in=-180, out=90] (51.center) to (48.center);
		\draw [style=ThickLine, in=90, out=0] (48.center) to (50.center);
		\draw [style=ThickLine, in=0, out=-90] (50.center) to (49.center);
		\draw [style=ThickLine, in=-90, out=-180] (49.center) to (51.center);
		\draw [style=ThickLine, in=-180, out=90] (55.center) to (52.center);
		\draw [style=ThickLine, in=90, out=0] (52.center) to (54.center);
		\draw [style=ThickLine, in=0, out=-90] (54.center) to (53.center);
		\draw [style=ThickLine, in=-90, out=-180] (53.center) to (55.center);
		\draw [style=ThickLine, in=-180, out=90] (59.center) to (56.center);
		\draw [style=ThickLine, in=90, out=0] (56.center) to (58.center);
		\draw [style=ThickLine, in=0, out=-90] (58.center) to (57.center);
		\draw [style=ThickLine, in=-90, out=-180] (57.center) to (59.center);
		\draw [style=ThickLine, in=-180, out=90] (63.center) to (60.center);
		\draw [style=ThickLine, in=90, out=0] (60.center) to (62.center);
		\draw [style=ThickLine, in=0, out=-90] (62.center) to (61.center);
		\draw [style=ThickLine, in=-90, out=-180] (61.center) to (63.center);
		\draw [style=ThickLine, in=-180, out=90] (67.center) to (64.center);
		\draw [style=ThickLine, in=90, out=0] (64.center) to (66.center);
		\draw [style=ThickLine, in=0, out=-90] (66.center) to (65.center);
		\draw [style=ThickLine, in=-90, out=-180] (65.center) to (67.center);
		\draw [style=ThickLine, in=-180, out=90] (71.center) to (68.center);
		\draw [style=ThickLine, in=90, out=0] (68.center) to (70.center);
		\draw [style=ThickLine, in=0, out=-90] (70.center) to (69.center);
		\draw [style=ThickLine, in=-90, out=-180] (69.center) to (71.center);
		\draw [style=ThickLine, in=-180, out=90] (75.center) to (72.center);
		\draw [style=ThickLine, in=90, out=0] (72.center) to (74.center);
		\draw [style=ThickLine, in=0, out=-90] (74.center) to (73.center);
		\draw [style=ThickLine, in=-90, out=-180] (73.center) to (75.center);
		\draw [style=ThickLine, in=-180, out=90] (79.center) to (76.center);
		\draw [style=ThickLine, in=90, out=0] (76.center) to (78.center);
		\draw [style=ThickLine, in=0, out=-90] (78.center) to (77.center);
		\draw [style=ThickLine, in=-90, out=-180] (77.center) to (79.center);
		\draw [style=ThickLine, in=-180, out=90] (83.center) to (80.center);
		\draw [style=ThickLine, in=90, out=0] (80.center) to (82.center);
		\draw [style=ThickLine, in=0, out=-90] (82.center) to (81.center);
		\draw [style=ThickLine, in=-90, out=-180] (81.center) to (83.center);
		\draw [style=ThickLine, in=-180, out=90] (87.center) to (84.center);
		\draw [style=ThickLine, in=90, out=0] (84.center) to (86.center);
		\draw [style=ThickLine, in=0, out=-90] (86.center) to (85.center);
		\draw [style=ThickLine, in=-90, out=-180] (85.center) to (87.center);
		\draw [style=ThickLine, in=-180, out=90] (91.center) to (88.center);
		\draw [style=ThickLine, in=90, out=0] (88.center) to (90.center);
		\draw [style=ThickLine, in=0, out=-90] (90.center) to (89.center);
		\draw [style=ThickLine, in=-90, out=-180] (89.center) to (91.center);
		\draw [style=DottedLine] (93.center) to (99.center);
		\draw [style=DottedLine] (92.center) to (97.center);
		\draw [style=DottedLine] (94.center) to (98.center);
		\draw [style=DottedLine] (95.center) to (96.center);
		\draw [style=ThickLine, in=-180, out=90] (101.center) to (100.center);
		\draw [style=ThickLine, in=90, out=0] (100.center) to (103.center);
		\draw [style=ThickLine, in=0, out=-90] (103.center) to (102.center);
		\draw [style=ThickLine, in=-90, out=180] (102.center) to (101.center);
		\draw [style=ThickLine, in=-180, out=90] (115.center) to (112.center);
		\draw [style=ThickLine, in=90, out=0] (112.center) to (114.center);
		\draw [style=ThickLine, in=0, out=-90] (114.center) to (113.center);
		\draw [style=ThickLine, in=-90, out=-180] (113.center) to (115.center);
		\draw [style=ThickLine, in=-180, out=90] (135.center) to (132.center);
		\draw [style=ThickLine, in=90, out=0] (132.center) to (134.center);
		\draw [style=ThickLine, in=0, out=-90] (134.center) to (133.center);
		\draw [style=ThickLine, in=-90, out=-180] (133.center) to (135.center);
		\draw [style=ThickLine, in=-180, out=90] (155.center) to (152.center);
		\draw [style=ThickLine, in=90, out=0] (152.center) to (154.center);
		\draw [style=ThickLine, in=0, out=-90] (154.center) to (153.center);
		\draw [style=ThickLine, in=-90, out=-180] (153.center) to (155.center);
		\draw [style=ThickLine, in=-180, out=90] (175.center) to (172.center);
		\draw [style=ThickLine, in=90, out=0] (172.center) to (174.center);
		\draw [style=ThickLine, in=0, out=-90] (174.center) to (173.center);
		\draw [style=ThickLine, in=-90, out=-180] (173.center) to (175.center);
		\draw [style=DottedLine] (193.center) to (199.center);
		\draw [style=DottedLine] (192.center) to (197.center);
		\draw [style=DottedLine] (194.center) to (198.center);
		\draw [style=DottedLine] (195.center) to (196.center);
		\draw [style=DottedLine] (244.center) to (243.center);
	\end{pgfonlayer}
\end{tikzpicture}
}
\caption{The Sen limit of the resolution $(T^4/\Z_2)'$ to an elliptic model with four $D_4$ singularities (right). The exceptional fibers in $(T^4/\Z_2)'$ are four copies of the resolution of a Kodaira type $I_0^*$ fiber which we degenerate to the associated elliptic surface singularity by shrinking all but one of the edge nodes in each exceptional fiber of the lefthand side.}
\label{fig:T4Z22}
\end{figure}

 The Sen limit with respect to the fibration $\pi_\alpha $ and a choice of section $F_\beta\equiv F_{I_\beta'}$ with fixed $\beta$ is now described at the level of topology by contracting all exceptional curves except $e_{I_\alpha,I_\beta'}$ for fixed $I_\beta'$. The 4 curves $F_{I_\alpha}$ which are of finite volume in $T^4/\Z_2$ are also contracted. This results in an elliptic fibration $X_\mathbb{E} \rightarrow  \P^1_\alpha$ and 4 degenerate fibers topologically given by the curves $e_{I_\alpha,I_\beta'}$. The 4 sections $F_{I_\beta}$ are collapsed onto the section $F_\beta$. Each curve $e_{I_\alpha,I_\beta'}$ contains a single singularity of ADE type $D_4$, more precisely it is a Kodaira type $I_0^*$ singular fiber. See figure \ref{fig:T4Z22}.

\begin{figure}
\centering
\scalebox{0.7}{\begin{tikzpicture}
	\begin{pgfonlayer}{nodelayer}
		\node [style=none] (0) at (-1.25, -1) {};
		\node [style=none] (1) at (-1.25, 1) {};
		\node [style=none] (2) at (1.25, 1) {};
		\node [style=none] (3) at (1.25, -1) {};
		\node [style=Circle] (4) at (-1.25, 1) {};
		\node [style=Circle] (5) at (1.25, 1) {};
		\node [style=Circle] (6) at (1.25, -1) {};
		\node [style=Circle] (7) at (-1.25, -1) {};
		\node [style=none] (8) at (1.75, -1.5) {$(0,1)$};
		\node [style=none] (9) at (-1.75, -1.5) {(0,0)};
		\node [style=none] (10) at (1.75, 1.5) {(1,1)};
		\node [style=none] (11) at (-1.75, 1.5) {(1,0)};
		\node [style=none] (12) at (-2.25, 0) {$F_{13}^{(1)},F_{14}^{(1)}$};
		\node [style=none] (13) at (2.25, 0) {$F_{13}^{(2)},F_{14}^{(2)}$};
		\node [style=none] (14) at (0, 2.5) {};
		\node [style=none] (15) at (0, -2.5) {};
		\node [style=none] (16) at (4, 0) {};
		\node [style=none] (17) at (-4, 0) {};
		\node [style=none] (18) at (-5, 0) {$\P_{12}\,:$};
		\node [style=none] (19) at (0, -1.5) {$F_{23}^{(1)},F_{24}^{(1)}$};
		\node [style=none] (20) at (0, 1.5) {$F_{23}^{(2)},F_{24}^{(2)}$};
	\end{pgfonlayer}
	\begin{pgfonlayer}{edgelayer}
		\draw [style=ThickLine] (1.center) to (2.center);
		\draw [style=ThickLine] (2.center) to (3.center);
		\draw [style=ThickLine] (3.center) to (0.center);
		\draw [style=ThickLine] (1.center) to (0.center);
		\draw [style=ThickLine, in=-90, out=0] (15.center) to (16.center);
		\draw [style=ThickLine, in=0, out=90] (16.center) to (14.center);
		\draw [style=ThickLine, in=90, out=180] (14.center) to (17.center);
		\draw [style=ThickLine, in=-180, out=-90] (17.center) to (15.center);
	\end{pgfonlayer}
\end{tikzpicture}
}
\caption{Cycles $F_\gamma^{(i)}$ projected to the base $\P^1_{\alpha}$ with $\alpha=(1,2)$. The black dots are the projection of the four $I_0^*$ fibers and labelled by their values for $I_\alpha\in \Z_2^2$. This follows studying the intersection of the lifts $F_\gamma^{(i)}$ to the smooth geometry with the curves $E_I$. }
\label{fig:FdegCycles}
\end{figure}
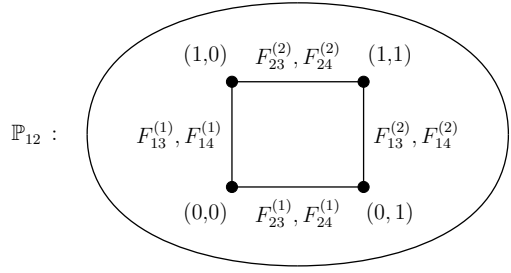

We now determine the gauge group associated with this geometry by M-theory. For this we track the 24 curves $F_{I_\gamma}$ through the deformation from $X'$ to $  X_\mathbb{E}$. These determine torsional cycles following \eqref{eq:KeyRelation2}. Let us concretely consider the case $\alpha=(1,2)$ and take the Sen limit labelled by $I_\beta'=(0,0)$, all other cases follow from relabelling. Then we have the degenerations
\be\ba\label{eq:LeftOverCurves}
\gamma=(12)\,: \qquad& \lbbb F_{I_\gamma}\rbbb \rightarrow \,\emptyset \\
\gamma=(13),(23),(24),(14)\,: \qquad& \lbbb F_{I_\gamma}\rbbb \rightarrow \lbbb F_\gamma^{(1)} ,F_\gamma^{(2)}  \rbbb \\
\gamma=(34)\,: \qquad& \lbbb F_{I_\gamma}\rbbb \rightarrow \lbbb F_\beta \rbbb  \\
\ea\ee
where $I_\gamma$ with $\gamma=(12)$ indexes the vertical curves collapsed by the limit, with $\gamma=(34)$ indexes the four sections which are degenerated to the zero section and other values of $\gamma$ gives mixed cases. Torsional 2-cycles of $X_\mathbb{E}$ for this new lattice of vanishing cycles are computed as in \eqref{eq:KeyRelation2} as differences of the curves on the righthand side in \eqref{eq:LeftOverCurves}. The relevant combinations are $F_\gamma^{(1)} -F_\gamma^{(2)}$ of which there are four and of which two are independent (the analysis proceeds via a coefficient matrix as in \eqref{eq:CoeffMatrix} and its Smith normal form). We deduce
\be
\Tor\,H_2(X_\mathbb{E})\cong \langle t_1, t_2\rangle \cong \Z_2^2\,.
\ee
The torsional cycles $F_\gamma^{(1)} -F_\gamma^{(2)}$ stretches diagonally between all $D_4$ singularities (orientations can be neglected as we are working over $\Z_2$). See figure \ref{fig:FdegCycles}. The non-Abelian gauge group is now
\be
G_\mathrm{loc} =Spin(8)^4 /\Z_2^2
\ee
where $\Z_2^2$ is the diagonal of the $Spin(8)$ factors.

The free 2-cycles follow immediately, we have the section $F_\beta$, the fiber class (equivalently  $e_{I_\alpha,I_\beta'}$ for some $I_\alpha$, different choices are homologous) and four classes $ F_\gamma^{(i)} $ (for some fixed $i$, different choices differ by torsion) with $\gamma=(13),(23),(24),(14)$, therefore
\be
H_2(X_\mathbb{E})\cong \Z^6 \oplus \Z_2^2\,.
\ee
The boundary map $\partial_2$ also follows in much the same way. The fiber class and $F_\beta$ are in the kernel. Conventions can be chosen such that the cycle $F_\gamma^{(i)}$ is mapped for $\gamma=(13),(24)$ and $\gamma=(23),(14)$ to the spinor and cospinor thimble of the relevant $D_4$ singularity. There are thus four factors of $\Z_2$ (two spinor (s), two cospinor (c)) sitting diagonally in the center $\Z_2^{(\rm s)}\times \Z_2^{(\rm c)}$ of pairs of $Spin(8)$ factors which can be undone by $U(1)$ rotations. We therefore find overall
\be
G_\textnormal{full}= \frac{Spin(8)^4 /\Z_2^2\times U(1)^4}{\Z_2^4} \times U(1)^2
\ee
in M-theory.

We can also move via circle compactification to F-theory. The gauge group becomes
\be
G_\textnormal{full,\,F-theory}=\lb Spin(8)^4 /\Z_2^2\times U(1)^4 \rb / \Z_2^4 \,,
\ee
i.e., we lost the Abelian KK factor associated via the zero section to the metric and the Abelian factor associated with the fiber which becomes a 2-form gauge field. These results match those derived in \cite{Cvetic:2021sxm, Guralnik:2001jh} via string junction analysis of the Mordell-Weil group of the elliptic model.

Finally, let us give the sequence \eqref{eq:MVS} for the above discussion
\be \ba
0~&\xrightarrow[\text{}]{} ~\big \langle T^2_\gamma \big\rangle  ~\xrightarrow[\text{}]{\;\jmath_2\;}  ~\langle  T^2_\beta, F_\beta, F_\gamma^{(i)} , t_1, t_2 \rangle ~\xrightarrow[\text{}]{\; \partial_2 \;}  ~   \bigoplus_{I_\alpha\;\! \in \;\! \Z_2^2} \Big\langle (\Tp_{\!c})^{(\rm s)}_{I_\alpha},(\Tp_{\!c})^{(\rm c)}_{I_\alpha} \Big\rangle~\xrightarrow[\text{}]{\;\imath_1\;} ~H_1(X^\circ)  ~\xrightarrow[\text{}]{}  ~0
\ea \ee
where $ T^2_\beta=\mathbb{E}$ is the elliptic fiber and $(\Tp_{\!c})^{(\rm c,s)}$ are compact representatives of the spinor and cospinor thimbles (which we write by isomorphisms as \eqref{eq:ref}). This gives the exact sequence
\be \ba
0~&\xrightarrow[\text{}]{} ~\Z^2\oplus \Z^4  ~\xrightarrow[\text{}]{\;\jmath_2\;}  ~\Z^2\oplus  \Z^4\oplus \Z^2_2 ~\xrightarrow[\text{}]{\; \partial_2 \;}  ~  (\Z_2^{2})^4 ~\xrightarrow[\text{}]{\;\imath_1\;} ~\Z_2^2  ~\xrightarrow[\text{}]{}  ~0\,.
\ea \ee

\section{M-theory on a Calabi-Yau Threefold: 5D Vacua} \label{sec:5d}

We now turn to examples of 5D vacua as obtained from M-theory on a Calabi-Yau threefold. There are many possible compact models, as well as local canonical singularities. In the 5D effective theory, such singularities can be interpreted as 5D SCFT sectors. It is worth noting that while there has been substantial progress in classifying 5D SCFTs (see e.g., \cite{Jefferson:2017ahm, DelZotto:2017pti, Jefferson:2018irk, Apruzzi:2019vpe, Apruzzi:2019opn, Apruzzi:2019enx, Xie:2022lcm} and \cite{Argyres:2022mnu} for a review), there is as yet no systematic classification of canonical singularities for threefolds available. As a general comment, in what follows we again focus on contributions to defects and symmetry operators obtained from wrapped branes. As such, contributions from discrete isometries will generally be deferred to future work.

With this in mind, we shall mainly focus on some simple illustrative examples of Calabi-Yau threefolds with 5D SCFT sectors. One case of interest which readily embeds in compact models is the local geometry $\mathbb{C}^3 / \mathbb{Z}_3$. An example of this sort, and the non-decoupling between different sectors in a 5D global model was considered in \cite{Heckman:2019bzm}. Our aim here will be to show how various global symmetry structures of each canonical singularity couple to gravity and are trivialized.

As another example, we consider singularities of the form $\mathbb{C}^{3} / \mathbb{Z}_4$. In this case we have both codimension 4 and 6 singularities, the latter arising as enhancement along the former, and the former supported on a non-compact cone $\mathbb{C} / \mathbb{Z}_2 \subset \mathbb{C}^{3} / \mathbb{Z}_4$. In any global model the locus of codimension 4 singularities is necessarily compactified whereby the corresponding non-Abelian flavor symmetry is gauged. The local model $\mathbb{C}^3/\Z_4$ engineers a 5D SCFT with a 2-group symmetry and embedding in a compact model allows us to track the fate of this generalized symmetry once coupled to gravity.

As a last class of examples we consider elliptic Calabi-Yau threefolds $\pi: X\rightarrow \mathbb{F}_n$ with Hirzebruch surface base and tune a non-Higgsable cluster on the base curve of self-intersection $-n$. Such geometries can exhibit a non-trivial Mordell-Weil group $\textnormal{MW}(\pi)$ and engineer 5D gauge theories, and it is well known that $\textnormal{MW}(\pi)$ determines the global form of the gauge group. The torsional, discrete subgroup thereof, in these particualr examples, was discussed in \cite{Apruzzi:2020zot} and here we reproduce and extend their results via cutting and gluing arguments.

Before proceeding to these specific examples, let us make a few additional comments on some features of these 5D models which apply in general.

Much as in our discussion of 7D vacua, we can also analyze possible emergent global symmetries as gravity is decoupled. Indeed, we can start from the compact geometry $X$ and consider the decompactification limit where the 5D Newton's constant $G_{N} \sim \mathrm{Vol}(X)^{-1} \rightarrow 0$. There are various ways we can achieve this when there is more than one K\"ahler modulus, and this is indeed the case in the models we consider. We can of course fully decouple each of the local models, and when we do so, we observe the emergence of local models, with their corresponding symmetry TFTs. In each symmetry TFT we encounter terms in the 6D action of the schematic form $B_2 \wedge d C_3$, as dictated by the braiding statistics of wrapped M2- and M5-branes wrapped on internal cycles. There are additional terms, as obtained from reduction of the topological terms of the 11D supergravity action, see \cite{Apruzzi:2021nmk} for additional discussion. One can also consider multiple sectors which are coupled to one another, as specified by including non-topological boundary conditions, much as in our analysis of section \ref{sec:7d}.

Another interesting feature of these 5D supergravity models is the generic appearance of 5D Abelian Chern-Simons terms (see Appendix \ref{app:5DCS}). Following \cite{Damia:2022bcd}, it was found that in a pure field theory limit with such terms, there is a global 2-form symmetry which suffers from an ABJ-type anomaly. Constructing the associated symmetry operators, there is a close analog to the analysis of \cite{Cordova:2022ieu, Choi:2022jqy} where the associated symmetry operators can support a fractional quantum Hall state. The resulting fusion rules thus yield a non-invertible symmetry. On general grounds, one expects such non-invertible symmetries to be absent in gravity, if only because one does not expect global symmetries.\footnote{Additionally, it is expected that graviational instantons explicitly break non-invertible symmetries to a maximal invertible symmetry \cite{McNamara:2021cuo}.}

It is nevertheless informative to see how this works in practice. Although one might wish to contemplate an ``approximate'' global symmetry which appears in the infrared, the divergence of the 2-form current is now of the schematic form $d \ast j_{2} \sim F\wedge F +  \mathrm{Tr} (\mathcal{R} \wedge \mathcal{R} )+ \:\!...$ where $\mathrm{Tr}(\mathcal{R} \wedge \mathcal{R})$ refers to the Pontryagin density of the tangent bundle and the ``...'' may also include various non-abelian gauge group factors. Such factors do not permit the construction of the same sort of ``fractionalized states'' used in \cite{Cordova:2022ieu, Choi:2022jqy, Damia:2022bcd}. Additionally, there is not even a clean decoupling limit where this 2-form symmetry really emerges; in the limit where gravity decouples the $U(1)$ factors again become non-dynamical anyway, so there is no 2-form current we can construct in the first place.\footnote{Gravity giveth and gravity taketh away.}

That being said, the appearance of these Chern-Simons couplings also generically leads to the appearance of a continuous (gauged) 2-group symmetry. This is because we have gauged $U(1)$ 1-form and 2-form symmetries from reduction of the (magnetic) M-theory 6-form potential on harmonic 4-forms and 3-forms on $X$. The Chern-Simons terms produce a non-trivial 2-group structure which one might formally wish to call a ``3-group'' (even though the structure is isomorphic to a 2-group).

Having stated these general considerations, we now turn to some explicit examples.

\subsection{Example: $ T^6/\Z_3$}\label{ssec:T6Z3}

The first Calabi-Yau threefold example we consider is $X=T^6/\Z_3$. M-theory on this Calabi-Yau orbifold engineers 27 copies of the 5D Seiberg SCFT $E_0$ \cite{Seiberg:1996bd}, each localized at codimension 6 (i.e., pointlike) singularities in $X$. We now discuss how the higher symmetry structures of these 27 local SCFT sectors trivializes and correlate when completed to the global model $X$. In the process we determine the extra, delocalized $U(1)$ symmetries stretching between the local models.

\subsubsection{Local Models}

In order to prepare the discussion let us briefly supply some further details on the geometry $T^6/\Z_3$, which we will be cutting apart, and establish notation. Much computational detail is postponed to appendix \ref{app:T6Z3} to which we make reference throughout.

To formulate the quotient split the six-torus as
\be\label{eq:Conventions}
T^6=T^2_1\times T^2_2\times T^2_3\,,
\ee
where each two-torus of complex structure $\tau_k=\textnormal{exp}({2\pi i/3})$ is parameterized by the complex coordinate $z_k$ where $k=1,2,3$. The $\Z_3$ group action is $z_i\mapsto \omega z_i$ with root of unity $\omega^3=1$ preserving each of the three two-tori. Overall there are $27$ fixed points in $T^6$ which projected to $X$ give 27 singularities. Each singularity is modelled locally on $\mathscr{X}=\mathbb{C}^3/\Z_3$. We note that the orbifold admits a crepant resolution $X'\rightarrow X$ in which each local copy of $\mathbb{C}^3/\mathbb{Z}_3$ is replaced with the line bundle $\mathcal{O}(-3) \rightarrow \mathbb{P}^2$ \cite{joyce2000CompactMW}.

We begin by discussing the generalized global symmetries of the SCFT sectors associated with each local model $\mathscr{X}=\mathbb{C}^3/\mathbb{Z}_3$ whose smooth, asymptotic boundary is $\partial \mathscr{X}=S^5/\Z_3$. The homology groups of the latter are
\be\label{eq:BdryHom}
H_n(S^5/\Z_3)\cong \{ \Z , \Z_3, 0, \Z_3, 0, \Z\}\,.
\ee
We focus on the contributions to the defect group obtained from M2-branes wrapped on cones of torsional 1-cycles, and their magnetic duals obtained from M5-branes wrapped on cones of torsional 3-cycles:
\begin{equation}\label{eq:DefectGpT6Z3}
\mathbb{D} \supset  (\mathbb{Z}_{3})^{(1)}_{\rm M2} \oplus (\mathbb{Z}_{3})^{(2)}_{\rm M5}\,,
\end{equation}
where the raised index denotes the spacetime dimensionality of the defect and the lowered index the membrane used to construct it. One can in principle also consider M5-branes wrapped on the torsional 1-cycle, which would result in a defect charged under a discrete 4-form symmetry $(\mathbb{Z}_{\mathrm{M5}})^{(4)}$, as well as over the entire space $\mathscr{X}$, which would result in a defect charged under a continuous 0-form symmetry $U(1)_{\mathrm{M5}}^{(0)}$. The 4-form symmetry is dual to a $-1$-form symmetry as associated with a discrete parameter which is decoupled from the SCFT. Likewise, the $U(1)$ 0-form symmetry is also decoupled. In what follows, we therefore ignore these contributions. As a passing remark, there is also a $\mathbb{Z}_3$ symmetry which comes from permuting the three coordinates of $\mathbb{C}^3$. This also decouples, and so we again ignore it in both the local and global models.

Restricting attention to the 1-form and magnetic dual 2-form symmetries, observe that the corresponding operators for the 1-form symmetries are constructed (see \cite{Heckman:2022muc}) from M5-branes wrapping the torsional 3-cycles ``at infinity'', while the operators for the 2-form symmetries are constructed from M2-branes wrapping the torsional 1-cycles ``at infinity''.

With defect and symmetry operators in hand the next step lies in fixing a choice of polarization, as specified by a maximal collection of mutually local subsets of defects or, equivalently, a maximal subset of commuting symmetry operators. In principle there can be anomalies in the field theory which obstruct certain choices, and a convenient way to catalog the possible consistent choices is to consider the associated symmetry TFT.

Indeed, the $E_0$ SCFT has a pure 1-form symmetry anomaly, which follows from the topological terms of 11D supergravity and the topological invariant:
\be \label{eq:anomaly}
\ell (S^3/\Z_3 \cdot S^3/\Z_3 , S^3/\Z_3 )=\frac{1}{3}
\ee
where $\ell$ and $ ``\cdot"$ denote the linking form and intersection pairing on $S^5/\Z_3$ respectively. Here the intersection computes as $S^3/\Z_3\cdot S^3/\Z_3=S^1/\Z_3$ which is the generator of $H_1(\partial \mathscr{X})$ and \eqref{eq:anomaly} follows. Alternatively, crepantly resolving $\mathscr{X}$, this anomaly can be recast as the triple intersection of the fractional divisor $\mathbb{P}^2/3$ which yields (see \cite{Apruzzi:2021nmk}):
\be
\mathbb{P}^2/3 \cdot \mathbb{P}^2/3 \cdot \mathbb{P}^2/3 =\frac{1}{3}\,.
\ee
It follows that for a single copy of the $E_0$ theory only electric polarizations are consistent.

Multiple copies of the $E_0$ theory however contain diagonal, anomaly free subgroups and in such cases mixed polarizations\,/\,gaugings of 1-form symmetry subgroups are possible. We will see that the global geometry $X$ naturally specifies such diagonal subgroups among its 27 local sectors.

\subsubsection{Gluing to Global Models}

We now proceed to glue together various local models to each a compact model, again making heavy use of the Mayer-Vietoris sequence.
For this we consider local, contractible neighbourhoods $U_I$ in $X$ centered on the 27 singularities, labelled by $I\in \mathbb{Z}^3_3$, and introduce our notion of local model
\be
\Xl=\bigcup_I \,U_I\,, \qquad X^\circ=X\setminus \Xl\,.
\ee
The sequence is now formulated with respect to the covering $X=\Xl \cup X^\circ$. In particular we extract two exact subsequences:
\be \label{eq:MVST6}\ba
0~&\xrightarrow[\text{}]{} ~H_4(X^\circ )  ~\xrightarrow[\text{}]{\;\jmath_4\;}  ~H_4(X) ~\xrightarrow[\text{}]{\;\partial_4\;}  ~   H_3(\partial \Xl) ~\xrightarrow[\text{}]{\;\imath_3\;} ~\Tor\,H_3(X^\circ)  ~\xrightarrow[\text{}]{}  ~0
\\
0~&\xrightarrow[\text{}]{} ~H_2(X^\circ )  ~\xrightarrow[\text{}]{\;\jmath_2\;}  ~H_2(X) ~\xrightarrow[\text{}]{\;\partial_2\;}  ~   H_1(\partial \Xl) ~\xrightarrow[\text{}]{\;\imath_1\;} ~\textnormal{Tor}\,H_1(X^\circ)  ~\xrightarrow[\text{}]{}  ~0\,.
\ea \ee

This parallels the K3 analysis, which resulted in \eqref{eq:MVS}, however, now we derive a pair of dual sequences, rather than a self-dual sequence: the symmetry operators appearing in one sequence act on defects appearing in the other.

Before analyzing the physics of \eqref{eq:MVST6} we evaluate the sequences. First note that clearly $H_n(\partial \Xl )\cong \Z_3^{27}$ for both $n=1,3$ from the 27 singularities $\mathbb{C}^3/\Z_3$ by \eqref{eq:BdryHom}. We compute
\be \label{eq:T6Z3}
H_n(T^6/\Z_3)=\begin{cases}   ~\Z \qquad\qquad ~~~~\;\!\;\!n=6 \\  ~0 \qquad\qquad ~~~~\;\!\;\!\,n=5 \\  ~\Z^9\oplus \Z_{3}^{4} \qquad ~\,\,~\!n=4 \\  ~\Z^2 \qquad\qquad ~~~~\;\!\;\!n=3 \\  ~\Z^9\oplus \Z_{3}^{17} ~~~~~~~\;\! n=2 \\ ~0 \qquad\qquad ~~~~\;\!\;\!\, n=1 \\ ~\Z\qquad\qquad ~~~~\;\!\;\! n=0\,, \end{cases}
\ee
in Appendix \ref{app:T6Z3} generalizing the discussion of section \ref{sec:FurtherExamples}. There we also argue
\be
H_2(X^\circ)\cong \Z^9\,, \qquad H_4(X^\circ)\cong \Z^9\,,
\ee
and compute $\textnormal{coker}\,\jmath_4=\Z_3^{10}$ and $\textnormal{coker}\,\jmath_2=\Z_3^{23}$. Putting everything together we thus find the pair of sequences to evaluate as
\be \label{eq:MVST6Z3}\ba
0~&\xrightarrow[\text{}]{} ~H_4(X^\circ )  ~\xrightarrow[\text{}]{\;\jmath_4\;}  ~H_4(X) ~\xrightarrow[\text{}]{\;\partial_4\;}  ~   H_3(\partial \Xl) ~\xrightarrow[\text{}]{\;\imath_3\;} ~\Tor\,H_3(X^\circ)  ~\xrightarrow[\text{}]{}  ~0
\\
0~&\xrightarrow[\text{}]{} ~\Z^9 ~\xrightarrow[\text{}]{\;\jmath_4\;}  ~\Z^9\oplus \Z_3^4 ~~\xrightarrow[\text{}]{\;\partial_4\;} ~ \Z_3^{27} ~ \xrightarrow[\text{}]{\;\imath_3\;} ~ \Z_3^{17} ~\xrightarrow[\text{}]{}  ~0\\[0.7em]
0~&\xrightarrow[\text{}]{} ~H_2(X^\circ )  ~\xrightarrow[\text{}]{\;\jmath_2\;}  ~H_2(X) ~\xrightarrow[\text{}]{\;\partial_2\;}  ~   H_1(\partial \Xl) ~\xrightarrow[\text{}]{\;\imath_1\;} ~\textnormal{Tor}\,H_1(X^\circ)  ~\xrightarrow[\text{}]{}  ~0
\\
0~&\xrightarrow[\text{}]{} ~\Z^9 ~\xrightarrow[\text{}]{\;\jmath_2\;}  ~ \Z^9\oplus \Z_3^{17}~\xrightarrow[\text{}]{\;\partial_2\;} ~ \Z_3^{27}  ~ \xrightarrow[\text{}]{\;\imath_1\;} ~\Z_3^{4} ~~\xrightarrow[\text{}]{}  ~0\,.
\ea \ee
These results pass the duality check
\be\ba\label{eq:DualityCheck}
\textnormal{Tor}\,H_1(X^\circ)&\cong (\textnormal{Tor}\,H_4(X))^\vee \cong \textnormal{Tor}\,H_4(X) \\   \textnormal{Tor}\,H_3(X^\circ) &\cong (\Tor\, H_2(X))^\vee  \cong \Tor\, H_2(X)
\ea\ee
which follows from the long exact sequence in relative homology, excision, deformation retraction and Poincar\'e-Lefschetz duality, much as in \eqref{eq:TorRelation}. Equivariant cohomology computations are consistent with these results.

The duality relations \eqref{eq:DualityCheck} make the physics strikingly clear. In the field theory sector of the compactification we have 27 candidate 1-form and 2-form symmetry operators resulting from M5- and M2-branes wrapped on $H_3(\partial \Xl)$ and $H_1(\partial \Xl)$ respectively. Upon gluing the local sectors to the bulk $X^\circ$ we introduce dynamical particles and strings via M2-branes and M5-branes wrapping the cycles
\be \label{eq:GaugingAndBreaking}
\text{M2's:}\,\,\, H_2(X)\cong \Z^9\oplus \Z_3^{17}\,, \qquad \text{M5's:}\,\,\, H_4(X)\cong \Z^9\oplus \Z_3^{4}.
\ee
The torsional factors break 17 copies of the candidate 1-form symmetries and 4 copies of the candidate 2-form symmetries. The free factors absorb another $\Z_3^6$ each. We therefore produce the 0-form gauge group
\be\label{eq:G0}
{G}_{\textnormal{full}}=\frac{\Z_3^{27}/\Z_3^{4}\times U(1)^9}{\Z_{3}^{6}}\cong \frac{\Z_3^{23}\times U(1)^9}{\Z_{3}^{6}} \cong \Z^{17}_3\times U(1)^{9}\,,
\ee
which involves an overall quotient by $\Z_3^{10}\cong \textnormal{Im}\,\partial_4 $. The $\Z_3^4\cong \textnormal{Tor}\,H_4(X)$ quotient is generated by cycles which stretch between 18 (divisible by 3) singularities and are therefore non-anomalous and can be readily gauged (as predicted by the geometry). The maps $\jmath_2,\jmath_4$ again determine normalizations. Lastly, there is also a $\mathbb{Z}_3$ permutation symmetry of the different $T^2$ factors which we have suppressed.

For completeness, let us now enumerate the various $p$-form gauge groups in the compact model. The case $p=0$ was already covered in \eqref{eq:G0} and we now discuss the cases $p= 1,2$. The respective gauge symmetry groups are read off straight from geometry and reduction of the 6-form potential on harmonic representatives of $H^4(X)$ and $H^3(X)$:
\begin{align}
     G^{(1)}_{\textnormal{full}}&=\Z^{4}_3\times U(1)^{9}\, \\
     G^{(2)}_{\textnormal{full}}&=U(1)^{2}\,.
\end{align}
The objects charged under the 1-form symmetry are effective strings (1-branes) obtained from M5-branes wrapped on $H_4(X)$, and the objects charged under the 2-form symmetry are membranes (i.e., 2-branes) obtained from M5-branes wrapped on $H_3(X)$.

We now confirm that in the model coupled to gravity, all of the above symmetries are indeed gauged or broken. For this, start by considering 27 decoupled $E_0$ theories with electric polarization and overall $\Z_3^{27}$ 1-form symmetry. Now, extend the field theory sector according to the gluing prescription above and introduce matter breaking the 1-form symmetry subgroup $\Z_3^{17}$ leaving a 1-form symmetry group $\Z_3^{10}$. Further, we are also asked to introduce a $U(1)^9$ gauge theory sector and this absorbs a $ \Z_3^6$ 1-form symmetry group exactly as in the discussion around \eqref{eq:gauginginFieldtheory}. The remaining $\Z^4_3$ 1-form symmetry is gauged and dualized to a quantum magnetic $\Z^4_3$ 2-form symmetry. The geometry then accounts for the absence of this symmetry as well since it is explicitly broken by the dynamical strings given by M5-branes wrapping $\mathrm{Tor}\,H_4(X)=\Z^4_3$. Overall no global symmetries remain.

\subsection{Example: $T^6/\Z_4$}

\begin{figure}[t!]
\centering
\includegraphics[scale=0.30, trim = {0cm 0cm 0cm 3cm}]{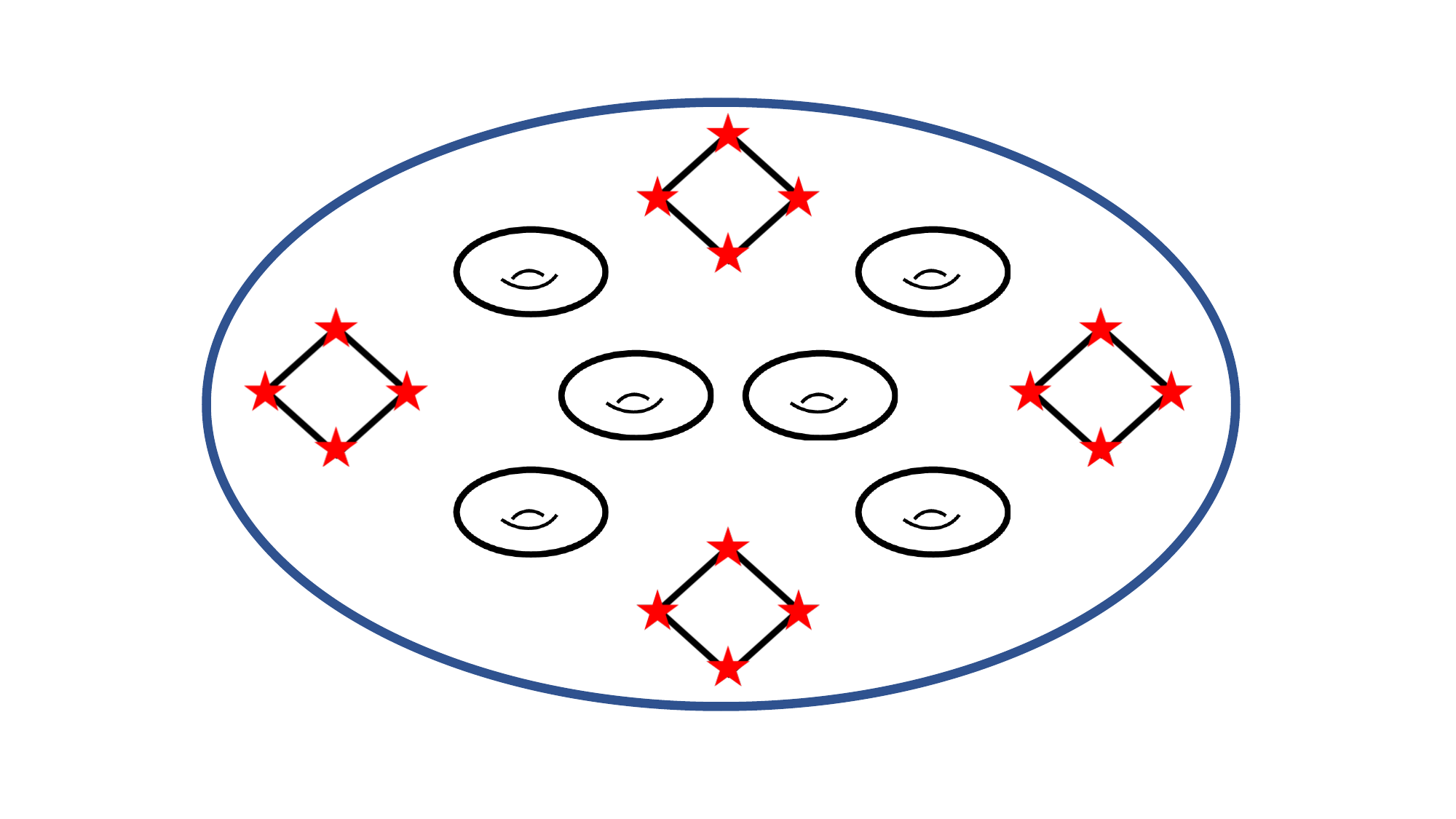}
\caption{Illustration of $T^6/\Z_4$ (blue oval). There are six $A_1$ singularities with topology $T^2$, and four $A_1$ singuarities with topology $T^2/\Z_2\cong \mathbb{P}^1$ (black diamonds). Each of the latter codimension 4 singularities have four enchanced singularities locally of the form $\mathbb{C}^3/\Z_4$ (red stars). Overall there are thus 16 5D SCFT sectors throughout the geometry. }
\label{fig:T6Z4}
\end{figure}

In this section we analyze the orbifold $T^6/\Z_4$. Singular strata now occur in both codimension 4 and 6, the structure of which is summarized in figure \ref{fig:T6Z4}. M-theory on $T^6/\Z_4$ engineers 16 5D SCFT sectors with local geometry $\mathbb{C}^3/\Z_4$. These exhibit four $\mathfrak{su}(2)^{4}$ flavor symmetries and the embedding into the global geometries compactifies the flavor branes of each $\mathfrak{su}(2)^{4}$ group into a single locus gauging it in the process. Further there are 6 copies of 7D SYM theory with Lie algebra $\mathfrak{su}(2)$ compactified to 5D on a $T^2$. We will again be interested in how the global model trivializes and correlates the higher symmetry structures of the local SCFT sectors. Further, we again determine the extra, delocalized $U(1)$ symmetries stretching between the local models.

To frame the analysis to follow let us lay out notation and introduce various local models. As before
split the six-torus $T^6$ into three two-tori $T^2_k$ here with complex structure $\tau_k=i$ and complex coordinate $z_k$ where $k=1,2,3$. The $\Z_4$ group action is
\be
(z_1,z_2,z_3) \mapsto (\omega z_1, \omega z_2, \omega^{-2} z_3)
\ee
with $\omega=i$. The orbifold $T^6/\Z_4$ contains six two-tori $T^2$ worth of $A_1$ singularities and four spheres $T^2/\Z_2$ worth of $A_1$ singularities which enhance at four points to codimension 6 singularities (see e.g., \cite{Joyce1999} for additional discussion). There are therefore a total of 16 singularities modelled on $\mathbb{C}^3/\Z_4$. With this we find the local models centered on codimension 4 singularities to take one of two forms:
\be
 \mathscr{X}_4^{(1)}= (\mathbb{C}^2\times T^2)/\Z_4= (\mathbb{C}^2/\Z_2\times T^2)/\Z_2\,, \qquad \mathscr{X}_4^{(2)}=\mathbb{C}^2/\Z_2\times T^2
\ee
while the local models centered on the codimension 6 singularities are
\be
\mathscr{X}_6=\mathbb{C}^3/\Z_4\,.
\ee
Four copies of $\mathscr{X}_6$ admit a gluing to one copy of $\mathscr{X}_4^{(1)}$.

\subsubsection{Local Models}

We now discuss the physics of the various local models.


\paragraph{Codimension Six Local Models: $\mathbb{C}^3/\Z_4$}\mbox{}\medskip \\  The building block $\mathscr{X}_6$ is the local model for the 16 codimension 6 singularities contained in $T^6/\Z_4$. In M-theory $\mathscr{X}_6$ engineers a (relative) 5D SCFT which we denote as
\be
\mathcal{T}_{\C^3/\Z_4}-[\mathfrak{su}(2)]\,,
\ee
and on its Coulomb branch, choosing an electric polarization, this theory admits a presentation as the 5D gauge theory $SU(2)_0$. The SCFT has flavor symmetry group $SO(3)$ (above we denoted its algebra $\mathfrak{su}(2)\cong \mathfrak{so}(3)$ which participates in the 2-group symmetry:\footnote{In principle there are can be an additional correlated action on the R-symmetry group. This is known to occur in other related examples such as various 6D and 4D SCFTs \cite{Heckman:2022suy}.}
\begin{equation}\label{eq:2gpSym}
0 \rightarrow \mathbb{Z}_2 \rightarrow \mathbb{Z}_4 \rightarrow SU(2) \rightarrow SO(3) \rightarrow 1\,.
\end{equation}
The geometry detects the centers of these groups via:
\be \label{eq:geometrification} \ba
&0 \rightarrow H_1(\partial \mathscr{X})^\vee \rightarrow H_1(\partial \mathscr{X}^\circ)^\vee \rightarrow H_1(\partial T_{K})^\vee\rightarrow  H_2(\partial \mathscr{X}^\circ )^\vee \rightarrow 0\\
&0 \rightarrow \Z_2 \rightarrow \Z_4 \rightarrow \Z_2 \rightarrow 0 \rightarrow 0\\
\ea \ee
where $\vee$ denotes the Pontryagin dual, as homology groups characterize defects and symmetry groups are the dual objects acting on defects. Here $\partial \mathscr{X}^\circ$ denotes the local model boundary minus the codimension 4 singularities $K\subset \partial \mathscr{X}$ and $T_K\subset \partial \mathscr{X}$ is a tubular neighbourhood of the codimension 4 singularities. The sequence $0 \rightarrow \Z_2 \rightarrow \Z_4 \rightarrow \Z_2 \rightarrow 0$ therefore states that taking two copies of a 5D defect is not trivial, rather it gives a 7D flavor defect \cite{Lee:2021crt}. For further details see \cite{DelZotto:2022fnw, Cvetic:2022imb}.

\paragraph{Codimension Four Local Models: $(\mathbb{C}^2/\Z_2 \times T^2) / \mathbb{Z}_2$}\mbox{}\medskip \\ We begin by discussing the physics of the building block $\mathscr{X}_4^{(1)}$. In M-theory $\mathscr{X}_4^{(1)}$ engineers a (relative) theory which we denote as
\be
\begin{array}{c}
\mathcal{T}_{\C^3/\Z_4}\\
\textnormal{\scriptsize $|$} \\
\mathcal{T}_{\C^3/\Z_4}-(\mathfrak{su}(2))-\mathcal{T}_{\C^3/\Z_4}\\
\textnormal{\scriptsize $|$}\\
\mathcal{T}_{\C^3/\Z_4}
\end{array}
\ee
where the four codimension 4 loci contained in four copies of $\mathscr{X}_6=\C^3/\Z_4$ glue to a single compact locus $T^2/\Z_2$ of $A_1$ singularities. The flavor branes in each copy of $\mathscr{X}_6$ are compactified and the flavor symmetry is therefore gauged,
resulting in the central node above.

Let us derive the spectrum of defects from geometry. The asymptotic boundary of $\mathscr{X}_4^{(1)}$ can be considered as a fibration over the singular locus
\be \label{eq:FibrationCo4}
\partial \mathscr{X}_4^{(1)}=(S^3/\Z_2\times T^2)/\Z_2 ~\rightarrow ~  T^2/\Z_2
\ee
with generic fibers given by lens spaces $S^3/\Z_2$. There are four orbifold points in $ T^2/\Z_2
$ where the fiber folds to $S^3/\Z_4$ and going around any of these the fiber is twisted by $-1\in \Z_2$. From here we compute via a Mayer-Vietoris sequence
\be\label{eq:BdryHomo}\ba
H_n( \partial \mathscr{X}_4^{(1)})&=\{ \Z, \Z_4\oplus \Z_2^3,\Z, \Z \oplus \Z_4\oplus \Z_2^3,0,\Z\}\,.
\ea
\ee

Let us identify the 1-cycle generators of $H_1( \partial \mathscr{X}_4^{(1)})$ in order to visualize the above result. Consider the four exceptional fibers $(S^3/\Z_4)_i$ which contain the Hopf fibers $\gamma_i=(S^1/\Z_4)_i$, where $i=1,2,3,4$. Each of these generates a copy of $\Z_4$ within its exceptional fiber.
Now, for all $i$, the double $2\gamma_i$ is homologous to the Hopf fiber of the generic lens space fiber $S^3/\Z_2$ and therefore we have $2\gamma_i-2\gamma_j=0$ in the boundary. With this it follows
\be\label{eq:BdryHomoAsQuotient}
H_1( \partial \mathscr{X}_4^{(1)})\cong \langle \gamma_1,\gamma_2,\gamma_3,\gamma_4 \rangle / \langle  2\gamma_i-2\gamma_j\rangle \cong  \Z_4\oplus \Z_2^3
\ee
immediately giving the result.

Now, the full defect group straightforwardly follows from \eqref{eq:BdryHomo}. We again focus on the torsional
contributions to the 1-form and 2-form symmetries:
\begin{equation}
\mathbb{D} \supset (\mathbb{Z}_{4}\oplus \Z_2^3)^{(1)}_{\rm M2} \oplus \lb  \mathbb{Z}_{4}\oplus \Z_2^3\rb^{(2)}_{\rm M5}\,,
\end{equation}
with notation as in line \eqref{eq:DefectGpT6Z3}.

Let us now focus on line defects and discuss the consequence of the 2-group symmetry \eqref{eq:2gpSym} when gluing four copies of $\mathscr{X}_6$ to $\mathscr{X}_4^{(1)}$. For this, we trace the geometric data of \eqref{eq:geometrification} through the gluing. The first observation is that the boundary $\partial \mathscr{X}_4^{(1)}$ does not glue from the boundaries of $(\partial \mathscr{X}_6)_i$, as this would contain singular loci, but rather
\be \label{eq:gluing2}
(\partial\mathscr{X}_6^\circ)_1 \cup (\partial\mathscr{X}_6^{\circ})_2 \cup (\partial\mathscr{X}_6^{\circ})_3 \cup (\partial \mathscr{X}_6^{\circ})_4 =\partial \mathscr{X}_4^{(1)}\,.
\ee
To make this manifest we can decompose the base $T^2/\Z_2$ of \eqref{eq:FibrationCo4} into 4 patches which each contain an orbifold point. The fibration restricted to each patch is then topologically a copy of $\partial\mathscr{X}_6^\circ$. The 1-cycles $\gamma_i$ generate $H_1((\partial\mathscr{X}_6^\circ)_i )$ and we can thus rewrite \eqref{eq:BdryHomoAsQuotient} as
\be
H_1( \partial \mathscr{X}_4^{(1)})\cong \bigoplus_{i=1}^4 H_1((\partial\mathscr{X}_6^\circ)_i )/ \langle  2\gamma_i-2\gamma_j\rangle\,.
\ee
This also immediately suggests a physically more intuitive presentation. Let us introduce the 5D line operator, constructed via an M2-brane wrapping as
\be
W=\textnormal{M2}(\textnormal{Cone}(2\gamma_i))
\ee
which is independent of the index $i$. This is the Wilson line in the fundamental representation
of the gauged flavor symmetry. We also introduce the lines
\be
L_i=\textnormal{M2}(\textnormal{Cone}(\gamma_i))
\ee
which are the line defects localized to each codimension 6 singularity. From the perspective of any codimension 6 singularity the line $W$ is the 7D flavor line obtained as $W=2L_i$. However, after the gluing, the flavor symmetry is gauged and $W$ is a genuine 5D line defect.

The 2-group structure now tells us how the flavor line defects extend 5D line defects $L_i$. Indeed, observe that we have:
\be\label{eq:subgroupstuff} \ba
\langle W, L_1,L_2,L_3,L_4 \rangle &\cong \Z_4\oplus \Z_2\oplus \Z_2\oplus \Z_2 \\
 &\neq \Z_2\oplus \Z_2 \oplus \Z_2\oplus \Z_2\oplus \Z_2,
\ea \ee
that is, we get a $\mathbb{Z}_4$ summand as obtained from the extension $0 \rightarrow \mathbb{Z}_2 \rightarrow \mathbb{Z}_4 \rightarrow \mathbb{Z}_2 \rightarrow 0$ rather than just a $\mathbb{Z}_2 \oplus \mathbb{Z}_2$ summand, as indicated in the bottom line.

\paragraph{Codimension Four Local Models: $\mathbb{C}^2 / \mathbb{Z}_2 \times T^2$}\mbox{}\medskip \\ The physics of the local model $\mathscr{X}^{(2)}_4=\C^2/\Z_2\times T^2$ is simply that of 7D $\mathfrak{su}(2)$ SYM theory compactified on $T^2$. All defects follow from
\be
H_n(\partial \mathscr{X}^{(2)}_4)\cong \{\Z, \Z^2\oplus \Z_2,\Z\oplus \Z_2^2,\Z\oplus \Z_2,\Z^2, \Z \}
\ee
which follows from the K\"unneth formula.

\subsubsection{Gluing Local Models}

Let us combine all codimension 4 local models into
\be
\Xl=\bigcup_{k=1}^4 \mathscr{X}^{(1)}_{4,k} ~\cup~ \bigcup_{\ell=1}^6 \mathscr{X}^{(2)}_{4,\ell}
\ee
where the contribution from the $\mathscr{X}^{(1)}_{4,k}$ collects the 16 codimension 6 singularities modelled on $\mathbb{C}^3/\Z_4$ into groups of four. Codimension six singularities are grouped according to which connected component of the codimension 4 locus they embed into. For $X=T^6/\Z_4$ we now consider the Mayer-Vietoris sequence for the covering $X=\Xl\cup X^\circ$, where $X^\circ$ is the complement of all singularities.

Above we computed the homology groups for $\Xl$. In appendix \ref{app:t6z4} we discuss in detail the fibration
\be
T^4/\Z_2 ~\hookrightarrow~ T^6/\Z_4 ~\rightarrow~ T^2/\Z_2\,,
\ee
which is induced by projection onto the third two-torus. This fibration presents $T^6/\Z_4$ as a fibration of Kummer surfaces away from four orbifold points where the Kummer surface $T^4/\Z_2$ folds to $T^4/\Z_4$. Using the homology groups for these fibers as computed in section \ref{sec:FurtherExamples}, we derive in Appendix \ref{app:t6z4} the homology groups
{\renewcommand{\arraystretch}{1.225}
\be\label{eq:T6Z4}
H_n(T^6/\Z_4)=\begin{cases}   ~\Z \\  ~0  \\  ~\Z^5\oplus \Z_4 \oplus \Z_2^4  \\  ~\Z^4\oplus \Z_2^4  \\  ~\Z^5 \oplus\Z_4\oplus  \Z_2^{17}  \\ ~0  \\ ~\Z \end{cases}
 \begin{array}{c} n=6\\ n=5 \\ n=4 \\ n=3 \\ n=2\\ n=1 \\ n=0 \end{array}
\ee}

\noindent with Betti-numbers $b_n=\{1,0,5,4,5,0,1 \}$. Finally, we also require the homology groups of $X^\circ$ to complete all entries of the Mayer-Vietoris sequence. For this we consider duality relations. First, note that, by Poincar\'e-Lefschetz duality and deformation retraction,
\be
H_n(X^\circ)\cong H^{6-n}(X^\circ, \partial X^\circ) \cong H^{6-n}(X, \textnormal{Sing}(X))
\ee
for all $n$. Here the singular set $\textnormal{Sing}(X)$ consists of six two-tori $T^2$ and four spheres $T^2/\Z_2$. We thus obtain:
\be
H_n( \textnormal{Sing}(X))\cong\{\Z^6\oplus \Z^4, \Z^{12}\oplus 0, \Z^6\oplus \Z^4, 0,0,0,0\}
\ee
and from here we deduce, via the long exact sequence in relative homology for the pair $(X, \textnormal{Sing}(X))$ and the universal coefficient theorem,
\be  \label{eq:dualityrelation}
\textnormal{Tor}\, H_n(X^\circ) \cong (\textnormal{Tor}\, H_{5-n}(X))^\vee\cong \textnormal{Tor}\, H_{5-n}(X)\,.
\ee

We emphasize that line \eqref{eq:dualityrelation} implies, as in all cases analyzed prior and via identical arguments, that no global symmetries of the field theory sector remain after gluing.

We now determine the gauge group $G_\textnormal{full}$. First let us give the naive answer:
\be
\widetilde{G}_\textnormal{full}^{\,\star}=[\mathbb{Z}_2^{3} \times \Z_4 \times SU(2)]^4 \times SU(2)^6\times U(1)^5\,,
\ee
where the $U(1)^5$ follows from $b_4=5$. Note that we have already identified via \eqref{eq:subgroupstuff} the subgroup $\Z_2\subset \Z_4$ as the center of $SU(2)$ and hence we can improve our naive starting point
\be
\widetilde{G}_\textnormal{full}=[\mathbb{Z}_2^{3} \times (\Z_4 \times SU(2))/\Z_2]^4 \times SU(2)^6\times U(1)^5\,.
\ee
Next, we utilize the trivialization of all global symmetries as derived from \eqref{eq:dualityrelation}. From a field theory perspective, starting in the local models, we have that completing the local models to a global model introduces magnetic strings which break a magnetic 2-form symmetry group
\be
\textnormal{Tor}\,H_4(X)\cong \Z_4\oplus \Z_2^4
\ee
and therefore we improve the naive answer to\smallskip
\be
\frac{\widetilde{G}_\textnormal{full}}{\textnormal{Tor}\,H_4(X)} =\frac{[\mathbb{Z}_2^{3} \times (\Z_4 \times SU(2))/\Z_2]^4 \times SU(2)^6}{\textnormal{Tor}\,H_4(X)} \times U(1)^5\,.
\ee
Similarly the compact geometry adds dynamical electric particles breaking a 1-form symmetry subgroup
\be
\textnormal{Tor}\,H_2(X)\cong \Z_4\oplus \Z_2^{17}
\ee
and hence we have an additional quotient
\be
G_\textnormal{full}=\Big( G_{\textnormal{loc}} \times U(1)^5\Big)/\mathcal{C}_\textnormal{Extra}\,, \qquad G_\textnormal{loc}=\frac{[\mathbb{Z}_2^{3} \times (\Z_4 \times SU(2))/\Z_2]^4 \times SU(2)^6}{\Z_4\oplus \Z_2^4}
\ee
such that the center of $G_{\textnormal{loc}}/\mathcal{C}_\textnormal{Extra}$ is isomorphic to $\textnormal{Tor}\,H_2(X)$, thus $\mathcal{C}_\textnormal{Extra}\cong \Z_4^2\oplus \Z_2^3$. The extension problem in this computation is the same as that of the Mayer-Vietoris sequence at the entries $\textnormal{Tor}\,H_2(X)$ and $H_1(\partial \Xl)$ and is solved there.

In addition to the gauge group (0-form symmetry), we also have various $p$-form gauge symmetries.
These follow straightforwardly from geometry. We have the 1-form and 2-form gauge groups:
\begin{align}
     G^{(1)}_{\textnormal{full}}&=\Z_4\times \Z^{4}_2\times U(1)^{5},\\
     G^{(2)}_{\textnormal{full}}&=\Z^4_2\times U(1)^{4}.
\end{align}
The objects charged under the 1-form symmetry are M5-branes wrapped on 4-cycles (1-branes, i.e., strings) and those charged under the 2-form symmetry are M5-branes wrapped on 3-cycles (2-branes, i.e., membranes).

One can again explicitly check that there are no global symmetries remaining in the model with gravity switched on.

\subsection{Example: Elliptic Calabi-Yau Threefolds}

\begin{figure}
\centering
\scalebox{0.7}{
\begin{tikzpicture}
	\begin{pgfonlayer}{nodelayer}
		\node [style=none] (0) at (-2, 3) {};
		\node [style=none] (1) at (-2, -3) {};
		\node [style=none] (2) at (-3, 2) {};
		\node [style=none] (3) at (3, 2) {};
		\node [style=none] (4) at (2, 3) {};
		\node [style=none] (5) at (2, -3) {};
		\node [style=none] (6) at (3, -2) {};
		\node [style=none] (7) at (-3, -2) {};
		\node [style=none] (8) at (0.15, 1.5) {\Large $b_+$};
		\node [style=none] (9) at (0.15, -1.5) {\Large $b_-$};
		\node [style=none] (10) at (-1.375, -0.05) {\Large $f_+$};
		\node [style=none] (11) at (1.375, -0.05) {\Large $f_-$};
		\node [style=none] (12) at (-0.225, -2.5) {\Large $-n$};
		\node [style=none] (13) at (0, 2.5) {\Large $n$};
		\node [style=none] (14) at (2.5, 0) {\Large $0$};
		\node [style=none] (15) at (-2.5, 0) {\Large $0$};
		\node [style=none] (16) at (0, -3) {};
	\end{pgfonlayer}
	\begin{pgfonlayer}{edgelayer}
		\draw [style=ThickLine] (4.center) to (5.center);
		\draw [style=ThickLine] (0.center) to (1.center);
		\draw [style=ThickLine] (2.center) to (3.center);
		\draw [style=ThickLine] (6.center) to (7.center);
	\end{pgfonlayer}
\end{tikzpicture}
}
\caption{Curves of the Hirzebruch surface $\mathbb{F}_n$.}
\label{fig:Hirzebruchcurves}
\end{figure}
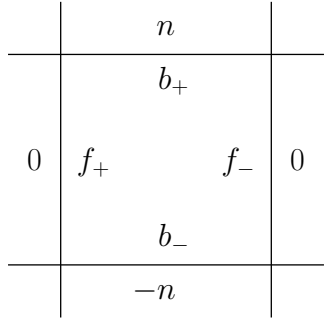

We now consider M-theory on elliptically fibered Calabi-Yau threefolds $\pi:X\rightarrow \mathbb{F}_n$ where the base is a Hirzebruch surface, presented as the projectivization of a holomorphic rank 2 bundle,
\be
\mathbb{F}_n =\P\lb \mathcal{O}_{\P^1}(0) \oplus \mathcal{O}_{\P^1}(-n) \rb\,.
\ee
From here it follows immediately that we can view the Hirzebruch surface as glued from two holomorphic line bundles, cut off at some finite radius $R$,
\be \label{eq:basedecomp}
\mathbb{F}_n= \mathcal{O}^{}_{\P^1}(-n)|_{r\;\!\leq\;\! R} \cup_{S^3/\Z_n} \mathcal{O}_{\P^1}(n)|_{r\;\!\leq\;\! R}
\ee
along their common lens space boundary.
The curves of the geometry are $b_\pm, f_\pm$ with base curves $b_\pm$ of $\mathcal{O}_{\P^1}(\pm n)$ and one-point compactifications of the fiber classes $f_\pm$, respectively. These curves are related as
\be
b_+\cdot b_- =0\,, \quad f_\pm \cdot b_\pm =1\,,\quad f_\pm\cdot f_\pm=0 \,,  \quad f_+=f_-\,, \quad b_+=b_-+n f_-
\ee
which is often presented as in figure \ref{fig:Hirzebruchcurves}.

 We now focus on the cases $n=3,4,6,8$ where Kodaira type $IV, I_0^*,IV^*, III^* $ singularities can be realized along the curve $b_-$ \cite{Aspinwall:1998xj, Apruzzi:2020zot}. These singularities are necessarily accompanied by a Kodaira singularities which project to the $ \mathcal{O}_{\P^1}(n)$ half of the base away from $b_-$. These may intersect, locally enhancing. Generically, however, we obtain pairs of degenerations $(\Phi_-,\Phi_+)$ of Kodaira type $\Phi_\pm$  localized to the two halves of \eqref{eq:basedecomp} given by:
\be\label{eq:ellipticsingularities}
(IV,I_1\oplus I_3)\,, \quad (I_0^*,I_2\oplus I_2\oplus I_2)\,, \quad (IV^*,I_1\oplus I_3)\,, \quad (III^*,I_1\oplus I_2)
\ee
giving rise to the non-abelian pairs of gauge algebras\footnote{All Kodaira fibers are split, so we get simply laced Lie algebras.}
 \be\label{eq:LieAlgs}
 (\mathfrak{su}(3),\mathfrak{su}(3))\,, \quad  (\mathfrak{so}(8),\mathfrak{su}(2)^3)\,, \quad  (\mathfrak{e}_6,\mathfrak{su}(3))\,, \quad  (\mathfrak{e}_7,\mathfrak{su}(2))\,.
 \ee
The monodromy about the discriminant loci $\Phi_\pm$ necessarily agree, we denote the common monodromy on 1-cycles as $M_n$. In particular the singularities $\Phi_\pm$ are not mutually local as 7-brane loci in F-theory.

Let us now define our notation for the local model and the bulk geometries:
\be \ba
\Xl&=  \pi^{-1}\lb  \mathcal{O}_{\P^1}(n)|_{r\;\!\leq\;\! R'< R}\rb \amalg   \pi^{-1}\lb  \mathcal{O}_{\P^1}(-n)|_{r\;\!\leq\;\! R'< R}\rb \equiv \Xl_+ \amalg \Xl_-\\[3pt] X^\circ &= X\setminus \Xl\,,
\ea \ee
which is defined in terms of the natural decomposition of the base $\mathbb{F}_n$ lifted to the total space via the projection $\pi$. The radius $R'<R$ is chosen such that the singular loci supporting $\Phi_\pm$ are fully contained in $\Xl_\pm$ respectively. Note that this decomposition differs slightly from those chosen throughout the paper, instead of taking tubular neighbourhoods of the singularities, we have taken the preimage of contractible base patches containing discriminant components. The physical reasoning is that the 5D local gauge theory physics in each component of $\Xl$ is well-understood.

It now follows that $X^\circ$ deformation retracts to an elliptic fibration over $S^3/\Z_n$ with monodromy $M_n$ along the Hopf circle of $S^3/\Z_n$. Further $X^\circ, \partial \Xl_\pm$ are topologically equivalent. As demonstrated in \cite{Hubner:2022kxr,Cvetic:2022imb},
this is sufficient to compute:
\be\label{eq:bdryHomo}
H_k(\partial \Xl_\pm)\cong H_k( X^\circ)\cong \{ \Z, \Z_n \oplus \Gamma_n, \Z, \Z \oplus \Z_n \oplus \Gamma_n,0,\Z\}
\ee
where $\Gamma_n\cong \Z_3, \Z_2^2,\Z_3,\Z_2$ for $n=3,4,6,8$ respectively.

We first focus on the torsional subgroups of \eqref{eq:bdryHomo}. Upon inclusion into the bulk $\Xl$ the 1- and 3-cycles generating $\Gamma_n$ trivialize due to monodromy effects, while the 1- and 3-cycles generating $\Z_n$ trivialize as these are generated by the Hopf circle of $S^3/\Z_n$. We thus obtain the discrete defect and symmetry operators which act on these defects. First, we have 1-form symmetries from M5-branes wrapped on torsional 3-cycles of $H_{3}(X^{\mathrm{loc}}_{\pm})$, and line defects charged under these symmetries from M2-branes wrapped on relative homology cycles of $H_{2}(X^{\mathrm{loc}}_{\pm}, \partial X^{\mathrm{loc}}_{\pm})$. Additionally, one can also consider the magnetic dual 2-form symmetry operators from M2-branes wrapping torsional 1-cycles of $H_{1}(\partial X^{\mathrm{loc}}_{\pm})$ and the corresponding heavy defects from M5-branes wrapping relative homology cycles of
$H_{4}(X^{\mathrm{loc}}_{\pm}, \partial X^{\mathrm{loc}}_{\pm})$. Clearly, defects and symmetry operators constructed from wrappings of $\Gamma_n$ are associated with the gauge theory dynamics of $\Phi_\pm$, in addition to these we have operators constructed from the universal $\Z_n$ contribution.

Next, consider free classes in \eqref{eq:bdryHomo} of degree 2 and 3. The universal factor of $\Z$ is generated in degree 2 by the fiber class and in degree 3 by the zero section (restricted to the boundary). These give rise, in particular, to 2- and 3-form $U(1)$ symmetries whose symmetry operators are $F4$ fluxbranes wrapped on these free homology cycles following \cite{Cvetic:2023plv}, while M5-branes wrapped over the zero section restricted to cycles in $\Xl_\pm$ give defect operators charged under these symmetries.


We now consider the global model and how defects compactify, see figure \ref{fig:Comp}. For this note the isomorphisms
\be \label{eq:double}
H_k(\partial \Xl_+)\cong H_k(X^\circ) \cong H_k(\partial \Xl_-)
\ee
which implies that we can glue any pair of defects with identified boundaries to compact cycles in $X$. To make this concrete,
consider the Mayer-Vietoris sequence with covering $X=\Xl \cup X^\circ$. Let us first focus on the entries:
\be \ba
\dots~\xrightarrow[\text{}]{}& ~H_4(X^\circ )\oplus H_4(\Xl)  ~\xrightarrow[\text{}]{\;\iota_4\;}  ~H_4(X) ~\xrightarrow[\text{}]{\;\partial_4\;}  ~   H_3(\partial \Xl) \\
~\xrightarrow[\text{}]{\;(\imath_3,\jmath_3)\;}& ~H_3(X^\circ)\oplus H_3(\Xl)   ~\xrightarrow[\text{}]{}  ~\dots
\ea \ee
One consequence of \eqref{eq:double} and the preceding discussion is that the mapping
\be
\partial_4\,: ~ H_4(X) \rightarrow H_3(\partial \Xl)\cong H_3(\partial \Xl_+)\oplus H_3(\partial \Xl_-)
\ee
has diagonal image with
\be \label{eq:GlobalStructure}
\textnormal{Im}\,\partial_4 =\Z \oplus \Z_n \oplus \Gamma_n \cong H_3( X^\circ) \,.
\ee
In order to discuss the effect of this on the gauge group $G_\textnormal{full}$, define $\widetilde{G}_-$ to be the simply connected Lie groups with algebra $\mathfrak{g}_-$. Define $\widetilde{G}_+$ to be the gauge group of the local model $\Xl_+$, in an electric polarization, which can have additional Abelian contributions beyond the non-Abelian contribution made explicit in \eqref{eq:LieAlgs}.

Next, note that the discrete group $\Gamma_n$ is the center of $\widetilde{G}_-$. With this, line \eqref{eq:GlobalStructure} now implies how the gauge groups of $\Xl_\pm$ combine, overall we have the gauge symmetry
\be
G_\textnormal{full}= \frac{\widetilde{G}_+\times \widetilde{G}_-}{\Gamma_n} \times \Z_n\times U(1)\,.
\ee
These results are closely related to results in \cite{Apruzzi:2020zot, Braun:2021sex} which focus on the torsional subgroup of the Mordell-Weil group $\textnormal{Tor}\,\textnormal{MW}(\pi)$, and in the cases above, was computed to be isomorphic to $\Gamma_n$ giving an alternative argument for the $\Gamma_n$ quotient. Further, for example when $n=3$, the homology class of the torsional $\Z_3$ section $\tau$ was computed in \cite{Apruzzi:2020zot}, modulo 1, to:
\be \ba
[\tau]&= -(\mathfrak{D}^{(\mathbb{Z}_3)}_++\mathfrak{D}^{(\mathbb{Z}_3)}_-)\\
\mathfrak{D}^{(\mathbb{Z}_3)}_\pm &= \frac{1}{3}(D_\pm^{(1)}+ 2D_\pm^{(2)})
\ea \ee
where $D_\pm^{(i)}$ are Cartan divisors. We note that this result generalizes \eqref{eq:sumofthimbles} and we see that torsional sections admit representations as a fractional linear combination of center divisors, where the latter were introduced in \cite{Hubner:2022kxr}. Given our previous analysis we can also note that there is no quotient involving the $U(1)$ factor as this is associated with the zero section which does not intersect the singularities.

Let us now proceed to studying the gauging and breaking of symmetries. For this consider the terms
\be \ba
\dots~\xrightarrow[\text{}]{}& ~H_2(X^\circ )\oplus H_2(\Xl)  ~\xrightarrow[\text{}]{\;\iota_2\;}  ~H_2(X) ~\xrightarrow[\text{}]{\;\partial_2\;}  ~   H_1(\partial \Xl) \\
~\xrightarrow[\text{}]{\;(\imath_1,\jmath_1)\;}& ~H_1(X^\circ)\oplus H_1(\Xl)   ~\xrightarrow[\text{}]{}  ~\dots
\ea \ee
and we conclude $\textnormal{Im}\,\partial_2\cong \Gamma_n\oplus \Z_n $ via similar arguments as above. We also have $H_1(X^\circ) \cong \Gamma_n\oplus \Z_n$ and so conclude from via duality relations similar to \eqref{eq:DualityCheck} that all discrete 1-form and 2-form symmetries in the local models are either gauged or explicitly broken by particles\,/\,strings resulting from M2-branes wrapped on compact 2-cycles or M5-branes wrapped on compact 4-cycles respectively.

\section{Conclusions} \label{sec:CONC}

Global symmetries provide important constraints on the structure of quantum field theories. In the context of stringy realizations of quantum field theories, it is natural to ask about the fate of these structures in models coupled to gravity. There are general expectations from the Swampland program which suggest that specific degrees of freedom must enter in order to be compatible with constraints from quantum gravity.
In this paper we have explained how local structures associated with generalized symmetries such as the spectrum of defects and topological symmetry operators embed in compact models. We have used this to extract the global form of the gauge group in 7D and 5D M-theory vacua, as well as how global symmetries can emerge in the deep infrared. When the compactification geometry has an elliptic fibration, these same backgrounds can also be used to define F-theory vacua in one higher dimension, and our results agree with methods based on the structure of the Mordell-Weil group of the elliptic model. In the case of 5D vacua, we have also seen how to consistently gauge\,/\,glue together various local models, tracking the fate of higher symmetries including examples with $0$-form, $1$-form and $2$-group structures. We now conclude by discussing some avenues for future investigation.

In our analysis we did not consider frozen singularities (see e.g. \cite{Witten:1997bs, deBoer:2001wca, Tachikawa:2015wka}), but in principle our analysis can be extended to these cases as well. These situations are also interesting because they generate 7D $\mathcal{N} = 1$ vacua with gauge group rank smaller than $22$. It would be interesting to apply the techniques of our paper to obtain the global form of the gauge groups for these cases as well.

While our analysis has primarily focused on quotients of tori and elliptically fibered Calabi-Yau manifolds, it would be natural to consider more general Calabi-Yau spaces with local singularities, as well as brane probes of these singularities. For example, some aspects of generalized symmetries, and the interplay with Verlinde's metastable monopole \cite{Verlinde:2006bc} were recently studied in \cite{Cvetic:2023plv}. It would be quite interesting to analyze the same higher symmetries in the related global models.

Additional physical and geometric phenomena will occur when the compactification manifold has more than six real dimensions. For example, it would be interesting to consider $G_2$-holonomy backgrounds. Examples include quotients of $T^7$, so our methods will likely apply to these situations as well. Further, twisted connected sum constructions of $G_2$-holonomy spaces (see e.g., \cite{Kovalev:2001zr, Corti2012A , Corti2012B}) naturally lend themselves to our cutting and gluing analysis as their construction runs precisely via such operations. In the case of Calabi-Yau fourfolds we can also expect to encounter terminal singularities (see e.g., \cite{Morrison1984, Anno2003}). This will likely give rise to new features in both M- and F-theory vacua which would be interesting to explore in detail.

One of the main assumptions made throughout this work is that our compactification results in asymptotic Minkowski spacetime, as will occur when compactifying on a manifold of special holonomy. More broadly, however, one can consider the consequences of switching on various fluxes, which results in more general backgrounds. Consistent truncations of such warped compactifications often result in AdS vacua. It would be interesting to investigate related gluing constructions of higher symmetries in this more general setting.

\section*{Acknowledgements}

We thank B.S. Acharya, F. Baume, M. Del Zotto, M. Dierigl, L. Lin, M. Montero,
A.P. Turner, T. Weigand, X. Yu, and H.Y. Zhang for helpful discussions.
MC is supported in part by the Slovenian Research Agency (ARRS No. P1-0306)
and Fay R. and Eugene L. Langberg Endowed Chair funds. The work of MC, JJH, and ET is supported by DOE (HEP) Award DE-SC0013528.
The work of MC, JJH, and MH is supported in part by a University Research Foundation grant at the University of Pennsylvania.
The work of MC and MH is also supported by the Simons Foundation Collaboration grant
\#724069 on ``Special Holonomy in Geometry, Analysis and Physics''. We thank the organizers of the meeting ``Geometry, Topology and Singular Special Holonomy Spaces'' held in Freiburg University in June 2022 for kind hospitality during part of this work.
JJH, MH and ET thank the Simons Summer Workshop on Physics 2023 held at the Simons Center for Geometry and Physics (SCGP) for hospitality during the completion of this work. MC, JJH and MH thank the Institute for Basic Science (IBS), Daejeon, for hospitality during the completion of this work. Additionally, MH thanks the Korea Advanced Institute of Science \& Technology (KAIST), Daejeon, for hospitality during the completion of this work.

\newpage

\appendix

\section{Resolutions and Combinatorics for $T^4 / \mathbb{Z}_n$}
\label{app:morequotients}

In this Appendix we discuss in further detail the geometry of the orbifolds $T^{4} / \mathbb{Z}_n$ discussed in section \ref{sec:7dExamp}. The goal of this appendix is to formulate the general homology analysis using methods of resolution, as made explicit for the example $T^4/\Z_2$ in section \ref{sec:Kummer}. One result of this analysis is again that thimbles constitute the fundamental building blocks to keep track of. This simplifies the combinatorial problem and straightforwardly allows to formulate results directly in the singular geometry, as demonstrated starting from \eqref{eq:Planes}. Some more technical computations in this analysis were separated into appendix \ref{app:AdditionalCyclicStuff}.

\paragraph{Resolutions}\mbox{}\medskip \\
We pick up the analysis initiated in section \ref{sec:Kummer} by first arguing that the rational curves contained in an exceptional fiber belonging to ${S}_k'$, introduced below \eqref{eq:fib}, are homologous to the generic fiber when weighted by the Kac labels of the associated extended Dynkin diagram. For $z\in T^2_k/\Z_n$ with stabilizer subgroup of order $m_z$ we have
\be\label{eq:SingularHomology1}
\mathcal{F}_k= m_{z\!\;}\mathcal{F}_{k,z} \,.
\ee
In the crepantly resolved geometry this relation is corrected by exceptional curves. Let $z_i$ denote the singular points of $ \mathcal{F}_{k,z} $ which lift to $m_{z_i}$ points
\be\label{eq:Relevant}
\mathcal{F}_k= m_{z}   \P^1_{k,z}+ \sum_{i} m_{z_i} e_{z_i}
\ee
where $ \P^1_{k,z}$ is the proper transform of $\mathcal{F}_{k,z}$ and $e_{z_i}$ denotes a sum of exceptional curves. If $e_{z_i}$ contracts to an A-type singularity of rank $N_i$ then we have, explicitly studying the resolution as in \cite{Wendland:2000ry},
\be\label{eq:thimble}
e_{z_i}=\sum_{\ell=1}^{N_i}\ell e_{i,\ell}
\ee
 where the curves $\lbbb  e_{i,\ell} \rbbb$ are the standard set of exceptional curves with self-intersection $(-2)$ intersecting with adjacency matrix of A-type of rank $N_i$. The final curve $e_{i,N_i}$ is the exceptional curve intersecting $  \P^1_{k,z}$ once. The numbers $m_z$ and $ m_{z_i} \ell $ are the Kac labels for the extended Dynkin diagram associated with the exceptional fiber in $S'_k$ projecting to $z$.

Note that $e_{z_i}/N_i$ mod $1$ is an A-type thimble. This pattern holds true more generally, that is \eqref{eq:SingularHomology1} is extended by collections of exceptional curves organized by thimbles.

The key relation in the above discussion is \eqref{eq:Relevant} which runs over all points $z,z'\in \P^1_k$ lifting to singular fibers listed in \eqref{eq:ExceptionalCurves}. From there we derive that fractional linear combinations of exceptional divisors
\be\label{eq:RationalCollection}
\frac{1}{\textnormal{gcd}(m_z,m_{z'})}\lb m_z\P^1_{k,z}-m_{z'}\P^1_{k,z'}\rb=\frac{1}{\textnormal{gcd}(m_z,m_{z'})}\sum_{i} \lb m_{z_i} e_{z_i} - m_{z_i'} e_{z_i'} \rb
\ee
are integral classes of the K3 lattice of $(T^4/\Z_n)'$. The righthand side descends to a collection of compact representatives for A-type thimbles when reducing coefficients modulo 1.

The rational curves contained in exceptional fibers or equivalently the vertical curves of the elliptic double pencil only span a subset of the K3 lattice. It was argued in \cite{Wendland:2000ry} that global sections of the elliptic double pencil complete the previously discussed set of curves to a generating set of the full K3 lattice.

We now consider the cases $n=3,4,6$. The case $n=2$ is highly symmetric and studied in detail in section \ref{sec:Kummer} and \cite{Nikulin_1975}. Begin by noting that the linear equation
\be\label{eq:LinearEq}
z_i=\alpha \bar z_j+\beta
\ee
with $i,j=1,2$ and $i\neq j$, is well-defined on $T^4$ whenever $\alpha$ is an integer power of the complex structure parameter $\tau$ or vanishing.\footnote{The case $\alpha=0$ leads to vertical curves for one of the elliptic pencils previously discussed.} For arbitrary $\beta$ the equation cuts out a two-torus parameterized by either $z_i$ or $z_j$. This two-torus is fixed by $\Z_n$ whenever $\beta$ is a fixed point of the $\Z_n$ action and descends to a curve in $T^4/\Z_n$. However, whenever the stabilizer of $\beta$ is non-trivial then we find collections of such curves which are grouped into orbits by $\Z_n$. Curves grouped into a single orbit now descend to a single curve in $T^4/\Z_n$. We denote the order of the stabilizer subgroup of $\beta$ by $N_\beta$ and introduce
\be\label{eq:KeyRelation3}
C^{\;\!ij}_{\alpha\beta}=\lbbb (z_i,z_j) \in T^4/\Z_k \,|\, z_i=\alpha \bar z_j+\beta\,, ~\alpha=\tau^m\,,~m\in \Z\,, ~ N_\beta\neq 1 \rbbb
\ee
where $\omega$ is the primitive $n$-th root of unity generating the $\Z_n$ group action. It was shown in \cite{Wendland:2000ry} that the proper transform of $C^{\;\!ij}_{\alpha\beta}$ can be oriented such that intersections with the curves \eqref{eq:thimble} are non-negative and that such curves form the center nodes of affine Dynkin diagrams, as previously the case for $\P^1_k$. Further, the collections of curves contributing to the same Dynkin diagram of curves, weighted by the corresponding Kac labels, are homologous when the slopes $\alpha$ of the central nodes agree but constants $\beta$ differ. This follows by tracing the corresponding homotopy in $T^4$ (shifts of $\beta$) through the $\Z_n$ quotient. Such homotopic collection of curves can again be equated in homology, as in \eqref{eq:RationalCollection} for cases with $\alpha=0$, and solving for the central nodes $C^{\;\!ij}_{\alpha\beta}$ we again find fractional linear combinations of exceptional curves which are integral classes of the K3 lattice.

Let $L_E$ denote the sublattice of the K3 lattice $H_2 ( (T^4/\Z_n)^{\prime};\Z)$ spanned by all exceptional curves and $\overline{L}_E$ the smallest primitive sublattice containing $L_E$. The set of exceptional curves together with fractional linear combinations of these derived from the elliptic double pencil as in \eqref{eq:RationalCollection} together with the fractional linear combinations discussed in the previous paragraph generate $\overline{L}_E$. We have
\be
\textnormal{Tor}\,H_2(T^4/\Z_n;\Z) \cong \overline{L}_E / L_E\,.
\ee
The elements of $ \overline{L}_E / L_E$ are linear combinations of A-type thimbles, as for example already noted in \eqref{eq:RationalCollection}. The generators of $\textnormal{Tor}\,H_2(T^4/\Z_n;\Z)$ are glued together from thimbles.

We now step through a number of different examples. We begin by revisiting the orbifold $T^4 / \mathbb{Z}_2$, and then turn to $T^4 / \mathbb{Z}_3$. The examples $T^4/\Z_n$ with $n=4,6$ can be analyzed in a similar fashion but carry less symmetry as $n$ is not prime.

\paragraph{Combinatorics and Affine Structures}\mbox{}\medskip \\
We now formulate the results of the computation sketched above in the singular geometry. The connection follows immediately from \eqref{eq:RationalCollection} whose right hand side is a difference of integer multiples of thimbles. There is one thimble per singularity and hence, including the multiplicities, these constitute the geometric units of the computation. Passing to the non-compact presentation of the thimbles, see \eqref{eq:ResolvedThimble}, we can then make statements in the singular geometry with contracted exceptional curves. 


\medskip

\noindent {\bf Case $X=T^4/\Z_2$\,:}

The fixed points carry the structure of an affine vector space modeled on $\Z_2^4$. Upon a choice of origin we can therefore label the $SU(2)$ factors and their centers by $I\in \Z_2^4$. Next, we introduce the set of planes
\be \label{eq:Planes}
\mathcal{H}=\lbbb H_{ij}^{kl}\subset \Z_2^4\,|\, I\in H_{ij}^{kl} ~\Leftrightarrow ~ \textnormal{positions }i,j \textnormal{ of $I$ have values }k,l \textnormal{ respectively} \rbbb
\ee
where $k,l\in \{0,1\}$ and $i,j\in \lbbb 1,2,3,4 \rbbb$. Each $H_{ij}^{kl}$ contains four indices. There are $24=4\times 6$ planes. Two planes are parallel whenever their values for $i,j$ agree but values for $k,l$ differ.

We discuss the quotients from this affine perspective. The quotient in \eqref{eq:SugraGroup} on non-Abelian factors is then formulated as
\be
\Z_2^5 \cong \Big\langle \diag \lb  \Z_2^H,  \Z_2^{H'}\rb \subset SU(2)^{16} \,|\, H,H' \in \mathcal{H}\textnormal{ parallel planes}\Big\rangle
\ee
where $\Z_2^H,\Z_2^{H'}$ is the diagonal subgroup of $SU(2)$ centers labelled by $I\in H,H'$ respectively. The Abelian factors admit a labelling by classes $[H]$ of parallel planes as determined by \eqref{eq:U1Generator}. The second quotient in \eqref{eq:SugraGroup} is by
\be \label{eq:one}
\Z_2^6 \cong  \Big\langle  \diag \lb \Z_2^{[H]} , \Z_2 \rb \subset (SU(2)^{16} / \Z_2^5)\times U(1)_{[H]}  \,|\, H\in \mathcal{H} \Big\rangle \,.
\ee

Note that quotients of $G_{\mathrm{loc}}$ always involve 8 center subgroups of the $SU(2)$ factors, i.e., the torsional cycles depicted in figure \ref{fig:Comp} stretch between 8 singularities. Quotients on the $U(1)$'s involve 4 center subgroups. \medskip

\noindent {\bf Case $X=T^4/\Z_3$\,:}

The fixed points carry the structure of an affine vector space modeled on $\Z_3^2$. Upon a choice of origin we can therefore label the $SU(3)$ factors and their centers by $I\in \Z_3^2$. Next, we introduce the set of lines
\be
\mathcal{L}=\lbbb L \subset \Z_3^2\,|\, L \textnormal{ is the graph of an affine linear equation} \rbbb
\ee
of which there are $12=3\times 4$. Each $L$ contains three indices. Two lines are parallel if they do not intersect.

We discuss the quotients from this affine perspective. The quotient in \eqref{eq:SugraGroup2} on non-Abelian factors is by
\be \label{eq:Tor2}
\Z_3^3 \cong \Big\langle \diag \lb  \Z_3^L,  -\Z_3^{L'}\rb \subset SU(3)^{9} \,|\, L,L' \in \mathcal{L}\textnormal{ parallel lines}\Big\rangle
\ee
where $\Z_3^L,\Z_2^{L'}$ is the diagonal subgroup of the center of the $SU(3)$ factors labelled by $I\in L,L'$ respectively. Here $-\Z_3^{L'}$ denotes the conjugate of $\Z_3^{L'}$. The Abelian factors admit a labelling by classes $[L]$ of parallel lines up to a redundancy we now explain.

Consider the Abelian factors $U(1)_{[L]}$ naively labelled by a class of parallel lines. Then we derive a quotient by
\be  \label{eq:two}
\Z_3^4 \cong  \Big\langle  \diag \lb \Z_3^{[L]} , \Z_3 \rb \subset (SU(3)^{9} / \Z_3^3)\times U(1)_{[L]}  \,|\, L\in \mathcal{L} \Big\rangle \,.
\ee
which does not match the $\Z_3^3$ quotient in $\eqref{eq:SugraGroup2}$. However, we have $\Z_3^6\cong \textnormal{coker}\,\jmath_2$ and therefore there exists a subgroup $\Z_3'\subset \Z_3^4$ (the diagonal) which is trivial in the center of $ SU(3)^{9}$ mod \eqref{eq:Tor2}. Therefore $\Z_3'$ is removed only from the Abelian factors and $U(1)^4_{[L]}/\Z_3'=U(1)^4$ where the $U(1)^4$ is the Abelian factor appearing in \eqref{eq:SugraGroup2}.

Note that quotients of $G_\mathrm{loc}$ always involve 6 center subgroups of the $SU(3)$ factors, i.e., the torsional cycles depicted in figure \ref{fig:Comp} stretch between 6 singularities. Quotients on the $U(1)$'s involve 3 center subgroups.

\medskip

\noindent {\bf Case $X=T^4/\Z_4$\,:} The fixed points do not carry the structure of an
affine vector space, however we can find $12=3\times 4$ planes, which we group into parallel triplets
\be\ba\label{eq:triplets}
\mathcal{G}_1&=  \lbbb \lbbb z_1=f \rbbb \subset X  \, |\, f\in F_0 \rbbb\,,  && \mathcal{G}_2=  \lbbb \lbbb z_2=f\rbbb \subset  X \, |\, f\in F_0\rbbb\,,\\
\mathcal{G}_3&=  \lbbb \lbbb z_1+z_2^*=f\rbbb \subset X\, |\, f\in F_0\rbbb\,, ~ && \mathcal{G}_4=  \lbbb \lbbb z_1+\tau z_2^*=f\rbbb \subset X \, |\, f\in F_0 \rbbb \\
\ea\ee
where $F_0$ is the set consisting of the three orbifold points of $T^2/\Z_4$. Again parallel planes do not share points. Here $\tau=i$ and the group action maps as $(z_1,z_2)\mapsto (iz_1,-iz_2)$ whereby each plane is well-defined.

The quotient by $\Z_4\oplus \Z_2^2$ on the non-Abelian group $G_\mathrm{loc} = SU(4)^4\times SU(2)^6$ is follows from the rows of the matrix $M_{T^4/\Z_4}$ given in \eqref{eq:T4Z4Matrix}. Note from there we can conclude that $\Z_4\oplus \Z_2^2$ embeds purely into the center of the $SU(4)^4$ factor:
\be
G_{\textnormal{full}}=\frac{(SU(4)^4/\Z_4\times \Z_2^2)\times SU(2)^{6}\times U(1)^4}{ \Z_4^2\times \Z_2^2}\,.
\ee
The $U(1)$ factors are in correspondence with classes of parallel planes $\mathcal{G}_i$. Planes are subsets of $\mathcal{G}_i$ with fixed $f\in F_0$, i.e. in the corresponding fibration with base $T^2/\Z_4$ they project to the same $f\in T^2/\Z_4$. Two planes are parallel if they are subsets of the same $\mathcal{G}_i$ but with respect to distinct fixed points in $F_0$. The $\Z_4^2\oplus \Z_2^2$ quotient involving the Abelian factors derives by considering any one representative of a parallel class and evaluating \eqref{eq:Relevant}.

Note this is the first instance with non-trivial multiplicities $m_z$ as introduced in \eqref{eq:RationalCollection}, see Appendix \ref{app:AdditionalCyclicStuff} for further details. Further, the quotients of $G_\ADE$ involve either 2 or 4 center subgroups of the $SU(4)$ factors, i.e., the torsional cycles depicted in figure \ref{fig:Comp} stretch between 2 or 4 singularities. Quotients on the $U(1)$'s involve 3 or 4 center subgroups of $SU(2)^6\times SU(4)^4$, see \eqref{eq:U1T4Z4}.

\medskip

\noindent {\bf Case $X=T^4/\Z_6$\,:} The fixed points do not carry the structure
of an affine vector space, however we can find $12=3\times 4$ planes, which we
group into parallel triplets which again take the form \eqref{eq:triplets} where, however,
$F_0$ is replaced with the set of the three orbifold points of $T^2/\Z_6$ and now $\tau=\exp(2\pi i/6)$.

The quotient by $\Z_6^3\oplus \Z_2$ on the non-Abelian group $SU(6) \times SU(3)^{4} \times SU(2)^{5}$ follows from the rows of the matrix $M_{T^4/\Z_6}$ given in \eqref{eq:T4Z6Matrix}. For the Abelian quotient similar remarks hold as in the $T^4/\Z_4$ example.

Again note the multiplicities $m_z$ are relevant as discussed in Appendix \ref{app:AdditionalCyclicStuff}. Further, the quotients of $G_\ADE$ involve either 5 or 6 center subgroups, i.e., the torsional cycles depicted in figure \ref{fig:Comp} stretch between 5 or 6 singularities. Quotients on the $U(1)$'s involve 3 or 4 center subgroups.

\section{Computational Details on $T^4/\mathbb{Z}_n$}
\label{app:AdditionalCyclicStuff}
In this Appendix we give details on the computations described in appendix \ref{app:morequotients}. See also Appendix \ref{app:nonAbelian} where computations are revisted and checked from an equivariant perspective.
Let us begin with the cases $n=4,6$. These cases are technically more involved as they do not exhibit the affine structures present for cases $n=2,3$ which rely on $n$ being prime. We start with labelling conventions for the ADE singularities.

We label the $A_3^4\oplus A_1^6$  singularities of $T^4/\Z_4$ as
\be \ba
&(A_3)_{IJ}\\
&(A_1)_{IJ}, (A_1)_+, (A_1)_-
\ea \ee
where $I,J\in \mathbb{Z}_2$, i.e., $I,J=0,1$ (see figure \ref{Fig:T4/Z4}).
\begin{figure}[]
\centering
\scalebox{0.7}{
\begin{tikzpicture}
	\begin{pgfonlayer}{nodelayer}
		\node [style=none] (0) at (-2.5, 7.5) {\Large$\frac{1}{2}$};
		\node [style=none] (1) at (-7.5, 8.5) {};
		\node [style=none] (2) at (-5.5, 8.5) {};
		\node [style=none] (3) at (-3.5, 8.5) {};
		\node [style=none] (4) at (-1.5, 8.5) {};
		\node [style=none] (5) at (0.5, 8.5) {};
		\node [style=none] (6) at (2.5, 8.5) {};
		\node [style=none] (9) at (-7.5, 6.5) {};
		\node [style=none] (10) at (-7.5, 4.5) {};
		\node [style=none] (11) at (-7.5, 2.5) {};
		\node [style=none] (12) at (-7.5, 0.5) {};
		\node [style=none] (13) at (-7.5, -1.5) {};
		\node [style=none] (16) at (-5.5, 6.5) {};
		\node [style=none] (17) at (-3.5, 6.5) {};
		\node [style=none] (18) at (-1.5, 6.5) {};
		\node [style=none] (19) at (0.5, 6.5) {};
		\node [style=none] (20) at (2.5, 6.5) {};
		\node [style=none] (23) at (-5.5, 4.5) {};
		\node [style=none] (24) at (-3.5, 4.5) {};
		\node [style=none] (25) at (-1.5, 4.5) {};
		\node [style=none] (26) at (0.5, 4.5) {};
		\node [style=none] (27) at (2.5, 4.5) {};
		\node [style=none] (30) at (-5.5, 2.5) {};
		\node [style=none] (31) at (-3.5, 2.5) {};
		\node [style=none] (32) at (-1.5, 2.5) {};
		\node [style=none] (33) at (0.5, 2.5) {};
		\node [style=none] (34) at (2.5, 2.5) {};
		\node [style=none] (37) at (-5.5, 0.5) {};
		\node [style=none] (38) at (-3.5, 0.5) {};
		\node [style=none] (39) at (-1.5, 0.5) {};
		\node [style=none] (40) at (0.5, 0.5) {};
		\node [style=none] (41) at (2.5, 0.5) {};
		\node [style=none] (44) at (-5.5, -1.5) {};
		\node [style=none] (45) at (-3.5, -1.5) {};
		\node [style=none] (46) at (-1.5, -1.5) {};
		\node [style=none] (47) at (0.5, -1.5) {};
		\node [style=none] (48) at (2.5, -1.5) {};
		\node [style=none] (65) at (-6.25, 8) {\Large$z_1$};
		\node [style=none] (66) at (-7, 7.25) {\Large$z_2$};
		\node [style=none] (68) at (-6.5, 1.5) {\Large $\frac{i}{2}$};
		\node [style=none] (69) at (-0.5, 7.5) {\Large $\frac{i}{2}$};
		\node [style=none] (70) at (-6.5, 5.5) {\Large 0};
		\node [style=none] (71) at (-4.5, 7.5) {\Large 0};
		\node [style=none] (72) at (1.5, 7.5) {\Large $\frac{1+i}{2}$};
		\node [style=none] (73) at (-6.5, -0.5) {\Large $\frac{1+i}{2}$};
		\node [style=none] (78) at (-4.5, 5.5) {\Large $4$};
		\node [style=none] (79) at (-2.5, 5.5) {\Large $2$};
		\node [style=none] (80) at (1.5, 5.5) {\Large 4};
		\node [style=none] (81) at (1.5, 3.5) {\Large 2};
		\node [style=none] (82) at (-0.5, 5.5) {\Large $2$};
		\node [style=none] (85) at (-2.5, 3.5) {\Large 2};
		\node [style=none] (86) at (-0.5, 3.5) {\Large 2};
		\node [style=none] (87) at (-2.5, 1.5) {\Large 2};
		\node [style=none] (88) at (-0.5, 1.5) {\Large 2};
		\node [style=none] (89) at (-4.5, 3.5) {\Large 2};
		\node [style=none] (90) at (-4.5, 1.5) {\Large 2};
		\node [style=none] (91) at (1.5, -0.5) {\Large 4};
		\node [style=none] (102) at (1.5, 1.5) {\Large 2};
		\node [style=none] (105) at (-4.5, -0.5) { \Large 4};
		\node [style=none] (106) at (-2.5, -0.5) {\Large 2};
		\node [style=none] (107) at (-0.5, -0.5) {\Large 2};
		\node [style=none] (114) at (-5, 6) {};
		\node [style=none] (115) at (-5, 5) {};
		\node [style=none] (116) at (-4, 5) {};
		\node [style=none] (117) at (-4, 6) {};
		\node [style=none] (118) at (-3, 5) {};
		\node [style=none] (119) at (-3, 6) {};
		\node [style=none] (120) at (0, 6) {};
		\node [style=none] (121) at (0, 5) {};
		\node [style=none] (122) at (1, 6) {};
		\node [style=none] (123) at (1, 5) {};
		\node [style=none] (126) at (-5, 4) {};
		\node [style=none] (127) at (-4, 4) {};
		\node [style=none] (128) at (-4, 1) {};
		\node [style=none] (129) at (-5, 1) {};
		\node [style=none] (130) at (-5, 0) {};
		\node [style=none] (131) at (-4, 0) {};
		\node [style=none] (134) at (-3, 3) {};
		\node [style=none] (135) at (-2, 4) {};
		\node [style=none] (136) at (0, 2) {};
		\node [style=none] (137) at (-1, 1) {};
		\node [style=none] (138) at (-1, 4) {};
		\node [style=none] (139) at (0, 3) {};
		\node [style=none] (140) at (-2, 1) {};
		\node [style=none] (141) at (-3, 2) {};
		\node [style=none] (142) at (1, 4) {};
		\node [style=none] (143) at (1, 1) {};
		\node [style=none] (146) at (-3, 0) {};
		\node [style=none] (147) at (0, 0) {};
		\node [style=none] (150) at (1, 0) {};
		\node [style=none] (162) at (-4.5, 8) {};
		\node [style=none] (163) at (-6.5, 3.5) {\Large $\frac{1}{2}$};
		\node [style=none] (164) at (-5, -1) {};
		\node [style=none] (165) at (-4, -1) {};
		\node [style=none] (166) at (-3, -1) {};
		\node [style=none] (167) at (0, -1) {};
		\node [style=none] (168) at (1, -1) {};
		\node [style=none] (169) at (2, -1) {};
		\node [style=none] (170) at (2, 0) {};
		\node [style=none] (171) at (2, 1) {};
		\node [style=none] (172) at (2, 4) {};
		\node [style=none] (173) at (2, 5) {};
		\node [style=none] (174) at (2, 6) {};
		\node [style=none] (175) at (-1.5, 5.5) {\scriptsize \color{red} $(A_1)_{00}$};
		\node [style=none] (176) at (-4.5, 2.5) {\scriptsize \color{red} $(A_1)_{01}$};
		\node [style=none] (177) at (1.5, 2.5) {\scriptsize \color{red} $(A_1)_{10}$};
		\node [style=none] (178) at (-1.5, -0.5) {\scriptsize \color{red} $(A_1)_{11}$};
		\node [style=none] (179) at (-3.75, 0.25) {\scriptsize \color{red} $(A_3)_{01}$};
		\node [style=none] (180) at (-3.75, 4.75) {\scriptsize \color{red} $(A_3)_{00}$};
		\node [style=none] (181) at (0.75, 4.75) {\scriptsize \color{red} $(A_3)_{10}$};
		\node [style=none] (182) at (0.75, 0.25) {\scriptsize \color{red} $(A_3)_{11}$};
		\node [style=none] (183) at (-2.25, 4.5) {\scriptsize \color{red} $(A_1)_{+}$};
		\node [style=none] (184) at (-1, 4.5) {\scriptsize \color{red} $(A_1)_{-}$};
	\end{pgfonlayer}
	\begin{pgfonlayer}{edgelayer}
		\draw [style=ThickLine]  (1.center) to (16.center);
		\draw [style=ThickLine]  (130.center) to (131.center);
		\draw [style=ThickLine]  (128.center) to (127.center);
		\draw [style=ThickLine]  (127.center) to (126.center);
		\draw [style=ThickLine]  (126.center) to (129.center);
		\draw [style=ThickLine]  (129.center) to (128.center);
		\draw [style=ThickLine]  (114.center) to (115.center);
		\draw [style=ThickLine]  (115.center) to (116.center);
		\draw [style=ThickLine]  (116.center) to (117.center);
		\draw [style=ThickLine]  (117.center) to (114.center);
		\draw [style=ThickLine]  (119.center) to (118.center);
		\draw [style=ThickLine]  (118.center) to (121.center);
		\draw [style=ThickLine]  (121.center) to (120.center);
		\draw [style=ThickLine]  (120.center) to (119.center);
		\draw [style=ThickLine]  (122.center) to (123.center);
		\draw [style=ThickLine]  (135.center) to (136.center);
		\draw [style=ThickLine]  (137.center) to (134.center);
		\draw [style=ThickLine]  (138.center) to (141.center);
		\draw [style=ThickLine]  (139.center) to (140.center);
		\draw [style=ThickLine]  [bend right=270, looseness=1.50] (140.center) to (141.center);
		\draw [style=ThickLine]  [bend left=90, looseness=1.50] (138.center) to (139.center);
		\draw [style=ThickLine]  [bend left=270, looseness=1.50] (135.center) to (134.center);
		\draw [style=ThickLine]  [bend left=90, looseness=1.50] (136.center) to (137.center);
		\draw [style=ThickLine] (142.center) to (143.center);
		\draw [style=ThickLine] (146.center) to (147.center);
		\draw [style=ThickLine] (9.center) to (20.center);
		\draw [style=ThickLine] (2.center) to (44.center);
		\draw [style=ThickLine] (122.center) to (174.center);
		\draw [style=ThickLine] (174.center) to (173.center);
		\draw [style=ThickLine] (173.center) to (123.center);
		\draw [style=ThickLine] (142.center) to (172.center);
		\draw [style=ThickLine] (172.center) to (171.center);
		\draw [style=ThickLine] (171.center) to (143.center);
		\draw [style=ThickLine] (150.center) to (168.center);
		\draw [style=ThickLine] (168.center) to (169.center);
		\draw [style=ThickLine] (169.center) to (170.center);
		\draw [style=ThickLine] (170.center) to (150.center);
		\draw [style=ThickLine] (147.center) to (167.center);
		\draw [style=ThickLine] (167.center) to (166.center);
		\draw [style=ThickLine] (166.center) to (146.center);
		\draw [style=ThickLine] (131.center) to (165.center);
		\draw [style=ThickLine] (165.center) to (164.center);
		\draw [style=ThickLine] (130.center) to (164.center);
	\end{pgfonlayer}
\end{tikzpicture}
}
\caption{We table pairs of fixed points of $T^2/\Z_4$. In $T^4/\Z_4$ their stabilizers are subgroups of $\Z_4$, we list the orders of these subgroups. These pairs are grouped into orbits as indicated.}
\label{Fig:T4/Z4}
\end{figure}
\begin{figure}[]
\centering
\scalebox{0.7}{
\begin{tikzpicture}
	\begin{pgfonlayer}{nodelayer}
		\node [style=none] (0) at (-6.5, 10.5) {\Large $\frac{e^{\pi i/6}}{\sqrt{3}}$};
		\node [style=none] (1) at (-11.5, 11.5) {};
		\node [style=none] (2) at (-9.5, 11.5) {};
		\node [style=none] (3) at (-7.5, 11.5) {};
		\node [style=none] (4) at (-5.5, 11.5) {};
		\node [style=none] (5) at (-3.5, 11.5) {};
		\node [style=none] (6) at (-1.5, 11.5) {};
		\node [style=none] (7) at (0.5, 11.5) {};
		\node [style=none] (8) at (2.5, 11.5) {};
		\node [style=none] (9) at (-11.5, 9.5) {};
		\node [style=none] (10) at (-11.5, 7.5) {};
		\node [style=none] (11) at (-11.5, 5.5) {};
		\node [style=none] (12) at (-11.5, 3.5) {};
		\node [style=none] (13) at (-11.5, 1.5) {};
		\node [style=none] (14) at (-11.5, -0.5) {};
		\node [style=none] (15) at (-11.5, -2.5) {};
		\node [style=none] (16) at (-9.5, 9.5) {};
		\node [style=none] (17) at (-7.5, 9.5) {};
		\node [style=none] (18) at (-5.5, 9.5) {};
		\node [style=none] (19) at (-3.5, 9.5) {};
		\node [style=none] (20) at (-1.5, 9.5) {};
		\node [style=none] (21) at (0.5, 9.5) {};
		\node [style=none] (22) at (2.5, 9.5) {};
		\node [style=none] (23) at (-9.5, 7.5) {};
		\node [style=none] (24) at (-7.5, 7.5) {};
		\node [style=none] (25) at (-5.5, 7.5) {};
		\node [style=none] (26) at (-3.5, 7.5) {};
		\node [style=none] (27) at (-1.5, 7.5) {};
		\node [style=none] (28) at (0.5, 7.5) {};
		\node [style=none] (29) at (2.5, 7.5) {};
		\node [style=none] (30) at (-9.5, 5.5) {};
		\node [style=none] (31) at (-7.5, 5.5) {};
		\node [style=none] (32) at (-5.5, 5.5) {};
		\node [style=none] (33) at (-3.5, 5.5) {};
		\node [style=none] (34) at (-1.5, 5.5) {};
		\node [style=none] (35) at (0.5, 5.5) {};
		\node [style=none] (36) at (2.5, 5.5) {};
		\node [style=none] (37) at (-9.5, 3.5) {};
		\node [style=none] (38) at (-7.5, 3.5) {};
		\node [style=none] (39) at (-5.5, 3.5) {};
		\node [style=none] (40) at (-3.5, 3.5) {};
		\node [style=none] (41) at (-1.5, 3.5) {};
		\node [style=none] (42) at (0.5, 3.5) {};
		\node [style=none] (43) at (2.5, 3.5) {};
		\node [style=none] (44) at (-9.5, 1.5) {};
		\node [style=none] (45) at (-7.5, 1.5) {};
		\node [style=none] (46) at (-5.5, 1.5) {};
		\node [style=none] (47) at (-3.5, 1.5) {};
		\node [style=none] (48) at (-1.5, 1.5) {};
		\node [style=none] (49) at (0.5, 1.5) {};
		\node [style=none] (50) at (2.5, 1.5) {};
		\node [style=none] (51) at (-9.5, -0.5) {};
		\node [style=none] (52) at (-7.5, -0.5) {};
		\node [style=none] (53) at (-5.5, -0.5) {};
		\node [style=none] (54) at (-3.5, -0.5) {};
		\node [style=none] (55) at (-1.5, -0.5) {};
		\node [style=none] (56) at (0.5, -0.5) {};
		\node [style=none] (57) at (2.5, -0.5) {};
		\node [style=none] (58) at (-9.5, -2.5) {};
		\node [style=none] (59) at (-7.5, -2.5) {};
		\node [style=none] (60) at (-5.5, -2.5) {};
		\node [style=none] (61) at (-3.5, -2.5) {};
		\node [style=none] (62) at (-1.5, -2.5) {};
		\node [style=none] (63) at (0.5, -2.5) {};
		\node [style=none] (64) at (2.5, -2.5) {};
		\node [style=none] (65) at (-10.25, 11) {\Large $z_1$};
		\node [style=none] (66) at (-11, 10.25) {\Large $z_2$};
		\node [style=none] (67) at (-10.5, 6.5) {\Large $\frac{e^{\pi i/6}}{\sqrt{3}}$};
		\node [style=none] (68) at (-10.5, 4.5) {\Large $\frac{2e^{\pi i/6}}{\sqrt{3}}$};
		\node [style=none] (69) at (-4.5, 10.5) {\Large $\frac{2e^{\pi i/6}}{\sqrt{3}}$};
		\node [style=none] (70) at (-10.5, 8.5) {\Large 0};
		\node [style=none] (71) at (-8.5, 10.5) {\Large 0};
		\node [style=none] (72) at (-2.5, 10.5) {\Large $\frac{1}{2}$};
		\node [style=none] (73) at (-10.5, 2.5) {\Large $\frac{1}{2}$};
		\node [style=none] (74) at (-10.5, 0.5) {\Large $\frac{e^{\pi i/3}}{2}$};
		\node [style=none] (75) at (-0.5, 10.5) {\Large $\frac{e^{\pi i/3}}{2}$};
		\node [style=none] (76) at (1.5, 10.5) {\Large $\frac{1+e^{\pi i/3}}{2}$};
		\node [style=none] (77) at (-10.5, -1.5) {\Large $\frac{1+e^{\pi i/3}}{2}$};
		\node [style=none] (78) at (-8.5, 8.5) {\Large $6$};
		\node [style=none] (79) at (-6.5, 8.5) {\Large 3};
		\node [style=none] (80) at (-2.5, 8.5) {\Large 2};
		\node [style=none] (81) at (-2.5, 6.5) {\Large 1};
		\node [style=none] (82) at (-4.5, 8.5) {\Large 3};
		\node [style=none] (83) at (-0.5, 8.5) {\Large 2};
		\node [style=none] (84) at (1.5, 8.5) {\Large 2};
		\node [style=none] (85) at (-6.5, 6.5) {\Large 3};
		\node [style=none] (86) at (-4.5, 6.5) {\Large 3};
		\node [style=none] (87) at (-6.5, 4.5) {\Large 3};
		\node [style=none] (88) at (-4.5, 4.5) {\Large 3};
		\node [style=none] (89) at (-8.5, 6.5) {\Large 3};
		\node [style=none] (90) at (-8.5, 4.5) {\Large 3};
		\node [style=none] (91) at (-2.5, 2.5) {\Large 2};
		\node [style=none] (92) at (-0.5, 2.5) {\Large 2};
		\node [style=none] (93) at (1.5, 2.5) {\Large 2};
		\node [style=none] (94) at (-2.5, 0.5) {\Large 2};
		\node [style=none] (95) at (-0.5, 0.5) {\Large 2};
		\node [style=none] (96) at (1.5, 0.5) {\Large 2};
		\node [style=none] (97) at (-2.5, -1.5) {\Large 2};
		\node [style=none] (98) at (-0.5, -1.5) {\Large 2};
		\node [style=none] (99) at (1.5, -1.5) {\Large 2};
		\node [style=none] (100) at (-0.5, 6.5) {\Large 1};
		\node [style=none] (101) at (1.5, 6.5) {\Large 1};
		\node [style=none] (102) at (-2.5, 4.5) {\Large 1};
		\node [style=none] (103) at (-0.5, 4.5) { \Large 1};
		\node [style=none] (104) at (1.5, 4.5) {\Large 1};
		\node [style=none] (105) at (-8.5, 2.5) {\Large 2};
		\node [style=none] (106) at (-6.5, 2.5) {\Large 1};
		\node [style=none] (107) at (-4.5, 2.5) {\Large 1};
		\node [style=none] (108) at (-8.5, 0.5) {\Large 2};
		\node [style=none] (109) at (-6.5, 0.5) {\Large 1};
		\node [style=none] (110) at (-4.5, 0.5) {\Large 1};
		\node [style=none] (111) at (-8.5, -1.5) {\Large 2};
		\node [style=none] (112) at (-6.5, -1.5) {\Large 1};
		\node [style=none] (113) at (-4.5, -1.5) {\Large 1};
		\node [style=none] (114) at (-9, 9) {};
		\node [style=none] (115) at (-9, 8) {};
		\node [style=none] (116) at (-8, 8) {};
		\node [style=none] (117) at (-8, 9) {};
		\node [style=none] (118) at (-7, 8) {};
		\node [style=none] (119) at (-7, 9) {};
		\node [style=none] (120) at (-4, 9) {};
		\node [style=none] (121) at (-4, 8) {};
		\node [style=none] (122) at (-3, 9) {};
		\node [style=none] (123) at (-3, 8) {};
		\node [style=none] (124) at (2, 8) {};
		\node [style=none] (125) at (2, 9) {};
		\node [style=none] (126) at (-9, 7) {};
		\node [style=none] (127) at (-8, 7) {};
		\node [style=none] (128) at (-8, 4) {};
		\node [style=none] (129) at (-9, 4) {};
		\node [style=none] (130) at (-9, 3) {};
		\node [style=none] (131) at (-8, 3) {};
		\node [style=none] (132) at (-9, -2) {};
		\node [style=none] (133) at (-8, -2) {};
		\node [style=none] (134) at (-7, 6) {};
		\node [style=none] (135) at (-6, 7) {};
		\node [style=none] (136) at (-4, 5) {};
		\node [style=none] (137) at (-5, 4) {};
		\node [style=none] (138) at (-5, 7) {};
		\node [style=none] (139) at (-4, 6) {};
		\node [style=none] (140) at (-6, 4) {};
		\node [style=none] (141) at (-7, 5) {};
		\node [style=none] (142) at (-3, 7) {};
		\node [style=none] (143) at (-3, 4) {};
		\node [style=none] (144) at (2, 4) {};
		\node [style=none] (145) at (2, 7) {};
		\node [style=none] (146) at (-7, 3) {};
		\node [style=none] (147) at (-4, 3) {};
		\node [style=none] (148) at (-4, -2) {};
		\node [style=none] (149) at (-7, -2) {};
		\node [style=none] (150) at (-3, 3) {};
		\node [style=none] (155) at (-3, -2) {};
		\node [style=none] (158) at (2, -2) {};
		\node [style=none] (161) at (2, 3) {};
		\node [style=none] (162) at (-1, 3) {};
		\node [style=none] (163) at (-3, 1) {};
		\node [style=none] (164) at (-3, -1) {};
		\node [style=none] (165) at (1, 3) {};
		\node [style=none] (166) at (2, 2) {};
		\node [style=none] (167) at (-2, -2) {};
		\node [style=none] (168) at (0, -2) {};
		\node [style=none] (169) at (2, 0) {};
		\node [style=none] (170) at (-8.5, 11) {};
		\node [style=none] (171) at (-7.75, 7.75) {\scriptsize \color{red} $(A_5)$};
		\node [style=none] (172) at (0.5, -0.5) {\scriptsize \color{red}  $(A_1)_+$};
		\node [style=none] (173) at (-1.5, 1.5) {\scriptsize \color{red}  $(A_1)_-$};
		\node [style=none] (174) at (0.5, 1.5) {\scriptsize \color{red}  $(A_1)_0$};
		\node [style=none] (175) at (-8.5, 1.5) {\scriptsize \color{red} $(A_1)_{z_2}$};
		\node [style=none] (176) at (-1.5, 8.5) {\scriptsize \color{red} $(A_1)_{z_1}$};
		\node [style=none] (177) at (-5.5, 8.5) {\scriptsize \color{red} $(A_1)_{z_2}$};
		\node [style=none] (178) at (-8.5, 5.5) {\scriptsize \color{red} $(A_2)_{z_1}$};
		\node [style=none] (179) at (-6.5, 7.5) {\scriptsize \color{red} $(A_2)_+$};
		\node [style=none] (180) at (-4.5, 7.5) {\scriptsize \color{red} $(A_2)_-$};
	\end{pgfonlayer}
	\begin{pgfonlayer}{edgelayer}
		\draw [style=ThickLine] (2.center) to (58.center);
		\draw [style=ThickLine] (9.center) to (22.center);
		\draw [style=ThickLine] (1.center) to (16.center);
		\draw [style=ThickLine] (131.center) to (133.center);
		\draw [style=ThickLine] (133.center) to (132.center);
		\draw [style=ThickLine] (132.center) to (130.center);
		\draw [style=ThickLine] (130.center) to (131.center);
		\draw [style=ThickLine] (128.center) to (127.center);
		\draw [style=ThickLine] (127.center) to (126.center);
		\draw [style=ThickLine] (126.center) to (129.center);
		\draw [style=ThickLine] (129.center) to (128.center);
		\draw [style=ThickLine] (114.center) to (115.center);
		\draw [style=ThickLine] (115.center) to (116.center);
		\draw [style=ThickLine] (116.center) to (117.center);
		\draw [style=ThickLine] (117.center) to (114.center);
		\draw [style=ThickLine] (119.center) to (118.center);
		\draw [style=ThickLine] (118.center) to (121.center);
		\draw [style=ThickLine] (121.center) to (120.center);
		\draw [style=ThickLine] (120.center) to (119.center);
		\draw [style=ThickLine] (122.center) to (123.center);
		\draw [style=ThickLine] (123.center) to (124.center);
		\draw [style=ThickLine] (124.center) to (125.center);
		\draw [style=ThickLine] (125.center) to (122.center);
		\draw [style=ThickLine] (135.center) to (136.center);
		\draw [style=ThickLine] (137.center) to (134.center);
		\draw [style=ThickLine] (138.center) to (141.center);
		\draw [style=ThickLine] (139.center) to (140.center);
		\draw [style=ThickLine] [bend right=270, looseness=1.50] (140.center) to (141.center);
		\draw [style=ThickLine] [bend left=90, looseness=1.50] (138.center) to (139.center);
		\draw [style=ThickLine] [bend left=270, looseness=1.50] (135.center) to (134.center);
		\draw [style=ThickLine] [bend left=90, looseness=1.50] (136.center) to (137.center);
		\draw [style=ThickLine] (142.center) to (143.center);
		\draw [style=ThickLine] (143.center) to (144.center);
		\draw [style=ThickLine] (144.center) to (145.center);
		\draw [style=ThickLine] (145.center) to (142.center);
		\draw [style=ThickLine] (147.center) to (148.center);
		\draw [style=ThickLine] (148.center) to (149.center);
		\draw [style=ThickLine] (146.center) to (149.center);
		\draw [style=ThickLine] (146.center) to (147.center);
		\draw [style=ThickLine] (150.center) to (161.center);
		\draw [style=ThickLine] (161.center) to (158.center);
		\draw [style=ThickLine] (158.center) to (155.center);
		\draw [style=ThickLine] (155.center) to (150.center);
		\draw [style=ThickLine] (162.center) to (163.center);
		\draw [style=ThickLine] (165.center) to (164.center);
		\draw [style=ThickLine] (166.center) to (167.center);
		\draw [style=ThickLine] (169.center) to (168.center);
	\end{pgfonlayer}
\end{tikzpicture}
}
\caption{We list pairs of fixed points of $T^2/\Z_6$. In $T^4/\Z_6$ their stabilizers are subgroups of $\Z_6$, we list the orders of these subgroups. These pairs are grouped into orbits as indicated.}
\label{Fig:T4/Z6}
\end{figure}

We label the $A_5\oplus A_3^4\oplus A_1^5$ singularities of $T^4/\Z_6$ as
\be\ba
&(A_5) \\
&(A_2)_{z_1=0} , (A_2)_{z_2=0}, (A_2)_{+}, (A_2)_{-}\\
&(A_1)_{z_1=0} , (A_1)_{z_2=0}, (A_1)_{+}, (A_1)_{0},(A_1)_{-}\,.
\ea \ee
See figure \ref{Fig:T4/Z6}.

Next we determine maximal sets of singularities which are parallel. We say that a set of singularities is maximal if there exists a fibration $\pi:T^4/\Z_n\rightarrow T^2/\Z_n$ such that all singularities are contained in the same fiber and, conversely, all singularities in that fiber are contained in the set. Two maximal sets are called parallel if they are maximal sets with respect to the same fibration $\pi$ and disjoint. Maximal sets are of ADE type, i.e. resolving the singularities we find the structures depicted in figure \ref{fig:ExtendedDynkin}.

The maximal sets of $E_7$ type for $T^4/\mathbb{Z}_4$ are
\be\label{eq:E71}\ba
z_1=[0]\,:&\qquad(A_3)_{00}, (A_3)_{01} , 2(A_1)_{01}\\
z_2=[0]\,:&\qquad (A_3)_{00}, (A_3)_{10} , 2(A_1)_{00}\\
z_1+\bar z_2=[0]\,:&\qquad (A_3)_{00}, (A_3)_{11} , 2(A_1)_{+}\\
z_1+\tau \bar z_2=[0]\,:&\qquad (A_3)_{00}, (A_3)_{11} , 2(A_1)_{-}\\
\ea\ee
and
\be\label{eq:E72} \ba
z_1=[(1+i)/2]\,:&\qquad(A_3)_{10}, (A_3)_{11} , 2(A_1)_{10}\\
z_2=[(1+i)/2]\,:&\qquad (A_3)_{01}, (A_3)_{11} , 2(A_1)_{11}\\
z_1+\bar z_2=[(1+i)/2]\,:&\qquad (A_3)_{10}, (A_3)_{01} , 2(A_1)_{-}\\
z_1+\tau \bar z_2=[(1+i)/2]\,:&\qquad (A_3)_{10}, (A_3)_{01} , 2(A_1)_{+}\\[3pt]
\ea\ee
while the maximal sets of $D_4$ type for $T^4/\mathbb{Z}_4$ are
\be\label{eq:D4}\ba
z_1= [1/2]\,:&\qquad 2(A_1)_{00}, 2(A_1)_{11}, 2(A_1)_{+}, 2(A_1)_{-}\\
z_2= [1/2]\,:&\qquad 2(A_1)_{01}, 2(A_1)_{10}, 2(A_1)_{+}, 2(A_1)_{-}\\
z_1+\bar z_2= [1/2]\,:&\qquad 2(A_1)_{01}, 2(A_1)_{01}, 2(A_1)_{00}, 2(A_1)_{11} \\
z_1+\tau \bar z_2= [1/2]\,:&\qquad 2(A_1)_{01}, 2(A_1)_{01}, 2(A_1)_{00}, 2(A_1)_{11} \,.
\ea\ee

The maximal sets of $E_8$ type for $T^4/\mathbb{Z}_6$ are
\be\label{eq:E8}\ba
z_1=[0]\,:&\qquad(A_5), 2(A_2)_{z_1=0} , 3(A_1)_{z_1=0}\\
z_2=[0]\,:&\qquad (A_5), 2(A_2)_{z_2=0} , 3(A_1)_{z_2=0}\\
z_1+\bar z_2=[0]\,:&\qquad (A_5), 2(A_2)_{+} , 3(A_1)_{+}\\
z_1+\tau \bar z_2=[0]\,:&\qquad (A_5), 2(A_2)_{-} , 3(A_1)_{-}\\
\ea\ee
 and maximal sets of $E_6$ type for $T^4/\mathbb{Z}_6$ are
\be\label{eq:E6}\ba
z_1= \Big[e^{\pi i/6}/\sqrt{3}\Big]\,:&\qquad 2(A_2)_{z_2=0} , 2(A_2)_{+}, 2(A_2)_{-}\\
z_2=\Big[e^{\pi i/6}/\sqrt{3}\Big]\,:&\qquad  2(A_2)_{z_1=0} , 2(A_2)_{-}, 2(A_2)_{+}\\
z_1+\bar z_2=\Big[e^{\pi i/6}/\sqrt{3}\Big]\,:&\qquad 2(A_2)_{z_1=0}, 2(A_2)_{z_2=0} , 2(A_2)_{-}\\
z_1+\tau \bar z_2=\Big[e^{\pi i/6}/\sqrt{3}\Big]\,:&\qquad 2(A_2)_{z_1=0}, 2(A_2)_{z_2=0} , 2(A_2)_{+}\\
\ea\ee
and maximal sets of $D_4$ type for $T^4/\mathbb{Z}_6$ are
\be\label{eq:D42}\ba
z_1=[1/2]\,:&\qquad 3(A_1)_{z_2=0}, 3(A_1)_{+} , 3(A_1)_{0} , 3(A_1)_{-} \\
z_2=[1/2]\,:&\qquad 3(A_1)_{z_1=0}, 3(A_1)_{+} , 3(A_1)_{0} ,3(A_1)_{-} \\
z_1+\bar z_2=[1/2]\,:&\qquad 3(A_1)_{z_1=0},  3(A_1)_{z_2=0} , 3(A_1)_{0} , 3(A_1)_{-}\\
z_1+\tau \bar z_2=[1/2]\,:&\qquad 3(A_1)_{z_1=0}, 3(A_1)_{z_2=0} , 3(A_1)_{+}, 3(A_1)_{0}  \,.
\ea\ee

In both cases we included multiplicities given by the orbit size, i.e. $n/r$ where $r$ is the rank of the singularity.

Next we organize the above into matrices. We pick two ordered bases
\be \ba
B_4&=\{ (A_3)_{00}, (A_3)_{01}, (A_3)_{10}, (A_3)_{11}, (A_1)_{00}, (A_1)_{01}, (A_1)_{10}, (A_1)_{11}, (A_1)_{+}, (A_1)_{-} \}\\
B_6&=\{ (A_5), (A_2)_{z_1 = 0}, (A_2)_ {z_2 = 0}, (A_2)_+,  (A_2)_-, (A_1)_ {z_1 = 0},  (A_1)_ {z_2 = 0},  (A_1)_+, (A_1)_0, (A_1)_-\}
\ea \ee
with respect to which the above maximal sets \eqref{eq:E71}, \eqref{eq:E72}, \eqref{eq:D4} respectively take the form
\be \label{eq:U1T4Z4}\ba
E_7^{(1)}[T^4/\Z_4]&=\left(
\begin{array}{cccccccccc}
 1 & 1 & 0 & 0 & 0 & 2 & 0 & 0 & 0 & 0 \\
 1 & 0 & 1 & 0 & 2 & 0 & 0 & 0 & 0 & 0 \\
 1 & 0 & 0 & 1 & 0 & 0 & 0 & 0 & 2 & 0 \\
 1 & 0 & 0 & 1 & 0 & 0 & 0 & 0 & 0 & 2 \\
\end{array}
\right) \\
E_7^{(2)}[T^4/\Z_4]&=\left(
\begin{array}{cccccccccc}
 0 & 0 & 1 & 1 & 0 & 0 & 2 & 0 & 0 & 0 \\
 0 & 1 & 0 & 1 & 0 & 0 & 0 & 2 & 0 & 0 \\
 0 & 1 & 1 & 0 & 0 & 0 & 0 & 0 & 0 & 2 \\
 0 & 1 & 1 & 0 & 0 & 0 & 0 & 0 & 2 & 0 \\
\end{array}
\right)\\
D_4[T^4/\Z_4]&=\left(
\begin{array}{cccccccccc}
 0 & 0 & 0 & 0 & 2 & 0 & 0 & 2 & 2 & 2 \\
 0 & 0 & 0 & 0 & 0 & 2 & 2 & 0 & 2 & 2 \\
 0 & 0 & 0 & 0 & 2 & 2 & 2 & 2 & 0 & 0 \\
 0 & 0 & 0 & 0 & 2 & 2 & 2 & 2 & 0 & 0 \\
\end{array}
\right)
\ea \ee
while the maximal sets \eqref{eq:E8}, \eqref{eq:E6}, \eqref{eq:D42} respectively take the form
\be \ba
E_8[T^4/\Z_6]&=\left(
\begin{array}{cccccccccc}
 1 & 2 & 0 & 0 & 0 & 3 & 0 & 0 & 0 & 0 \\
 1 & 0 & 2 & 0 & 0 & 0 & 3 & 0 & 0 & 0 \\
 1 & 0 & 0 & 2 & 0 & 0 & 0 & 3 & 0 & 0 \\
 1 & 0 & 0 & 0 & 2 & 0 & 0 & 0 & 0 & 3 \\
\end{array}
\right) \\
E_6[T^4/\Z_6]&=\left(
\begin{array}{cccccccccc}
 0 & 0 & 2 & 2 & 2 & 0 & 0 & 0 & 0 & 0 \\
 0 & 2 & 0 & 2 & 2 & 0 & 0 & 0 & 0 & 0 \\
 0 & 2 & 2 & 0 & 2 & 0 & 0 & 0 & 0 & 0 \\
 0 & 2 & 2 & 2 & 0 & 0 & 0 & 0 & 0 & 0 \\
\end{array}
\right)
\\
D_4[T^4/\Z_6]&=\left(
\begin{array}{cccccccccc}
 0 & 0 & 0 & 0 & 0 & 0 & 3 & 3 & 3 & 3 \\
 0 & 0 & 0 & 0 & 0 & 3 & 0 & 3 & 3 & 3 \\
 0 & 0 & 0 & 0 & 0 & 3 & 3 & 0 & 3 & 3 \\
 0 & 0 & 0 & 0 & 0 & 3 & 3 & 3 & 3 & 0 \\
\end{array}
\right)\,.
\ea \ee

Now we subtract parallel maximal sets, taking multiplicity into account. Multiplicities are determined by the ratio between the generic fiber volume and the exceptional fibers of ADE type. This generalizes \eqref{eq:CoeffMatrix}. We have the coefficient matrices
\be \ba
M_{T^4/\Z_4}=\textnormal{Join}\Big\{~&E_7^{(1)}[T^4/\Z_4] -E_7^{(2)}[T^4/\Z_4] ,\\
&2 E_7^{(1)}[T^4/\Z_4] - D_4[T^4/\Z_4] , \\ &2 E_7^{(2)}[T^4/\Z_4] - D_4[T^4/\Z_4] ~\Big\}\\
M_{T^4/\Z_6}=\textnormal{Join}\Big\{~&2E_8[T^4/\Z_6] -E_6[T^4/\Z_6] , \\ &3 E_8[T^4/\Z_6] - D_4[T^4/\Z_6] ~\Big\}\,.
\ea \ee
Here join refers to the operation of stacking all rows into a single matrix with the same number of columns. Explicitly we have
\be\label{eq:T4Z4Matrix}
\scalebox{0.8}{$\textnormal{\large $M_{T^4/\Z_4}$}=\left(
\begin{array}{cccccccccc}
(A_3)_{00} & (A_3)_{01} & (A_3)_{10} & (A_3)_{11} & (A_1)_{00} & (A_1)_{01} & (A_1)_{10} & (A_1)_{11} & (A_1)_{+} & (A_1)_{-} \\[3pt] \hline
 1 & 1 & -1 & -1 & 0 & 2 & -2 & 0 & 0 & 0 \\
 1 & -1 & 1 & -1 & 2 & 0 & 0 & -2 & 0 & 0 \\
 1 & -1 & -1 & 1 & 0 & 0 & 0 & 0 & 2 & -2 \\
 1 & -1 & -1 & 1 & 0 & 0 & 0 & 0 & -2 & 2 \\
 2 & 2 & 0 & 0 & -2 & 4 & 0 & -2 & -2 & -2 \\
 2 & 0 & 2 & 0 & 4 & -2 & -2 & 0 & -2 & -2 \\
 2 & 0 & 0 & 2 & -2 & -2 & -2 & -2 & 4 & 0 \\
 2 & 0 & 0 & 2 & -2 & -2 & -2 & -2 & 0 & 4 \\
 0 & 0 & 2 & 2 & -2 & 0 & 4 & -2 & -2 & -2 \\
 0 & 2 & 0 & 2 & 0 & -2 & -2 & 4 & -2 & -2 \\
 0 & 2 & 2 & 0 & -2 & -2 & -2 & -2 & 0 & 4 \\
 0 & 2 & 2 & 0 & -2 & -2 & -2 & -2 & 4 & 0 \\
\end{array}
\right)$}
\ee
and
\be\label{eq:T4Z6Matrix}
\scalebox{0.8}{$ {\textnormal{\large $M_{T^4/\Z_6}$}}=\left(
\begin{array}{cccccccccc}
(A_5) & (A_2)_{z_1 = 0} & (A_2)_ {z_2 = 0} & (A_2)_+ &  (A_2)_- & (A_1)_ {z_1 = 0}&  (A_1)_ {z_2 = 0} &  (A_1)_+& (A_1)_0& (A_1)_- \\[3pt] \hline
 2 & 4 & -2 & -2 & -2 & 6 & 0 & 0 & 0 & 0 \\
 2 & -2 & 4 & -2 & -2 & 0 & 6 & 0 & 0 & 0 \\
 2 & -2 & -2 & 4 & -2 & 0 & 0 & 6 & 0 & 0 \\
 2 & -2 & -2 & -2 & 4 & 0 & 0 & 0 & 0 & 6 \\
 3 & 6 & 0 & 0 & 0 & 9 & -3 & -3 & -3 & -3 \\
 3 & 0 & 6 & 0 & 0 & -3 & 9 & -3 & -3 & -3 \\
 3 & 0 & 0 & 6 & 0 & -3 & -3 & 9 & -3 & -3 \\
 3 & 0 & 0 & 0 & 6 & -3 & -3 & -3 & -3 & 9 \\
\end{array}
\right)$}
\ee
where we appended our labelling conventions. Computing the Smith normal form we find
\be\ba
\textnormal{SNF}(M_{T^4/\Z_4})&=\lb \begin{array}{cc} M_4 & {\mathbf 0}_{7\times 3} \\  {\mathbf 0}_{5 \times 7} &  {\mathbf 0}_{5 \times 3} \end{array} \rb\,,\quad && M_4=\textnormal{diag}(1,2,2,4,4,4,4)\\[5pt]
\textnormal{SNF}(M_{T^4/\Z_6})&=\lb M_6 ~ {\mathbf 0}_{8\times 2}  \rb \,, &&  M_6=\textnormal{diag}(1,6,6,6,6,6,6,6)\\
\ea\ee
where $M_n$ are quadratic sub-blocks and ${\mathbf 0}_{n\times m}$ denotes a block of zeros of dimension $n\times m$. From here we divide by $n$, take the result modulo 1 and conclude
\be
\textnormal{Tor}\,H_2(T^4/\Z_4)\cong \Z_4\oplus \Z_2^2\,, \qquad \textnormal{Tor}\,H_2(T^4/\Z_6)\cong \Z_6\,.
\ee
The attentive reader has noticed that we have not considered all curves of the form \eqref{eq:KeyRelation3}. Rather, we have only taken a minimal independent set of such curves in order to not overload presentation.

Finally, let us give, for completeness and in less detail, the matrix for the case $T^4/\Z_3$ where no multiplicities have to be taken into account. It reads
\be\label{eq:T4Z3Matrix} \scalebox{0.8}{$
{\textnormal{\large $M_{T^4/\Z_3}$}}=\left(
\begin{array}{ccccccccc}
 1 & 1 & 1 & -1 & -1 & -1 & 0 & 0 & 0 \\
 1 & 1 & 1 & 0 & 0 & 0 & -1 & -1 & -1 \\
 0 & 0 & 0 & 1 & 1 & 1 & -1 & -1 & -1 \\
 1 & 0 & -1 & -1 & 1 & 0 & 0 & -1 & 1 \\
 1 & -1 & 0 & 0 & 1 & -1 & -1 & 0 & 1 \\
 0 & -1 & 1 & 1 & 0 & -1 & -1 & 1 & 0 \\
 1 & -1 & 0 & -1 & 0 & 1 & 0 & 1 & -1 \\
 1 & 0 & -1 & 0 & -1 & 1 & -1 & 1 & 0 \\
 0 & 1 & -1 & 1 & -1 & 0 & -1 & 0 & 1 \\
 1 & -1 & 0 & 1 & -1 & 0 & 1 & -1 & 0 \\
 1 & 0 & -1 & 1 & 0 & -1 & 1 & 0 & -1 \\
 0 & 1 & -1 & 0 & 1 & -1 & 0 & 1 & -1 \\
 1 & -1 & 0 & 0 & 1 & -1 & -1 & 0 & 1 \\
 1 & 0 & -1 & -1 & 1 & 0 & 0 & -1 & 1 \\
 0 & 1 & -1 & -1 & 0 & 1 & 1 & -1 & 0 \\
 1 & -1 & 0 & -1 & 0 & 1 & 0 & 1 & -1 \\
 1 & 0 & -1 & 0 & -1 & 1 & -1 & 1 & 0 \\
 0 & 1 & -1 & 1 & -1 & 0 & -1 & 0 & 1 \\
\end{array}
\right)$}
\ee
from which one computes $\textnormal{Tor}\,H_2(T^4/\Z_3)=\Z_3^3$.

\section{Non-Abelian Quotients $T^{4} / \Gamma$ and Equivariant Cohomology}\label{app:nonAbelian}

In this Appendix we consider quotients of the form $X=T^4 / \Gamma$ for $\Gamma$ a non-Abelian finite group and compute the homology groups appearing in Mayer-Vietoris sequence
\be \label{eq:MVSapp}\ba
0~&\xrightarrow[\text{}]{} ~H_2(X^\circ )  ~\xrightarrow[\text{}]{\;\jmath_2\;}  ~H_2(X) ~\xrightarrow[\text{}]{\;\partial_2\;}  ~   H_1(\partial \Xl)\cong \oplus_i \,H_1(\partial U_i) ~\xrightarrow[\text{}]{\;\imath_1\;} ~H_1(X^\circ)  ~\xrightarrow[\text{}]{}  ~0
\ea \ee
for all possible $\Gamma$ such that $X$ is K3 manifold. Rather than applying the algebro-geometric techniques we used to determine these homology groups when $\Gamma$ was Abelian, we present an alternative method that expresses each of the homology groups in the exact sequence \eqref{eq:MVSapp} as equivariant cohomology groups. We believe the methods employed here can be of wider use for calculating homology groups of orbifolds that arise from finite group quotients. After reviewing the classification of quotients for $T^4/\Gamma$, for $\Gamma$ non-Abelian, we employ equivariant methods to, in particular, express $H_2(X)$ in terms of (twisted) cohomology groups of the classifying space $B\Gamma$, and end this appendix with calculating the global form of the gauge groups for the 7D supergravity theories that arise from compactifying M-theory on $T^4/\Gamma$.

The most general possible finite groups $\Gamma$ such that $T^4/\Gamma$ is Calabi-Yau are \cite{19881}:
\begin{equation}
    \Gamma=\Z_2, \; \Z_3, \; \Z_4, \; \Z_6, \; \mathcal{D}_4, \; \mathcal{D}_5, \;  \mathcal{T}
\end{equation}
which, in addition to the cyclic quotients studied in the main body of this paper, include the binary dihedral groups $\mathcal{D}_N$ for $N = 4,5$ as well as the binary tetrahedral group $\mathcal{T}$.\footnote{In our conventions, $\mathcal{D}_N$ is an order $4N-8$ group isomorphic to $\mathbb{Z}_{2N-4} \rtimes \mathbb{Z}_2$ and is a double cover of the dihedral group, the symmetries of an $(N-2)$-gon. The group $\mathcal{T}$ is an order $48$ isomorphic to $D_4\rtimes \Z_3$ and is the double cover of the tetrahedral group, the symmetries of a tetrahedron.} A new feature relative to the Abelian cases is that depending on how $\mathcal{D}_4$ and $\mathcal{T}$ act on $T^4$, we may have different ADE singularities on $T^4/\Gamma$. The group $D_4$ has three possible actions\footnote{Up to isomorphisms and moving in connected regions of K3 moduli space that do not resolve the ADE singularities.} on $T^4$ so we denote them by $\mathcal{D}_4$, $\mathcal{D}_{4}^{\prime}$, and $\mathcal{D}_{4}^{\prime \prime}$, while $\mathcal{T}$ has two possible actions which we denote by $\mathcal{T}$ and $\mathcal{T}^{\prime}$. The ADE singularity structure is summarized in Table \ref{tab:ADEloci} (see Lemma 3.19 of \cite{19881} and Table 18 of \cite{deBoer:2001wca}).

{\renewcommand{\arraystretch}{1.225}
\begin{table}[t!]
\centering
\begin{tabular}{|c||c|}
\hline
 $X$ &  ADE Singularities  \\
 \hline \hline
$T^4/\mathcal{D}_4$ &  $D^2_4\oplus A^3_3\oplus A^2_1$  \\
\hline
$T^4/\mathcal{D}^{\prime}_4$ & $D^4_4\oplus  A^3_1$    \\
\hline
$T^4/\mathcal{D}^{\prime \prime}_4$ &  $A^6_3\oplus A_1$   \\
\hline
$T^4/\mathcal{D}_5$ & $D_5\oplus A^3_3\oplus A^2_2\oplus A_1$   \\
\hline
$T^4/\mathcal{T}$ & $E_6\oplus D_4\oplus A^4_2\oplus A_1$   \\
\hline
$T^4/\mathcal{T}^{\prime}$ &  $A_5\oplus A^2_3\oplus A^4_2$  \\ \hline
\end{tabular}
\caption{Collection of complex codimension 2 ADE singularties for each $T^4/\Gamma$. This immediately tells us the non-Abelian part of the gauge algebra of the 7D supergravity associated to M-theory compactified on $T^4/\Gamma$, with the full gauge algebra having an additional $\mathfrak{u}(1)^3$ factor in these cases so that the total rank is 22.}
\label{tab:ADEloci}
\end{table}}

Describing now the $\mathcal{D}_4$ actions on $T^4$, first let us write $T^4$ as $\mathbb{C}^2/\Lambda$ under the following identifications
\begin{equation}\label{eq:vanillaquotient}
  \mathbb{C}^2/\Lambda:  (z_1,z_2)\sim (z_1+1, z_2)\sim (z_1,z_2+1)\sim (z_1+i,z_2)\sim (z_1,z_2+i).
\end{equation}
Then the \textit{unprimed} $\mathcal{D}_4$ action on this $T^4$ is generated by the following elements:
\begin{equation}\label{eq:d4vanilla}
    \alpha_{4}: \; \; (z_1,z_2)\mapsto (iz_1,-iz_2), \quad\quad \delta: \; \; (z_1,z_2)\mapsto (-z_2,z_1).
\end{equation}
The primed $\mathcal{D}_4$ actions are on the four-tori $T^4$ with a different complex structure than in \eqref{eq:vanillaquotient}. $\mathcal{D}^{\prime}_4$ acts on $\mathbb{C}^2/\Lambda^{\prime}$ which is defined by the identifications
\begin{equation}
   \mathbb{C}^2/\Lambda^{\prime}: (z_1,z_2)\sim (z_1+1,z_2)\sim (z_1+\sqrt{2}i, z_2)\sim (z_1,z_2+\sqrt{2}i)\sim (z_1+\lambda,z_2+i\lambda) ,
\end{equation}
where $\lambda \equiv 1/\sqrt{2}+i$, while $\mathcal{D}^{\prime \prime}_4$ acts on $\mathbb{C}^2/\Lambda^{\prime \prime}$ which is defined by
\begin{equation}
   \mathbb{C}^2/\Lambda^{\prime\prime}:  (z_1,z_2)\sim (z_1+1,z_2)\sim (z_1,z_2+\sqrt{2})\sim (z_1, z_2+i)\sim (z_1+\lambda,z_2+i/\sqrt{2}).
\end{equation}
The action of $\mathcal{D}^{\prime}_4$ and $\mathcal{D}^{\prime \prime}_4$ can both be defined by the generators:
\begin{equation}\label{eq:d4strawberry}
    \alpha_{4}: \; \; (z_1,z_2)\mapsto (iz_1,-iz_2), \quad\quad \widetilde{\delta}: \; \; (z_1,z_2)\mapsto \left(-z_2+\frac{(1+i)}2,z_1\right).
\end{equation}
where $(z_1,z_2)$ are to be understood as coordinates on $\mathbb{C}^2/\Lambda^{\prime}$ and $\mathbb{C}^2/\Lambda^{\prime \prime}$ respectively.

The $\mathcal{D}_5$ action on on $T^4$ is relatively simple, as it acts on a $T^4$ with the complex structure $\mathbb{C}^2/\Lambda$ shown in \eqref{eq:vanillaquotient} and is given by
\begin{equation}
     \alpha_{6}: \; \; (z_1,z_2)\mapsto (\zeta z_1,\zeta^{-1} z_2), \quad\quad \delta: \; \; (z_1,z_2)\mapsto (-z_2,z_1).
\end{equation}
with $\zeta=\exp(2\pi i /6)$.

Moving on to the $\mathcal{T}$ and $\mathcal{T}^\prime$ actions on $T^4$, the former can be defined on $\mathbb{C}^2/\Lambda$ with complex structure \eqref{eq:vanillaquotient} with generators $\alpha_4$ and $\delta$ appearing in \eqref{eq:d4vanilla} along with the additional generator of order three
\begin{align}
    \mu: \; \frac{1}{1-i}\begin{pmatrix}
        1 & i \\
        1 & -i
    \end{pmatrix}
\end{align}
acting on $(z_1,z_2)^\mathrm{T}$. The $\mathcal{T}^\prime$ action can be defined on either the $\mathbb{C}^2/\Lambda^{\prime}$ or $\mathbb{C}^2/\Lambda^{\prime \prime}$ complex structures of $T^4$, and is generated by the $\alpha_4$ and $\widetilde{\delta}$ appearing in \eqref{eq:d4strawberry} as well as $\widetilde{\mu}$ which is defined as the composition of $\mu$ with an affine translation:
\begin{equation}
    \widetilde{\mu}:  \left[ (z_1,z_2)\mapsto \left(z_1+\frac{1+i}{2}, z_2+\frac{-1+i}{2} \right)  \right] \circ \mu.
\end{equation}
For more details on non-Abelian orientation-preserving actions of $T^4$ see section 3 of \cite{19881}.

We now move on to detailing a method for determining the groups appearing in \eqref{eq:MVSapp} for each non-Abelian quotient using equivariant cohomology. The advantage of this method is that it is computationally efficient as it reduces the homology computations to twisted cohomology of the classifying spaces $B\Gamma$, while on the other hand it leaves a hands-on definition of the homology classes of $T^4/\Gamma$ harder to obtain, an aspect which is beyond the scope of this Appendix.

To understand how the homology groups of $X=T^4/\Gamma$ and $X^\circ=({T^4}/\Gamma )^\circ$ can be determined from the equivariant cohomology groups associated to  $Y=T^4$, we first note that we can excise the fixed points of the $\Gamma$ action on $Y=T^4$ to give a space $Y^\circ=({T^4})^\circ$ which is acted on by $\Gamma$ freely. The equivariant cohomology groups of $Y^\circ$ with respect to $\Gamma$ are therefore isomorphic to the singular cohomology groups of the smooth quotient $X^\circ = Y^\circ/\Gamma$, we have:
\be
H_\Gamma^*(Y^\circ)=H^*(X^\circ)\,.
\ee

Now let us denote the collection of neighborhoods of fixed points in $Y$ of the $\Gamma$ action by $Y^{\textnormal{loc}}$. Topologically, $Y^{\textnormal{loc}}$ is a collection of balls centered on the fixed points. We now have via the excision property of (co)homology, equivalence of relative equivariant and singular cohomologies for free actions, and finally Poincar\'e-Lefschetz duality the three isomorphisms:\footnote{We use integer coefficients unless stated otherwise.}:
\be
H_\Gamma^*(Y,Y^{\textnormal{loc}})\cong  H_\Gamma^*(Y^\circ,\partial Y^{\textnormal{loc}})\cong H^*(X^\circ,\partial \Xl) \cong H_{4-*}(X^\circ)\,.
\ee
The relative cohomology group $H_\Gamma^*(Y,Y^{\textnormal{loc}})$ sits in the long exact sequence of equivariant relative homology for the pair $(Y,Y^{\textnormal{loc}})$ which is given by
\be
\dots ~\rightarrow ~ H^k_\Gamma( Y, Y^{\textnormal{loc}}) ~\rightarrow ~  H^k_\Gamma (Y) ~\rightarrow ~   H^k_\Gamma ( Y^{\textnormal{loc}}) ~\rightarrow ~ H^{k+1}_\Gamma( Y, Y^{\textnormal{loc}}) ~\rightarrow ~ \dots\,.
\ee
We simplify from here by noting that $Y^{\textnormal{loc}}$ deformation retracts onto fixed points and lifting this retraction to $(Y^{\textnormal{loc}} \times E \Gamma)/\Gamma$ we find
\be\label{eq:BG}
 H^k_\Gamma ( Y^{\textnormal{loc}})  = \bigoplus_{i} H^k( B\Gamma_i)
\ee
where $\Gamma_i\subset \Gamma$ is the subgroup of the action fixing the respective fixed point. As for the equivariant cohomology groups $H^k_\Gamma (Y) $, fortunately these were calculated in many cases in \cite{deBoer:2001wca} using the Serre spectral sequence for the fibration $(Y^{\textnormal{loc}} \times E \Gamma)/\Gamma\rightarrow B\Gamma$.

Combining all of the above we find the exact sequence
\be
\dots ~\rightarrow ~  \bigoplus_{i} H^{k-1}( B\Gamma_i) ~\rightarrow ~ H_{4-k}(X^\circ) ~\rightarrow ~  H^k_\Gamma (Y) ~\rightarrow ~    \bigoplus_{i} H^k( B\Gamma_i) ~\rightarrow ~  \dots\,.
\ee
where we let $H_\ell(X^\circ)=0$ if $\ell<0$. With this, the computation of the homology groups of $X^\circ$ is in many cases reduced to determining the maps of this sequence.

Note that $X^\circ$ deformation retracts to a three-dimensional space and therefore $H_\Gamma^k(Y,Y^{\textnormal{loc}}) =0$ for $k\geq 4$ which implies
\be
H^k_\Gamma(Y)\cong  \bigoplus_{i} H^k( B\Gamma_i)\qquad (k\geq 4)\,.
\ee
Further, for discrete subgroups of $SU(2)$ we have $H^k( B\Gamma_i)=0$ for odd $k$ and $H^2( B\Gamma_i)=\mathrm{Ab}(\Gamma_i)$ which splits the remaining exact subsequence into two pieces given by
\be\ba
0 ~\rightarrow ~ H^0_\Gamma (Y) ~\rightarrow ~  \Z^N ~\rightarrow ~ H_3(X^\circ) ~\rightarrow ~  H^1_\Gamma(Y)~\rightarrow ~ 0 \\
0 ~\rightarrow ~ H_2 (X^\circ) ~\rightarrow ~ H^2_\Gamma(Y) ~\rightarrow ~  \bigoplus_{i}  \Gamma^{\textnormal{ab}}_i ~\rightarrow ~ H_1(X^\circ)  ~\rightarrow ~ H^3_\Gamma(Y) ~\rightarrow ~ 0
\ea \ee
where $N$ is the number of fixed points. This simplifies further by noting that $H^3_\Gamma(Y)=H^1_\Gamma(Y)=0$ for our torus quotients according to \cite{deBoer:2001wca}, so then have
\be
0 ~\rightarrow ~ H_2 (X^\circ) ~\rightarrow ~ H^2_\Gamma(Y) ~\rightarrow ~  \bigoplus_{i}  \Gamma^{\textnormal{ab}}_i ~\rightarrow ~ H_1(X^\circ)  ~\rightarrow  ~ 0
\ee
matching \eqref{eq:MVS} in all entries but $H^2_\Gamma(Y)$ where instead it reads $H_2(X)$. We therefore propose that
\be\label{eq:Cool1}
H^2_\Gamma(Y)\cong H_2(X)
\ee
which agrees with the available results in \cite{deBoer:2001wca,Wendland:2000ry}.

We now provide a basic summary of our equivariant cohomology calculations, most of which already appears in \cite{deBoer:2001wca} with the exception of the case $\Gamma=\mathcal{T}$. The $E_2$-page of the Serre spectral sequence $(Y^{\textnormal{loc}} \times E \Gamma)/\Gamma\rightarrow B\Gamma$ is given by
\begin{equation}\label{eq:equivariantSerrespec}
  E_2^{p,q}=H^p(B\Gamma, (H^q(T^4,\mathbb{Z}))_\Gamma)
\end{equation}
where the subscript $\Gamma$ indicates the $\Gamma$-module structure inherited from the action on $T^4$. A crucial feature to make note of is that Lemma 3.4 of \cite{19881} implies that the $\mathcal{D}_4$-module structure of $H^q(T^4,\Z)$ has identical  $\mathcal{D}^{\prime}_4$- and $\mathcal{D}^{\prime \prime}_4$-module structure, and likewise its $\mathcal{T}$- and $\mathcal{T}^{\prime}$-module structures are identical. This is not surprising since the extra affine transformations in the primed actions do not affect the $k$-forms on $T^4$.

Next, note that for $p+q=2$, we have $E_2^{p,q}=E_\infty^{p,q}$ because for $\Gamma$ any ADE subgroup of $SU(2)$, $H^k(B\Gamma,\mathbb{Z})=0$ for odd $k$ and $H^k(B\Gamma, \mathbb{Z}^n_\Gamma)=0$ for $k$ even if $\mathbb{Z}^n_\Gamma$ is a non-trivial irreducible $\Gamma$-module. So from (\ref{eq:Cool1}) we have that
\begin{equation}\label{eq:homologyequivcoh}
 H_2(T^4/\Gamma)= H^0(B\Gamma, (\mathbb{Z})^6_\Gamma)\oplus H^1(B\Gamma, (\mathbb{Z})^4_\Gamma) \oplus  H^2(B\Gamma, \mathbb{Z} ).
\end{equation}
We list these cohomology groups with (un)twisted coefficients in the table in \eqref{tab:specseqgrps}. Detailing now how these groups are derived, first notice that the group $H^2(B\Gamma,\Z)$, where $\mathbb{Z}$ is a trivial $\Gamma$-module, is simply $\mathrm{Ab}(\Gamma)$. The group $H^0(B\Gamma, (\mathbb{Z}^6)_\Gamma)$ on the other hand consists of $\Gamma$-invariant elements of $(\mathbb{Z}^6)_\Gamma$ where the $\Gamma$ action is the induced action on $H^2(T^4,\Z)$.

The groups $H^1(B\Gamma,(\mathbb{Z}^4)_\Gamma)$ require a little more discussion. If $\Gamma=\mathbb{Z}_k$, let us single out a generator $\alpha\in \Z_k$. Then if we consider the following (not necessarily exact) sequence of maps
\begin{equation}
   \Z^4 \; \xrightarrow[]{d_0} \; \Z^4 \; \xrightarrow[]{d_1} \; \Z^4
\end{equation}
where $d_0=1-\alpha$ and $d_1=1+\alpha+\alpha^2+...+\alpha^{k-1}$, then we have that
\begin{equation}\label{eq:h1twisted}
    H^1(B\Z_k,(\mathbb{Z}^4)_{\Z_k})=\frac{\mathrm{Ker} d_1}{\mathrm{Im}d_0}.
\end{equation}

For the groups $D_{4+n}$, we start with the fact that these can be presented as extensions of $\Z_2$
\begin{equation}
    0 \xrightarrow[]{} \; \Z_{4+2n} \; \xrightarrow[]{} \; D_{4+n} \; \xrightarrow[]{} \; \Z_2 \; \xrightarrow[]{}  0\,.
\end{equation}
This implies a fibration structure on $BD_{4+n}$ of the form
\begin{equation}
     B\Z_{4+2n} \; \hookrightarrow \; BD_{4+n} \; \xrightarrow[]{} \; B\Z_2
\end{equation}
so we can calculate $H^1(BD_{4+n},(\Z)^4_{D_{4+n}})$ using the Serre spectral sequence which collapses at the second page to
\begin{equation*}\ba\label{eq:h1twistdtype}
    H^1(BD_{4+n},(\Z)^4_{D_{4+n}})&=H^1(B\Z_2,\left[ H^0(B\Z_{4+2n},(\Z)^4_{\mathbb{Z}_{4+2n}})\right]_{\Z_2})\\&~~~\,\oplus  H^0(B\Z_2,\left[ H^1(B\Z_{4+2n},(\Z)^4_{\mathbb{Z}_{4+2n}})\right]_{\Z_2})\,.
\ea\end{equation*}
Notice that the $\Z_{4+2n}$-module structure on $H_1(T^4,\Z)=\Z^4$ is induced from its $D_{4+n}$-module structure, and furthermore that
\be
H^1(B\Z_2,\left[ H^0(B\Z_{4+2n},(\Z)^4_{\mathbb{Z}_{4+2n}})\right]_{\Z_2})=0
\ee
because there are no $\Z_{4+2n}$ invariant generators of $\Z^4$. One may also calculate twisted cohomology groups for $BD_{4+2n}$ using a method similar to \eqref{eq:h1twisted}, see for instance equations (254) and (255) of \cite{deBoer:2001wca}.\footnote{We observe that our table entry $H^1(BD_4,(\Z)_{D_4}^4)=\Z^2_2$ does not match that found in Table (260) of \cite{deBoer:2001wca} (note that they use $U(1)$ coefficients for the $T^4$ cohomology groups). We argue that their result of ``$\Z_2$" is not possible since $D_4$ clearly acts on a basis of 1-cocycles given by $\{ dz_1,dz_2,d\overline{z}_1,d\overline{z_2} \}$ in a reducible manner (i.e. (anti)-holomorphicity conditions of forms preserved). This implies that $H^1(BD_4,(\Z)_{D_4}^4)$ must take the form $G^2$ for some Abelian group $G$.}

Moving on to the binary tetrahedral group $\mathcal{T}$, we can use the presentation
\begin{equation}
    0 \xrightarrow[]{} \; D_4\; \xrightarrow[]{} \; \mathcal{T} \; \xrightarrow[]{} \; \Z_3 \; \xrightarrow[]{}  0
\end{equation}
to see that we have the following fibration structure of classifying spaces
\begin{equation}
     BD_4 \; \hookrightarrow \; B\mathcal{T} \; \xrightarrow[]{} \; B\Z_3\,.
\end{equation}
We can again use the Serre spectral sequence, as well as the fact that $H^0(BD_4,(\Z)^4_{D_4})=0$, to conclude that
\begin{equation}\label{eq:h1twiste6}
H^1(B\mathcal{T},(\Z)^4_{\mathcal{T}})=H^0(B\Z_3, \left(  H^1(BD_4,(\Z)^4_{D_4})\right)_{\Z_3})\,.
\end{equation}
We can calculate the RHS of by noting that
\be
(H^1(BD_4,(\Z)^4_{D_4}))_{\Z_3}=(\Z^2_2)_{\Z_3}
\ee
where the $\Z_3$ module structure is induced by the well-known triality automorphism of the Klein group. There are no (non-trivial) elements in $\Z^2_2$ invariant under the $\Z_3$ automorphism, so therefore $H^1(B\mathcal{T},(\Z)^4_{\mathcal{T}})=0$.

We now briefly comment on how our equivariant results can be used to match with the homology groups of $T^6/\Z_3$ and $T^6/\Z_4$ in the main text. For the former case, we compute
\begin{equation}
\ba
    H_4(T^6/\mathbb{Z}_3)=H^1_{\Z_3}(T^6)&=H^2(B\Z_3, \Z) \oplus H^1(B\Z_3,[H^1(T^6)]_{\Z_3})\oplus H^0(B\Z_3,[H^2(T^6)]_{\Z_3})\\&=\Z_3\oplus \Z^3_3 \oplus 0\,,
\ea \end{equation}
while in the latter we have
\begin{equation}
\ba
   H_4(T^6/\Z_4)=H^2_{\Z_4}(T^4)&=H^2(B\Z_4,\Z)\oplus H^1(B\Z_4,[H^1(T^6)]_{\Z_4})\oplus H^0(B\Z_4,[H^2(T^6)]_{\Z_4})\\&=\Z_4\oplus \Z^4_2\oplus \Z^5\, .
\ea \end{equation}
These follow from the collapsing of the Serre spectral sequence in low cohomological degree and indeed match our results for $H_4(T^6/\Z_3)$ and $H_4(T^6/\Z_4)$ in the main text. However, understanding non-trivial differentials in the spectral sequence \eqref{eq:equivariantSerrespec} are required to compare with $H_2(T^6/\mathbb{Z}_3)$ and $H_3(T^6/\Z_3)$ in \eqref{eq:T6Z3} which is beyond the scope of this work.

{\renewcommand{\arraystretch}{1.225}
\begin{table}
\centering
\begin{tabular}{|c||c|c|c|}
\hline
  $T^4/\Gamma$ & $H^0(B\Gamma, (\mathbb{Z})^6_\Gamma)$ & $H^1(B\Gamma, (\mathbb{Z})^4_\Gamma)$ & $H^2(B\Gamma, \mathbb{Z} )$ \\ \hline \hline
 $T^4/\mathbb{Z}_2$ & $\Z^6$ & $\Z^4_2$ & $\Z_2$ \\  \hline
 $T^4/\mathbb{Z}_3$ & $\Z^4$ & $\Z^2_3$ & $\Z_3$ \\  \hline
  $T^4/\mathbb{Z}_4$ & $\Z^4$ & $\Z^2_2$ & $\Z_4$ \\  \hline
   $T^4/\mathbb{Z}_6$ & $\Z^3$ & $0$ & $\Z_6$ \\  \hline
 $T^4/\mathcal{D}_4$ & $\Z^3$ & $\Z^2_2$ & $\Z^2_2$ \\  \hline
  $T^4/\mathcal{D}_5$ & $\Z^3$ & $0$ & $\Z_4$ \\  \hline
  $T^4/\mathcal{T}$ & $\Z^3$ & $0$ & $\Z_3$ \\  \hline
\end{tabular}
\caption{Cohomology groups of $B\Gamma$ with twisted coefficients given by $\Gamma$-modules $(\Z)^6_\Gamma \equiv H^2(T^4,\mathbb{Z})$ and $(\Z)^4_\Gamma \equiv H^1(T^4,\mathbb{Z})$ whose module structure is induced from the $\Gamma$ action on $T^4$. The sum of the groups each row is equal to $H_2(T^4/\Gamma,\Z)$. The module structure of the unprimed actions of $\mathcal{D}_4$ and $\mathcal{T}$ are identical to the primed actions which is why we do not distinguish these in the table. }
\label{tab:specseqgrps}
\end{table}}

\paragraph{7D Gauge Groups for non-Abelian $\Gamma$}\mbox{}\medskip \\
We are now in a position to write down the global form of the 7D gauge groups for M-theory compactified on $T^4/\Gamma$ for all possible non-Abelian actions $\Gamma$. This is equivalent to stating the collection of representations of massive charged particles and charged monopole 3-branes. Since $H_1(T^4/\Gamma)=0$ in all cases, these two types of massive excitations are guaranteed to be mutually local. Even though the equivariant method does not give a hands-on definition of the 2-cycles in $T^4/\Gamma$, we find that the maps in \eqref{eq:MVSapp} can be fixed by exactness of the sequence.

\noindent {\bf Case $X=T^4/ \mathcal{D}_4$\,:} We find the exact sequence
\be \ba
0~&\xrightarrow[\text{}]{\;\imath_2\;} ~\Z^3  ~\xrightarrow[\text{}]{\;\jmath_2\;}  ~\Z^3\oplus \Z_2^4 ~\xrightarrow[\text{}]{\; \partial_2 \;}  ~ (\Z_2\oplus\Z_2)^2 \oplus \Z_4^3 \oplus \Z_2^2 ~\xrightarrow[\text{}]{\;\imath_1\;} ~ \Z_2^4  ~\xrightarrow[\text{}]{\;\jmath_1\;}  ~0\,.
\ea \ee
The continuous gauge group is
\be
G_{\textnormal{full}}=\frac{([Spin(8)^2 \times SU(4)^3 \times SU(2)^{2}]/\Z_2^4)\times U(1)^3}{\Z_2^2\times \Z_4}\,.
\ee

\noindent {\bf Case $X=T^4/ \mathcal{D}_4'$\,:} We find the exact sequence
\be \ba
0~&\xrightarrow[\text{}]{\;\imath_2\;} ~\Z^3  ~\xrightarrow[\text{}]{\;\jmath_2\;}  ~\Z^3\oplus \Z_2^4 ~\xrightarrow[\text{}]{\; \partial_2 \;}  ~ (\Z_2\oplus\Z_2)^4 \oplus \Z_2^3  ~\xrightarrow[\text{}]{\;\imath_1\;} ~ \Z_2^4  ~\xrightarrow[\text{}]{\;\jmath_1\;}  ~0\,.
\ea \ee
The continuous gauge group is
\be
G_{\textnormal{full}}=\frac{([Spin(8)^4 \times SU(2)^{3}]/\Z_2^4)\times U(1)^3}{\Z_2^3}\,.
\ee

\noindent {\bf Case $X=T^4/ \mathcal{D}_4^{\prime \prime}$\,:} We find the exact sequence
\be \ba
0~&\xrightarrow[\text{}]{\;\imath_2\;} ~\Z^3  ~\xrightarrow[\text{}]{\;\jmath_2\;}  ~\Z^3\oplus \Z_2^4 ~\xrightarrow[\text{}]{\; \partial_2 \;}  ~ \Z^6_4\oplus \Z_2   ~\xrightarrow[\text{}]{\;\imath_1\;} ~ \Z_2^4  ~\xrightarrow[\text{}]{\;\jmath_1\;}  ~0\,.
\ea \ee
The continuous gauge group is
\be
G_{\textnormal{full}}=\frac{([SU(4)^6 \times SU(2)]/\Z_2^4)\times U(1)^3}{\Z^2_4\times\Z_2}\,.
\ee

\noindent {\bf Case $X=T^4/ \mathcal{D}_5$\,:} We find the exact sequence
\be \ba
0~&\xrightarrow[\text{}]{\;\imath_2\;} ~\Z^3  ~\xrightarrow[\text{}]{\;\jmath_2\;}  ~\Z^3\oplus \Z_4 ~\xrightarrow[\text{}]{\; \partial_2 \;}  ~ \Z_4\oplus\Z_4^3 \oplus \Z_3^2 \oplus \Z_2 ~\xrightarrow[\text{}]{\;\imath_1\;} ~ \Z_4  ~\xrightarrow[\text{}]{\;\jmath_1\;}  ~0\,.
\ea \ee
The continuous gauge group is
\be
G_{\textnormal{full}}=\frac{([Spin(10) \times SU(4)^3 \times SU(3)^{2}\times SU(2)]/\Z_4)\times U(1)^3}{\Z_{12}^2\times \Z_2}\,.
\ee

\noindent {\bf Case $X=T^4/\mathcal{T}$\,:} We find the exact sequence
\be \ba
0~&\xrightarrow[\text{}]{\;\imath_2\;} ~\Z^3  ~\xrightarrow[\text{}]{\;\jmath_2\;}  ~\Z^3\oplus \Z_3  ~\xrightarrow[\text{}]{\; \partial_2 \;}  ~ \Z_3\oplus\Z_2^2 \oplus \Z_3^4 \oplus \Z_2 ~\xrightarrow[\text{}]{\;\imath_1\;} ~ \Z_3   ~\xrightarrow[\text{}]{\;\jmath_1\;}  ~0\,.
\ea \ee
The continuous gauge group is
\be
G_{\textnormal{full}}=\frac{([E_6 \times Spin(8) \times SU(3)^{4}\times SU(2)]/\Z_3)\times U(1)^3}{\Z^3_6}\,.
\ee

\noindent {\bf Case $X=T^4/\mathcal{T}^{\prime}$\,:} We find the exact sequence
\be \ba
0~&\xrightarrow[\text{}]{\;\imath_2\;} ~\Z^3  ~\xrightarrow[\text{}]{\;\jmath_2\;}  ~\Z^3\oplus \Z_3  ~\xrightarrow[\text{}]{\; \partial_2 \;}  ~ \Z_6\oplus \Z^2_4 \oplus \Z^4_3~\xrightarrow[\text{}]{\;\imath_1\;} ~ \Z_3   ~\xrightarrow[\text{}]{\;\jmath_1\;}  ~0\,.
\ea \ee
The continuous gauge group is
\be
G_{\textnormal{full}}=\frac{([SU(6) \times SU(4)^2 \times SU(3)^{4}]/\Z_3)\times U(1)^3}{\Z^2_{12}\times \Z_6}\,.
\ee

\section{Homology Groups  $H_n(T^6/\Z_3)$ and $H_n(T^6/\Z_4)$} \label{sec:T6Homo}

In this Appendix we give details on the computations of the homology groups $H_n(T^6/\Z_3)$ and $H_n(T^6/\Z_4)$.

\subsection{$H_n(T^6/\Z_3)$}
\label{app:T6Z3}
We now determine the homology groups $H_n(T^6/\Z_3)$. For this we first determine all fibrations of the form
\be
 T^6/\Z_3 ~\rightarrow~ T^2/\Z_3\,, \qquad T^6/\Z_3 ~\rightarrow ~T^4/\Z_3\,,
\ee
with generic two- and four-torus fiber respectively. In both cases the base is an invariant two- and four-torus folded by the $\Z_3$ action and the fibration is realized by orhogonal projection (the $\Z_3$ action is an isometry). For notational conventions see \eqref{eq:Conventions}.

We begin by studying the set of fibrations $ T^6/\Z_3 \rightarrow T^2/\Z_3$. The possible bases are lines
\be\label{eq:locus}
L^{(i)}_\alpha(f_j,f_k)\,: \qquad z_j=\alpha_j z_i + f_j\,, \qquad  z_k=\alpha_k z_i + f_k
\ee
parameterized by $z_i$ where $\{ i,j,k\}= \{1,2,3\} $ and $f_j,f_k$ are each one of the three fixed points on $T^2_j,T^2_k$ respectively. Further we have $\alpha_i,\alpha_j\in \{0,1,\tau, \tau^2\}$ with $\tau=\textnormal{exp}({2\pi i/3})$ and $\alpha=(\alpha_j,\alpha_k)$. Requiring the line $L^{(i)}_\alpha(f_j,f_k)$ to be invariant under the $\Z_3$ action fixes possible constant terms to be fixed points $f_j,f_k$. The parameters $\alpha$ are constrained by $z_i,z_j,z_k$ taking values modulo $1,\tau$. The map from labels $(i,\alpha, f_j,f_k)$ to lines is not injective.

At fixed points $z_i=f_i$ of $T^2_i$ the values for $z_j,z_k$ also compute to fixed points. To every line $L^{(i)}_\alpha(f_j,f_k)$ we thus associate a line in $\Z_3^3$ consisting of three fixed points. We denote such a line by $\mathcal{L}_{IJK}=\{I,J,K\}$ with $I,J,K\in \Z_3^3$ labelling the fixed points. For convenience let us label $L^{(i)}_\alpha(f_j,f_k)$ by the fixed the singularities it contains, that is write $L_{IJK}$ for the same object. Note that here lines include affine lines, they are not required to contain the origin $z_i=0$.

Let us consider a fixed fibration $T^6/\Z_3\rightarrow L_{IJK}$ for a given line $ L_{IJK}$. Let us denote the generic four-torus fiber by $T^4_{IJK}$. At the $i$-th fixed point of the line $L_{IJK}$ the four-torus fiber is folded to the divisor $D_{IJK}^{(i)}=T^4_{IJK}/\Z_3$ and in homology we have
\be\label{eq:Obvious}
T^4_{IJK} = 3D_{IJK}^{(i)}\,, \qquad i=1,2,3\,.
\ee
Now consider the crepant resolution $(T^6/\Z_3)'$. The lines intersect the exceptional divisors $\P^2_I$ in points and therefore we have the fibration $(T^6/\Z_3)'\rightarrow L_{IJK}$ for any line $ L_{IJK}$. Now resolution introduces exceptional curves and the relation \eqref{eq:Obvious} is extended to
\be\label{eq:Obvious2}
T^4_{IJK} = 3D_{IJK}^{(i)}+ \sum_{H\,\in\, D_{IJK}^{(i)}} \P^2_H\,, \qquad i=1,2,3
\ee
where $H$ runs over the 9 fixed points contained in $D_{IJK}^{(i)}\subset X$. In equation \eqref{eq:Obvious2} we more precisely have the proper transform of $D_{IJK}^{(i)}$ but to avoid additional notational elements we do not include this in the notation. The crucial part of equation \eqref{eq:Obvious2} is that the lefthand side is independent of $i$. We thus find the integral cycles
\be\label{eq:FractionalDivisor}
 \frac{1}{3}\lb \sum_{H\,\in\, D_{IJK}^{(i)}} \P^2_H -  \sum_{H'\,\in\, D_{IJK}^{(i')}} \P^2_{H'}\rb=D_{IJK}^{(i')}-D_{IJK}^{(i)} \in H_4(X')\,.
\ee
Note the analgous structure in \eqref{eq:KeyRelation2}.

We can rephrase this result as follows. Consider the planes in $\Z_3^3$ orthogonal to the lines $\mathcal{L}_{IJK}$. Planes are parallel whenever they do not share a point. Orthogonal to a given line there are three parallel planes. The relation \eqref{eq:FractionalDivisor} is then the difference of the exceptional divisors associated with fixed points in two parallel planes.

We now study the set of fibrations of the form $T^6/\Z_3 \rightarrow T^4/\Z_3$. The possible bases are the planes
\be\label{eq:locus2}
P_{\alpha,f}^{(k)} \,: \qquad  z_k=\alpha_i z_i + \alpha_j z_j+ f_k
\ee
parameterized by $z_i,z_j$ and otherwise with notation as introduced above. Most comments made for $T^6/\Z_3 \rightarrow T^2/\Z_3$ generalize and we do not repeat them.

The important observation is now that the computation proceeds as for the case $T^4/\Z_3$ (or $T^4/\Z_2$). We take the affine perspective reducing the problem to combinatorics. We compute
\be\ba
\textnormal{coker}\,\jmath_2&=\Z_3^{23}\,, \quad &&\Tor\,H_2(X)\cong \Z_3^{17}\,,\\
\textnormal{coker}\,\jmath_4&=\Z_3^{10}\,, \quad &&\Tor\,H_4(X)\cong \Z_3^4\,,
\ea\ee
where the maps $\jmath_n$ appear in the sequence \eqref{eq:MVST6}.

\subsection{$H_n(T^6/\Z_4)$}
\label{app:t6z4}

Here we discuss the homology groups $H_n(T^6/\Z_4)$. We compute these via a cutting and gluing construction. This computation differs from the computation of $H_n(T^6/\Z_3)$ as now codimension 4 and 6 singularities occur simultaneously and we are unaware of a reduction of the analysis to a counting problem, as in all other cases analyzed throughout this paper.

Let us setup the cutting and gluing problem. The $\Z_4$ action on $T^6=T^2_1\times T^2_2\times T^2_3$ with coordinates $(z_1,z_2,z_3)$ and complex structures $(i,i,i)$ respectively is $(z_1,z_2,z_3)\sim (iz_1,iz_2,-z_3)$. We consider the fibration
\be
\pi\,:~~ T^6/\Z_4 ~\rightarrow~ B= T^2_3/\Z_2
\ee
whose base is topologically a two-sphere with four orbifold points. The generic fiber $F=(T^2_1\times T^2_2)/\Z_2$ degenerates at these points to $F/\Z_2=(T^2_1\times T^2_2)/\Z_4$. Next we decompose the base $B$ into four disks
\be
B=D_1\cup D_2\cup D_3\cup D_4
\ee
each centered on a orbifold point and containing exactly one orbifold point. The disks are chosen such that they overlap in contractible patches. This decomposition lifts to a covering
\be
T^6/\Z_4=U_1\cup U_2\cup U_3\cup U_4\,, \qquad U_k=\pi^{-1}(D_k)
\ee
to which we now iteratively apply the Mayer-Vietoris sequence.

Before we set up this Mayer-Vietoris sequence we analyze the quotient map
\be
q\,:~~ F=(T^2_1\times T^2_{2\,})/\Z_2~\rightarrow~ F/\Z_2=(T^2_1\times T^2_{2\,})/\Z_4
\ee
for its properties feature repeatedly throughout the sequences. First, consider its lift $q_4$ to homology in degree 4, which is clearly given by multiplication by 2. Next, consider its lift $q_2$ to homology in degree 2, which is a map
\be \label{eq:QuotientMap}
q_2\,:~~ H_2(F)\cong \Z^6\oplus \Z_2^5 ~\rightarrow~H_2(F/\Z_2)\cong \Z^4\oplus \Z_4 \oplus \Z_2^2\,.
\ee
Here $\Z_2$ acts on $H_2(F)$ acting on free and torsional generators separately. Any two-sphere generating a free class in $H_2(F/\Z_2)$ is of the form $T^2/\Z_4$ while those in $H_2(F)$ are topologcially $T^2/\Z_2$. Therefore free generators are related by multiplication by 2 and thefore
\be
q_2|_{\textnormal{free}}\,:~~  \Z^6~\rightarrow~ \Z^4\,, \quad \textnormal{ker}\,q_2|_{\textnormal{free}}=\Z^2\,, \quad \textnormal{coker}\,q_2|_{\textnormal{free}}=\Z_2^4
\ee
The torsional subgroup $\textnormal{Tor}\, H_2(F)\cong  \Z_2^5$ is in turn generated by elements separately labelled by 8 of the 16 singularities in $F$. The $\Z_2$ action fixes 4 of the 16 singularities and permutes the other 12, see section \ref{sec:Kummer}. Writing out generators explicitly we conclude
\be
q_2|_{\textnormal{tor}}\,:~~  \Z_2^5~\rightarrow~  \Z_4 \oplus \Z_2^2\,, \quad \textnormal{ker}\,q_2|_{\textnormal{tor}}=\Z_2^2\,, \quad \textnormal{coker}\,q_2|_{\textnormal{tor}}=\Z_2\,,
\ee
and therefore find overall
\be
\textnormal{im}\,q_2\cong \Z^4 \oplus \Z_2^3\,,  \quad \textnormal{ker}\,q_2\cong \Z^2 \oplus \Z_2^2\,, \quad \textnormal{coker}\,q_2\cong \Z_2^5 \,.
\ee
The map $q_0$ is the identity map and $q_1,q_3$ are trivial.

Let us now consider the Mayer-Vietoris sequence gluing $U_1,U_2$ to the union $U_1\cup U_2$. The intersection $U_{12}=U_1\cap U_2$ retracts to $F$ while both $U_1,U_2$ retract to $F/\Z_2$. The patches $U_1,U_2$ are topologically identical and the map
\be
Q_n\,:~~ H_n(F) ~\rightarrow ~ H_n(U_1)\oplus H_n(U_2)
\ee
in the sequence is therefore diagonal and given by $Q_n=(q_n,q_n)$. Therefore we compute
\be
H_n(U_1\cup U_2)\cong \{ \Z , 0 ,\Z^4 \oplus\Z_4  \oplus \Z_2^7, \Z^2\oplus \Z_2^2, \Z \oplus \Z_2\}
\ee
and clearly we also have $H_n(U_1\cup U_2)\cong H_n(U_3\cup U_4)$.

Next we glue $U_{12}=U_1\cup U_2$ to $U_{34}=U_3\cup U_4$ along $U_{12}\cap U_{34}$. The latter retracts to a fibration of $F$ over a circle linking two of the four orbifold points in $B$. Encircling each orbifold point $F$ is glued to itself by a $\Z_2$ twist. The monodromy about two points is therefore trivial and the intersection therefore retracts to the direct product $F\times S^1$, therefore by K\"unneth's formula
\be
H_n(U_{12}\cap U_{34})\cong \{ \Z , \Z , \Z^6\oplus \Z_2^5,  \Z^6\oplus \Z_2^5, \Z , \Z \}\,.
\ee
The Mayer-Vietoris sequence with the above simplifications is
\be \ba
& &&\qquad~\;\! 0\oplus 0  && \xrightarrow[\text{}]{\;\jmath_6\;} ~H_6(T^6/\Z_4)  ~\xrightarrow[\text{}]{}
\\
&\xrightarrow[\text{}]{} ~H_4( F)\otimes H_1(S^1) &&\xrightarrow[\text{}]{\;\imath_5\;}  ~  0\oplus 0 && \xrightarrow[\text{}]{\;\jmath_5\;} ~H_5(T^6/\Z_4)  ~\xrightarrow[\text{}]{}
\\
&\xrightarrow[\text{}]{} ~H_4(F) &&\xrightarrow[\text{}]{\;\imath_4\;}  ~    H_4(U_1\cup U_2)\oplus H_4( U_3\cup U_4) && \xrightarrow[\text{}]{\;\jmath_4\;} ~H_4(T^6/\Z_4)   ~\xrightarrow[\text{}]{}
\\
&\xrightarrow[\text{}]{}~H_2(F) \otimes H_1(S^1)&&\xrightarrow[\text{}]{\;\imath_3\;}  ~    H_3(U_1\cup U_2)\oplus H_3( U_3\cup U_4 ) && \xrightarrow[\text{}]{\;\jmath_3\;} ~H_3(T^6/\Z_4)   ~\xrightarrow[\text{}]{}
\\
&\xrightarrow[\text{}]{}  ~H_2(F)&&\xrightarrow[\text{}]{\;\imath_2\;}  ~    H_2(U_1\cup U_2)\oplus H_2( U_3\cup U_4 ) && \xrightarrow[\text{}]{\;\jmath_2\;} ~H_2(T^6/\Z_4)   ~\xrightarrow[\text{}]{}
\\
&\xrightarrow[\text{}]{}  ~H_1(S^1)&&\xrightarrow[\text{}]{\;\imath_1\;}  ~  0\oplus 0 && \xrightarrow[\text{}]{\;\jmath_1\;} ~H_1(T^6/\Z_4)  ~\xrightarrow[\text{}]{}  ~0
\ea \ee
which immediately gives $H_0(T^6/\Z_4)\cong \Z$ and $H_1(T^6/\Z_4)\cong 0$ and $H_6(T^6/\Z_4)\cong \Z$. As before we have that $U_1\cup U_2$ and $U_3\cup U_4$ are topologically identical and the maps $\imath_n$ are therefore diagonal. Let us now study these maps in turn.

The map $\imath_2:\Z^6\oplus \Z_2^5\rightarrow (\Z^4\oplus \Z_4\oplus\Z_2^7)^2$ has image and kernel as the map $Q_2$, therefore
\be\label{eq:i2}
\textnormal{ker}\,\imath_2 \cong \Z^2 \oplus \Z_2^2\,, \qquad \textnormal{coker}\,\imath_2\cong  \Z^4 \oplus\Z_4\oplus  \Z_2^{17} \,,
\ee
and due to $S^1\subset B$ we conclude $H_2(T^6/\Z_4)\cong  \Z^5 \oplus\Z_4\oplus  \Z_2^{17} $. The map $\imath_4:\Z\rightarrow (\Z\oplus\Z_2)^2$ has image and kernel as the map $Q_4$, therefore
\be\label{eq:i4}
\textnormal{ker}\,\imath_4 \cong 0\,, \qquad \textnormal{coker}\,\imath_4\cong  \Z \oplus  \Z_2^{3} \,,
\ee
from which we conclude $H_5(T^6/\Z_4)\cong 0$. The map $\imath_3:\Z^6\oplus \Z_2^5\rightarrow (\Z^2\oplus \Z_2^2)^2$ follows from the monodromy action on $H_2(F)$ discussed following \eqref{eq:QuotientMap}. Denote by $\sigma\in H_2(F)$ a generator invariant under this action. The 3-cycle $\sigma\times S^1 $ can then be contracted and vanishes in $H_2(F) \otimes H_1(S^1)$. Denote by $\rho_i\in H_2(F)$ with $i=1,2$ two generators interchanged by the action. In this case $\rho_1-\rho_2$ is inverted and necessarily collapses at both orbifold points contained in either $U_1\cup U_2$ or $U_3\cup U_4$. Fibering $\rho_1-\rho_2$ over a line connecting the orbifold points constructs a 3-cycle. The two 3-cycles $\rho_i\times S^1$ both map to this 3-cycle (or the inverse thereof) upon contracting $S^1$ to a line running between the orbifold base points. We have the split
\be
H_2(F)=H_2(F)_+\oplus H_2(F)_-=(\Z^2\oplus \Z_2)\oplus(\Z^4\oplus \Z_2^4)
\ee
where the first, second summand are associated with $\sigma$'s, $\rho$'s respectively. We conclude
\be
\textnormal{im}\,\imath_3 \cong \Z^2\oplus \Z_2^2\,, \qquad \textnormal{ker}\,\imath_3 \cong \Z^4\oplus \Z_2^3\,, \qquad \textnormal{coker}\,\imath_3\cong   \Z^2\oplus \Z_2^2\,,
\ee
from which we derive, together with \eqref{eq:i2} and \eqref{eq:i4}, the two short exact sequences
\be\ba
&0~\rightarrow \Z \oplus \Z_2^3 ~ \rightarrow~ H_4(T^6/\Z_4 ) ~ \rightarrow~\Z^4\oplus \Z_2^3 ~\rightarrow~ 0\\
&0~\rightarrow \Z^2\oplus \Z_2^2 ~ \rightarrow~ H_3(T^6/\Z_4 ) ~ \rightarrow~\Z^2\oplus \Z_2^2 ~\rightarrow~ 0\,.
\ea\ee
Identifying the cycles we conclude $H_3(T^6/\Z_4)\cong \Z^4\oplus \Z_2^4$ and $H_4(T^6/\Z_4)\cong \Z^5\oplus \Z_4 \oplus \Z_2^4$. The latter matches with equivariant cohomology compuations. Notice that we have the duality check $H_4(T^6/\Z_4)\cong H_1((T^6/\Z_4)^\circ)$ with the latter readily computable by equivariant methods. Overall we find the result given in \eqref{eq:T6Z4}.

\section{5D Abelian Chern-Simons Couplings} \label{app:5DCS}

In this Appendix we collect the 5D Abelian Chern-Simons Couplings for 5D vacua obtained from M-theory on $T^6 / \mathbb{Z}_3$ and $T^6 / \mathbb{Z}_4$. These follow from reduction of the 11D supergravity coupling $C \wedge G \wedge G$. Let us note that we can also extract mixed $U(1)$-gravity-gravity Chern-Simons couplings from reduction of the $C_3 \wedge X_8(\mathcal{R})$ term.

\subsection{$T^6 / \mathbb{Z}_3$}

Consider first M-theory on $X = T^6 / \mathbb{Z}_3$.
The $U(1)^{9}$ gauge group factor in $G_{\textnormal{full}}$ is accompanied by Chern-Simons terms
\begin{equation}\label{eq:5dCS}
    \frac{K_{IJK}}{24\pi^2} \int_{5D} A_I F_{J} F_K\,, \qquad K_{IJK}\in \mathbb{Z}\,, \qquad I,J,K=1,\dots,9\,,
\end{equation}
where we have normalized such that $K_{IJK}$ are the integer-valued K-matrix coefficients. These follow from reduction of the 11D supergravity term:
\be
-\frac{1}{6(2\pi)^3}\int C_3\wedge G_4\wedge G_4
\ee
and are computed by triple intersection numbers among the generators of $\textnormal{Free}\,H_2(X)$. We can compute these intersection numbers noting that $H^2(X,\Z)=\mathbb{Z}^9$ correspond to $\mathbb{Z}_3$ invariant 2-forms of $H^2(T^6)$ which are given by
\be
\omega_{i\Bar{\jmath}}:=\frac{i}{2}dz_i\wedge d\Bar{z}_{\Bar{\jmath}}\,.
\ee
We relabel these 2-forms as $\omega_I$ with $I=1,\dots,9$ defined by $I=3(i-1)+\Bar{\jmath}$. The Chern-Simons levels are then
\be
K_{IJK}=\int_{T^6/\Z_3}\omega_I \wedge  \omega_J \wedge \omega_K\,.
\ee
They are computed using $[\mathrm{Vol}_{T^6}]= 3\in \Z \cong H^6(T^6/\Z_3,\Z)$, where $\mathrm{Vol}_{T^6}$ is the volume form on $T^6$, and explicitly read:
\begin{equation}
   K_{IJK}=\begin{cases}   ~3 \;  \; ~~~~\;\!\;\!(I,J,K)=(1,5,9),  (1,6,8),(2,4,9),(2,6,7), (3,4,8), (3,5,7)  \\
  \qquad   \; \qquad  \qquad \quad \; \; \, \textnormal{+ permutations} \\[3pt]
   ~0 \; \;  ~~~~\;\!\;\!\,\mathrm{Otherwise} \end{cases}
\end{equation}

\subsection{$T^6 / \mathbb{Z}_4$}

Consider next M-theory on $T^6 / \mathbb{Z}_4$. The Abelian Chern-Simons levels of the $U(1)^5$ factor can be computed in the same way as in the $T^6/\Z_3$ example, relying again on the free part $\Z^5\subset H^2(X)$ being generated by $\Z_4$ invariant 2-forms of $T^6$. If we define
\be \ba
\omega_1&=\frac{i}{2}dz_1\wedge d\Bar{z}_1\,, \\  \omega_2&=\frac{i}{2}dz_1\wedge d\Bar{z}_2\,, \\ \omega_3&=\frac{i}{2}dz_2\wedge d\Bar{z}_1\,, \\ \omega_4&=\frac{i}{2}dz_2\wedge d\Bar{z}_2\,,\\ \omega_5&=\frac{i}{2}dz_3\wedge d\Bar{z}_3\,,
\ea \ee
then we have a CS action \eqref{eq:5dCS} with K-matrix
\be
K_{IJK}=\int_{T^6/\Z_4}\omega_I\wedge \omega_J\wedge  \omega_K=\begin{cases}   ~4 \;  \; ~~~~\;\!\;\!(I,J,K)=(1,4,5), (2,3,5)\textnormal{ + permutations} \\
   ~0 \; \;  ~~~~\;\!\;\!\,\mathrm{Otherwise.} \end{cases}
\ee
where $I,J,K=1,\dots,5$.

\newpage

\bibliographystyle{utphys}
\bibliography{HigherSwamp}

\end{document}